%% file: tauspinner-application-studies.tex
\documentclass{article}
\usepackage[T1]{fontenc}
\usepackage[latin1]{inputenc}
\usepackage{graphics}
\usepackage{graphicx}
\usepackage{longtable}
\usepackage{color}
\usepackage{pslatex}
\usepackage{hyperref}
\usepackage{sverb}
\usepackage{subfigure}

\begin{document}
\begin{titlepage}
 
\begin{flushright} 
{ IFJPAN-IV-2013-19 
} 
\end{flushright}
 
\vskip 3 mm
\begin{center}
{\bf\huge  Application of {\tt TauSpinner } for studies on $\tau$-lepton polarization 
and spin correlations in $Z$, $W$ and $H$ decays at LHC}
\end{center}
\vskip 13 mm

\begin{center}
   {\bf A. Kaczmarska$^{a}$,   J. Piatlicki$^{b}$,
        T. Przedzi\'nski$^{b}$, E. Richter-W\c{a}s$^{c}$
 and Z. W\c{a}s$^{a,d}$  }\\
       {\em $^a$  Institute of Nuclear Physics, PAN,
        Krak\'ow, ul. Radzikowskiego 152, Poland}\\ 
       {\em $^b$ The Faculty of Physics, Astronomy and Applied Computer Science, \\ 
Jagellonian University, Reymonta 4, 30-059 Cracow, Poland}\\
       {\em $^c$ Institute of Physics, \\ 
Jagellonian University, Reymonta 4, 30-059 Cracow, Poland}\\
{\em $^d$ CERN PH-TH, CH-1211 Geneva 23, Switzerland }
\end{center}
\vspace{1.1 cm}
\begin{center}
{\bf   ABSTRACT  }
\end{center}

The $\tau$-lepton plays an important role in the physics program at the Large Hadron Collider 
(LHC). It offers a powerful probe in searches for New Physics.
Spin of $\tau$ lepton represents an interesting phenomenological quantity which can be used
for the sake of separation of signal from background or in measuring  properties
of New Particles decaying to $\tau$ leptons. 
A proper treatment of $\tau$ spin effects in the Monte Carlo simulations is important for
understanding  the detector acceptance as well as for the measurements of $\tau$
polarization and $\tau$ spin correlations.

The {\tt TauSpinner} package represents a tool which can be used to modify $\tau$ spin effects in any
sample containing $\tau$ leptons. 
Generated samples of events featuring $\tau$ leptons produced from intermediate state 
$W$, $Z$, Higgs bosons can be used as an input.
The information on the polarization and spin correlations is reconstructed from the kinematics of the 
$\tau$ lepton(s) (also $\nu_\tau$ in case of $W$-mediated processes) and $\tau$ decay products.  No other information stored in the event record is needed.
By calculating spin weights, attributed on the event-by-event basis, it enables numerical evaluation of 
the spin effects on experimentally measured distributions and/or modification of the spin effects.
With {\tt TauSpinner}, the experimental techniques developed over years since LEP 1 times may be 
used and extended for LHC applications.


We review a selection of simple distributions which can be used to monitor the $\tau$ spin effects
(polarization and spin correlations) in leptonic $\tau$ decays and hadronic $\tau$ decays with up to three pions. The main 
purpose is to provide basic benchmark 
distributions for validation of spin content of the user-prepared event sample
and to visualize significance of the $\tau$ lepton spin polarization and correlation effects. 
The utility programs, demonstration examples for use of {\tt TauSpinner} libraries, are prepared and documented. 
New methods, with respect to previous publications, for validation of such an approach are provided.
Other topics like methods to evaluate {\tt TauSpinner} systematic errors 
or sensitivity of experimental distributions to 
explore spin effects are also addressed, but are far from being exploited.
Results of semi-analytical calculations, and some  effects of QED bremsstrahlung, are shown as well.

This approach is of particular interest for estimation of the theoretical systematic 
errors for implementation of spin effects in so-called embedded $\tau$ lepton samples,
where $ Z \to \mu \mu$ events are selected from data and muons are replaced with simulated $\tau$ leptons.
Such embedding techniques are used in several analyses at LHC for estimating dominant background 
from $Z \to \tau \tau$ process to the Higgs boson $H \to \tau \tau$  searches. 


\vskip 1 cm


\vspace{0.2 cm}
 
\vspace{0.1 cm}
\vfill
{\small
\begin{flushleft}
{   IFJPAN-IV-2013-19
\\ February 2014
}
\end{flushleft}
}
 
\vspace*{1mm}
\footnoterule
\noindent
{\footnotesize \noindent  $^{\dag}$
This project is financed in part from funds of Polish National Science
Centre under decisions  DEC-2011/03/B/ST2/00107 (TP,JP), DEC-2011/03/B/ST2/02632 (ZW), DEC-2011/03/B/ST2/00220 (ERW)
and 2012/07/B/ST2/03680 (AK).
}
\end{titlepage}
\newpage
\tableofcontents

\newpage
\section{Introduction}

The successful research programme of LHC experiments requires the careful 
analysis of a multitude of different final states. 
A broad spectrum of interesting observables have been  
developed over the years \cite{Aad:2009wy,Chatrchyan:2008aa,Alves:2008zz}. One of the physics quantities, which can be used 
for such purpose, is the spin state of the produced $\tau$ leptons.
For the processes, like the charged or neutral Higgs boson production and 
their respective important backgrounds from single $W$ or single $Z$ production,
the spin effects can be measured \cite{Aad:2012cia,Deigaard:2012fqa}
and also used for optimising signal from background separation.  
The spin of the final state $\tau$ lepton
carries information on the  $\tau$ production processes
and manifests itself in  distributions of the $\tau$ decay products.
There is a multitude of $\tau$ decay channels which are accessible experimentally. The dominant ones are 
$\tau^\pm \to l^\pm \nu_l\nu_\tau$,  $\tau^\pm \to \pi^\pm \nu_\tau$, 
$\tau^\pm \to \rho^\pm \nu_\tau$. 
The decay channels listed above represent more than 2/3 of the total $\tau$ lepton decay width. 
In all these channels spin effects 
manifest themselves in the energy spectrum of the visible $\tau$ decay products, 
but for each channel differently.
In the past \cite{Harton:1995dj,Heister:2001uh}, the channel $\tau^\pm \to a_1^\pm \nu_\tau$ was also 
often proposed for the $\tau$ 
spin measurements, but for this case, more sophisticated distributions
were necessary.
We skip discussion of this channel
from the study presented here.

We focus our paper on describing strategy for validating $\tau$ spin effects\footnote{The transverse $\tau$ spin effects are not yet installed in {\tt TauSpinner}. Motivation for such natural 
extension is if directions of $ \pi^0$ can be separated from the one  of
$ \pi^\pm$. For introduction of such an extension, complete spin density matrix of the
produced $\tau$-pair has to be provided. This also requests some rather straightforward 
tests of kinematics. The other parts of the algorithm  are already prepared.}
in the analyses at LHC experiments, which can be performed with the help of 
the {\tt TauSpinner} \cite{Czyczula:2012ny,Banerjee:2012ez} program.
For that purpose we recall  simple  distributions 
used for 
evaluation of $\tau$ spin effects at the LEP time \cite{Heister:2001uh,jadach-was:1984,Eberhard:1989ve},
the fractions of $\tau$ lepton energy carried by its observable decay products,
and provide several methods to verify if for the 
particular sample the spin effects can be observed. These distributions can 
be used as a validation check if spin effects were properly transmitted 
to the generated sample or provide important information for feasibility studies 
in planning of the experimental analysis. One should keep in mind that at LHC
although  fractions of the $\tau$ lepton's energy 
carried by its observable decay products is not directly measurable  
and thus of a limited use for the experimental analyses,  it was
partly adapted to LHC applications already in \cite{Pierzchala:2001gc}.
The $\tau^+ \tau^-$ pairs (or   $\tau \nu_\tau$ pairs) carry only  small fraction of the colliding proton momenta and the $\tau$ leptons 
 energies
differ substantially from the beam energies, contrary to how it was in the case at LEP 1
where $\tau$ energies where strongly constrained by the beam energies.
One should however not underestimate their usefulness for different Monte Carlo
studies, thanks to their simplicity and direct sensitivity to the spin effects. 

The paper is organized as follows. We recall main properties of the 
{\tt TauSpinner} algorithm in Section~\ref{Sec:tauspinner} and in Section~\ref{Sec:samples} explain details on the
event samples used for providing numerical results.
In Section~\ref{Sec:motivation}, we describe the properties of $\tau$ lepton spin effects which may be of interest 
at LHC and how they are transmitted to $\tau$ decay products.
In  subsections we discuss the semi-analytical formula for spectra of leptons and single $\pi$'s from $\tau$ decays 
and the effects of QED bremsstrahlung on these spectra, which can be also described in semi-analytical form.
In the following subsections we describe  plots 
we propose for benchmarking spin effects
and provide  examples  and short discussion on the numerical results. 
In Section~\ref{Sec:Installation}, we describe technical details of installation of those example programs. The 
summary, Section~\ref{Sec:Summary}, closes the paper.
The complete set of automatically generated benchmarking plots is collected  in Appendices
of a preprint version of our paper.
This documents the output from new, more advanced set of example programs for using {\tt TauSpinner}
libraries.

\section{{\tt TauSpinner} brief description}\label{Sec:tauspinner}

The {\tt TauSpinner} is a program associated with {\tt Tauola++}, enabling calculation of weights for the previously 
generated or constructed by other means events, for example like with embedding technique,
where  $Z \to \mu^+\mu^-$ events are selected from data and 
muons are replaced by the $\tau$-leptons  with simulated decays \cite{embedding-atlas}. 
The events must  feature kinematics of  $\tau$ lepton production and decay products, but 
information on partons from which intermediate resonance decaying to  the $\tau$'s  was produced is 
assumed to be  unknown, and therefore is not used.
The algorithm calculates for each event,  from  this information alone,  a spin weight
 corresponding to a presumed  configuration, for example Higgs or $Z/\gamma^{*}$ production and decay.   
The part of the weight related to the production 
of $\tau$ lepton pair (or $\tau$ lepton and associated with its production  $\nu_\tau$ in case 
of e.g. $W$ mediated processes) is calculated only from the four-momenta of the  $\tau-\tau$ 
($\tau-\nu_\tau$) lepton pair. The information on flavours of the initial state partons,
quarks or gluons,  are assumed not to be available and are attributed stochastically on the
basis of matrix elements for parton level hard processes and parton density functions (PDFs) of the user choice.
As default, processes mediated by  single $W$, $Z/\gamma^{*}$ production are assumed. Alternative processes, like
Higgs-mediated processes, can be used as well, but then the intermediate state has to be explicit in the event record.
For each $\tau$ lepton the decay part of the weight is calculated from the matrix element 
of the corresponding $\tau$ decay channel, as classified by the algorithm. For this purpose, the matrix elements
of {\tt Tauola++} library \cite{Davidson:2010rw} are used. To calculate this weight the four-momenta of $\tau$ 
decay products need to be boosted to $\tau$ rest-frame. This requires careful treatment of possible rounding errors.

The weight constructed with the help of {\tt TauSpinner}, $WT$, is separated into 
multiplicative components: production ($wt_{\sigma_{prod}}$), decay 
($wt_{\Gamma_{decay}}^{\tau^\pm}$) and
spin correlation/polarization ($wt_{spin}$): 
\begin{eqnarray}
WT&=&wt_{\sigma_{prod}} wt_{\Gamma_{decay}}^{\tau^+}wt_{\Gamma_{decay}}^{\tau^-} wt_{spin} \nonumber \\
wt_{spin}&=& R_{i,j} h^i_{\tau^+} h^j_{\tau^-}.
\end{eqnarray}
In the present note we use only  the last  component  of the
weight $WT$, the spin weight $wt_{spin}$, leaving interesting discussion on the other ones ($wt_{\sigma_{prod}}$,  $wt_{\Gamma_{decay}}^{\tau^+}$ and $wt_{\Gamma_{decay}}^{\tau^-}$) aside
to other applications,
namely ref. \cite{Banerjee:2012ez}. The definition of the spin correlation matrix $R_{i,j}$ and polarimetric vectors 
for the decay of $\tau$ leptons  $h^i_{\tau^+}$, $h^j_{\tau^-}$ is rather lengthy and also well known. Therefore
 we refer the reader to our previous publications \cite{jadach-was:1984,Pierzchala:2001gc,Jadach:1993hs} for detailed definitions.
For the discussion presented here, it is important to recall only that   $h^i_{\tau^+}$, $h^j_{\tau^-}$ are defined completely 
from the kinematics of the corresponding $\tau$ decay products and  $R_{i,j}$ from the $\tau$ production kinematics. For every event  $0<wt_{spin}<4$
by construction.
The average  of $wt_{spin}$ taken over the unconstrained event sample,  
up to statistical error equals to 1.

With the $wt_{spin}$ weight one can evaluate on event-by-event basis spin effects transmitted from 
the production to the decay of $\tau$ leptons.
By definition  $wt_{spin}=1$ if those effects are omitted.
Consequently, reweighting each event with  $WT=1/wt_{spin}$  can remove spin effects from generated sample.
Also, the cases when only part 
of spin effects is taken into account,  more specifically the
spin correlation  but no effects due to 
vector and axial couplings to 
the intermediate $Z/\gamma^{*}$ state\footnote{To $\tau$ leptons
or to  incoming quarks only.}, can be corrected with the help of the appropriate  weights. On the other hand,
the spin effects can be also removed completely or the missing parts installed.


\section{Analysed event samples} \label{Sec:samples}

In this paper, we use the samples of events from $pp$ collision at 8~TeV center-of-mass energies, featuring
final states of $\tau$ lepton pairs with a mass close to that of the $Z$ or $W$, generated with 
{\tt Pythia8} Monte Carlo \cite{Sjostrand:2007gs}. These samples, each 
of 10M events, are  stored in   {\tt HepMC} format~\cite{Dobbs:2001ck}. 
Essentially default\footnote{%
The configuration parameters are detailed in Appendix \ref{Sec:InputFiles},
For the generation of 
multiphoton final state radiation in $Z$ and $W$ decays {\tt Photos} 
Monte Carlo \cite{Davidson:2010ew} was used.
}
initialisation parameters of {\tt Pythia}, are used
and no selection criteria are applied on the kinematics of outgoing $\tau$'s.
The decays of $\tau$ leptons are generated with {\tt Tauola++} initialized
with  standard options\footnote{In general QED bremsstrahlung was not taken into
account in  $\tau$ decays. Only for preparation of Fig.~\ref{Fig:QED},  $\tau$ leptons were 
re-decayed with  QED bremsstrahlung ON or OFF.}. 
The samples are generated including spin effects (polarisation and correlations): we will refer to these samples as {\it original (orig, pol, polarized)} samples.
Starting from the  {\it original} samples, depending on the studied effects, the spin effects are 
removed, with the help of {\tt TauSpinner} weights: {\it unpolarized (unweighted, unpol)} samples are obtained.  

For some of the presented results we have created events originating from the spin-0 resonance 
of the mass of $Z$ boson and couplings of the Higgs boson, denoted as $\Phi$ resonance. 
This was performed for convenience using $Z \to \tau \tau$ events generated with 
{\tt Pythia8} Monte Carlo and $\tau$ leptons decaying with {\tt Tauola++}
configured for the scalar resonance decay. 
Such events, denoted in this paper as $\Phi \to \tau \tau$ events,  
serve to illustrate spin effects between $\tau$ pairs originating  from the decay of vector or scalar boson of the same mass and width.
Please note that {\tt TauSpinner} weights could be as well used to reweight complete $Z \to \tau \tau$ events 
to represent  $\Phi \to \tau \tau$ events,
however this procedure introduces large statistical fluctuations due to the large spread of the weights
when reweighting for spin effects from vector to scalar resonance decays. 

As a very interesting example, we point to the case when spin effects are removed completely from 
the {\it original} sample and are reinstalled back with only spin correlations but not spin polarization effects.
While reintroducing only the spin correlations,  it is sufficient to use information on the
four-momenta of $\tau$ leptons and their decay products. Reintroducing effects from polarization
requires information on structure functions of partons forming decaying resonance. 
In the case of embedded $Z \to \tau \tau$ samples~\cite{embedding-atlas}, it means introducing theoretical uncertainty 
due to assumed PDF's parametrisation to the sample, which  appriory, was free from such uncertainties.
The theoretical systematic error for such approach can be assigned by comparing
spin effects calculated from approximation which rely on four-momenta of $\tau$ leptons
and their decay products with the one exploring full hard
scattering parton level amplitudes. 

For some auxiliary tests, discussed in Section~\ref{Sec:consistency-check}, we have also used 1M events 
from Drell-Yan $ pp \to Z/\gamma* \to \tau \tau $ process generated within the virtuality interval of $m_{\tau \tau}$ = 1.0-1.5~TeV. 

We would like to stress an important feature of this strategy,
for removing or reintroducing 
spin effects. It  represents a solution for the cases when the sample of events, which feature $\tau$ 
lepton decays, is generated and processed with CPU-intensive
simulation of the detector response. There is no need to 
prepare another reference sample with spin effects excluded, as this effect
can be introduced by  weights calculated by {\tt TauSpinner}.
The solution may be very helpful to estimate the  sensitivity of the sample to its spin content
as one can profit from using correlated events  to reduce the effects
from statistical fluctuations. 
Finally, with this strategy, at very low CPU-cost the spin effects can be evaluated  
to validate correctness of the generation of the sample under scrutiny.  

\section{Physics motivation of test observables and numerical results} \label{Sec:motivation}

In the analysis of experimental data it is important to evaluate effects 
due to particular theoretical phenomena, incorporated in tools 
used in preparation of the experimental 
distributions where at the same time all experimental effects are taken into account.
Only then, one can decide if the studied effects are sizeable and can be 
distinguished from  effects such as eg. background contamination.
Distortion of the energy spectra of decay products due to polarization of $\tau$ leptons 
and spin correlations are examples of such effect. 

Due to the short lifetime and their parity-violating decays, $\tau$
leptons are the only leptons whose spin information  
is transmitted to the observed decay products
kinematics. 
In the $\tau$ lepton decay the neutrino(s) escape detection, so complete kinematics of 
all decay products cannot be reconstructed experimentally. 
We assume however that $\tau$ decay channel can be correctly determined
and for the sake of definition of the test distribution,  that the fraction of $\tau$ energy carried 
by all observable decay products combined can be used. 
This leads to relatively simple semi-observables even if still does not explore all correlations
and energy fractions of the secondary decay products ($\rho$, $a_1$...). 

It would be optimal to measure energies of individual
$\tau$ decay products  and use all  of them simultaneously achieving then
substantial gain in the sensitivity to the spin.
That is the case, for example, in $\tau^\pm \to \rho^\pm \nu_\tau \to \nu_\tau \pi^\pm \pi^0$ decay channel.
The difference between $ \pi^\pm$ and $ \pi^0$ energies is determined by the spin of $\rho$ which 
carries information on the  spin of $\tau$.  
This type of constraints is desirable  to be included in any realistic studies, 
but substantially adds to the complexity of the $\tau$ decay response to its spin.  

Such effects 
are of course taken into account in {\tt TauSpinner} algorithms
but are not explored with  the distributions we study in this paper.
Let us remind that
precision tests of the Standard Model were performed at LEP~1, with significant
and well documented effort
on experimental, theoretical and computational 
levels \cite{Altarelli:1989hv,Altarelli:1989hx}. In particular, manifestation 
of $\tau$ lepton 
polarization in its all main decay channels was carefully explored
 \cite{Heister:2001uh}, in the context of  measuring the intermediate $Z$ boson
properties.
In our  discussion, we recall some of the  phenomenological and technical 
considerations 
of that time \cite{jadach-was:1984,Eberhard:1989ve}, which may be useful 
for the LHC applications as well, as shown in ref.~\cite{Pierzchala:2001gc}.
 In the LEP 1 analyses, observables 
were   at first limited 
to the fraction of $\tau$ energy carried by its observable decay products, $x$.
Such fractions could have been used directly at LEP~1 experiments because 
$\tau$ energy was essentially equal to the beam energy. In general,
 in $m_\tau \ll M_{Z,W}$ limit, the fraction $x$ is independent from the 
boost and remain  the same in the rest frame of intermediate
$Z$ (or $W$) or in the lab frame. That is why it is of potential interest
 as a first step in preparation 
of  the spin measurements at LHC experiments, or to validate 
the correctness of spin implementation in the generated samples.
Even though $x$ is not reconstructed experimentally, knowledge of
spin effect to  distributions in this variable can be used  rather straightforward to estimate how the
distributions, of the actual interest, will be modified. Because of the simplicity
and direct relation to properties of the decay matrix elements, the $x$  
variable is also a good choice for Monte Carlo benchmarks\footnote{
Our definitions   follow refs.~\cite{jadach-was:1984,Eberhard:1989ve,Pierzchala:2001gc}.
At LEP1 time the $x$ fraction was the actual observable; for the collisions of the 
center-of-mass energy close to 
 the $Z$ peak the $\tau$ lepton energies
were  close to the beam energies. The $x$ variable received a lot of phenomenological attention. }
on polarization and spin correlation effects.

\subsection{Semi-analytical formulae}

The pattern of the $\tau$ response to spin can be studied by the Monte Carlo methods through its decay, 
taking into account the complexity of multi-dimensional signatures. We return to this
solution later in the paper. 
In some cases, simple analytical formulae are nonetheless available. They can be quite helpful to 
visualize the effects in an intuitive way, even if some details  of the distributions would be neglected.

In case of $\tau^\pm \to \pi^\pm \nu_\tau$ and $\tau^\pm \to l^\pm\nu_l\nu_\tau$ decays 
formulae for energy spectra
for visible decay products, neglecting mass and QED bremsstrahlung effects have been known for a long time~\cite{jadach-was:1984}.
For $\tau^\pm \to \pi^\pm \nu_\tau$ it is
\begin{equation}
1 +P \times (2x-1) \label{pion}
\end{equation}
and for $\tau^\pm \to \ell^\pm\nu_l\nu_\tau$ ($ \ell = e, \mu$) it reads as
\begin{equation}
\frac{5}{3} -3x^2+\frac{4}{3}x^3
-P \times \Bigl(-\frac{1}{3}+3x^2-\frac{8}{3}x^3\Bigr),\label{lepton}
\end{equation}
where $P$ denotes $\tau$ polarization and $x$ is a fraction of $\tau^\pm$ 
energy carried by $\pi^\pm$ or $\ell^\pm$.

These analytic forms of the spectra can be extended to the case when effects 
of radiative corrections are 
taken into account. Such parametrization of the 
 spectra for decay products of polarized $\tau$ leptons are given in ref.~\cite{Eberhard:1989ve},
formulae\footnote{Unfortunately 
this review is known to have typing mistakes.} A3 and A4.

For tests presented here, we assume that bremsstrahlung in decays is not taken into 
account in the event generation or that its effect can be neglected. Also, that the mass terms (non neglible for muons) can be neglected.
Otherwise, the semi-analytical formulae would become much more complicated.
With time, as it was the case of LEP~\cite{Heister:2001uh}, these effects may become of interest as well. It is straightforward 
to introduce to formulae~(\ref{pion}),~(\ref{lepton}) effects due to bremsstrahlung; not only in $\tau$ decay itself, but also in decay 
of intermediate $Z$ or $W$ bosons. 

With above assumptions, simple formulae as~(\ref{pion}),~(\ref{lepton}) can be fitted to the histograms
for the energy  fractions $x$,
  to evaluate effective $P$ polarization and conclude if  the particular sample feature the spin effect 
and/or if this effect is 
big enough to be  statistically
significant. 

In case semi-analytical formula is not available for the distribution, or distribution is distorted by kinematical selection, 
one can use for fitting the linear combination of reference spectra corresponding to pure left-handed
 and right-handed $\tau$ leptons.
This  template fit technique was already used by LEP experiments~\cite{Heister:2001uh}. Such a reference spectra
can be  obtained with the help of Monte Carlo methods.

\subsection{Spin correlations and  polarization monitoring plots} \label{Sec:PlotCategories}

It is expected that in most cases of interest at the LHC,  $\tau$ leptons are produced through the decay of intermediate states
of $W$, $Z/\gamma^*$ or $H$ bosons. Because of the detector properties,  fraction
of $\tau^\pm$ energy carried by visible decay products, respectively $x_1$ ($\tau^{+}$) and $x_2$ ($\tau^{-}$), is 
a natural choice for the 
monitoring variables. 

\begin{figure}[h!]
\centering 
\subfigure[$\tau \to l \nu_l \nu_\tau $]{
\includegraphics[scale=0.43]{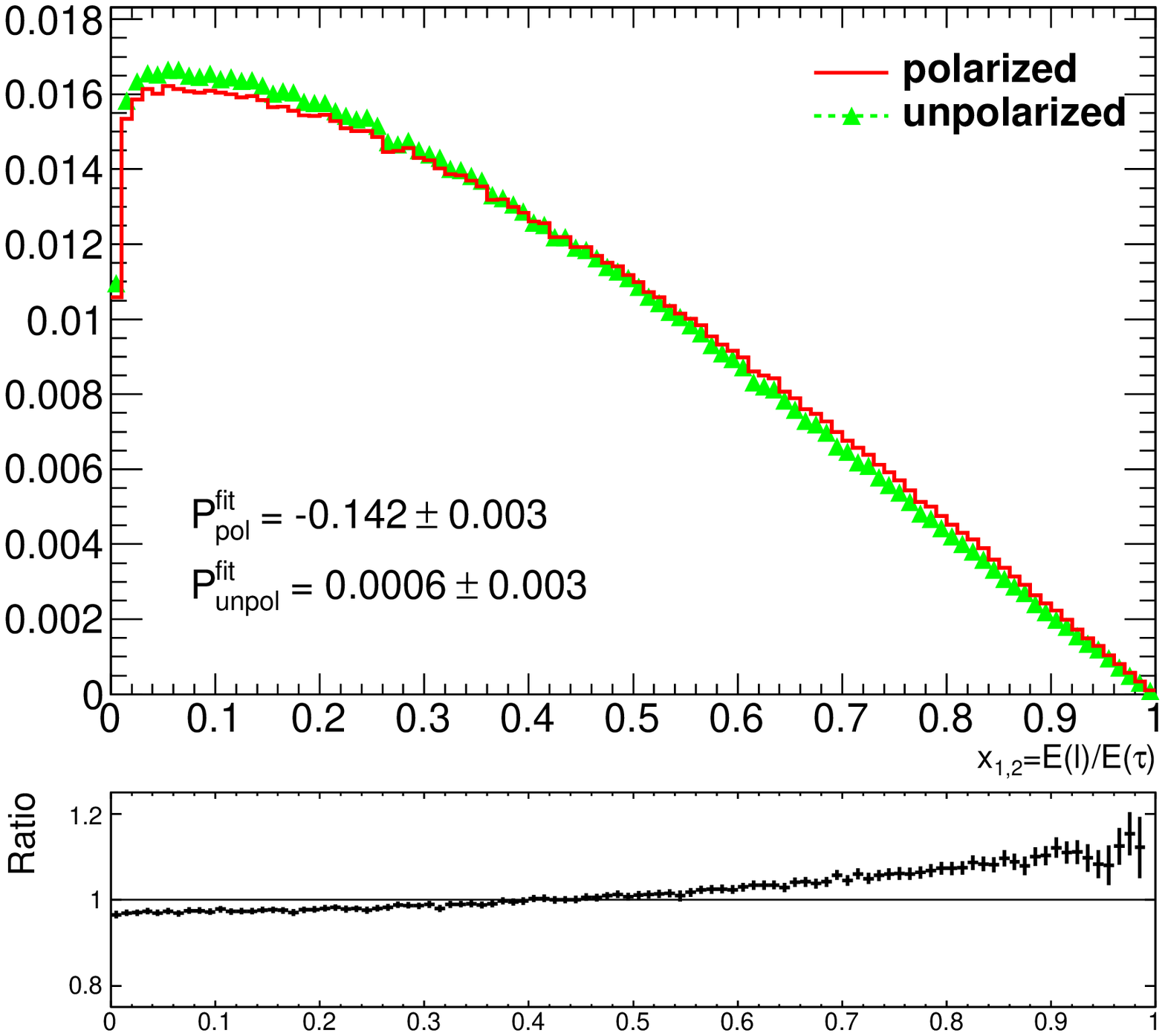}
}
\subfigure[$\tau \to  \pi \nu_\tau $]{
\includegraphics[scale=0.43]{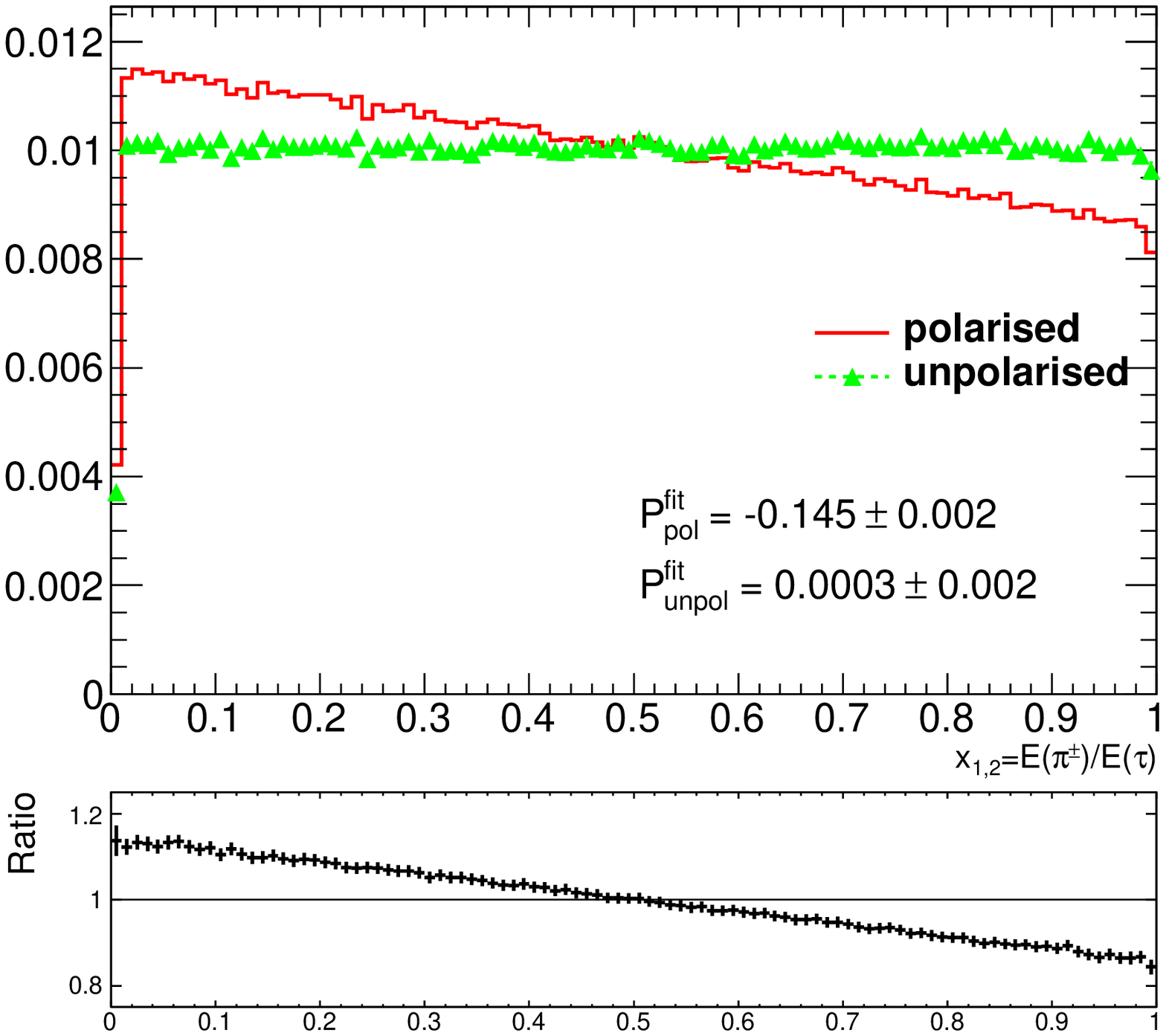}
}
\subfigure[$\tau \to  \rho \nu_\tau $]{
\includegraphics[scale=0.43]{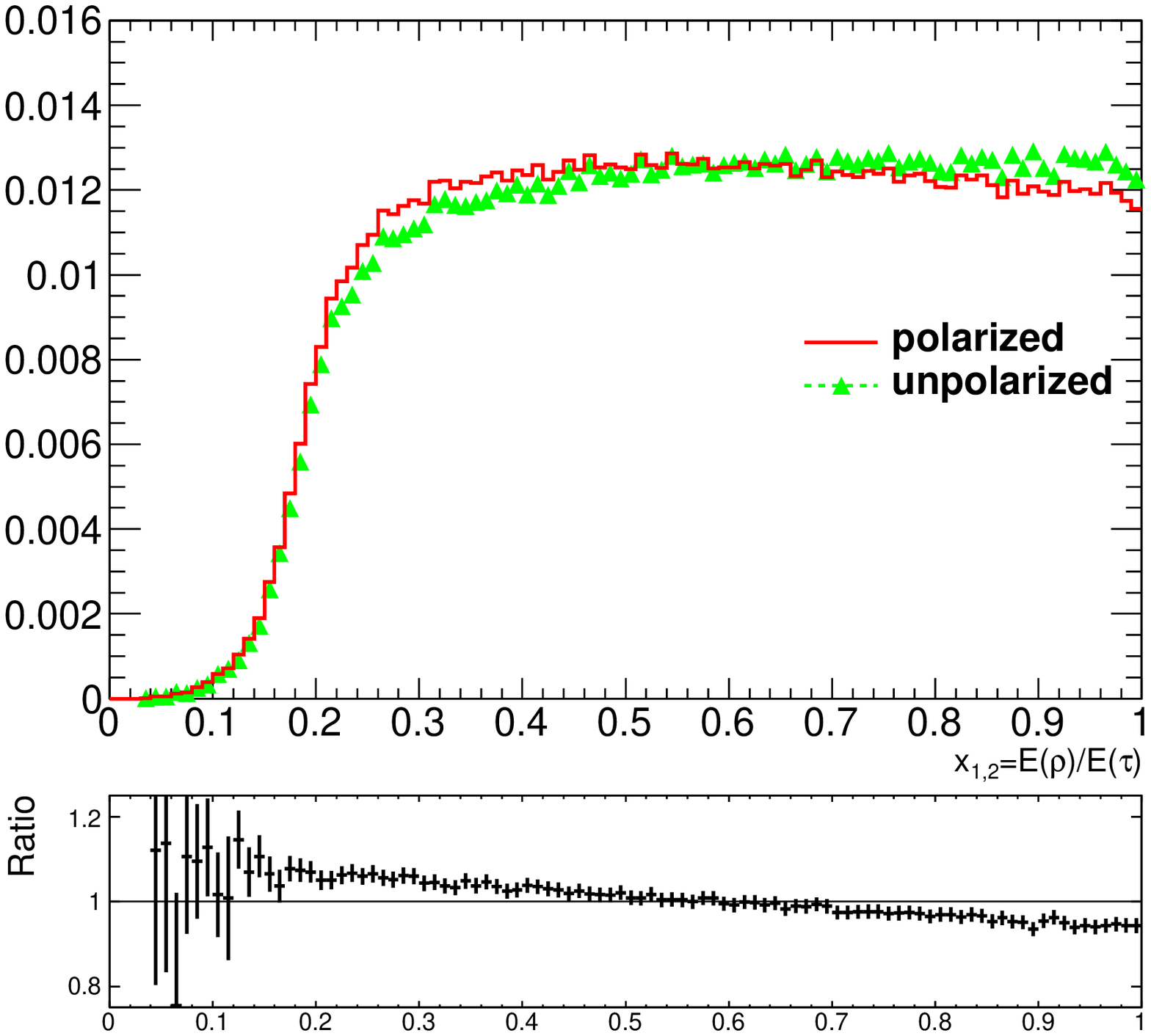}
}
 \caption{\label{Fig:spectra} 
 The spectra of visible $\tau$ decay  energy  normalized to $\tau$ energy, $x_{1,2}$. 
Spin effects included (red, solid line) and neglected
(green, dashed line with triangles). The  $\tau$ leptons are produced through 
$Z$ decay close to the mass peak. The $\tau$ polarization $P$ is obtained from the  fit to the distributions constructed from  $Z \to \tau \tau$ sample
for polarized and unpolarized (unweighted) cases. 
For the fit, the first bin in  $\tau \to \pi \nu$ case and first five bins in $\tau \to l \nu_l \nu_\tau $ case
where mass effects would be the largest,  were 
omitted.
  }
\end{figure}

The example spectra of $x_{1,2}$ for the specific
$\tau$ decay channels are shown in Fig.~\ref{Fig:spectra} for $Z \to \tau^+ \tau^-$ decays, separately 
for leptonic, single $\pi$ and $2 \pi$ decay channels. 
For construction
of these plots the {\it original } events sample, discussed in Section~\ref{Sec:samples}, was used.
The unpolarized spectra were obtained from the  {\it original } sample, using
weights calculated by {\tt TauSpinner}.  From the comparison of the two spectra the 
effect of polarization can be evaluated. As shown in Fig.~\ref{Fig:spectra}, the spectra and their sensitivity 
to the spin vary dramatically depending on the $\tau$  decay channel. 
Negative polarization leads to harder spectra in case of $\tau \to \ell \nu_\ell\nu_\tau$ decays,
and softer in case of $\tau \to  \pi \nu_\tau $ .  The effect is of 20\%  at the very end of the spectra. 

In Fig.~\ref{Fig:spectra} results from the fits to respectively
formulae~(\ref{pion}) and~(\ref{lepton}) are given:
$P_{\mathrm{orig}}=-0.142 \pm 0.003$ ($P_{\mathrm{unweighted}}= 0.0006 \pm 0.003$ unpolarized) for leptonic mode and 
$P_{\mathrm{orig}}=-0.145 \pm 0.002$ ($P_{\mathrm{unweighted}}= 0.0003 \pm 0.002$ unpolarized) for single $\pi$ decay mode.
Note that the differences between the results for the leptonic and $\pi$ modes, 
even though formula~(\ref{lepton}) is missing mass corrections for muons are small. This is because the first bins of the histograms were excluded from fitting.
The nominal average $\tau$ polarization of sample used, as
estimated by appropriate method of {\tt TauSpinner}, reads  $P=-0.144 \pm 0.001$.
The effect from the missing muon mass contributes less than -0.005 to the polarization
obtained from the fit to distribution in leptonic channel,
both for the original sample $P_{\mathrm{pol}}$ and for the spin unweighted sample
$P_{\mathrm{unpol}}$. 
Statistical errors on the fit results correspond to samples of 10M events, as discussed in Section~\ref{Sec:samples}.
Although fit results are not precise (they are biased by approximations
of analytic formulae~(\ref{pion}),~(\ref{lepton})), some distinguishing power between polarized 
and unpolarized
sample is demonstrated. Statistical errors on the fitted values are calculated by the {\tt ROOT} fitting 
package \cite{root-install-www}.

\begin{figure}[h!]
\centering 
\subfigure[$Z \to \tau^+ \tau^-;\; \tau^\pm \to \pi^\pm \nu_\tau$]{
\includegraphics[scale=0.43]{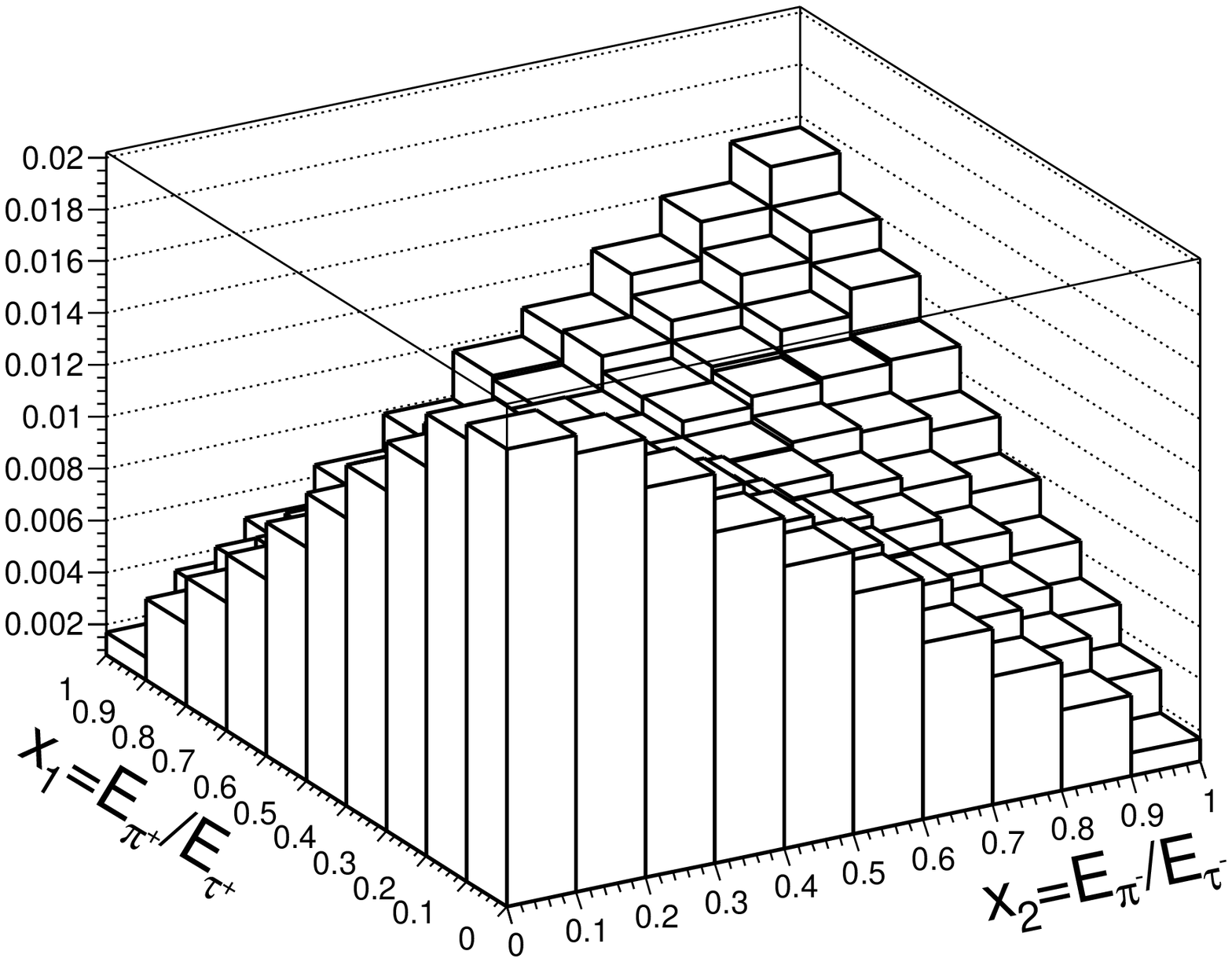}
}
\subfigure[$\Phi \to \tau^+ \tau^-;\; \tau^\pm \to \pi^\pm \nu_\tau$]{
\includegraphics[scale=0.43]{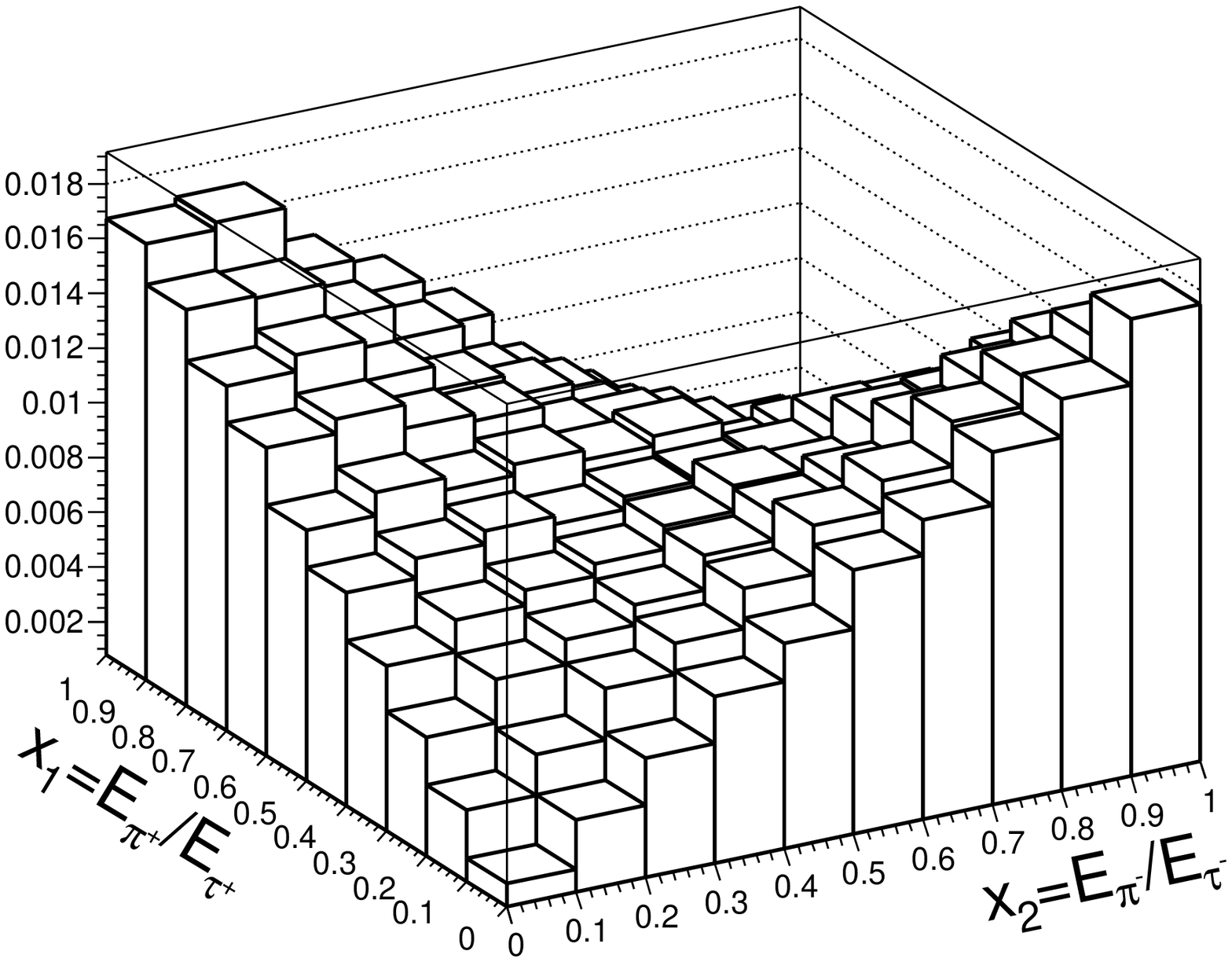}
}
\subfigure[$Z\to \tau^+ \tau^-;\; \tau^\pm \to \pi^\pm \nu_\tau $]{
\includegraphics[scale=0.43]{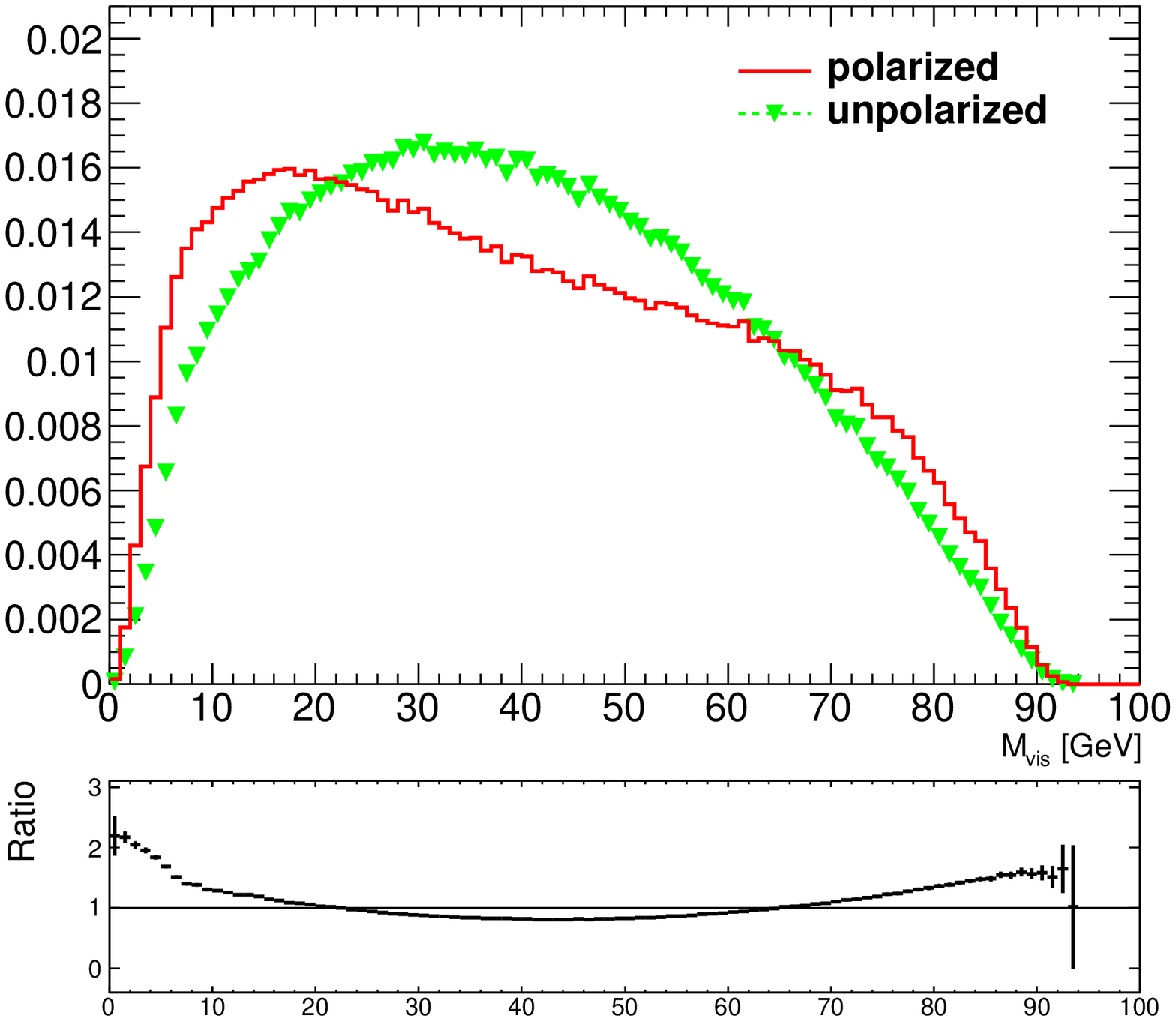}
}
\subfigure[$\Phi \to \tau^+ \tau^-;\; \tau^\pm \to \pi^\pm \nu_\tau$]{
\includegraphics[scale=0.43]{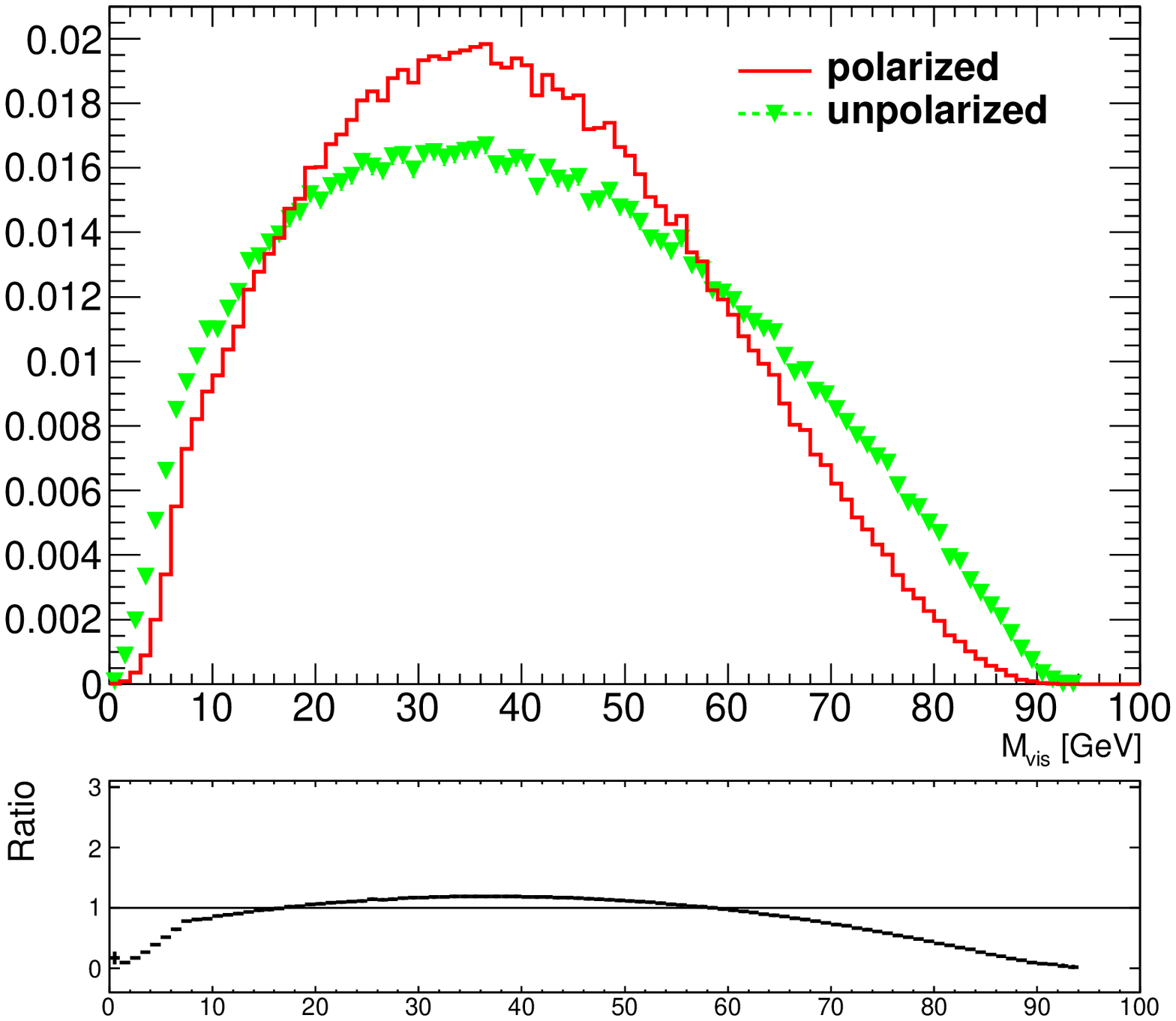}
}
 \caption{\label{Fig:scattergams}  
The case of $Z (\Phi)  \to \tau^+ \tau^- ; \tau^\pm \to \pi^\pm \nu_\tau$.
On plots (a, b): lego plots of  $E_{\pi^+}/E_{\tau^+} \times E_{\pi^-}/E_{\tau^-}$
 and
on  (c, d):  invariant mass distributions of  visible $\tau^+$ and $\tau^-$ decay products are shown. In case when  
spin effects are included red (solid line), otherwise
green (dashed line with triangles) is used. For plots (a, c): the  $\tau$ leptons produced through $Z$ decay are used.
For plots (b, d): $\tau^+\;\tau^-$ pairs from spin-0 state $\Phi$ are used.
All distributions are normalized to unity.}
\end{figure}

The spin effects show up differently depending on the particular $\tau$ decay channel,
as shown in Fig.~\ref{Fig:spectra}. As a consequence the spin correlation manifests itself differently 
depending on the particular cases of the  $\tau^+$ and $\tau^-$ decay channels.
In Fig.~\ref{Fig:scattergams}a and  Fig.~\ref{Fig:scattergams}b we provide 
the two-dimensional plots (lego plots) for $Z \to \tau \tau$ events with both  $\tau^{\pm} \rightarrow \pi^\pm \nu_\tau$
and the distributions of invariant mass $M_{vis}$ for the all visible  $\tau$ pairs decay products
combined, Fig.~\ref{Fig:scattergams}c and  Fig.~\ref{Fig:scattergams}d.
Note that to a good approximation $M_{vis}=Q x_1 x_2$ where $Q$ denotes mass of the 
$\tau$-pair.
In case of spin-0 state $\Phi$,
the fast-fast ($x_{1,2} > 0.5$) and slow-slow ($x_{1,2} < 0.5$)
pairs of $\pi^\pm$ are disfavoured, whereas in $Z/\gamma^{*}$ case the fast-slow and
slow-fast configurations are less populous.
Each configuration of  $\tau$ decay channels feature different spin response pattern. 
We refer reader to series of the plots in Appendices
collecting automatically created numerical results respectively for 
$W$, $Z/\gamma^{*}$ and  $\Phi$ cases and different configurations of the $\tau$ decay modes. 
The Appendices (attached to the preprint version of our paper) represent
examples of output from programs described in Section~\ref{Sec:Installation}.

\subsection{Fits to energy fractions, radiative corrections and experimental cuts}
\label{subsec:cuts}

If there is no kinematical selection with resulting correction, mass corrections are neglected and QED bremsstrahlung in decays of $\tau$ 
and $Z\to \tau\tau$ is not present, analytic formulae for distributions of $x_1$, $x_2$ are given by simple polynomial 
expressions; formulae~(\ref{pion}) and~(\ref{lepton}). 
These formulae can be used to fit the distributions and extract the value of polarization $P$.
The fit can be performed for both, polarized and unpolarized distributions (the second
ones constructed from appropriately weighted events).
The values  of the parameter $P$ obtained from the fit (average $\tau$ polarization) 
for the polarized and unpolarized distributions
enables simple diagnostic if a given $\tau$ decay channel is sensitive 
to the spin effects. Fit errors 
on the parameter $P$
provide an estimate on the statistical sensitivity to 
the spin effects.
One can evaluate statistical significance 
of the spin determination for the leptonic and single $\pi$ decay
 modes%
\footnote{It would be interesting to check how this observation is preserved
in case when experimental cuts are applied.} 
as can be seen from Fig.~\ref{Fig:spectra}.

\begin{figure}[h!]
\centering 
\subfigure[$\tau \to e \nu_e \nu_\tau (\gamma)$]{
\includegraphics[scale=0.43]{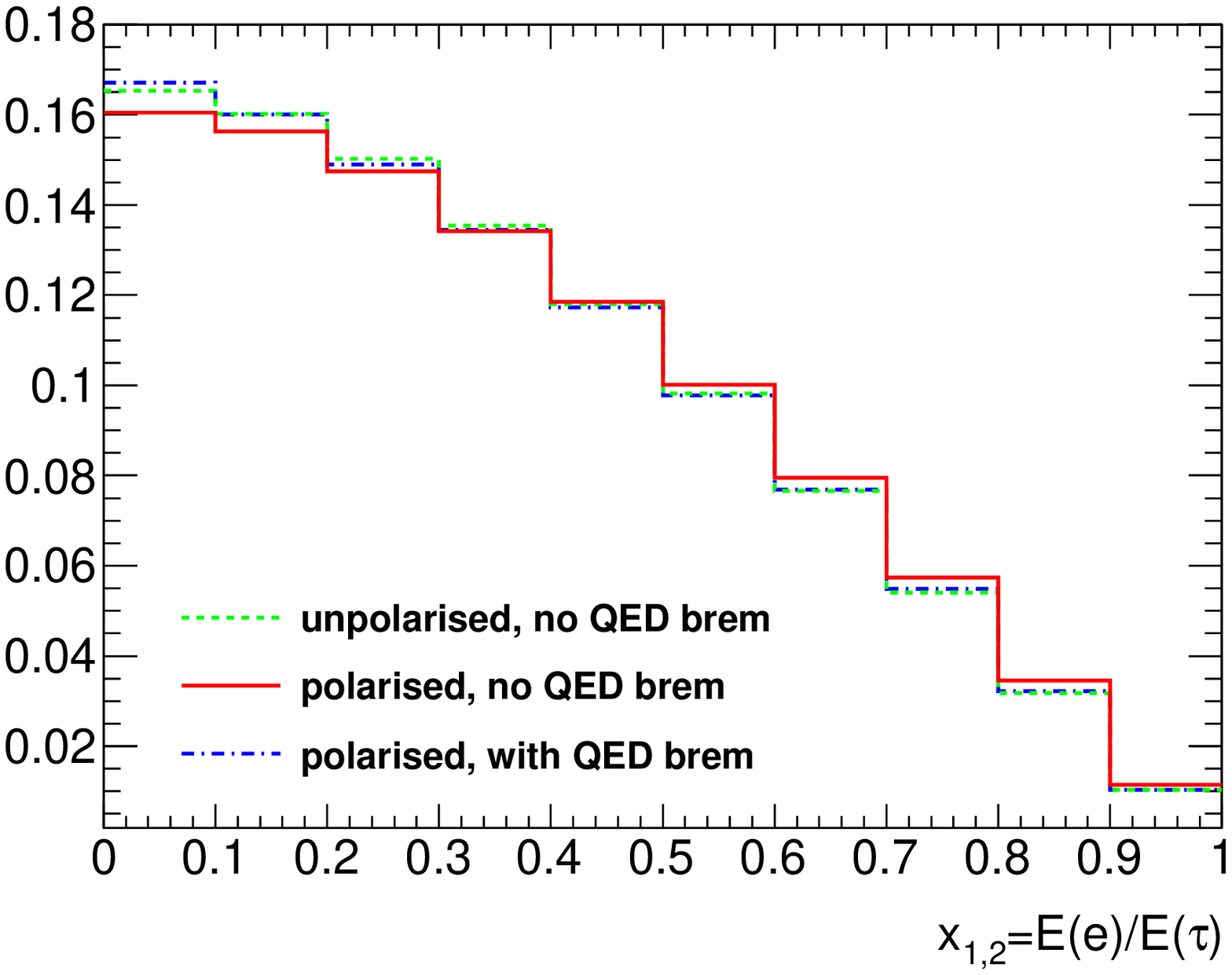}
}
\subfigure[$\tau \to \mu \nu_\mu \nu_\tau (\gamma)$]{
\includegraphics[scale=0.43]{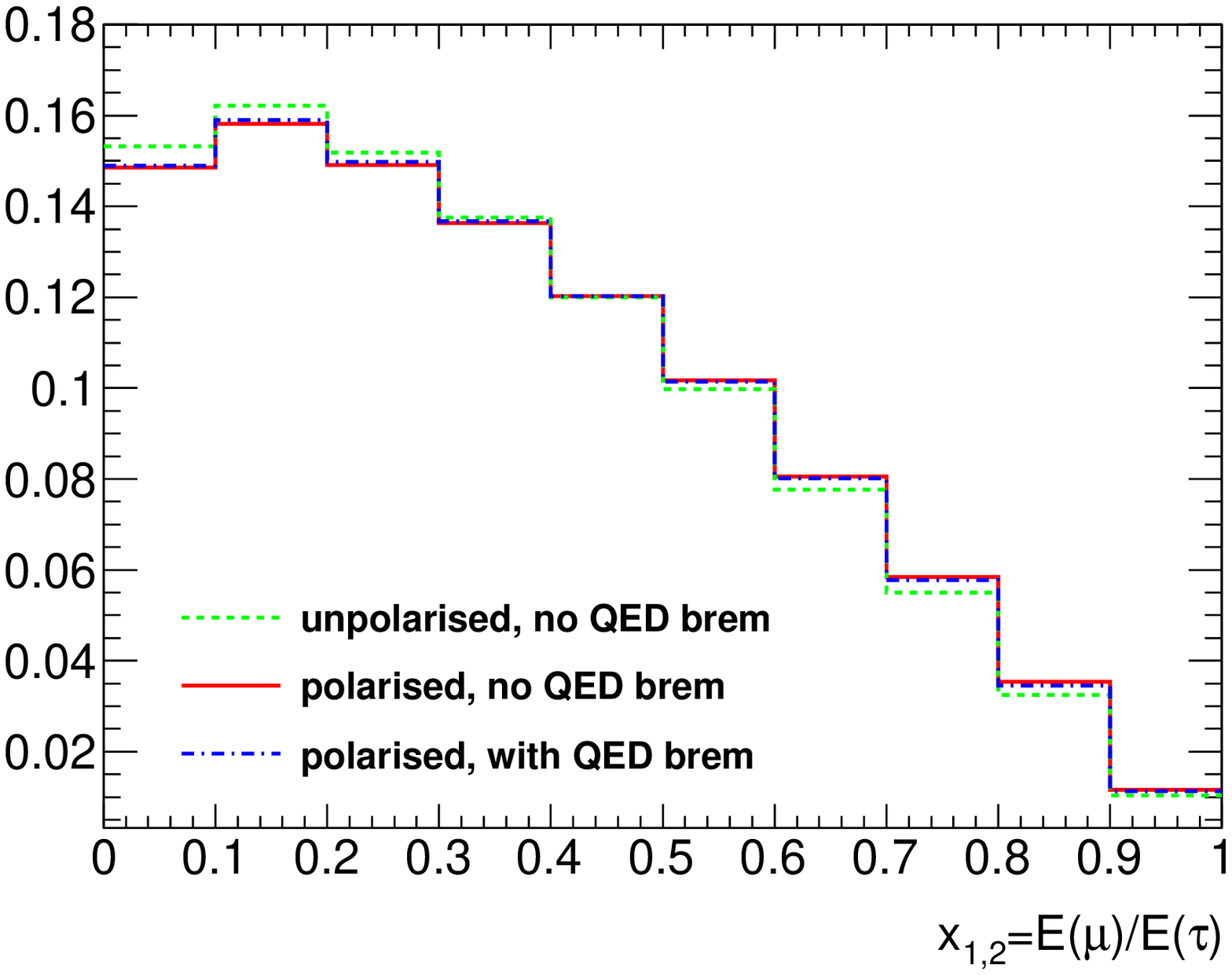}
}
 \caption{\label{Fig:QED} Example  plots for the effects of QED bremsstrahlung
in leptonic decays of  $\tau$.
 Spectra of visible $\tau$ decay  energy  normalized to $\tau$ energy, $x_{1,2}$, are 
shown. 
Spin effects and QED bremsstrahlung are excluded  for green (dashed line), for  
spin effects included but QED bremsstrahlung excluded red (solid line) and 
finally both spin effects  and  QED bremsstrahlung included 
blue (dotted line). Distributions constructed for $Z \to \tau \tau$ decays.
Left-hand plot is for $\tau$ decays to electron, right-hand plot for
$\tau$ decays to muons. For the plots, histograms were rebinned. 
For the fits of histograms,  all 100 bins  except 
the first five (that is exactly as  in fits for Fig. \ref{Fig:spectra})  were used.
The expression~(\ref{lepton}) was fitted to the spectra.
For the $e$ channel, 
$P= 0.004\pm 0.002, -0.141 \pm 0.002, -0.015 \pm 0.002$ 
respectively for
unpolarized, polarized and polarized with bremsstrahlung effect included cases.
For the $\mu$ channel, the  analogous result reads  
$P= -0.009 \pm 0.002, -0.153 \pm 0.002, -0.124 \pm 0.002$.   }
\end{figure}

As it was explained in ref.~\cite{Eberhard:1989ve}, deformation of the spectra due to  radiative 
corrections   can be as big as the effect of 
$\tau$ polarization itself. Also our fit results given in caption of Fig.~\ref{Fig:QED} support this observation. 
Depending on whether for calculation of  $x_{1,2}$ bremsstrahlung photons   are 
combined with the lepton or not,  effect on $\tau$ polarization obtained  from the 
fit of formulae~(\ref{lepton})  may substantially differ in size\footnote{
It is  important to verify if such spectra with radiative corrections
will be useful. Other effects, such as experimental cuts,
may change shapes  as well,
making such theoretical improvements of a minimal interest only. }. 
The main deformations are for $x_{1,2}$ close to 0 or 1.
In general case, when distributions are not sufficiently well described 
by formulae~(\ref{pion}) and~(\ref{lepton}), Monte Carlo methods can be used to obtain the spectra and dependence 
on the polarization $P$  similarly to how it was done for the   measurements  of the $Z$ couplings
performed at LEP~\cite{Heister:2001uh}. 

\subsection{From benchmarks toward realistic experimental distributions.}

\begin{figure}[h!]
\centering 
\subfigure[$W^- \to \tau^- \bar \nu_\tau; \tau^- \to \pi^- \pi^0\pi^0 \nu_\tau$]{
\includegraphics[scale=0.43]{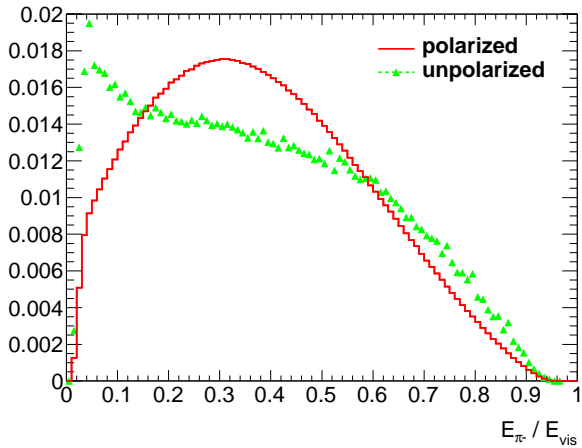}
}
\subfigure[$Z/\gamma^* \to \tau^- \tau^+; \tau^- \to \pi^- \pi^0\pi^0 \nu_\tau$]{
\includegraphics[scale=0.43]{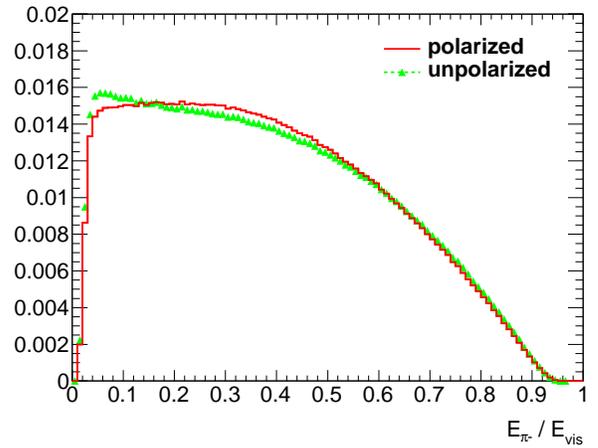}
}
 \caption{\label{Fig:A1} Example  plots for effects of spin
in  $\tau^- \to \pi^- \pi^0\pi^0 \nu_\tau$; the $E_{\pi^-} / E_{vis}$ distribution.
Case when spin effects are included is denoted by red (solid line), for 
spin effects excluded, green (dashed line with triangles) is used. Left hand side plot is for 
$W^- \to \tau^- \bar \nu_\tau$ production, right hand side plot 
for $Z/\gamma^* \to \tau^- \tau^+$. Note large statistical fluctuations for 
unpolarized distributions in $W$ case obtained with unweighting procedure. 
It is because in case of 100 \% polarization like for $W\to \tau \nu_\tau$ decays, 
the spin weight $wt_{spin}$ can approach zero. Its inverse
 used for unweighting polarization, can therefore become arbitrarily large, resulting in (integrable)
singularity of the distribution. These large fluctuations  indicate the limitation for 
use of weights method to remove spin effects from already generated events.}
\end{figure}

Numerical results presented above were prepared in an idealized case,
where no  experimental selection was applied to the analyzed samples. We have relied
on the unobservable 
fractions of visible $\tau^\pm$ energies, $x_1$ and $x_2$.  This is well suited
for testing Monte Carlo programs and detector simulation samples. 
In Fig.~\ref{Fig:A1}, we show the impact of spin effects on experimentally observable
and sensitive to spin quantity $E_{\pi^-} / E_{vis}$  for  $\tau^- \to \pi^- \pi^0\pi^0 \nu_\tau$ decay channel.
As one can conclude effects of the spin are sizeable. The effect
can be evaluated  using the {\tt TauSpinner} unweighting algorithm. The $a_1$ decay channel
is less suitable for testing the programs or simulations 
because of more complex interpretation of different spectra. It indicates further applications, more oriented to 
realistic studies than the ones we have collected in Appendices for technical purposes.
For applications in experimental data analysis, one should address possible complications
like background contamination or limited acceptance. 
The template fit technique  \cite{Heister:2001uh},  convenient at LHC, 
as demonstrated  in \cite{Aad:2012cia}, can be used for spin in this case
as well.

\newpage
\subsection{Consistency checks}\label{Sec:consistency-check}

For the  {\tt TauSpinner}   algorithm the questions of theoretical 
systematic error are of a  great importance. We do not plan to review 
this aspect of the program development now. Some results are already 
documented in \cite{Czyczula:2012ny,Banerjee:2012ez}, but more detailed studies
will be needed when the precision requirements will become more strict than 
presently. The work with explicit multi-leg QCD matrix elements of appropriate form, 
like in Ref.~\cite{vanHameren:2008dy}, will be mandatory.

It has been  known for a long time \cite{Mirkes:1992hu,Mirkes:1994eb}, that predictions for the Drell-Yan 
processes must lead to the dependency on the polar and azimuthal angles of outgoing leptons in the 
center-of-mass frame of decaying resonance in the form of second order spherical harmonics.
This feature leads to the broad spectrum of possible applications, from validating
implementations of higher order QCD corrections in the Monte Carlo programs,
to the indirect measurement of the mass of the W boson \cite{Aaltonen:2013wcp}.

For shown here new tests, 
it is important to notice  that in the process of preparing
spin weights, {\tt TauSpinner} calculates 
all ingredients of the effective Born parton level cross section, 
as described 
in \cite{jadach-was:1984,Eberhard:1989ve,Pierzchala:2001gc,Jadach:1993hs}. 
Predictions for other observables or quantities of phenomenological interest, 
such as quark level forward-backward asymmetry or probability of 
a given quark flavour 
to originate a particular hard process event, can be obtained when executing the code. 
Because of mentioned above properties of QCD,  formulae for polarization and 
other quantities, remain essentialy as at LEP.

If the studied sample is generated by the Monte Carlo
program and  {\it physics  history entries} (flavours and momenta of quarks 
entering hard process) are stored, 
one can directly use this information
to retrieve properties of the electroweak matrix elements and hadronic interactions of the studied
events sample to validate precision of the {\tt TauSpinner} algorithms. 
The four-momenta and flavours of the incoming quarks can be used to calculate
parton level forward-backward asymmetry or rate of production from distinct quark flavour.
These results can then be compared with the similar quantities estimated from the
weights calculated by the {\tt TauSpinner} algorithms using kinematics of the $\tau$ decay products only,
providing very interesting test on the precision of TauSpinner algorithms.

Unfortunately information on four momenta and flavours of incoming quarks is usually 
available only for Monte Carlo with parton showers based on the leading
logarithm approach. At the next to leading logarithm level
\cite{Kleiss:1990jv} such information may be available as well, but it is not  
necessarily the case. One should mention here that because of the spin-1 nature 
of objects decaying to  pair of leptons, the  angular distributions of $\tau$ leptons in the rest frame of $\tau$ pair are 
described by spherical harmonics, of at 
most  the order of two. This explains why higher order QCD corrections,
contributing higher than second order spherical harmonics,
must be  small \cite{Mirkes:1992hu}. 

\begin{figure}[h!]
\centering 
\subfigure[$d \bar d \to \tau \tau$]{
\includegraphics[scale=0.43]{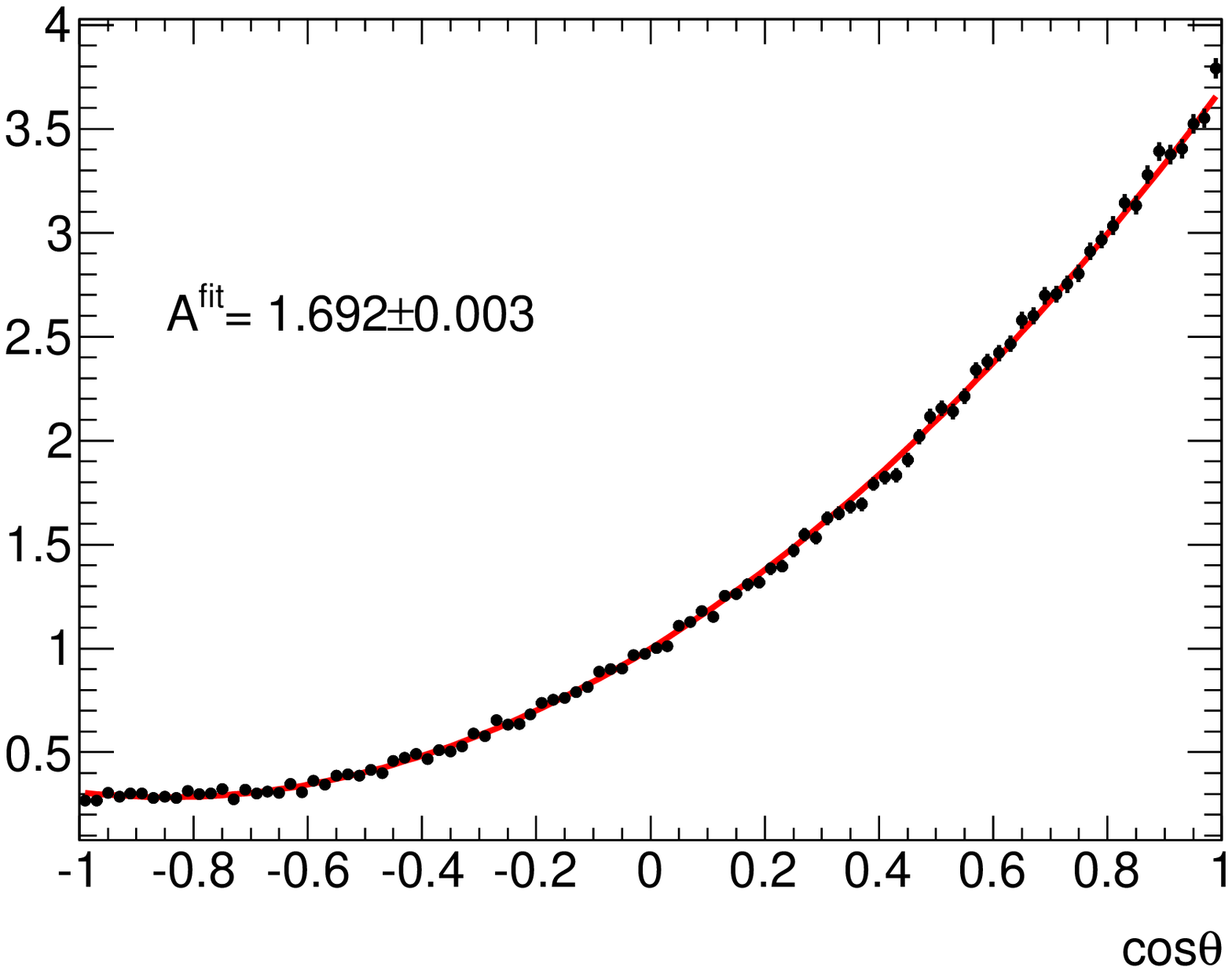}
}
\subfigure[$u \bar u \to \tau \tau$]{
\includegraphics[scale=0.43]{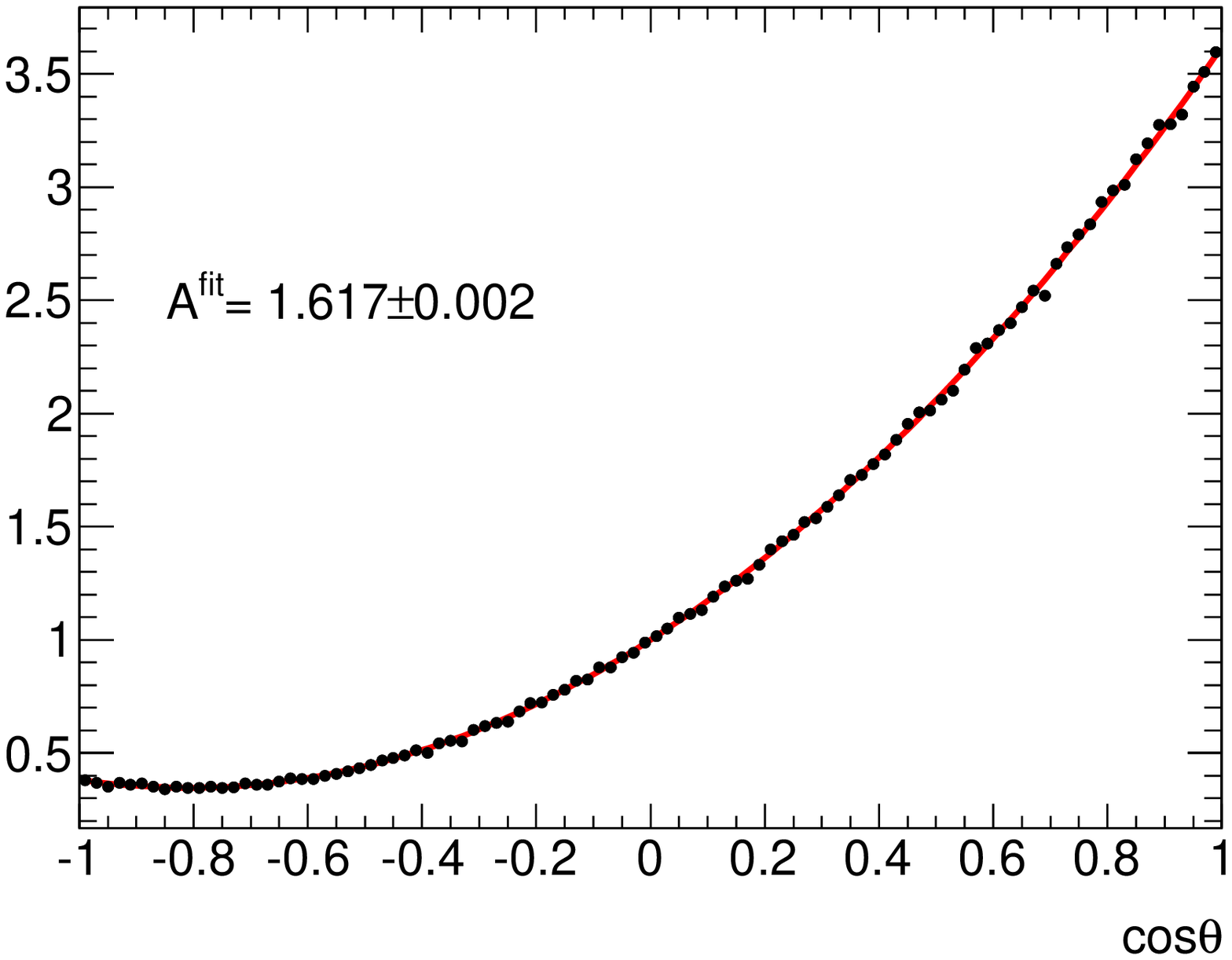}
}
 \caption{\label{A-fit}  Differential distribution in the hard scattering 
angle $\cos \theta$ calculated from {\it physics history entries} in the event record,
 the superimposed  fit of   ($1+\cos^2 \theta + A \cos \theta$) red line 
is shown.
Histograms are normalized to $\frac{8}{3}$ (the integral of fit function in range $[-1,1]$).
 The vituality of the $\tau \tau$ pair was restricted to the range 1-1.5 TeV.}
\end{figure}

We have prepared following tests, supplementary to the ones
of subsection~\ref{Sec:PlotCategories}, which exploit {\it physics history entries}.
These tests may be particularly interesting if some kind of inconsistency is 
found in the analyzed sample
and one is debugging its origin:
\begin{enumerate}
\item[A] {\bf Test of kinematic reconstruction.}
In  {\tt TauSpinner}, to evaluate $\tau$ scattering angle $\theta^*$, an algorithm described 
in \cite{Was:1989ce} is used. Resulting  $\cos \theta^*$ is compared with $\cos \theta$
of scattering angle calculated from {\it physics history entry} of the  event record. 
The difference of the two results is monitored.
\item[B] {\bf Test of electroweak Born cross section.}
For the sample featuring {\it physics history entries}, the scattering angle of the outgoing lepton in the hard process 
can be calculated and appropriate angular distribution
plotted separately for each flavour of incoming quarks. 
This distribution, in the leading log approximation have functional form
 ($1+\cos^2 \theta + A \cos \theta$). 
Coefficient in front  of  $\cos \theta$, defining size of forward-backward asymmetry $A_{FB}$, 
can be obtained from the fit of this function to $\cos \theta$ distribution obtained 
from the analysed sample. The same coefficient can be calculated with the help of 
{\tt TauSpinner} algorithm. This calculation uses as an input information of parton density functions (PDFs)
which is convoluted with the parton level matrix-element of the hard process. 
The average value of the coefficient $A$ (defining size of the forward-backward asymmetry $A_{FB}$) can be therefore obtained independently
from the algorithm responsible for calculating spin weights and in particular  scattering angles.
The comparison of the results obtained from the fit to $\cos \theta$
distribution constructed from {\it physics history entries} of the events on one side,
 and of {\tt TauSpinner} internal calculation of $A$ 
(when only PDFs and virtuality of $\tau$-pair is used) on the other side,
provides tests for effective Born parameters consistency in the analysed sample and {\tt TauSpinner} code. Results 
of this test depend also on the choice of PDFs and on the correctness of the {\tt TauSpinner} algorithm 
for reconstruction of PDFs arguments (fractions of proton momenta carried by partons)  from the kinematics of the $\tau$'s 
(used is virtuality  and pseudorapidity  of the $\tau \tau$ system). 

An example of such comparison is given in Fig.~\ref{A-fit} for the case of Drell-Yan events with virtuality
in the range of 1~-~1.5~TeV. The fit  gives A= 1.617 +/- 0.002 for up quarks and A= 1.692 +/- 0.003
for down quarks. From the {\tt TauSpinner} calculation using Born amplitude, value of the A parameter
averaged over the same sample (calculated from Born cross section) read respectively 
1.613 and 1.691 with negligible statistical error~\footnote{In this case 
we concentrate on matching the electroweak parameters in initialization of 
{\tt Pythia} and {\tt TauSpinner}, hence the initial state hadronic effects were
switched off. We have checked, that if more complete treatment is used,
quality of the  agreemet between $A$ and $A^{\mathrm {fit}}$ is
degraded by $\sim$0.01, but shapes of the distributions become more complicated.}.
For other choices of the virtualities range agreement was found to be of a similar quality. 

\item[C] {\bf Test of PDFs.}
To some level, previous test can be complemented with the direct test of PDFs.
Fraction of sample events (production rates) with hard process of particular incoming 
quark flavour can be compared with that fraction attributed by the {\tt TauSpinner} algorithm.
In the second case, every event contribute, but with the weight proportional to
quark level total cross sections multiplied by respective PDFs.
For the example shown in Fig.~\ref{A-fit} we obtained respectively:
rate of down quarks 0.217 ({\tt TauSpinner} 0.216) and 
rate of up quarks   0.766 ({\tt TauSpinner} 0.762).
\end{enumerate}

The results for tests [B] and [C] depend on the PDF's used internally by {\tt TauSpinner} and on effective
Born-level cross-section used at parton quark level. In case of [B] dominant contribution
comes from the odd power of axial couplings, whereas in case [C] even powers of 
the axial couplings dominate the result.
That is why the two tests are to a large extent independent.

\begin{enumerate}
\item[D]{\bf Partial polarization test.}
An option that the events sample features  spin correlation of the 
two $\tau$ leptons, but not fully the polarization effects due to production of intermediate 
state $Z/\gamma^{*}$ couplings is of practical interest when constructing so called $\tau$ embedded
sample from $Z \to \mu \mu$ events selected from data.  
For example the sample features dominant spin effect, due to vector nature of the
intermediate state, but is free of systematic error of 
the  electroweak effective Born and of incoming quark PDFs. If only angular dependence of the 
polarization is neglected, the systematic error due to PDFs on the spin effects  is reduced
and much smaller systematic error due to the effective electroweak Born parameters remain. In both cases,
relatively small neglected effects can be evaluated and introduced  with the help of {\tt TauSpinner}
weights, see Section~\ref{Sec:tauspinner-uses}.
\end{enumerate}

In the discussion of numerical results presented above, samples 
without any kinematical cuts were used. However, one may be interested
to test how our algorithm will perform if only a particular class 
of events, for example of high 
$p_T$ configurations only, is used.
All the tests listed above can be then performed using  sub-samples defined
by the particular set of cuts.
In such cases,  the validity of the  {\tt TauSpinner} algorithms and of the parton shower
algorithm used for the sample generation can be explored in more exclusive phase-space regions.

\subsection{Reference plots} \label{Sec:PlotAppendices}

Let us now discuss briefly the large collection of automatically created plots prepared by our testing programs\footnote{
These programs are included in the {\tt TauSpinner} distribution tar-ball. See Section~\ref{Sec:Installation} for more details.}. For the 
preprint version of our paper such plots are collected into rather lengthy  appendices for the $W$, $Z/\gamma^*$ and spin-0 resonance $\Phi$.
The distributions shown, depend on the particular sample used. 
We have grouped the figures for  each $\tau$ decay channel (case of $W$)
or for each pair of $\tau$ decay channels (case of $Z$) separately. 
For leptonic and single $\pi$ decay channels
results of the fits to spectra~(\ref{pion}) or~(\ref{lepton}) are given.
The input samples feature complete longitudinal spin effects. 
The {\tt TauSpinner} weights were used to unpolarize the sample. For the case 
of $\Phi \to \tau \tau$  events the $Z \to \tau \tau$ sample was used, but
$\tau$ decays were regenerated instead of reweighted for better numerical stability. 

For each type of decaying resonance, we give specification of the sample used
for the respective set of  plots, reporting the number of the analyzed events with the decomposition 
into particular (pair of)  $\tau$ decay channels and initialization of the generator used for sample preparation. 
We also plot the control distribution of invariant mass of $\tau\tau$ ($\tau\nu_\tau$) system. 
This sanity plot verifies if the sample consists
of events  at the resonance peak or if substantial contribution from low energy 
or very high mass tails is included in the sample as well.
Spin effects are different if events are taken at, above or below the $Z$ peak. 

Afterwards come collection of  
 plots to large extend following layouts of Fig.~\ref{Fig:spectra} and Fig.~\ref{Fig:scattergams}: 

\begin{itemize}
\item
In case of $W \to \tau \nu_\tau$ decay only the one-dimensional distribution of the energy fraction
carried by $\tau$ lepton is plotted, comparing the case of polarized and unpolarized samples. 
In the captions of the plots, similarly as in  Fig.~\ref{Fig:spectra},
the fitted value of the $\tau$ polarization $P$ is given as a measure of spin sensitivity of analyzed samples. 
\item
In the case of the $Z/\gamma^*$ mediated processes and for each particular combination of $\tau^+$ and $\tau^-$ decay 
channel, the  set of histograms are collected.
The first is the two-dimensional lego plot constructed from the fractions of energies of $\tau^+$ and $\tau^-$ 
carried by their corresponding  observable decay products.  Analogous lego plot is also shown for the case 
when spin effects are removed  with the help of weights calculated by {\tt TauSpinner}. 
Ratio of the two distributions is given in the  lego plot of the second row. 
It demonstrates the strength of the spin effect. On the right hand side of this 
lego plot
the one-dimensional histogram for invariant mass, of all visible products of $\tau^+$ and $\tau^-$ combined, is given.
It provides a convenient way of representing spin correlation effects in case of smaller samples, which may be insufficient
to fill the two-dimensional distributions.
The last two plots show single $\tau^+$ and $\tau^-$ decay product spectra respectively
(each plot containing original sample, sample with modifications
due to {\tt TauSpinner} weights and their ratio). Spectra are normalized to unity.
In the captions of the plots, see Fig.~\ref{Fig:spectra}, the
fitted value of $\tau$ polarization $P$ is given as a measure of spin sensitivity of analyzed samples. 
Note that the groups of plots for the cases when  decay channels for $\tau^+$ and $\tau^-$ are simply 
interchanged, coincide up to permutation of axes, unless some cuts are introduced by the user. 
\item
In case of $\Phi$-mediated process, set of plots analogous to  $Z/\gamma^*$-mediated processes
are given.
\end{itemize} 

The proposed set of benchmark plots can be extended further with 
the help of provided validation programs, in particular for the cases of
partial implementation of polarization and spin correlations effects. 
Respective 
systematic errors can be evaluated.

\section{Technical details} \label{Sec:Installation}

For the purpose of this paper, a directory {\tt TauSpinner/examples/applications}%
\footnote{In {\tt Tauola++ v1.1.4}, released on 12 Dec 2013, this directory was called {\tt TauSpinner/examples/tauspinner-validation}.
All subsequent directories and programs have been renamed following the new convention. 
In particular, directories: {\tt applications-plots-and-paper, applications-rootfiles, applications-fits} was respectively 
called {\tt tauspinner-validation-results, tauspinner-validation-plots, tauspinner-validation-fit} and programs 
{\tt applications-plots.cxx, applications-comparison.cxx, applications-fits.cxx} were called 
{\tt tauspinner-validation-plots.cxx, tauspinner-validation-comparison.cxx, tauspinner-validation-fit.cxx}.
While the naming of programs and subdirectories changed, the content of the programs remained the same.}
has been added to the previous distributions of {\tt Tauola++}. It contains several tools used to produce the plots
for this paper and to obtain necessary results. It was also extended with several tests that help validate {\tt TauSpinner}.
If {\tt Tauola++} is configured with all prerequisites needed to compile {\tt TauSpinner} package,
as well as {\tt TauSpinner} examples\footnote{Up-to-date instructions can be found on the {\tt Tauola++} website in the documentation
to the most-recent version of the package \cite{tauolaC++}.},
compiling these additional programs should not require any further setup and can be done by executing {\tt make} in {\tt applications} directory.

\subsection{ The {\tt applications} directory}

In the following subsection we will briefly describe the sub-directories for this package and their use.

\subsubsection{Generating plots}

The main program, {\tt applications-plots.cxx}, generates plots which are latter included in the pdf file (like of Appendix~\ref{sec:OutputTest}).
It uses the same algorithm as the one used in {\tt tau-reweight-test.cxx}; part of
the examples for {\tt TauSpinner} included in {\tt Tauola++} tar-ball starting
from version of November 2012.
In this example code, input file {\tt events.dat} is processed and for
each event $WT$ weight is calculated. The set of histograms is filled with weighted (to remove spin effects) and not weighted events,
separately for each $\tau$ decay mode or $\tau$ pair decay mode combination.
Histograming and plotting is done using the {\tt ROOT} library \cite{root-install-www} (also fits are performed with the help of
{\tt RooFit} library).

This program can be used to recreate plots in the Appendices. For this, a datafile with $W$ and $Z$ which
decay into $\tau$'s is needed. Note that since the template {\tt LaTeX} file is prepared
for both $W$ and $Z$ samples, this program can be executed on a single sample
file containing both types of events or on two samples with separate $W$
and $Z$ events\footnote{This program does not produce histograms stored in Appendix~\ref{sec:OutputH}. These plots
require change of the {\tt PDGID} of the $Z$ boson so {\tt TauSpinner} can calculate weight as if the intermediate boson is Higgs.
This change is omitted from the example provided with the distribution tar-ball for simplicity.}.
Only channels $\tau \to \mu \nu_\mu \nu$, $\tau \to e \nu_e \nu$, $\tau \to \pi \nu$
and $\tau \to \rho \nu$ are analysed. To run the program:

\begin{itemize}
\item make sure that {\tt ROOT} configuration is available through {\tt root-config},
\item execute {\tt make} in {\tt TauSpinner/examples/applications} directory,
\item verify that settings in file {\tt applications-plots.conf} are correct,
      including the path to input file\footnote{Note 
      that example file {\tt examples/events.dat} can be used to verify
                if the program compiles and runs correctly. However, it contains
                only a sample of 100 $Z \to \tau^+ \tau^- \to \pi^+ \pi^- \nu_\tau \nu_\tau$ events.},
\item execute {\tt ./applications-plots.exe applications-plots.conf}.
\end{itemize}

A set of plots will be generated in the directory indicated by the configuration file
(the default one is {\tt applications-plots-\-and-paper})
and a breakdown of the $\tau$ decay channels found in the sample will be written
at the end of running the program. If the input file contains both $W$ and $Z$
decays, two sets of plots will be generated, each accompanied with summary of the $W$ and $Z$ events  properties.  The
program also saves all histograms created during processing time to {\tt out.root} file. 
This file can be used to archive the results for further analysis or to add
fits to the plots.

\subsubsection{Adding fits}

The code for adding fits is provided in the subdirectory {\tt applications-fits}.
It is built along with other programs when executing {\tt make} in {\tt applications} directory.
This tool adds fits to the histograms generated by
{\tt applications-plots.exe}
using the formulae~(\ref{pion}) and~(\ref{lepton}), results of the fits are 
stored in the rootfile. See the {\tt README} file in this directory
for details on how input files are processed.

This program uses rootfiles from subdirectory {\tt applications-rootfiles}.
They are specified in the default configuration file {\tt app\-li\-cat\-ions-fits.conf} as the input files of this program.
The resulting plots, with added fit information on polarization, will be stored in
{\tt applications-plots-and-paper} directory. Previously generated plots will be overwritten.
This can be changed in the configuration file with path to the output directory.

As mentioned in Section~\ref{subsec:cuts}, the fit can be applied not to the whole
range but to the interval $(x_1,x_2)$, that is why an option to perform fits only in the limited range of [$x_{\min}$,$x_{max}$]
has been provided in the code and is controlled by the
configuration file.

\subsubsection{Recreating figures~\ref{Fig:QED} and~\ref{Fig:A1}}

The subdirectory {\tt applications-rootfiles} contains rootfiles of 
histograms necessary to reproduce all plots shown in our paper.
These rootfiles are used by {\tt applications-fits.exe}. Histograms
for the plots that are not part of the Appendices are also stored in the 
rootfiles.
Executing {\tt make} will invoke code to generate the plots for Figures~\ref{Fig:QED} and~\ref{Fig:A1}.
Note that generation of these rootfiles requires different setup and different data samples than for any other plots.
While necessary changes are straightforward, including such options would add to the already complex
structure of the validation programs, thus they were skipped in the distribution.

\subsubsection{Additional tests and tools}\label{Sec:Installation-tools}

Two additional subdirectories:
\begin{itemize}
\item 
{\tt test-bornAFB} 
\item {\tt test-ipol} 
\end{itemize}
were added for further  validation
of the {\tt TauSpinner} library. These tests are somewhat peripheral to the main topic of the paper,
thus they were only documented in the {\tt README} files of the corresponding sub-directories.
The result of the first test is briefly discussed in Section~\ref{Sec:consistency-check}, while
the second one has not been presented here. It is, however, included in the package as a validation test
of  {\tt TauSpinner} options ({\tt Ipol} = 0, 1, 2, 3).

The {\tt applications} directory contains additional programs:
\begin{itemize}
\item
The {\tt hepmc-tauola-redecay.cxx}, while not being an example for {\tt TauSpinner}, can be used to
process existing input file and remove $\tau$ decays substituting them with new ones generated by
{\tt Tauola++}. This tool can be used to generate unpolarized $\tau$ decays needed to verify different
{\tt TauSpinner} options (see Section~\ref{Sec:tauspinner-uses}). Note, that as with
{\tt Tauola++}, generation options are limited by the available information stored in the data files.
\item
The {\tt applications-comparison.cxx}, uses two input files.  First one  is considered  
as a reference. For the second one  {\tt TauSpinner} weights are used. 
The same set of histograms is produced for both input files and compared afterwards.
This program can be used to validate {\tt TauSpinner} options, as for example in case E described in Section~\ref{Sec:tauspinner-uses}.
\end{itemize}

Details of how to use both programs are described in {\tt README} of the directory.

\subsubsection{Generating pdf file}

The subdirectory {\tt applications-plots-and-paper} contains the {\tt LaTeX} files, as well as all other
files necessary to prepare Appendices of this paper. Executing {\tt make} in this directory generates the pdf file as of our paper.

Text of  Appendices is stored in  files: {\tt appendixA.tex} and {\tt appendixB.tex}.
The user can thus easily re-attach results of the program run to the documentation of his own project starting from the
template {\tt user-analysis.tex};
the {\tt make user-analysis} will include
 Appendices  into short {\tt user-analysis.pdf}. 

\subsubsection{Final remarks}
It is possible to  redo, for the sake of documenting results of one 
own  tests, 
 all figures and other numerical results of the Appendix A 
(that is also of {\tt user-analysis.pdf}).
In case the physics assumptions are substantially different than
the one used for  the present paper, shapes of the obtained distribution
may differ as well.  
In every case 
 the following step have to be followed:

\begin{enumerate}
\item generate a sample of $W$ and $Z$ decays to $\tau$; $\tau$ decaying to $\mu, e, \pi$ and $\rho$;
\item run {\tt applications-plots.exe} on this sample;
\item run {\tt applications-fits.exe} on the resulting rootfile and store the output in {\tt applications-plots-and-paper}
      subdirectory;
\item execute {\tt make} in {\tt applications-plots-and-paper} subdirectory.
\end{enumerate}

Further details on each of these steps, including more technical details on the output and input files, are given
in the distribution tar-ball and in the {\tt README} files located in {\tt TauSpinner/examples/applications}
directory and all of the sub-directories. 

The numerical results of whole paper can also be reconstructed. Scripts 
for most of the necessary operations are prepared and documented elsewhere 
in the paper or in {\tt README} files.

\subsection{Input file formats}
Essentially any {\tt HepMC} \cite{Dobbs:2001ck} file (saved in {\tt HepMC::IO\_GenEvent} format) can be processed~\footnote{Note however that
it is user responsability to verify that HepMC file contains events with correctly structured information for {\tt TauSpinner}
to find outgoing $\tau$ leptons and their decay products.}
Files with events stored in different format can be either converted to {\tt HepMC} or interfaced 
using  methods described in {\tt TauSpinner} documentation 
and used in the default example {\tt tau-reweight-test.cxx}.

Note that only the file {\tt applications-plots-and-paper/input-file-info.txt}
should be updated by the user with the information on the event sample processed.
All other text files will be updated by the appropriate tools described in previous section.
The content of these text files is included in the output file of pdf format, 
as shown in Appendix~\ref{Sec:InputFiles}.

\subsection{Rounding error recovering algorithm}
The $\tau$ leptons stored in data files can be ultra-relativistic. This may cause 
problems for the part of algorithm recalculating matrix elements 
for $\tau$ decays. For our example, there was no problem with errors from rounding numbers,
but in general such problems are expected.

The following correcting algorithm is prepared:
\begin{enumerate}
\item
  For each stable $\tau$ decay product its energy is recalculated from the mass
and momentum.
\item
The four-momentum of the $\tau$ is recalculated from the sum of four-momenta of its 
decay products.
\item
The algorithm performs check if resulting operation doesn't introduce sizeable
modifications, incompatible with rounding error recovery. If it does, a
 warning message is printed. 
 This may indicate other than rounding error,
 difficulty with the production file. 
For example, some decay products not stored (eg. expected as non-observable
soft photons).
\end{enumerate}

The algorithm is located in file {\tt applications/CorrectEvent.h}.
An example of its use is provided in {\tt applications/\-hepmc-\-tau\-ola-redecay.cxx}.
By default, this algorithm is turned off.

\subsection{Package use cases}\label{Sec:tauspinner-uses}

This package can be used to validate several {\tt TauSpinner} options representing different applications of
{\tt TauSpinner}. Such tests include, but are not limited to:
\begin{itemize}
\item[(A)] {\bf Applying longitudinal spin effects:} adding spin effect to an unpolarized sample
           using weights {\tt WT} calculated by {\tt TauSpinner}. For this purpose, set {\tt Ipol=0} in the configuration file.
\item[(B)] {\bf Removing spin effects:} removing spin effects from the polarized sample
           using weights calculated by {\tt TauSpinner}. This is the default option used for our figures.
           The weight {\tt 1/WT} instead of {\tt WT} should be used. 
\end{itemize}
Note, that  regardless of whether {\tt Ipol=0} or {\tt 1}, {\tt TauSpinner} works 
in the same manner.
The two options are distinguished  at the level of the  user program 
 only (use $1/WT$ instead of $WT$ to reweight events), as shown in our demo.
\begin{itemize}
\item[(C)] {\bf Working on the input file with spin correlations but without polarization:}
           initialize {\tt TauSpinner} with {\tt Ipol=2}. In this case {\tt WT} will represent
           correction necessary for implementation of the full longitudinal spin effects.
           Analogously, if the sample feature  $\tau$ polarization, but polarization
           is missing dependence on the $\tau$ leptons directions, {\tt TauSpinner} should be initialized with {\tt Ipol=3}
           and the missing dependence can be corrected with calculated weight  {\tt WT}.
\item[(D)] {\bf Replacing spin effects of $Z/\gamma^*$   with the Higgs-like spin-0 state spin correlations}: 
           This could be realized with weights ($\frac{1}{wt_{spin}}$ to remove spin effects of $Z/\gamma^*$ times $wt_{spin}^\Phi$ to introduce spin effects of $\Phi$)
           without modification of event kinematics. Due to large spread in the weights,
           this method introduces large statistical fluctuations.  
           Alternatively, this can be realised by regenerating  $\tau$ decays with {\tt Tauola++} 
           configured for the $\tau$ pair originating from the scalar state resonance.
\item[(E)] {\bf Validation:} test is similar to test [A], we apply spin effect to
           a sample without polarization. However, for this test we take the polarized sample and replace its $\tau$ decays
           by new, non-polarized ones using {\tt Tauola++}. \\
           This allows to test different {\tt Ipol} options as mentioned in test (C).
           It requires different setup and use of two input files. 
           The {\tt TauSpinner} should be executed on this new sample.
           The result should be then  compared with the ones from {\it original} sample.
           Tools required to perform these steps are described in Section~\ref{Sec:Installation-tools}.
\end{itemize}

The results of the tests B and D are presented in the Appendices of this paper.
The details of test C are described in {\tt README} of {\tt applications/test-ipol} subdirectory.
Tools, that can be adapted to perform tests A and E, have been provided as well (see Section~\ref{Sec:Installation-tools}).

We have successfully performed tests A-E on samples generated with {\tt Pythia8 + Photos++ + Tauola++} (in some cases  {\tt Pythia8 } alone). 
Satisfactory results, of similar quality as discussed in our paper, sections \ref{subsec:cuts} and \ref{Sec:consistency-check} were always found.
Further details for all of the cases listed above are given in the distribution tar-ball.

\newpage
\section{Summary} \label{Sec:Summary}

In this paper we presented the use of {\tt TauSpinner} libraries for testing effects resulting from spin correlations 
and polarization of $\tau$ leptons in processes at LHC featuring $W$, $Z$ and $H $ decays.
New example  programs  were developed and incorporated into program distribution tar-ball. 
The purpose of
these  programs is to analyze spin effects using information on the kinematics of  $\tau$ decay 
products of events  
stored in a file. Moreover, they provide a convenient tool for  
validating the {\tt TauSpinner} algorithms. As an important use case, this set of programs
provides a method to evaluate systematic error on spin effect
implementation in so called {\it embedded $\tau$} samples, an experimental technique used for 
analyses at LHC experiments.

For the purpose of presenting methodology, a set of kinematical distributions was selected and the physics properties  
of these distributions were  explained on some example plots. 
Event samples featuring $\tau$ lepton decays  of $W$ and $Z$ production  at LHC energies were generated.
The weights calculated by {\tt TauSpinner} algorithms were  used then to remove the effects due to polarization of decaying $\tau$.
The sample featuring no spin effects was also created on flight for comparisons.
For studying the spin effects of the spin-0 intermediate state $\Phi$, the $Z \to \tau \tau $ sample was modified, namely 
the $\tau$ leptons decay products were removed and the decays were generated again, using spin density matrix of $H \to \tau^+ \tau^- $ decay.
The complete set of benchmark plots from analyses of these samples, graphical output from 
our program,  is collected in the Appendices of the preprint version of our paper.
What is shown in Appendix~\ref{sec:OutputTest}  are plots  for $W$ and $Z$ 
decays, from executing our  program on a single event file. With the additional run,
one can prepare a set of plots shown in Appendix~\ref{sec:OutputH}, for the case when instead of spin effects from intermediate $Z/\gamma^*$ 
the  Higgs couplings were used for the preparation of events file. 
As expected, effects of removed polarisation are present and 
spin correlations are of opposite sign 
to that of the $Z/\gamma^*$ case.

The details of the program installation and use were  given. Our example provides test that
algorithms of {\tt TauSpinner} used for calculating spin weights are equivalent to the ones in {\tt Tauola++},
the $\tau$ decay library used for creation of initial samples.
 We have also demonstrated  how {\it physics history entries} of event samples can be used
to provide  validation tests for  algorithm of effective Born level kinematic reconstruction and cross-section calculations used in {\tt TauSpinner}\footnote{In our paper this method
was used for comparisons with
{\tt Pythia8 } results, which are of the  leading logarithm precision 
level similarly as  {\tt TauSpinner}. Further extensions of this
method is possible. Properties of factorization of exact QCD multiparton 
amplitudes need then to be used in the {\it physics history entries} of
events  stored by the reference generators. As one can see   
 \cite{vanHameren:2008dy}, such properties are present for  QCD amplitudes, 
as it 
was the case of multiphoton amplitudes of QED which were used for 
$\tau$ lepton spin effects in KKMC Monte Carlo \cite{Jadach:1999vf}
of LEP applications.}. 


The easiest case to understand are the spin effects  of $\tau^\pm \to \pi^\pm \nu_\tau$ decays. The spectra
are affected by spin as discussed in ref. \cite{jadach-was:1984,Pierzchala:2001gc}, that is why we have frequently used this decay channel for the
example plots  of the paper.
Due to $Z$ polarization  there is clearly identifiable slope for the  $\pi^\pm$'s energy spectrum.
The spin correlations of the two $\tau$'s disfavour configurations when one of the $\pi$ is hard and the other one is soft.
For the  $\Phi \to \tau \tau$ case the spin correlations effect is opposite.

For other $\tau$ decay channels, effects of spin are more complex and we have presented results in the main body of
the paper only for the $\tau$'s of polarization originating from $Z \to \tau \tau$ decays. 
Pattern of spin correlations and single $\tau$ polarization effect depends on the $\tau$ decay channel.
One can easily notice from the lego plots that it will be affected by the kinematical selection on the other $\tau$ decay products
as well, biasing the  observed $\tau$ polarization.

One should keep in mind when performing above tests that features of presented distributions depend strongly on the analyzed event sample. 
If  $\tau$ leptons predominantly originate
from the decays of $Z$, $W$, or $H$, of virtualities close to the resonance peaks,
and with the spin effects taken into account, the distributions should be similar to the
ones presented in this paper. However, it might not always be the case. For results presented in this 
paper  parameters of the electroweak interactions and the PDFs used were carefully 
tuned between events generation and TauSpinner analysis codes. If this is not the case 
particular patterns of discrepancies may appear. We will investigate this point in the future.

To evaluate sensitivity to the spin the average $\tau$ lepton polarization fits to the histograms of the simple analytic
distributions are provided. 
The effects of QED bremsstrahlung or mass corrections are not  incorporated into the functions  used for fits in case 
of leptonic $\tau$ decay.
At LEP  \cite{Eberhard:1989ve}, it was shown
that they may be of the similar size and shapes as polarization effects.
We have also discussed that the alternative to analytical formule, Monte Carlo based template distributions are useful for fits
and evaluation of spin effects.

From the discussion presented in this paper we left aside discussion on functionality of matrix-element re-weighting
like the one described in ref.~\cite{Banerjee:2012ez} and recently being upgraded even further. 
This approach may be helpful to evaluate if spin effects
present in a given sample can be helpful to distinguish different production mechanisms, ones combined with the effects
of production distributions.
In our present study, we however concentrated on discussing spin effects only.



\section*{Acknowledgments}
We thank Prof. Erez Etzion for inspiring comment that bremsstrahlung effects
 given in semi analytical 
form (as in the LEP time {\tt CALASY} program) may be of interest  for the LHC 
applications 
as well. We thank Dr. Ian Nugent for careful and critical reading of our manuscript and for discussions. We thank Dr. Will Davey for discussions as well. 

\providecommand{\href}[2]{#2}
\newpage
\appendix

\newcommand{\przedro}{}
\newcommand{\greenlineis}{is for modified sample after removing polarisation using {\tt TauSpinner} weights, }

\include{appendixA}
\include{appendixB}

\end{document}

%% file: appendixA.tex
\setcounter{figure}{0}

\renewcommand\thefigure{\thesection.\arabic{figure}}

\section{Benchmark results}\label{sec:OutputTest} 

The Appendix A with its subsections represents output%
\footnote{ Text is adopted for the sake of paper preparation in 
a minor way only.}
from a single execution of {\tt applications-plots.exe} program.
The numerical results of the Appendix  are obtained  
from the  events file  generated 
 by run of {\tt Pythia8} combined  with {\tt Tauola++}, details of initialization  
are given later in the Appendix. 
For the fits,  all 100 bins except 
the first five in $\tau \to l \nu_l \nu_\tau $ case  and first one in $\tau \to \pi \nu$ case, were used\footnotemark[18].
\footnotetext[18]{If bremsstrahlung in $\tau$ decays would be present, the result of the fits would differ.
For example, for leptonic channel (see Fig.3), the shift of $\sim$0.07 would be present.}

The figures in the main part of our 
paper are taken from the ones of Appendices, with somewhat improved graphic style 
for better readability. 
For the plots of Fig.~1,  the bottom-left plots 
 Figs.~\ref{Fig:spectra1},~\ref{Fig:spectra2}~and~\ref{Fig:spectra3}, were selected.
The lego plot (and visible mass plots) of Fig.~2 are shown as the top-left lego plots (and mid-right plots)
of Figs.~\ref{Fig:spectra2}~and~\ref{Fig:scattergrams2} respectively.


\subsection{Input files} \label{Sec:InputFiles}
The list of files and additional information on generation of events 
used for the plots: 
{\small \verbinput{input-file-info.txt} }

\vspace{1\baselineskip}

\noindent
Configuration file used by the program:

\vspace{1\baselineskip}

{\small \verbinput{input-config-file.txt} }

\noindent
Configuration file used by the fitting program:

\vspace{1\baselineskip}

{\small \verbinput{fit-config-file.txt} }

\newpage
\subsection{W decays}\label{Sec:W}
The invariant mass distribution and break-down on the $\tau$ decay channels are shown
for $\tau \nu_\tau$-pair originating from $W$ decay. The spin effects should not depend on the virtuality of the
$W$ intermediate state, but this may be not the case if New Physics samples are studied.
\vskip 3 mm

\begin{figure}[h!]
\centering 
\resizebox*{0.45\textwidth}{!}{\includegraphics{\przedro 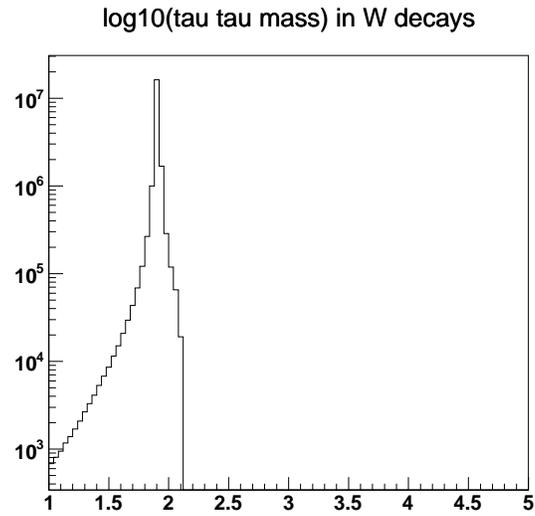}} \\
\caption{ Invariant mass distribution of $\tau \nu_\tau$-pair originating from $W$ decay.}
\end{figure}

{\small \verbinput{input-W-event-count.txt} }
\newpage

\subsubsection{ The energy spectrum: $\tau^\pm \to \mu^\pm, e^\pm$}
\vspace{1\baselineskip}

\begin{figure}[h!]
\centering 
\resizebox*{0.49\textwidth}{!}{\includegraphics{\przedro 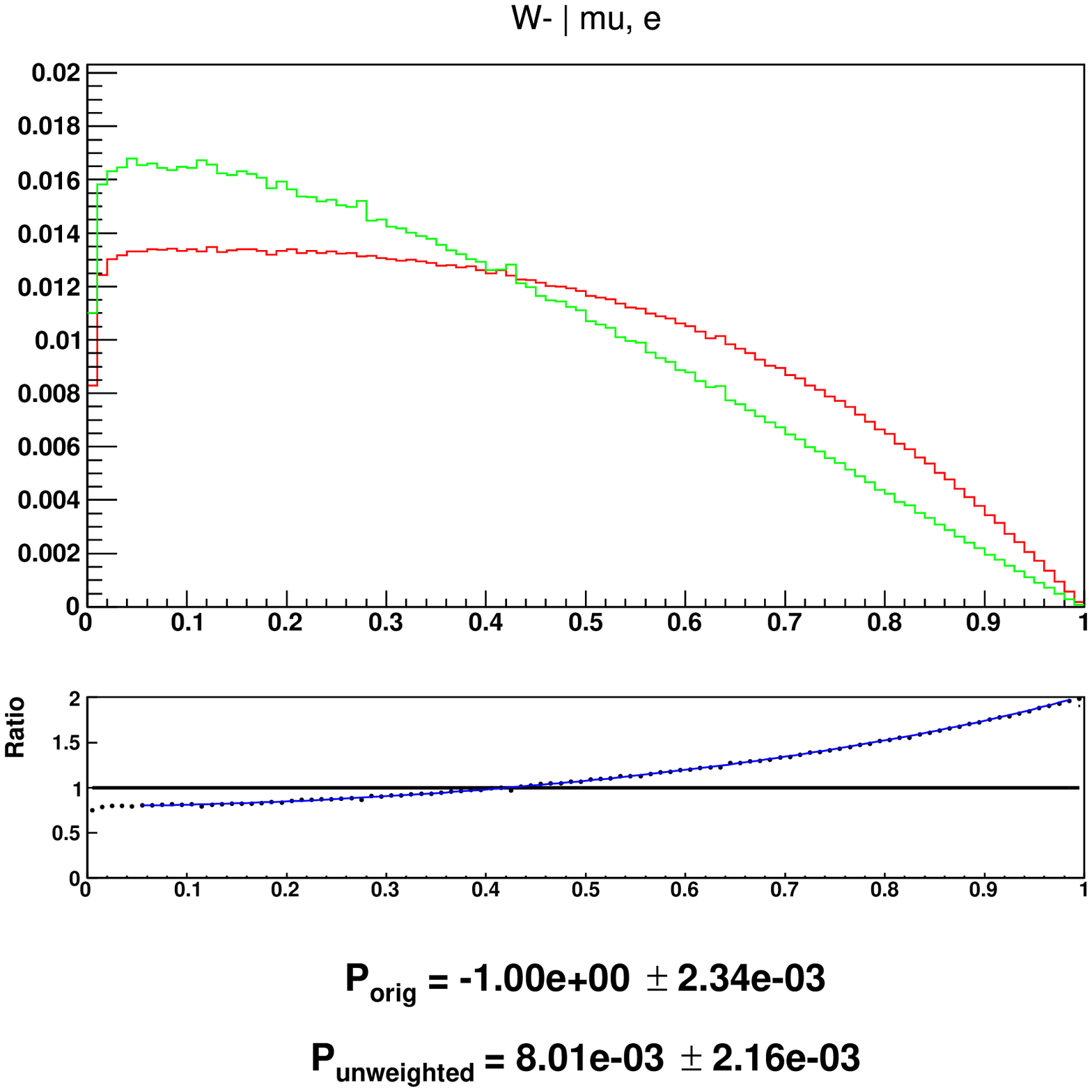}}
\resizebox*{0.49\textwidth}{!}{\includegraphics{\przedro 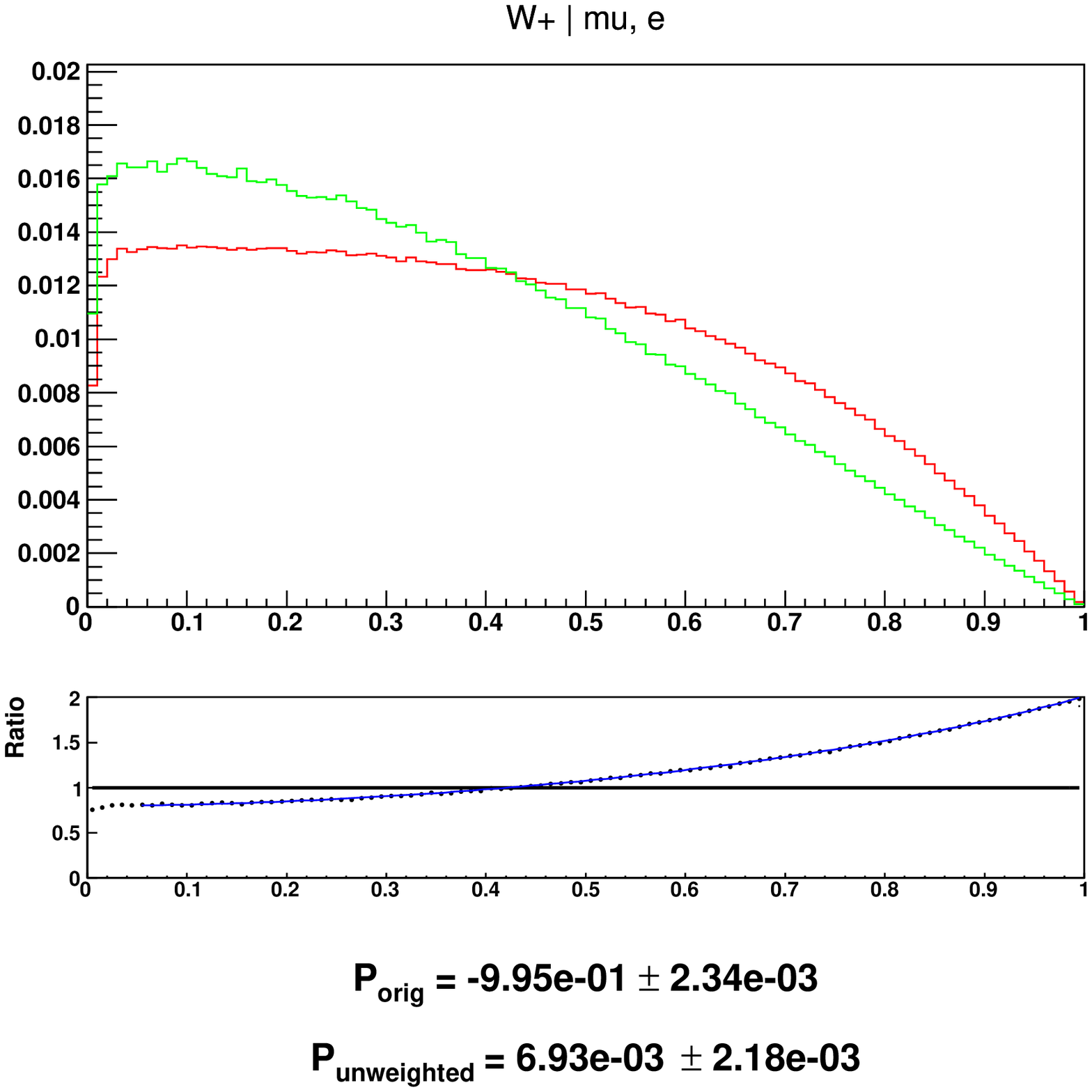}} \\
\caption{ Fraction of $\tau$ energy carried by its visible  decay products%
$^{18}$. 
\textcolor{red}{Red line} is for original sample,
\textcolor{green}{green line} \greenlineis
black line is ratio \textcolor{red}{original}/\textcolor{green}{modified}.
}
\end{figure}

\subsubsection{ The energy spectrum: $\tau^\pm \to \pi^\pm$}
\vspace{1\baselineskip}

\begin{figure}[h!]
\centering 
\resizebox*{0.49\textwidth}{!}{\includegraphics{\przedro 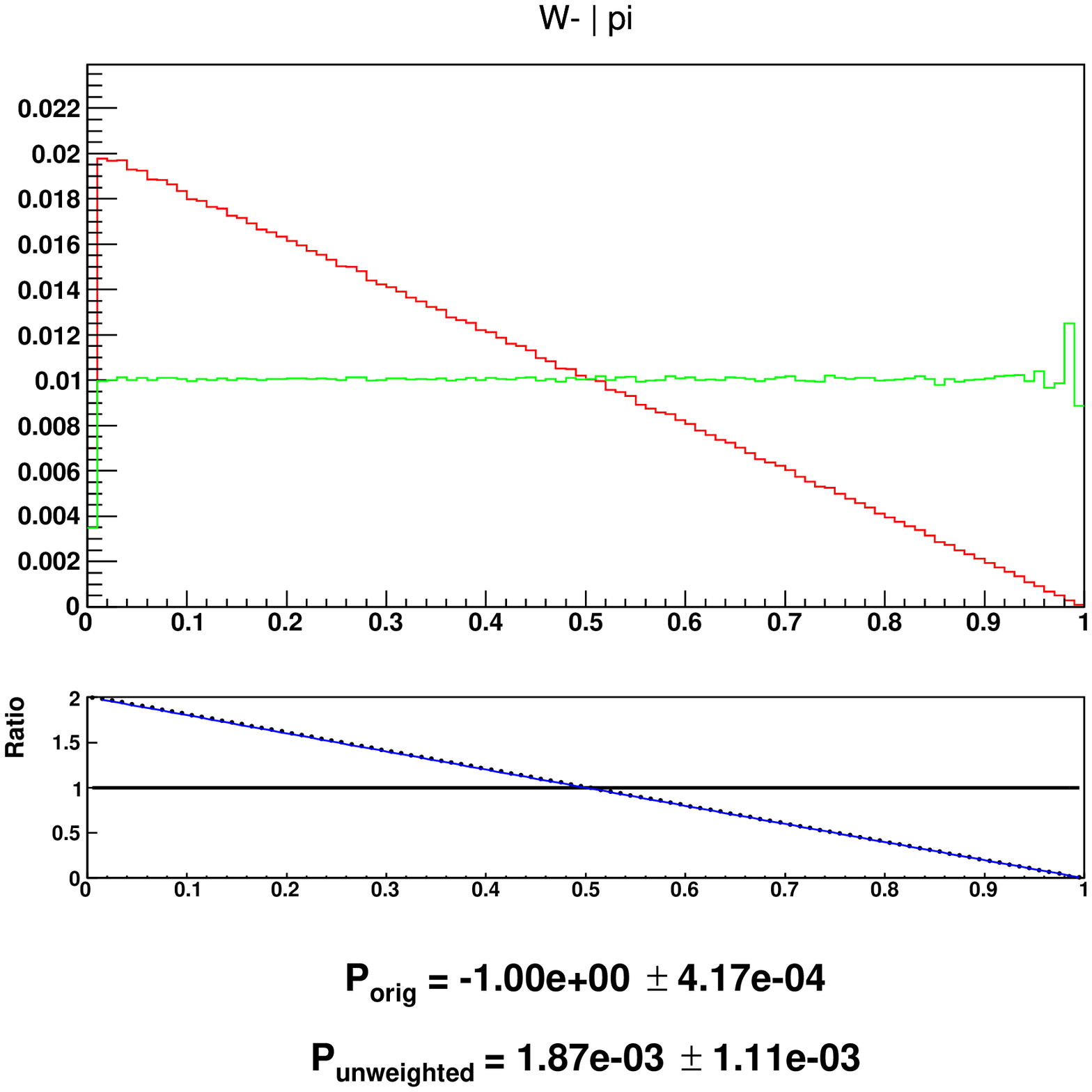}}
\resizebox*{0.49\textwidth}{!}{\includegraphics{\przedro 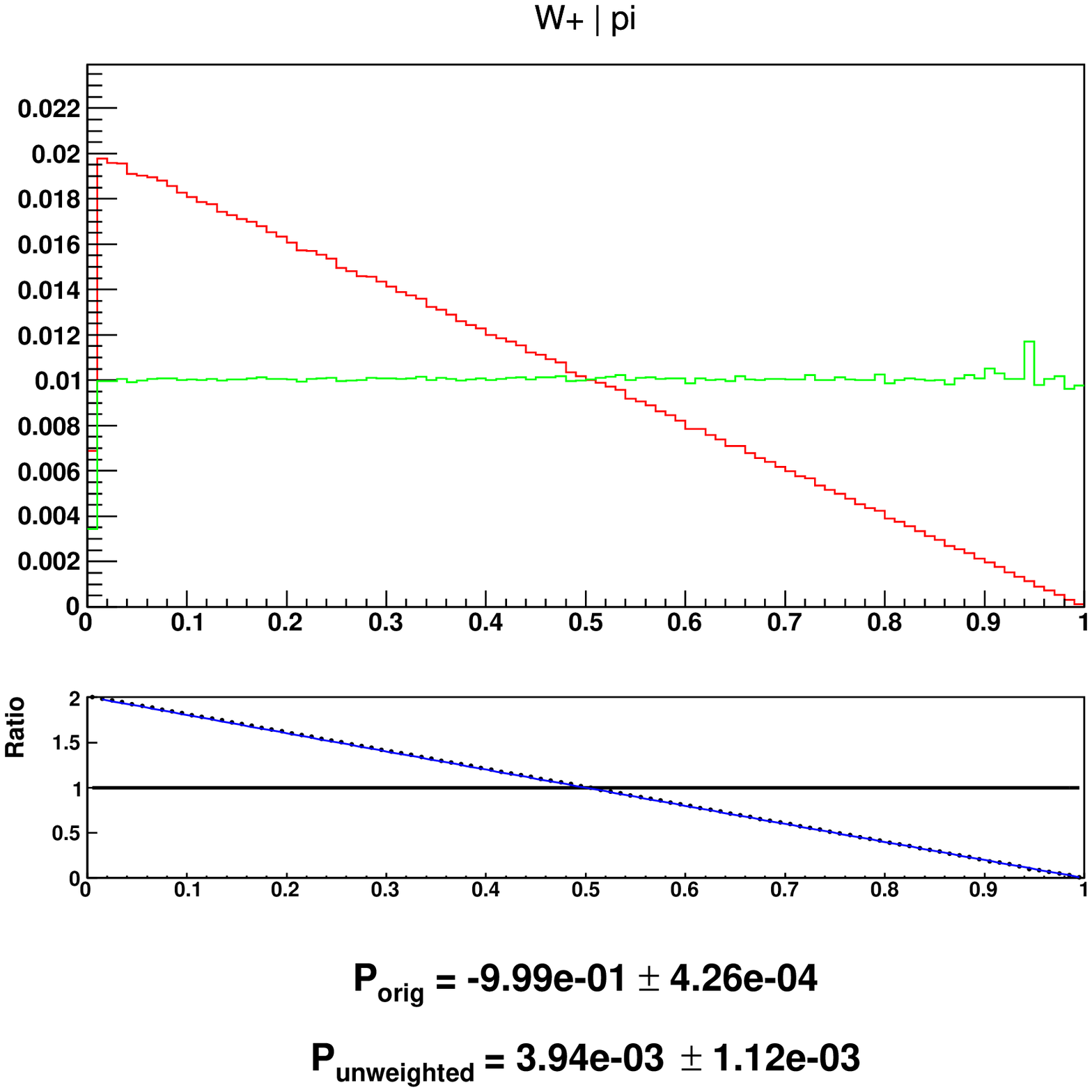}} \\
\caption{  Fraction of $\tau$ energy carried by its visible  decay products$^{18}$.
\textcolor{red}{Red line} is for original sample,
\textcolor{green}{green line} \greenlineis
black line is ratio \textcolor{red}{original}/\textcolor{green}{modified}.
}
\end{figure}

\newpage
\subsubsection{ The energy spectrum: $\tau^\pm \to \rho^\pm$}
\vspace{1\baselineskip}

\begin{figure}[h!]
\centering 
\resizebox*{0.49\textwidth}{!}{\includegraphics{\przedro 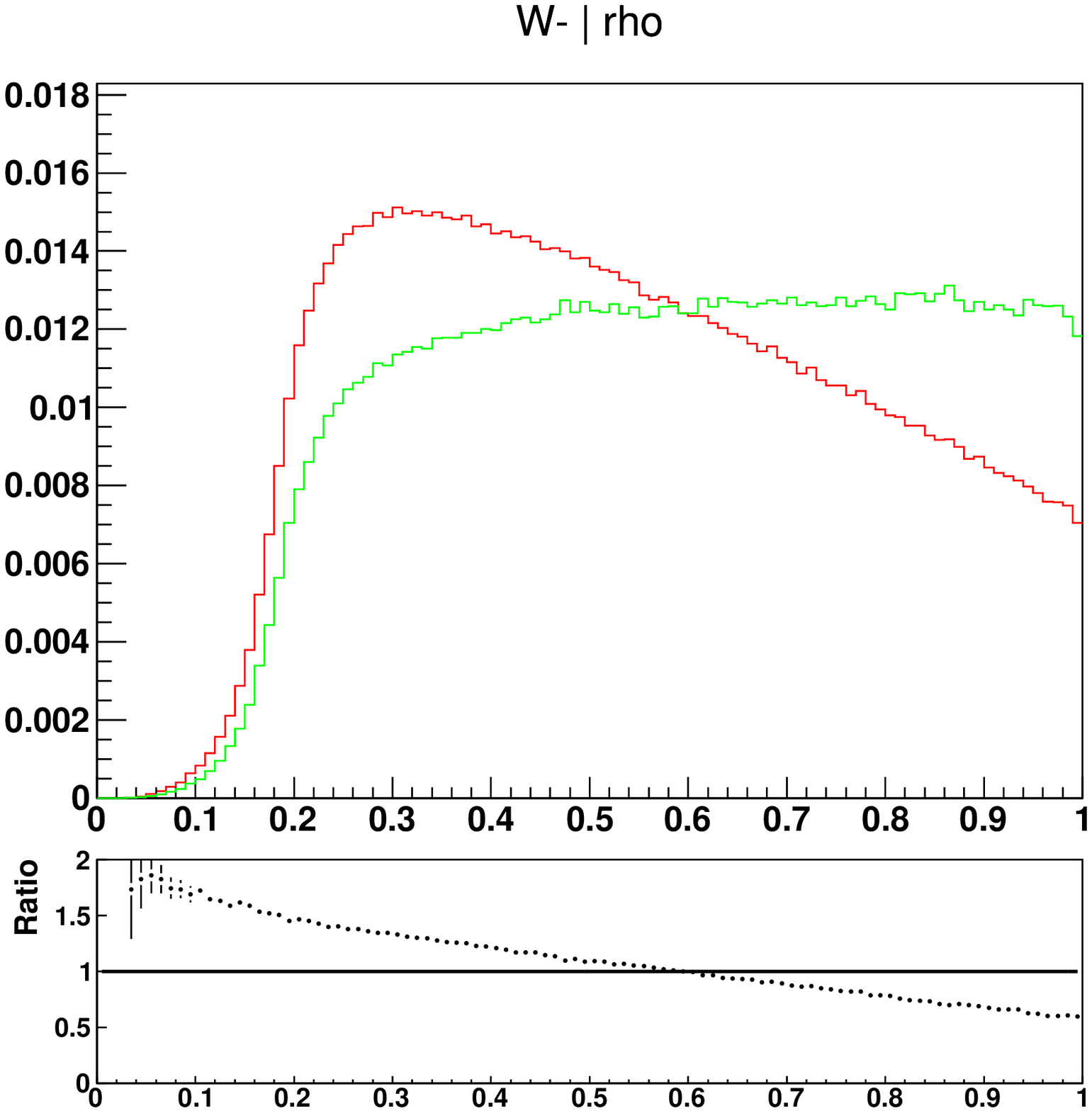}}
\resizebox*{0.49\textwidth}{!}{\includegraphics{\przedro 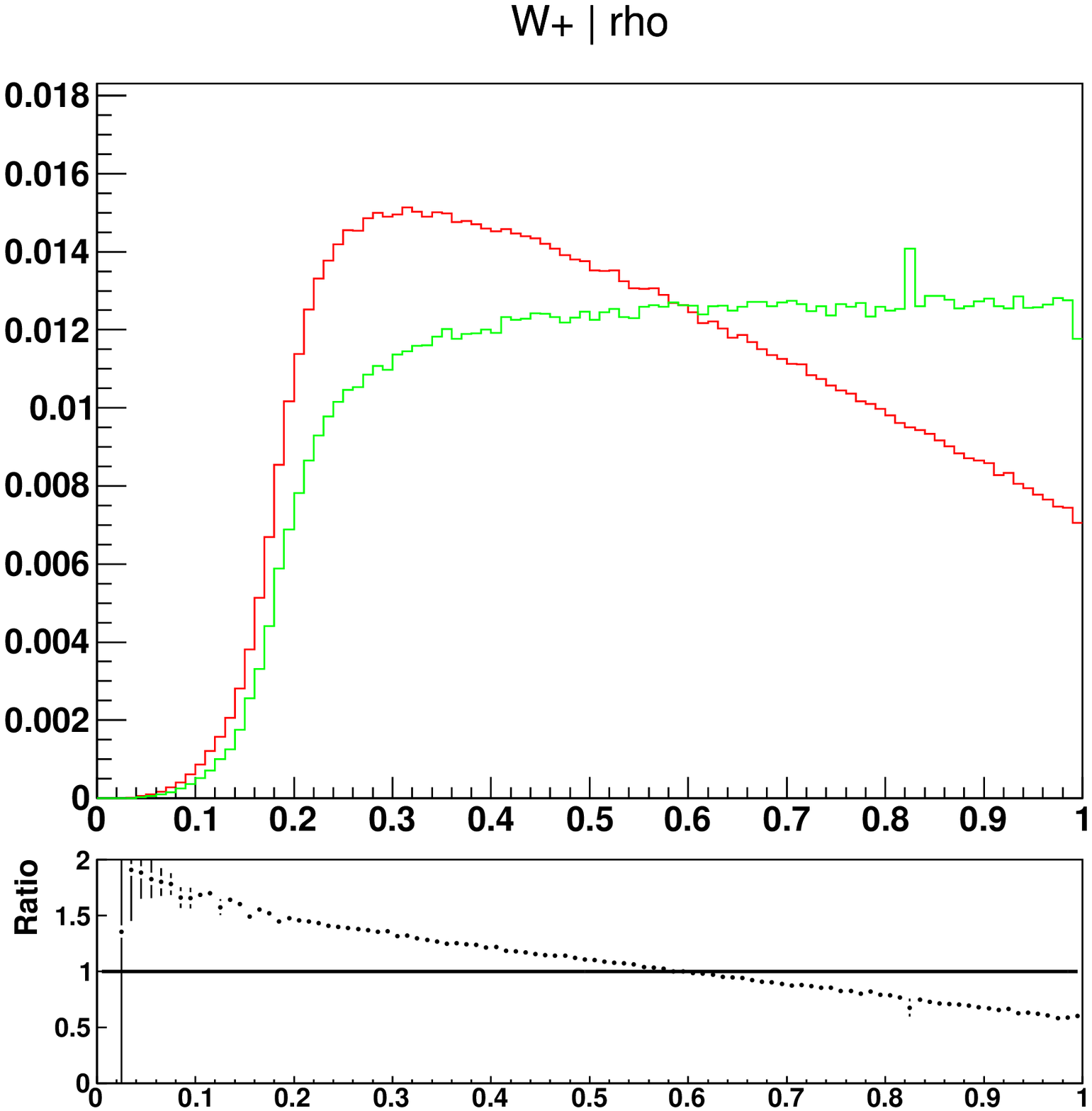}} \\
\caption{Fraction of $\tau$ energy carried by its visible  decay products.
\textcolor{red}{Red line} is  for original sample,
\textcolor{green}{green line} \greenlineis
black line is ratio \textcolor{red}{original}/\textcolor{green}{modified}.
}
\end{figure}

\newpage

\subsection{Z decays}\label{Sec:Z}

The invariant mass distribution and break-down on the $\tau$ decay channels are shown
for $\tau \tau$-pair originating from $Z$ decay.
The spin effects strongly depend on the virtuality of the
$Z/\gamma^*$ intermediate state. Events were generated explicitly requiring virtuality
of $Z\gamma^*$ within 88-94 GeV window. Minor 
contamination from some other process is nevertheless observed (we have not traced it back).
\vskip 3 mm

\begin{figure}[h!]
\centering 
\resizebox*{0.45\textwidth}{!}{\includegraphics{\przedro 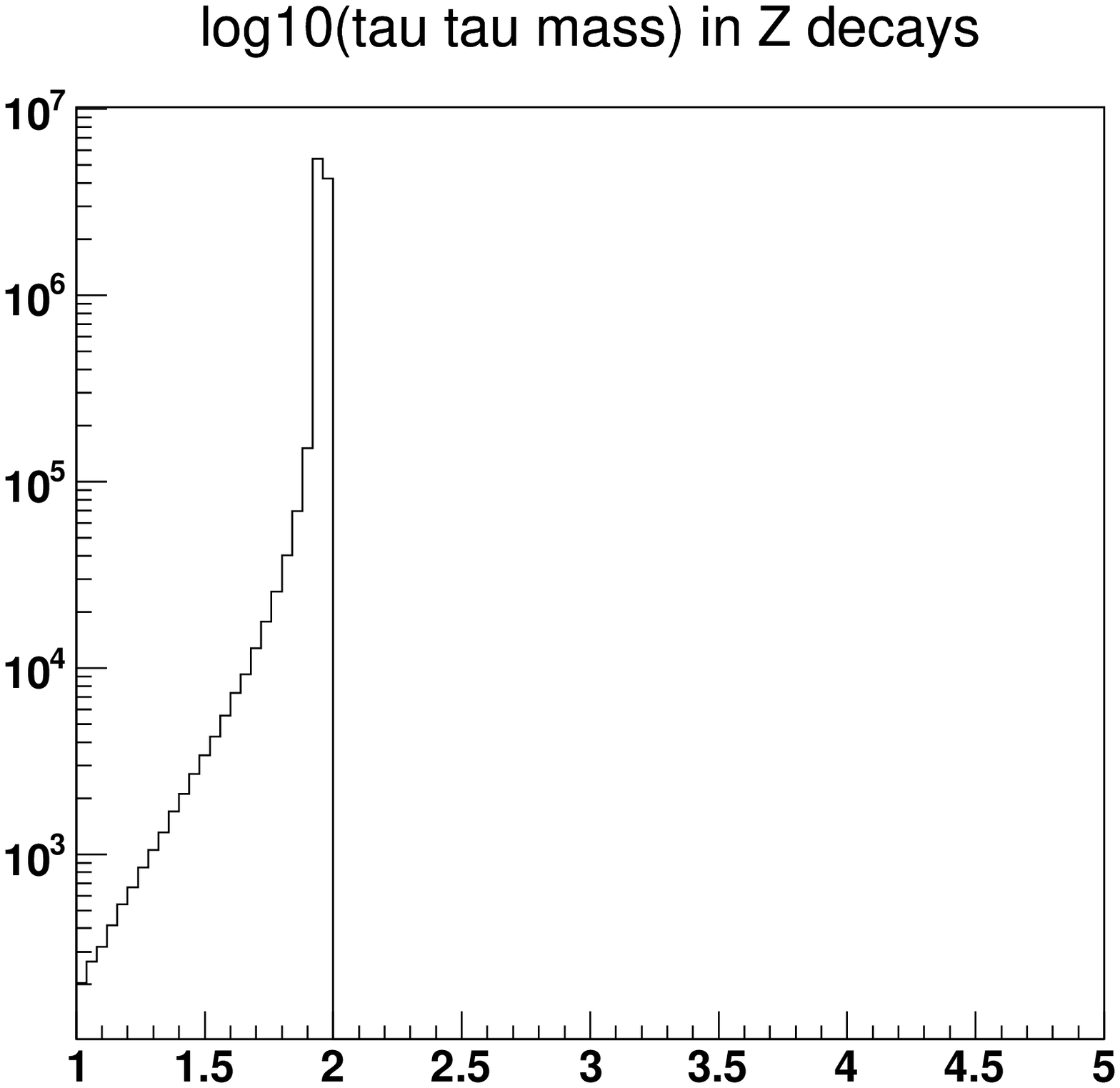}} \\
\end{figure}

{\small \verbinput{input-Z-event-count.txt} }

\newpage
\subsubsection{The energy spectrum: $\tau^- \to \mu^-, e^-$ {\tt vs } $\tau^+ \to \mu^+, e^+$}
\vspace{3\baselineskip}

\begin{figure}[h!]
\centering
\resizebox*{0.49\textwidth}{!}{\includegraphics{\przedro 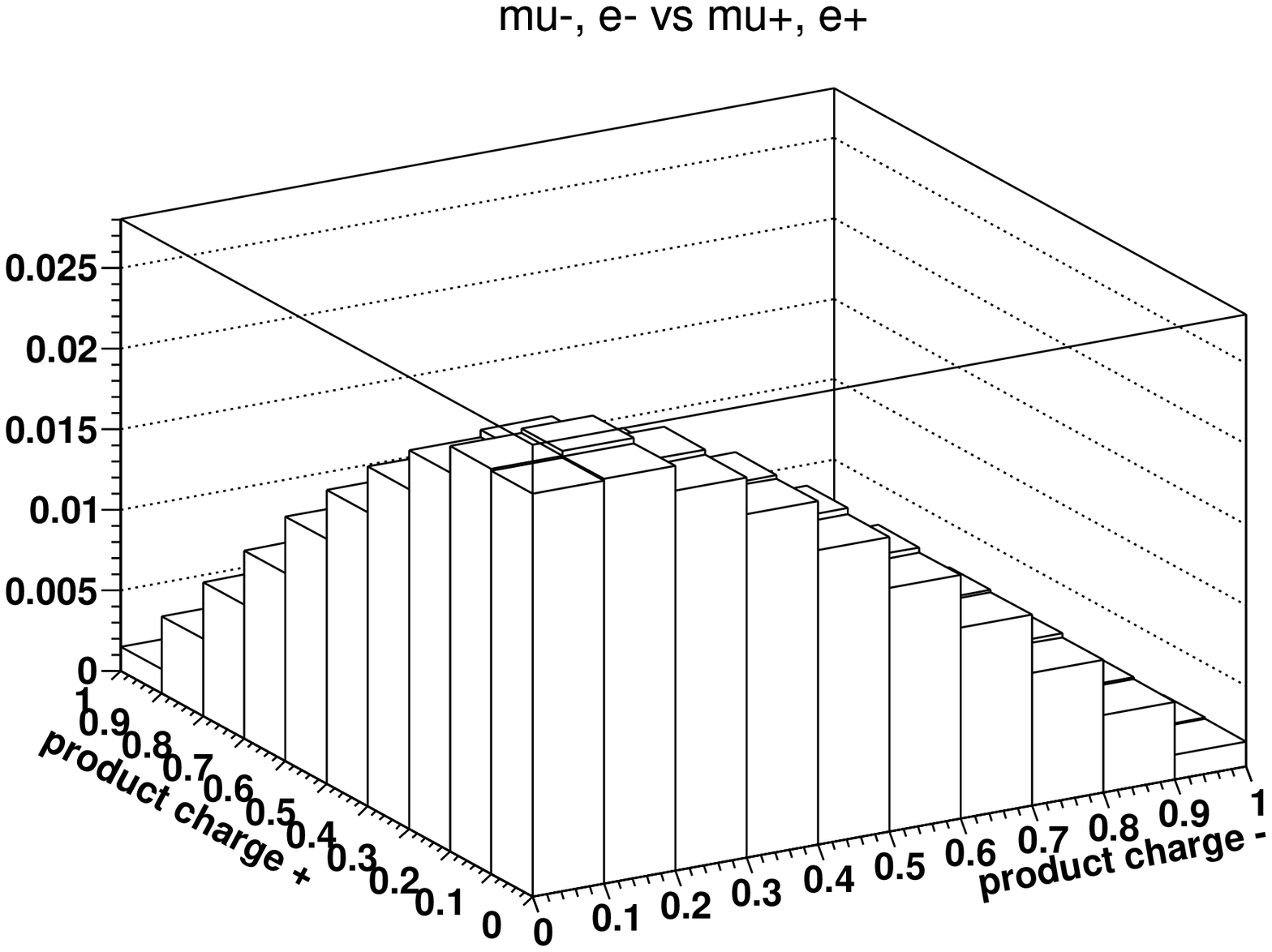}}
\resizebox*{0.49\textwidth}{!}{\includegraphics{\przedro 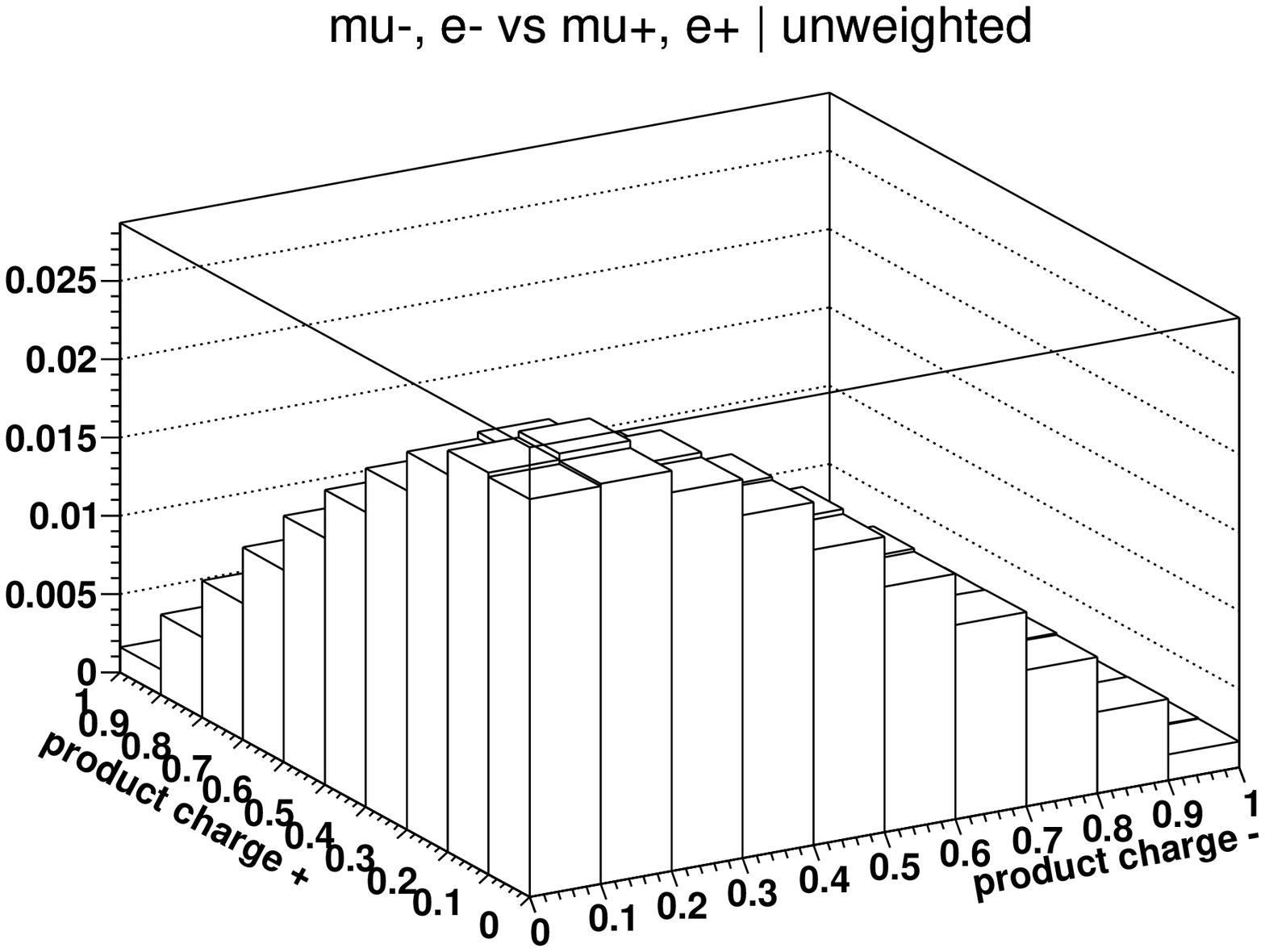}} \\
\resizebox*{0.49\textwidth}{!}{\includegraphics{\przedro 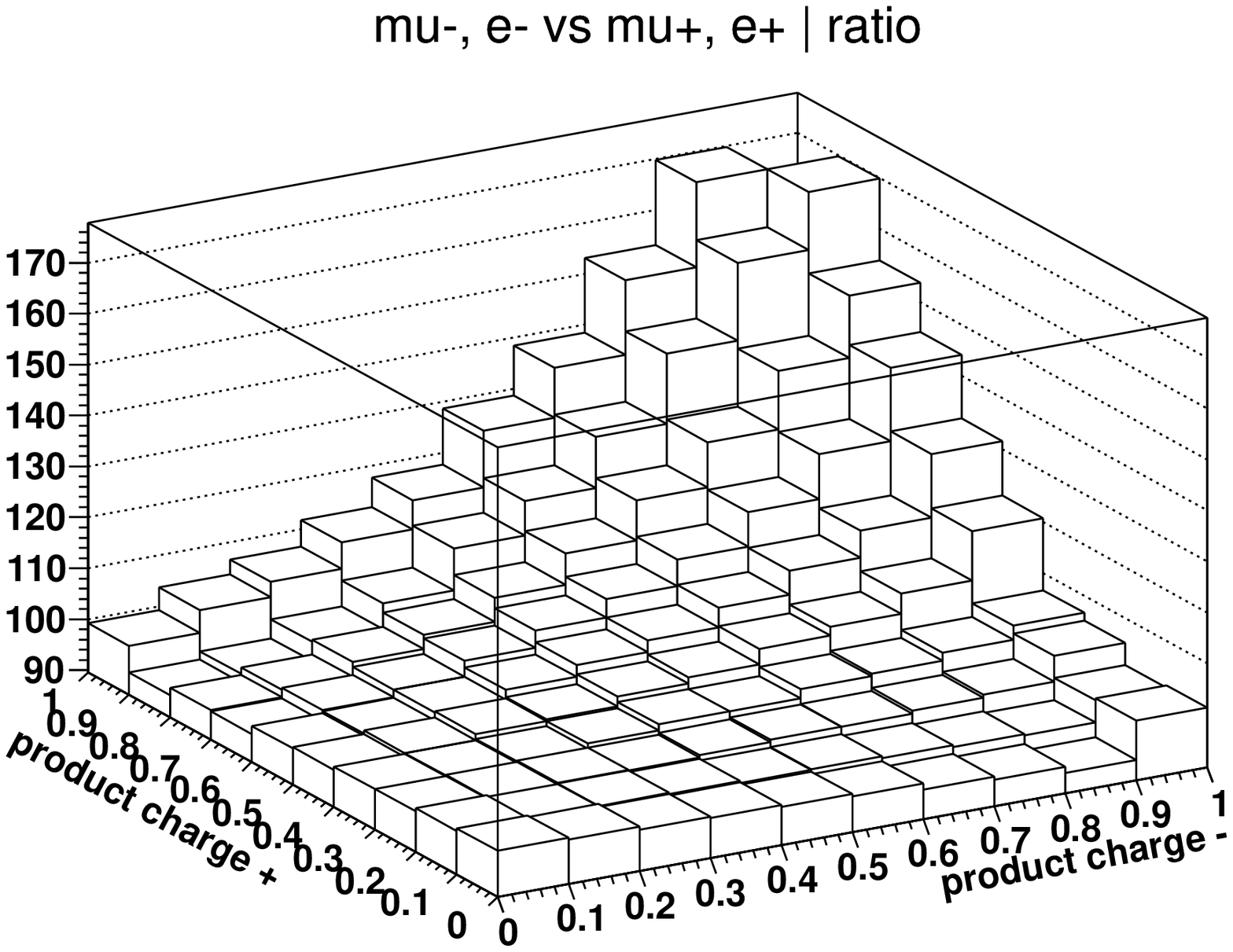}}
\resizebox*{0.49\textwidth}{!}{\includegraphics{\przedro 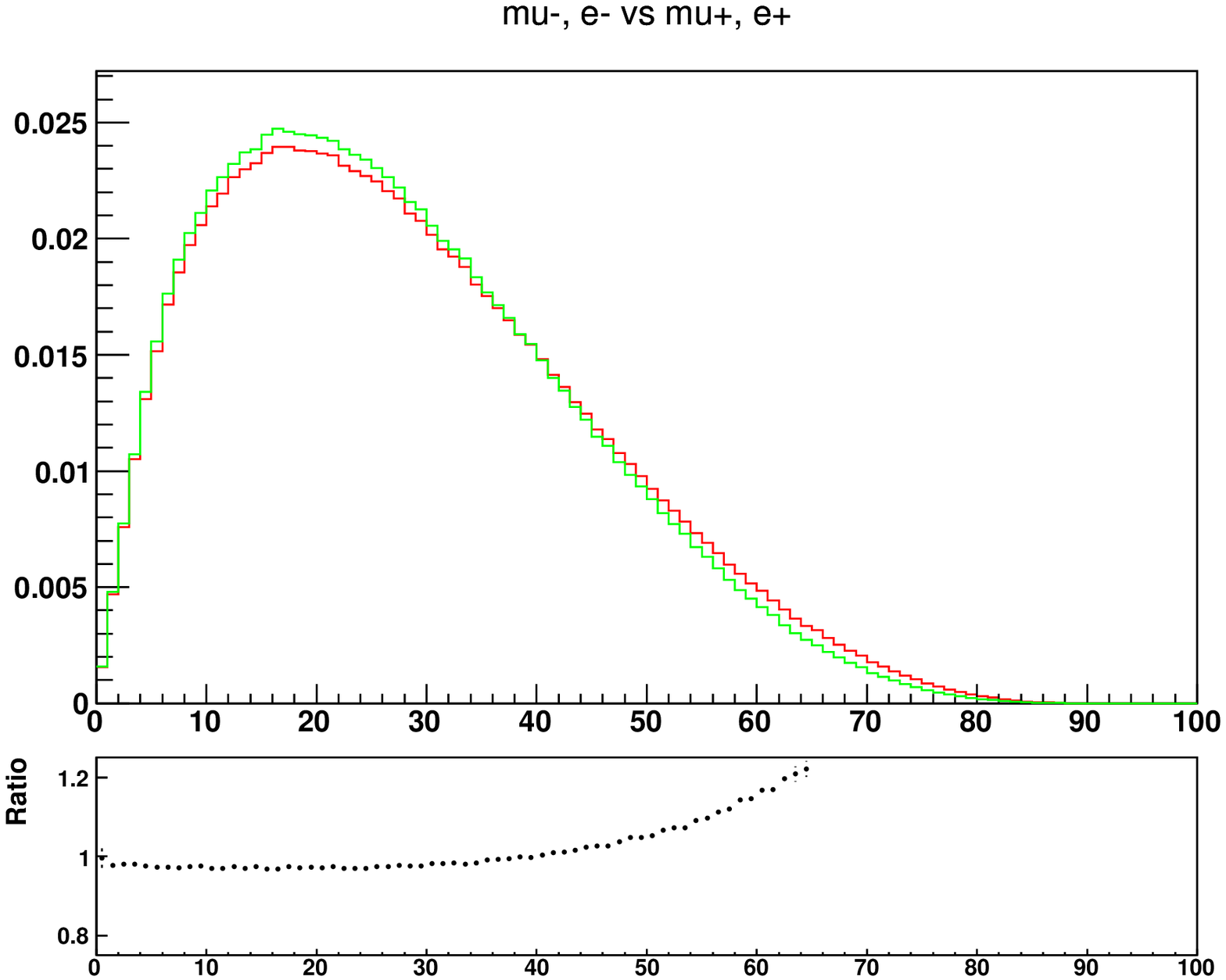}} \\
\resizebox*{0.49\textwidth}{!}{\includegraphics{\przedro 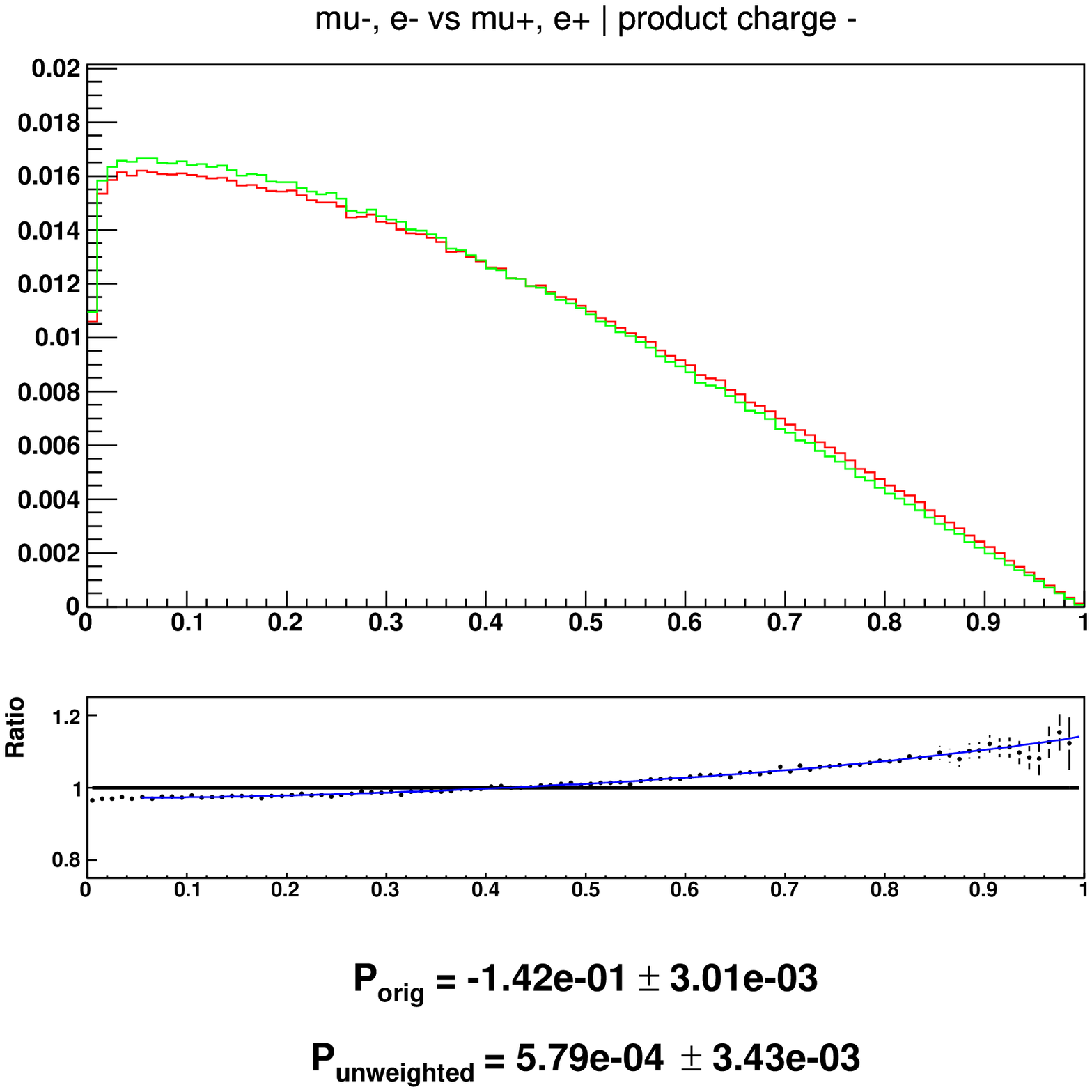}}
\resizebox*{0.49\textwidth}{!}{\includegraphics{\przedro 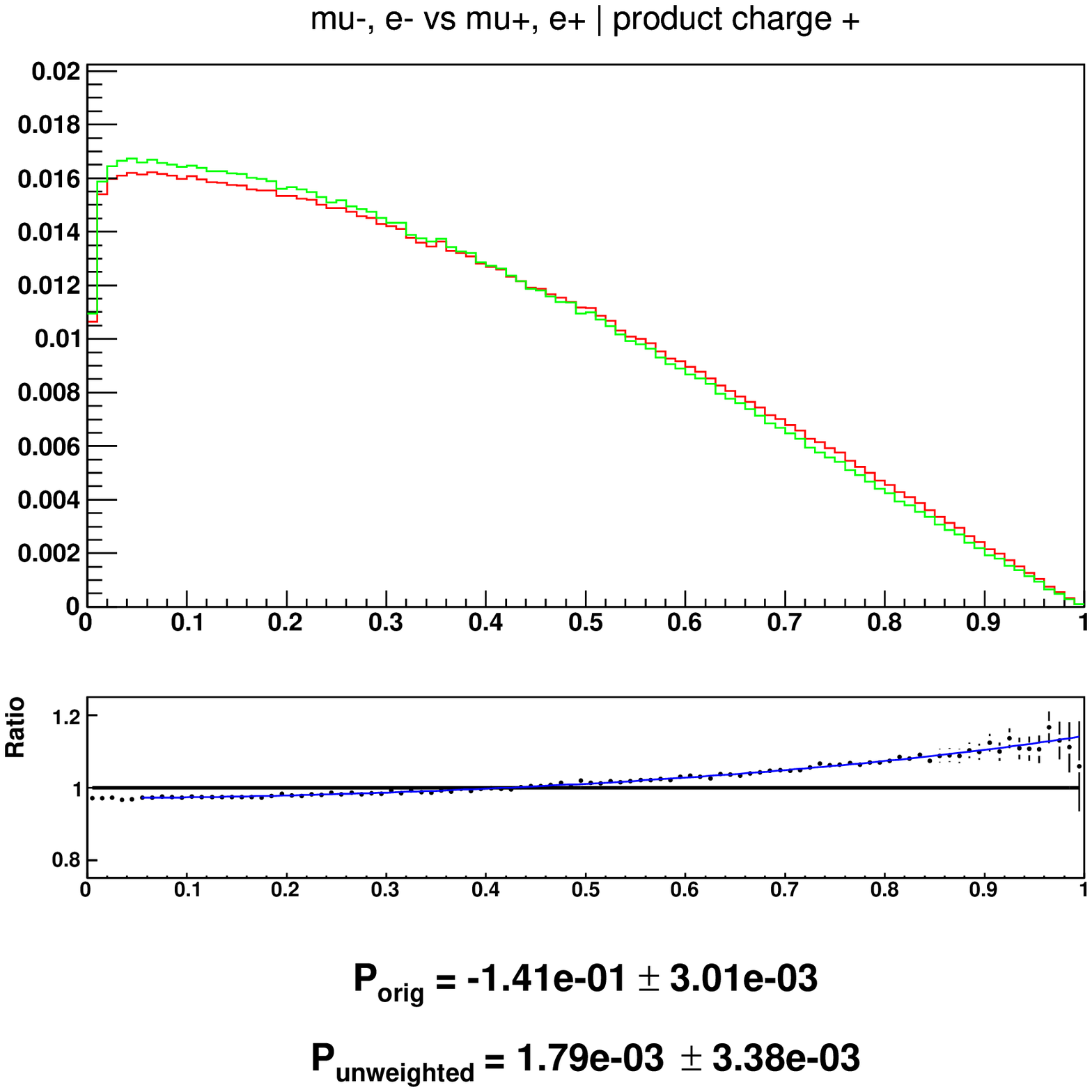}} \\
\caption{ \small Fractions of  $\tau^+$ and $\tau^-$ energies carried by their visible  decay products:
two dimensional lego plots and one dimensional spectra$^{18}$.
\textcolor{red}{Red line} is  for original sample,
\textcolor{green}{green line} \greenlineis
black line is ratio \textcolor{red}{original}/\textcolor{green}{modified} with whenever available superimposed result for the
fitted functions.
}\label{Fig:spectra1}
\end{figure}

\newpage
\subsubsection{The energy spectrum: $\tau^- \to \mu^-, e^-$ {\tt vs } $\tau^+ \to \pi^+$}
\vspace{3\baselineskip}

\begin{figure}[h!]
\centering
\resizebox*{0.49\textwidth}{!}{\includegraphics{\przedro 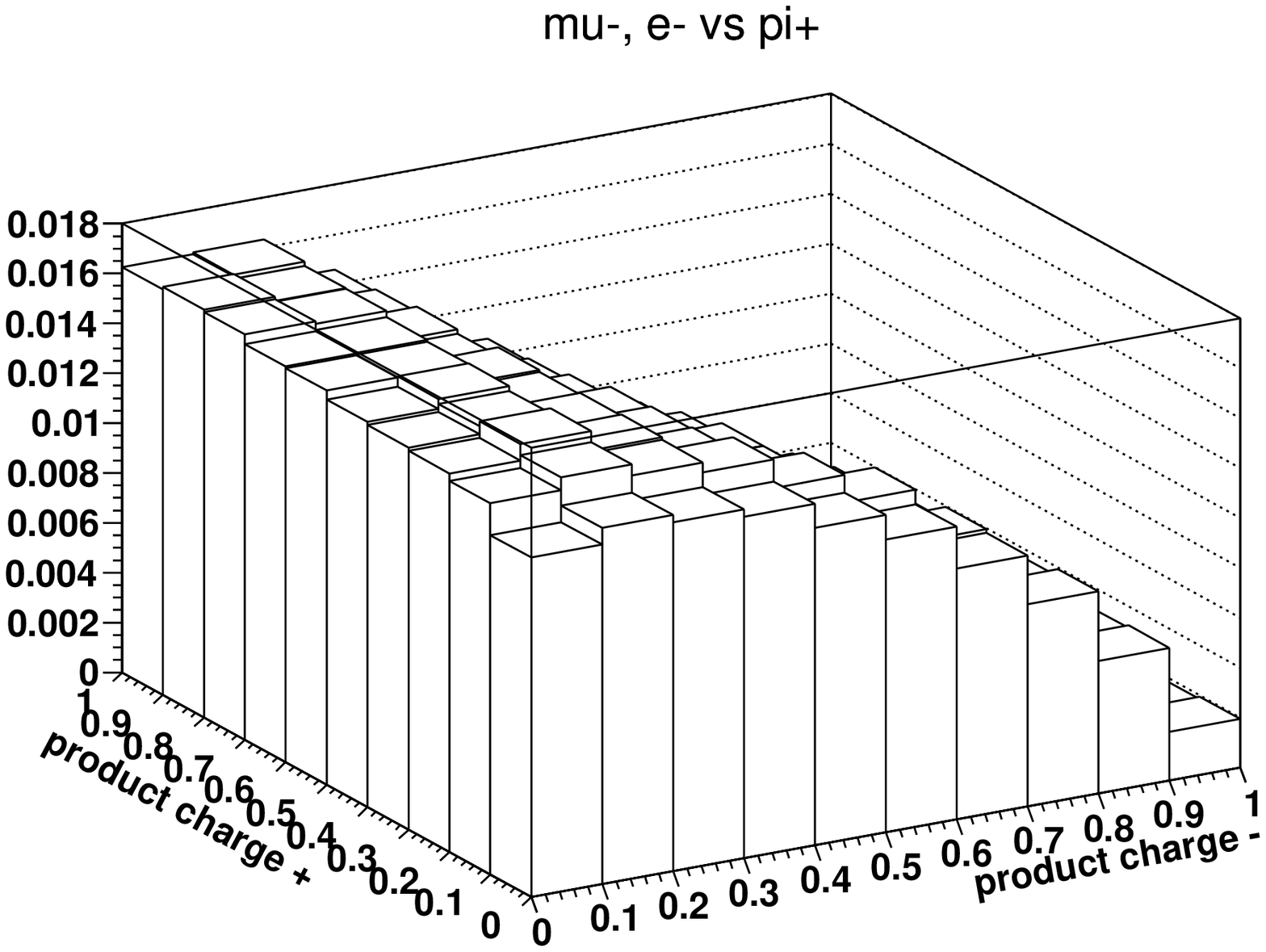}}
\resizebox*{0.49\textwidth}{!}{\includegraphics{\przedro 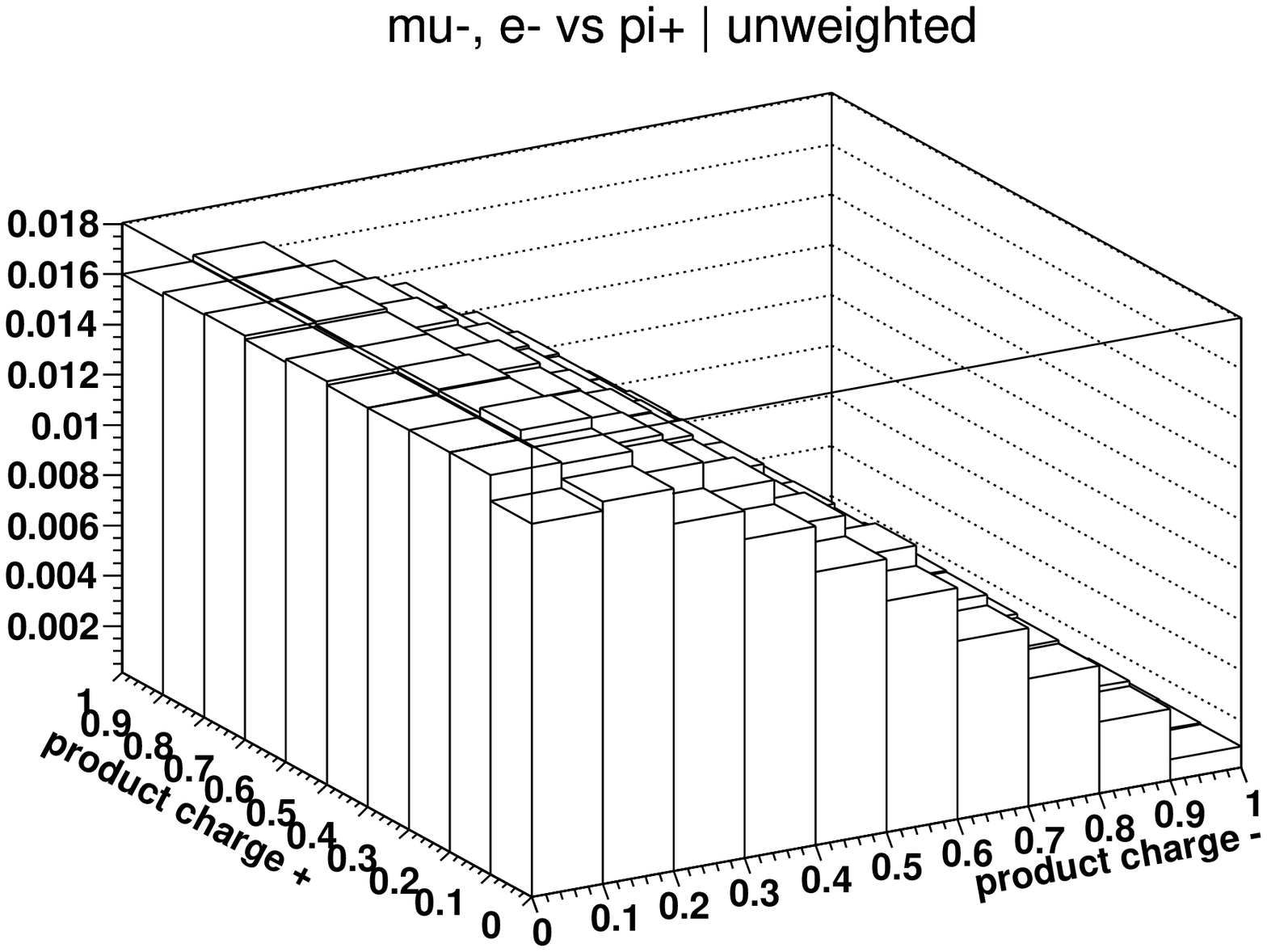}} \\
\resizebox*{0.49\textwidth}{!}{\includegraphics{\przedro 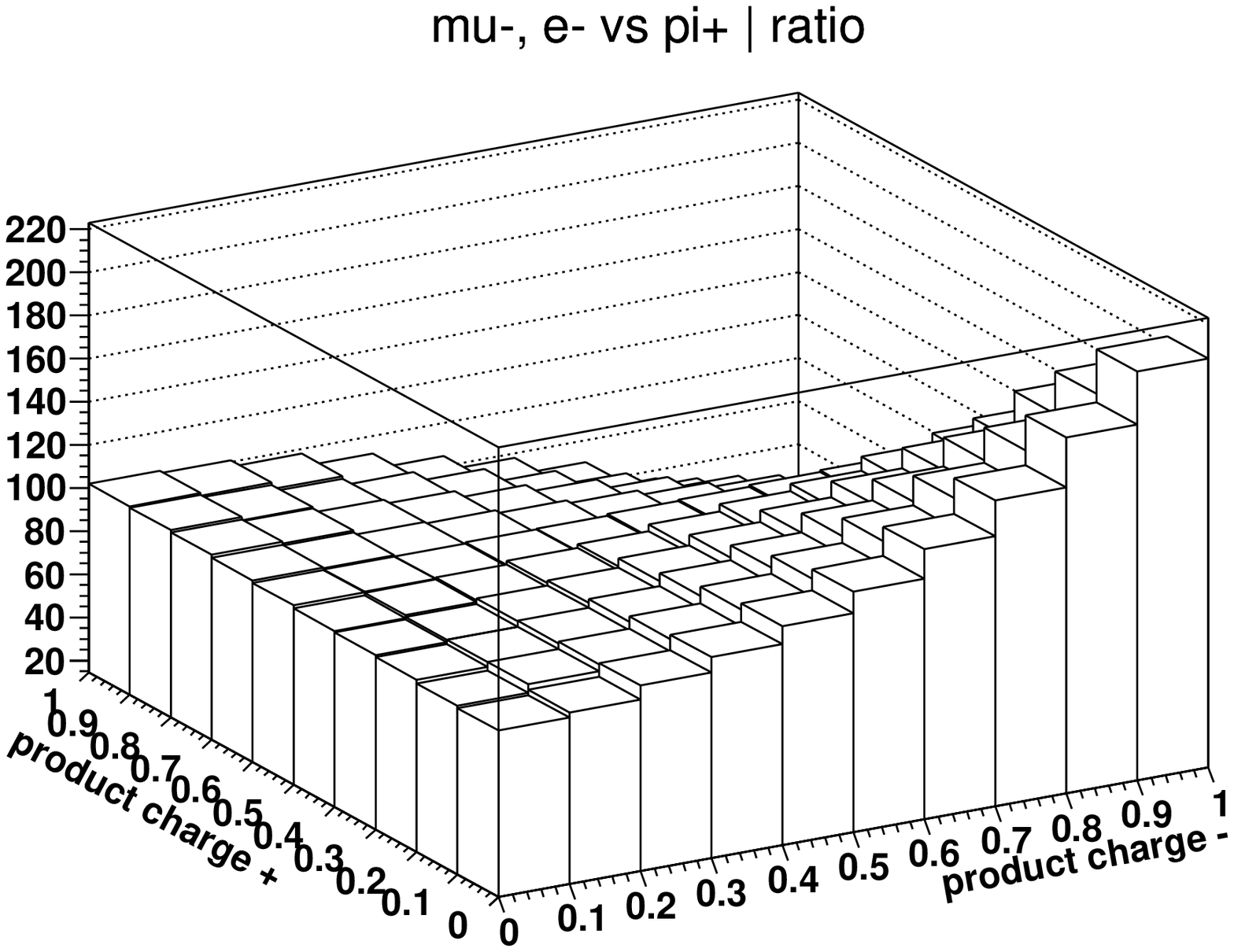}}
\resizebox*{0.49\textwidth}{!}{\includegraphics{\przedro 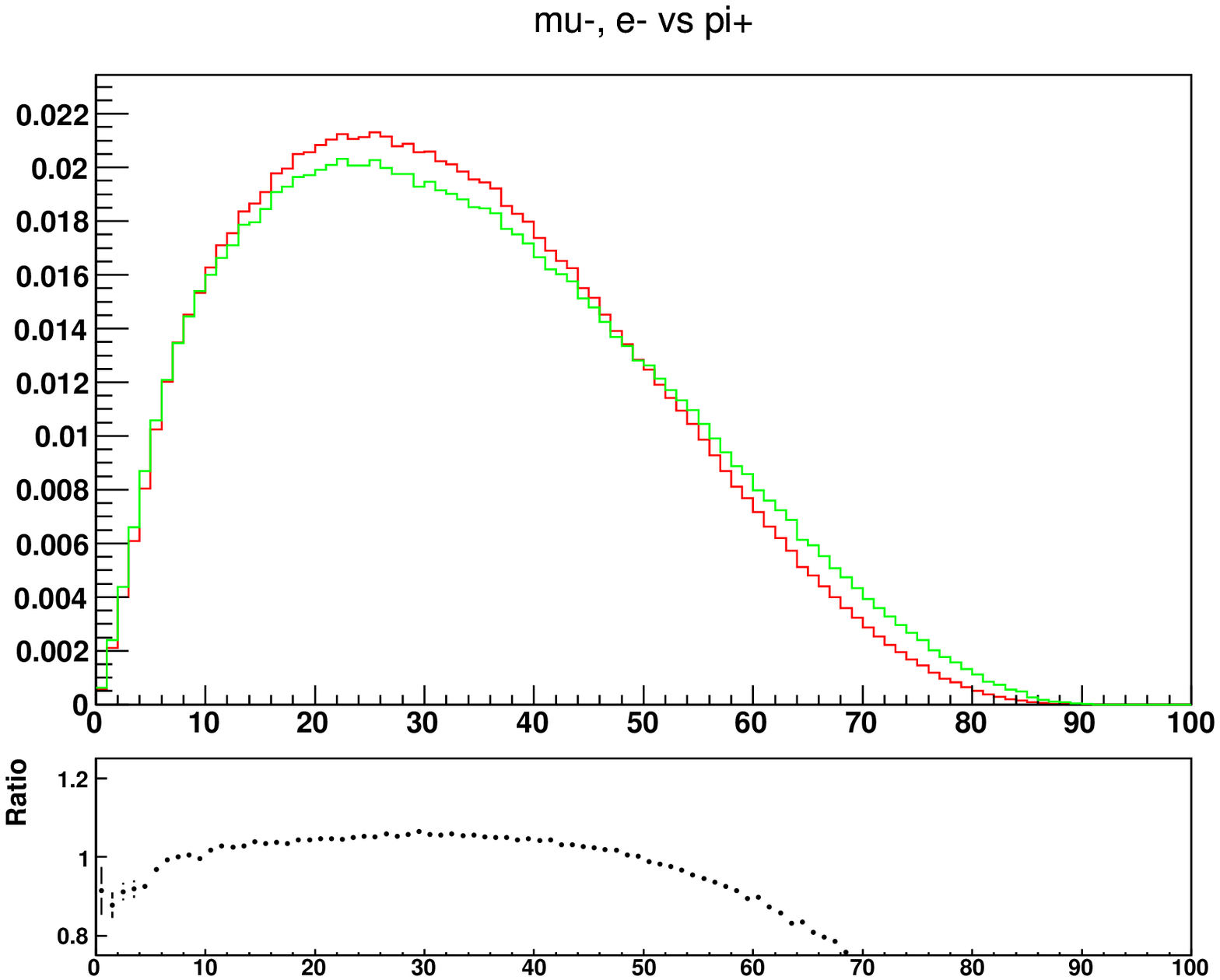}} \\
\resizebox*{0.49\textwidth}{!}{\includegraphics{\przedro 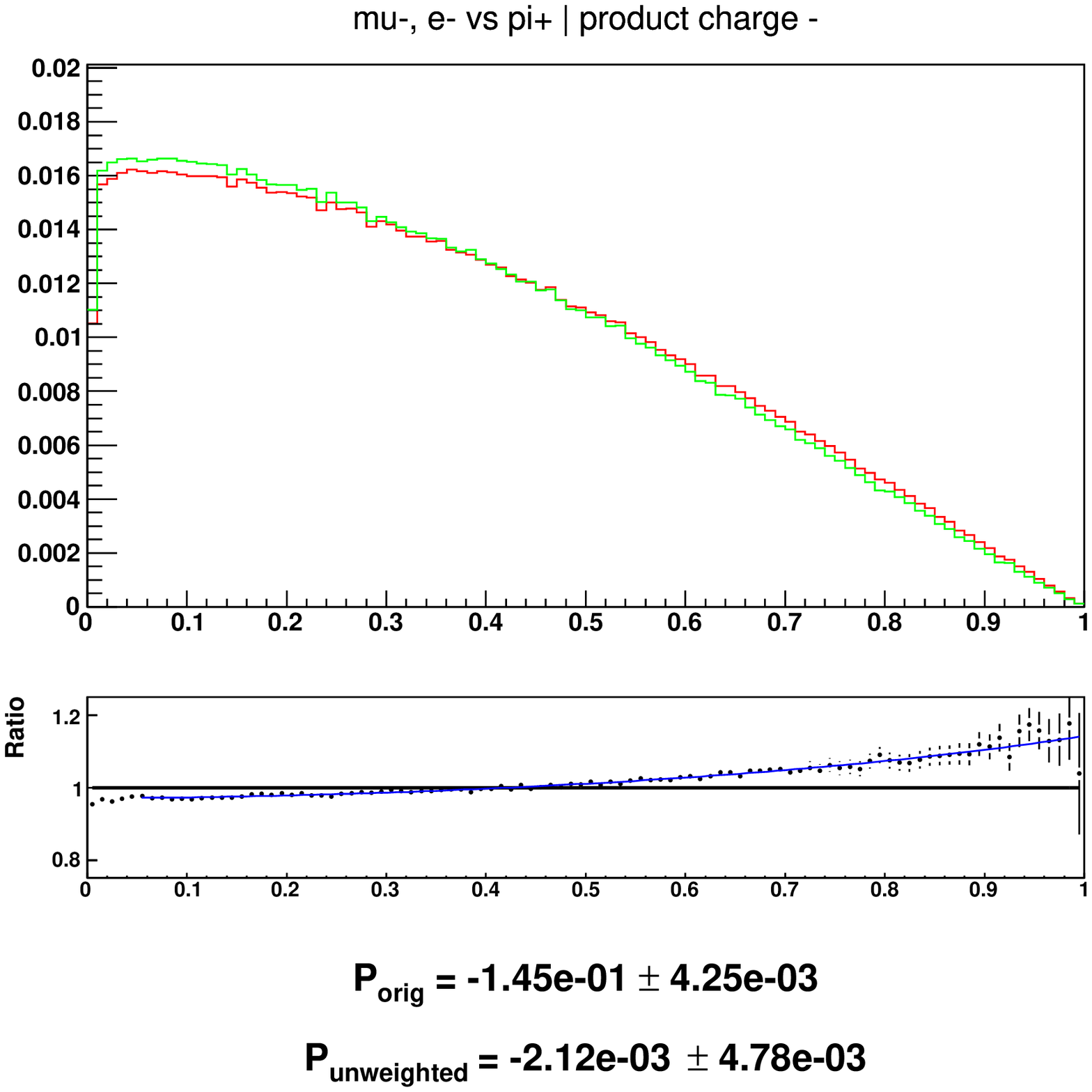}}
\resizebox*{0.49\textwidth}{!}{\includegraphics{\przedro 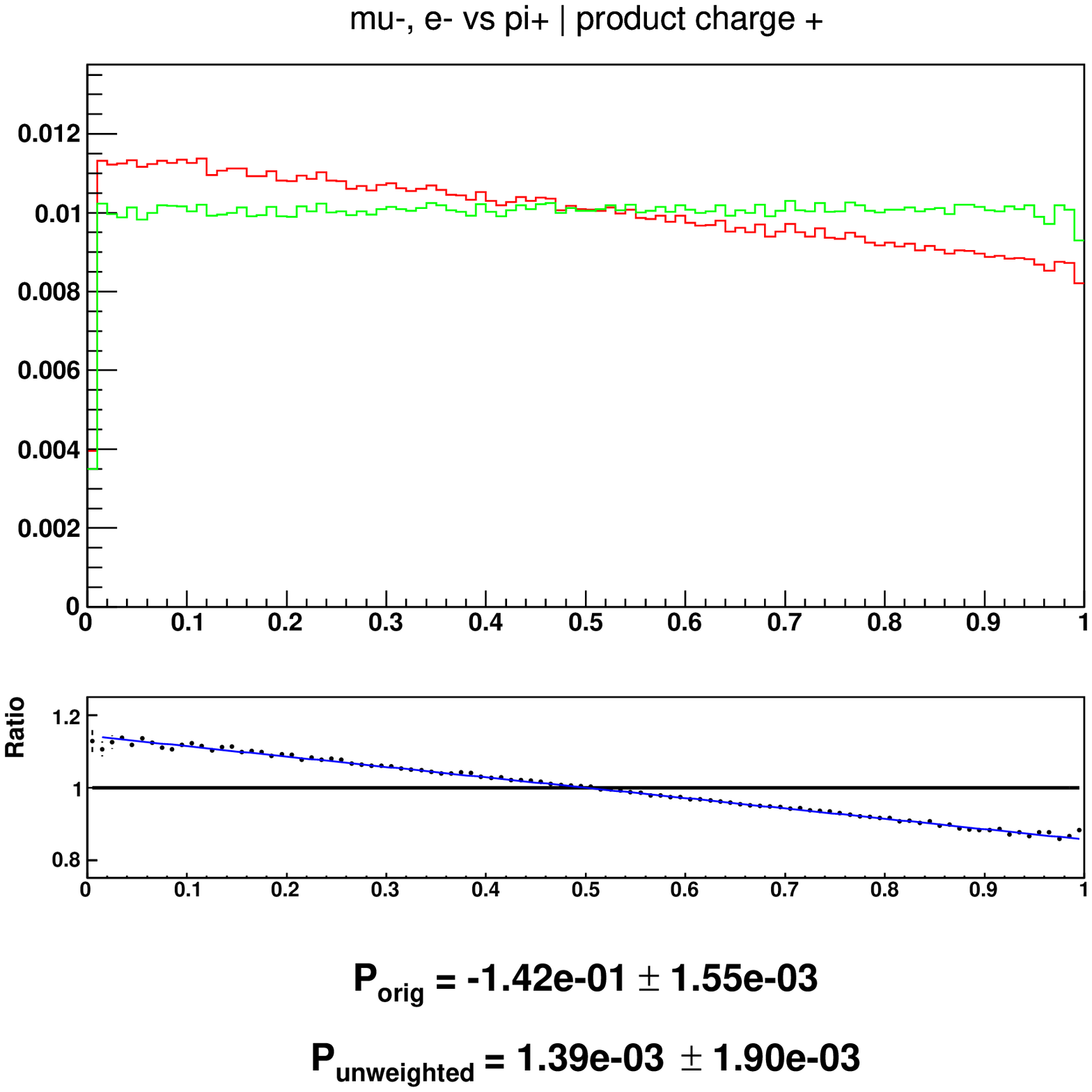}} \\
\caption{\small Fractions of  $\tau^+$ and $\tau^-$ energies carried by their visible  decay products:
two dimensional lego plots and one dimensional spectra$^{18}$.
\textcolor{red}{Red line} is  for original sample,
\textcolor{green}{green line} \greenlineis
black line is ratio \textcolor{red}{original}/\textcolor{green}{modified} with whenever available superimposed result for the
fitted functions.
}
\end{figure}

\newpage
\subsubsection{The energy spectrum: $\tau^- \to \pi^-$ {\tt vs } $\tau^+ \to \mu^+, e^+$}
\vspace{3\baselineskip}

\begin{figure}[h!]
\centering
\resizebox*{0.49\textwidth}{!}{\includegraphics{\przedro 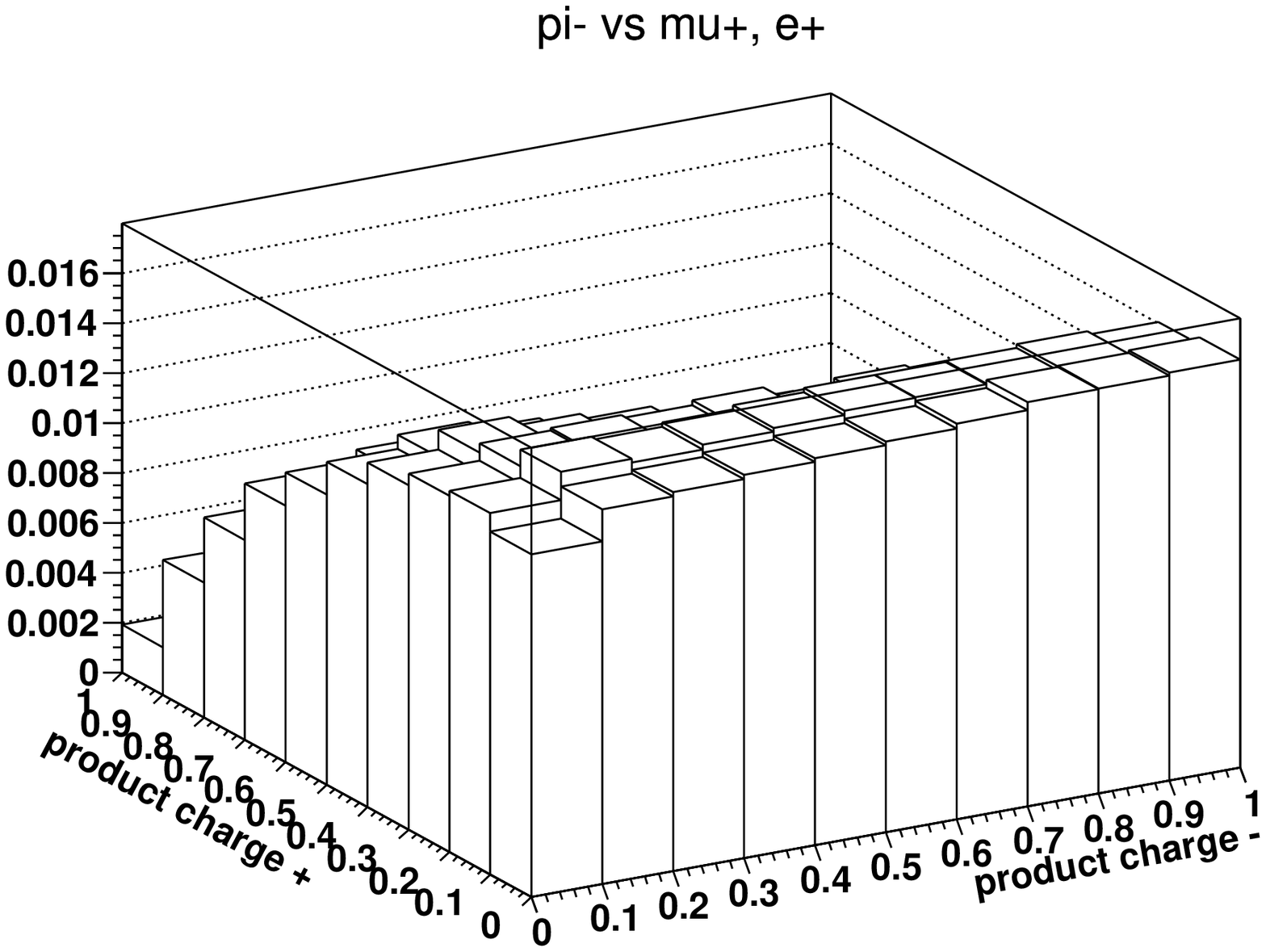}}
\resizebox*{0.49\textwidth}{!}{\includegraphics{\przedro 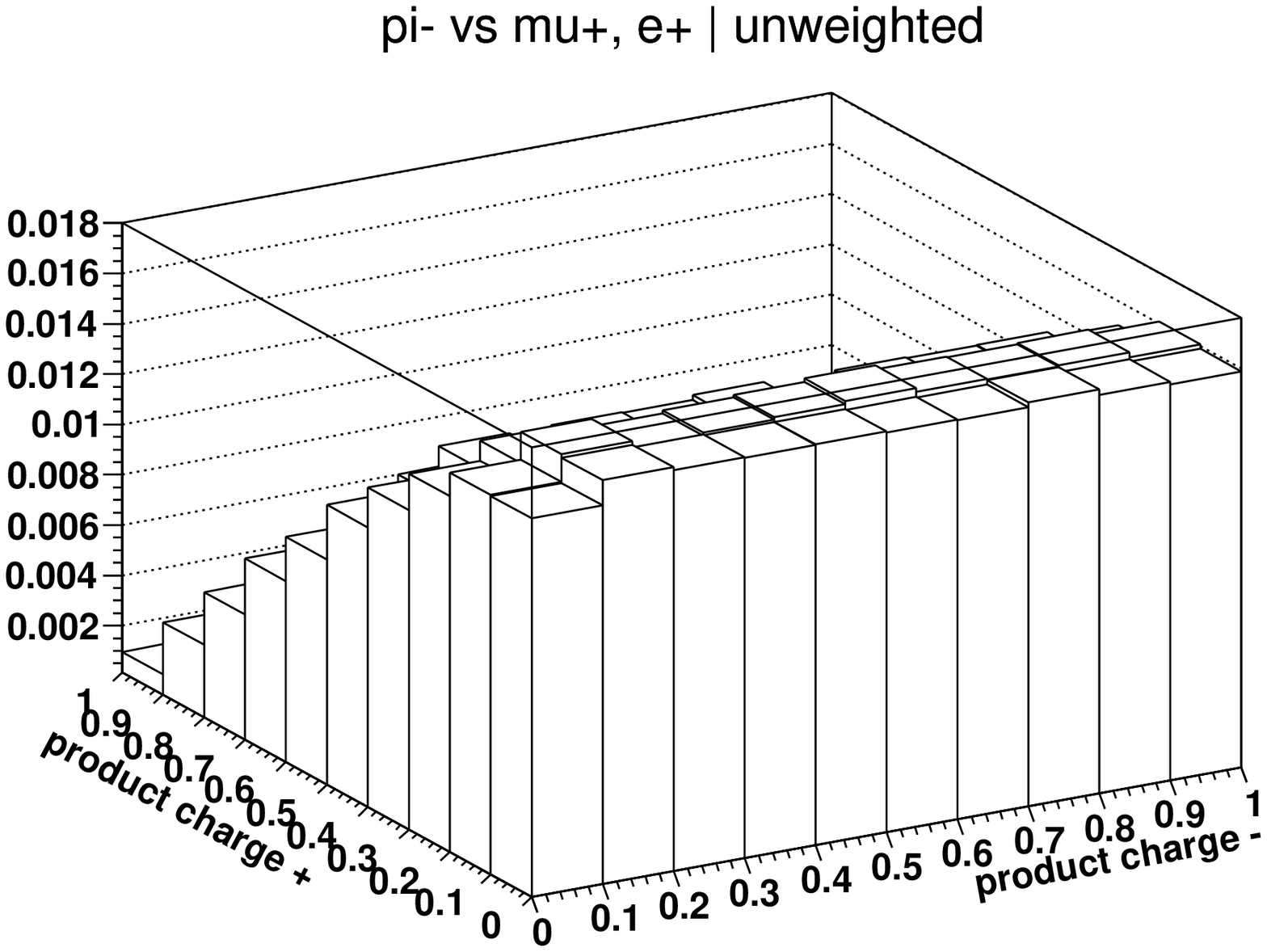}} \\
\resizebox*{0.49\textwidth}{!}{\includegraphics{\przedro 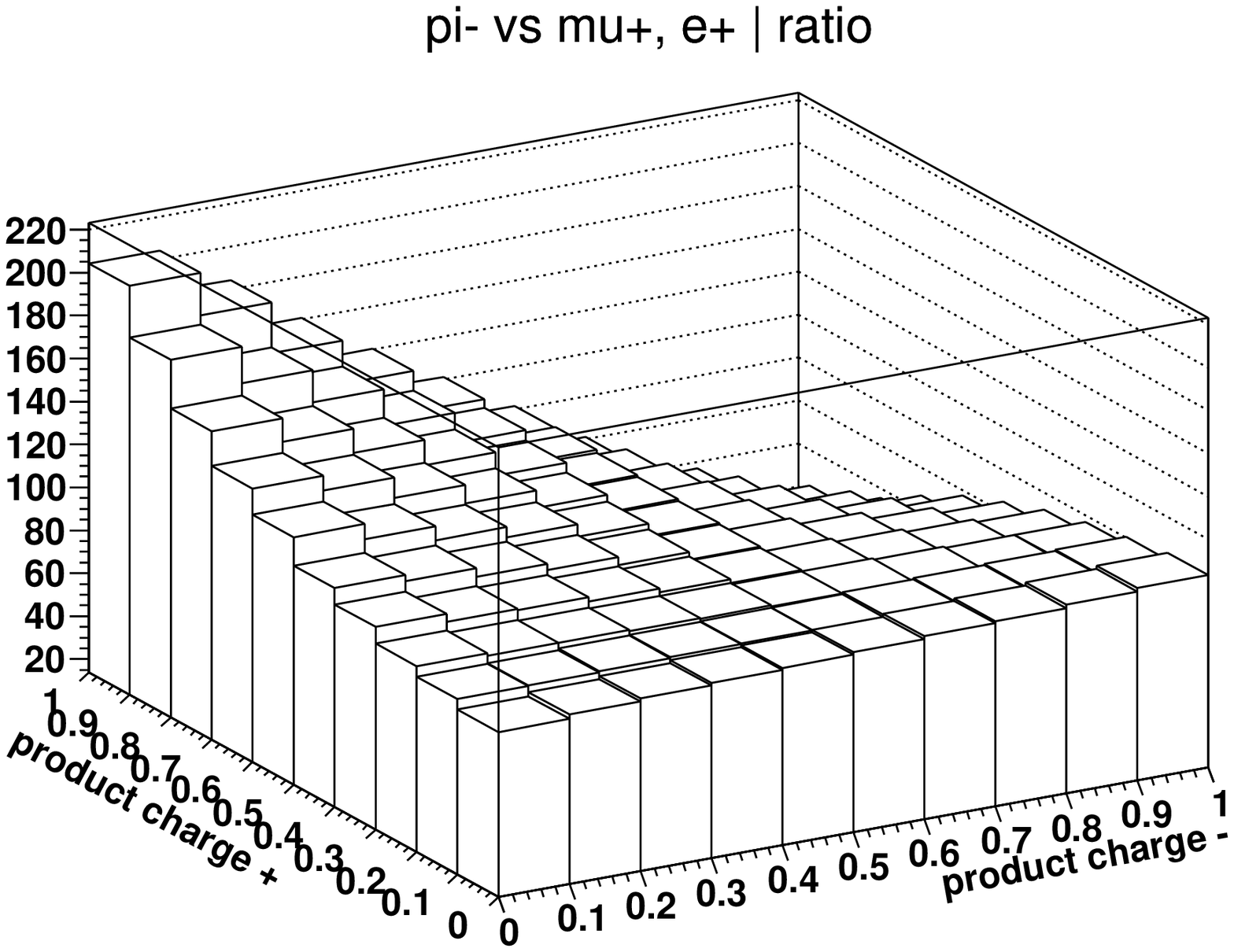}}
\resizebox*{0.49\textwidth}{!}{\includegraphics{\przedro 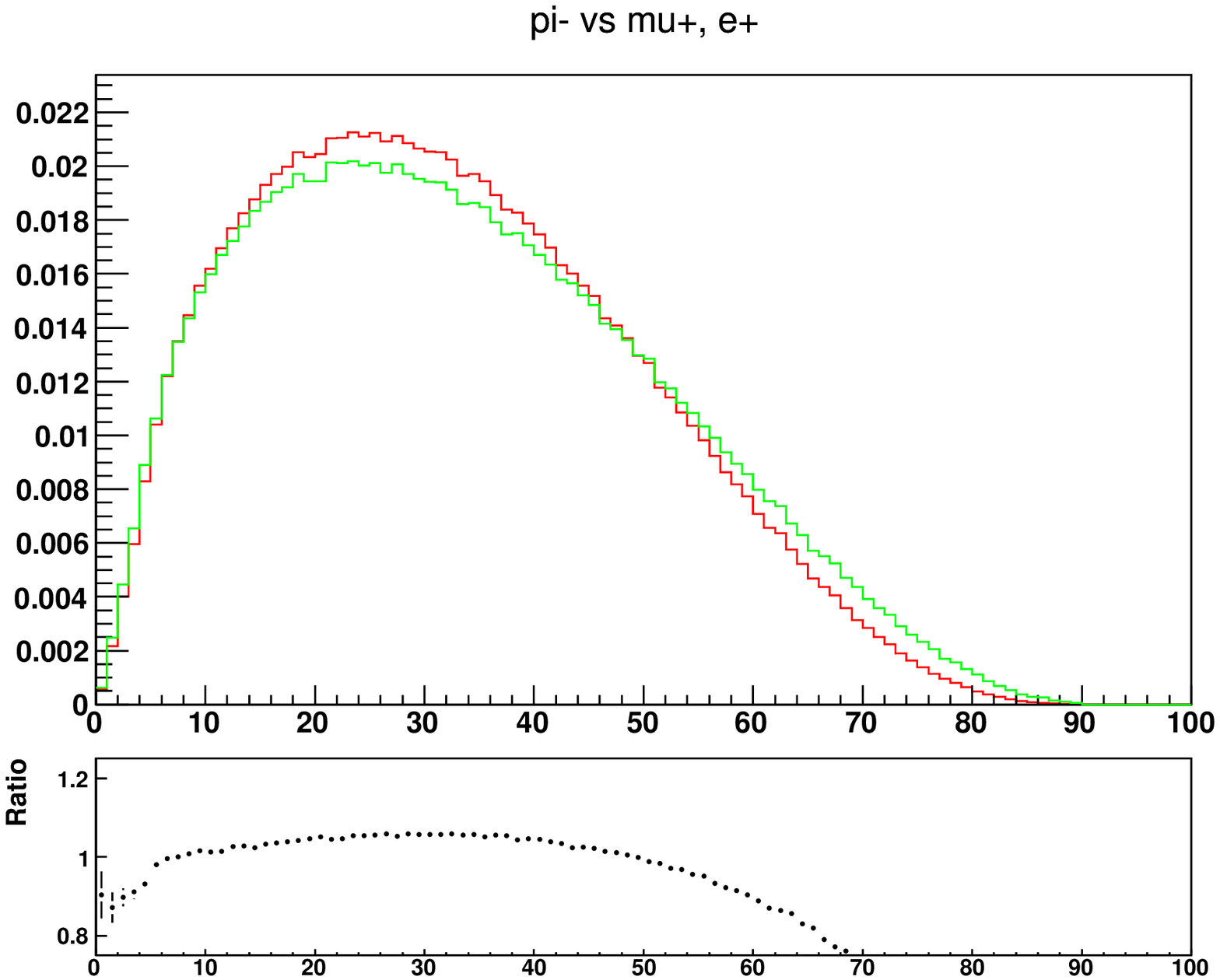}} \\
\resizebox*{0.49\textwidth}{!}{\includegraphics{\przedro 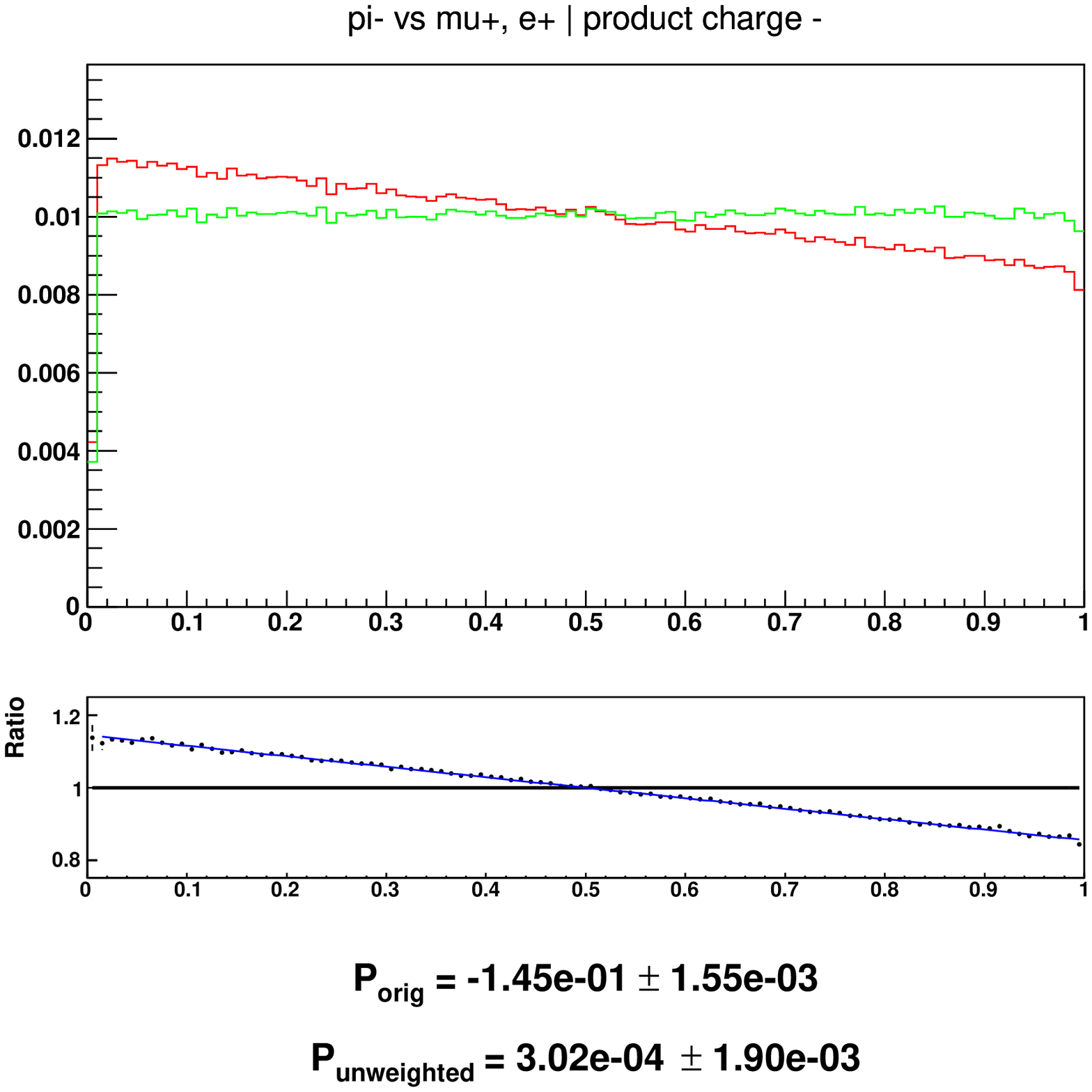}}
\resizebox*{0.49\textwidth}{!}{\includegraphics{\przedro 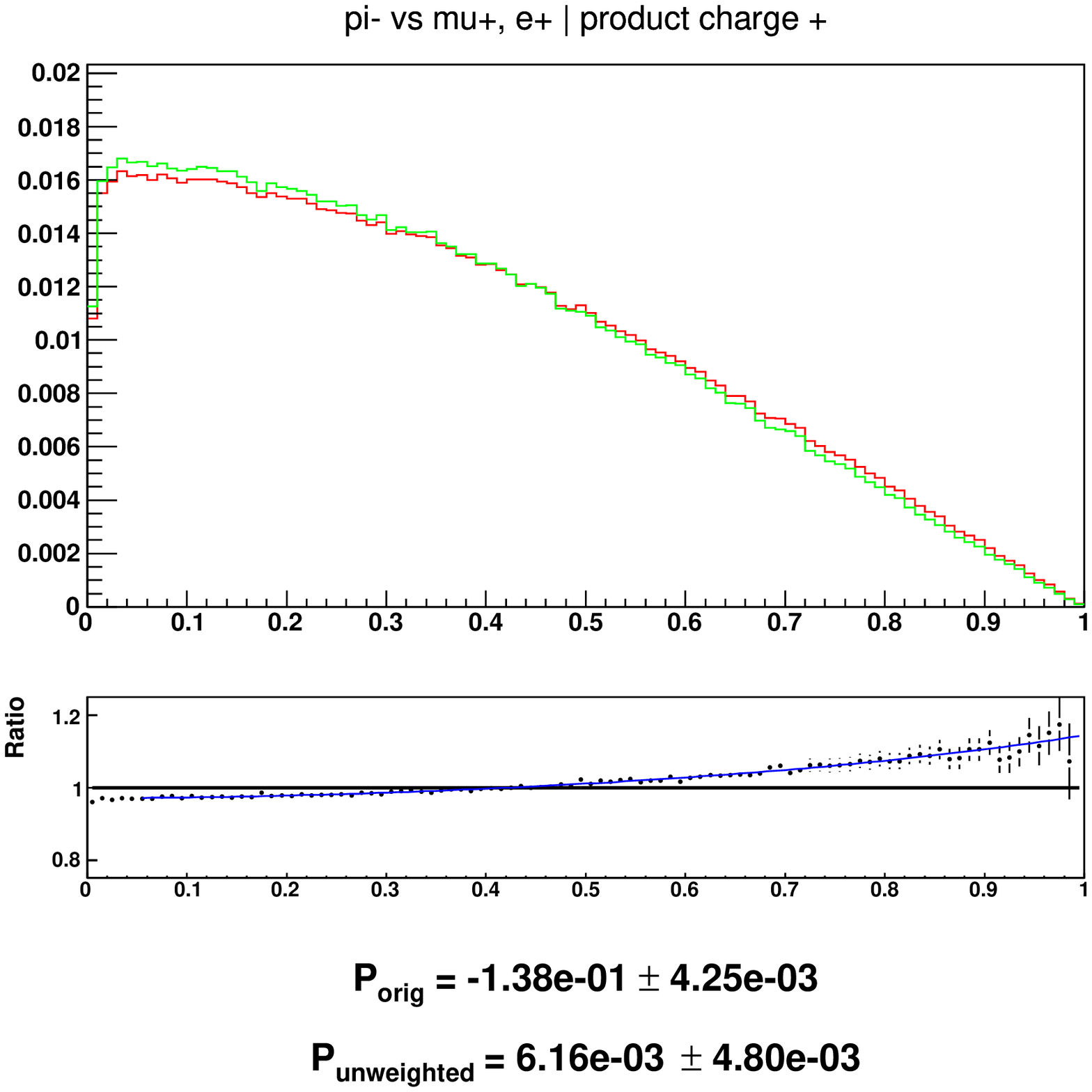}} \\
\caption{\small Fractions of  $\tau^+$ and $\tau^-$ energies carried by their visible  decay products:
two dimensional lego plots and one dimensional spectra$^{18}$.
\textcolor{red}{Red line} is  for original sample,
\textcolor{green}{green line} \greenlineis
black line is ratio \textcolor{red}{original}/\textcolor{green}{modified} with whenever available superimposed result for the
fitted functions.
}
\end{figure}

\newpage
\subsubsection{The energy spectrum: $\tau^- \to \pi^-$ {\tt vs } $\tau^+ \to \pi^+$}
\vspace{3\baselineskip}

\begin{figure}[h!]
\centering
\resizebox*{0.49\textwidth}{!}{\includegraphics{\przedro 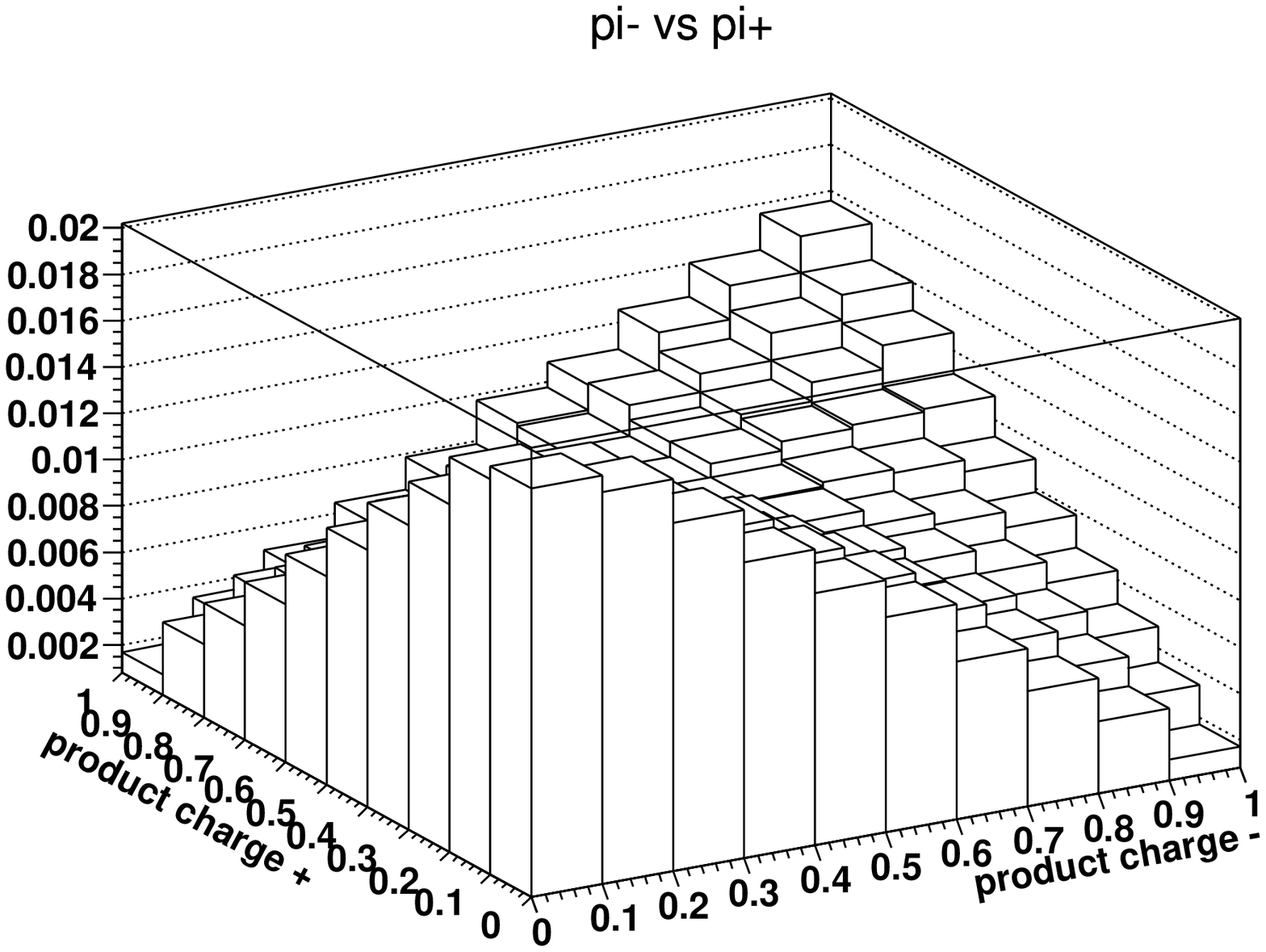}}
\resizebox*{0.49\textwidth}{!}{\includegraphics{\przedro 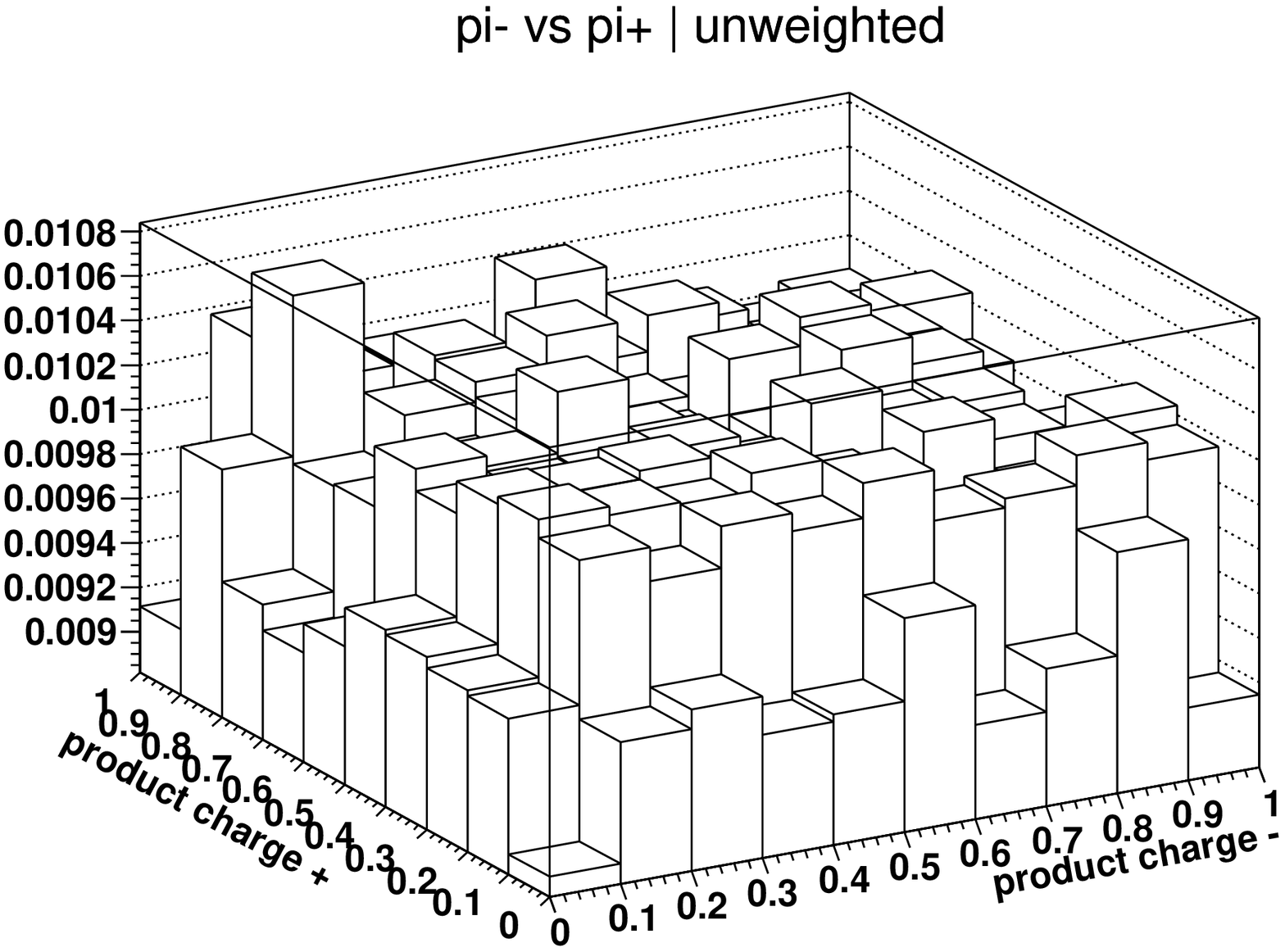}} \\
\resizebox*{0.49\textwidth}{!}{\includegraphics{\przedro 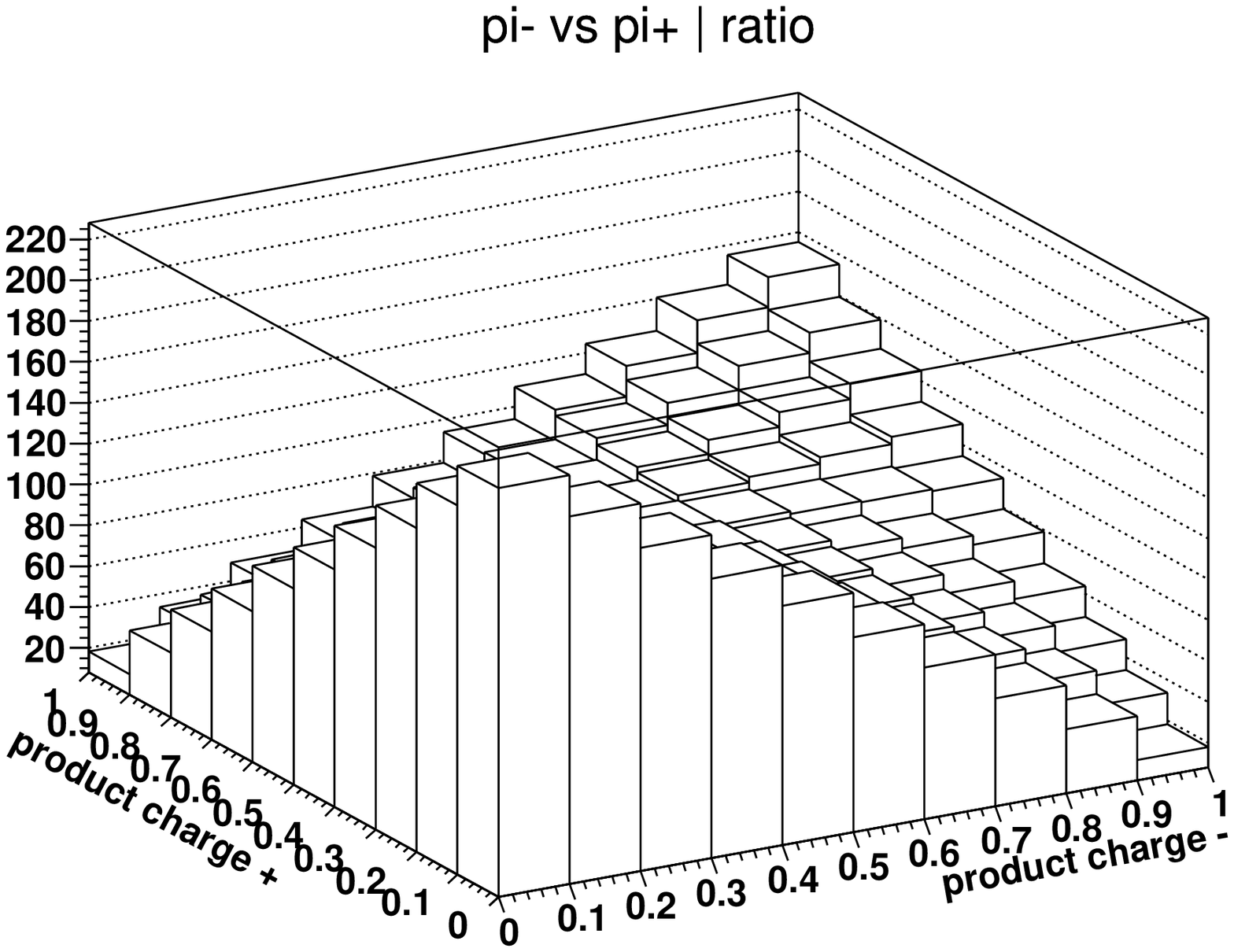}}
\resizebox*{0.49\textwidth}{!}{\includegraphics{\przedro 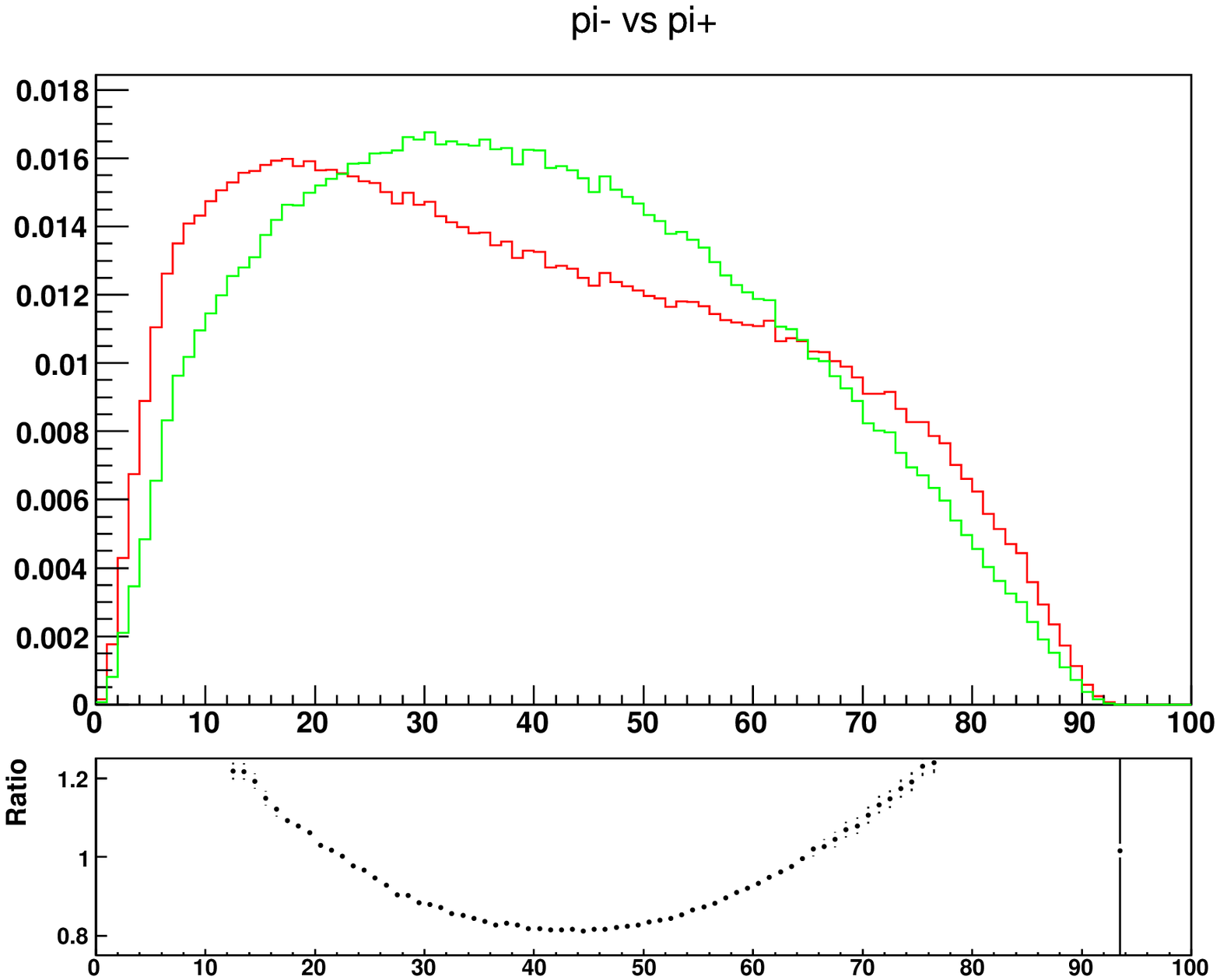}} \\
\resizebox*{0.49\textwidth}{!}{\includegraphics{\przedro 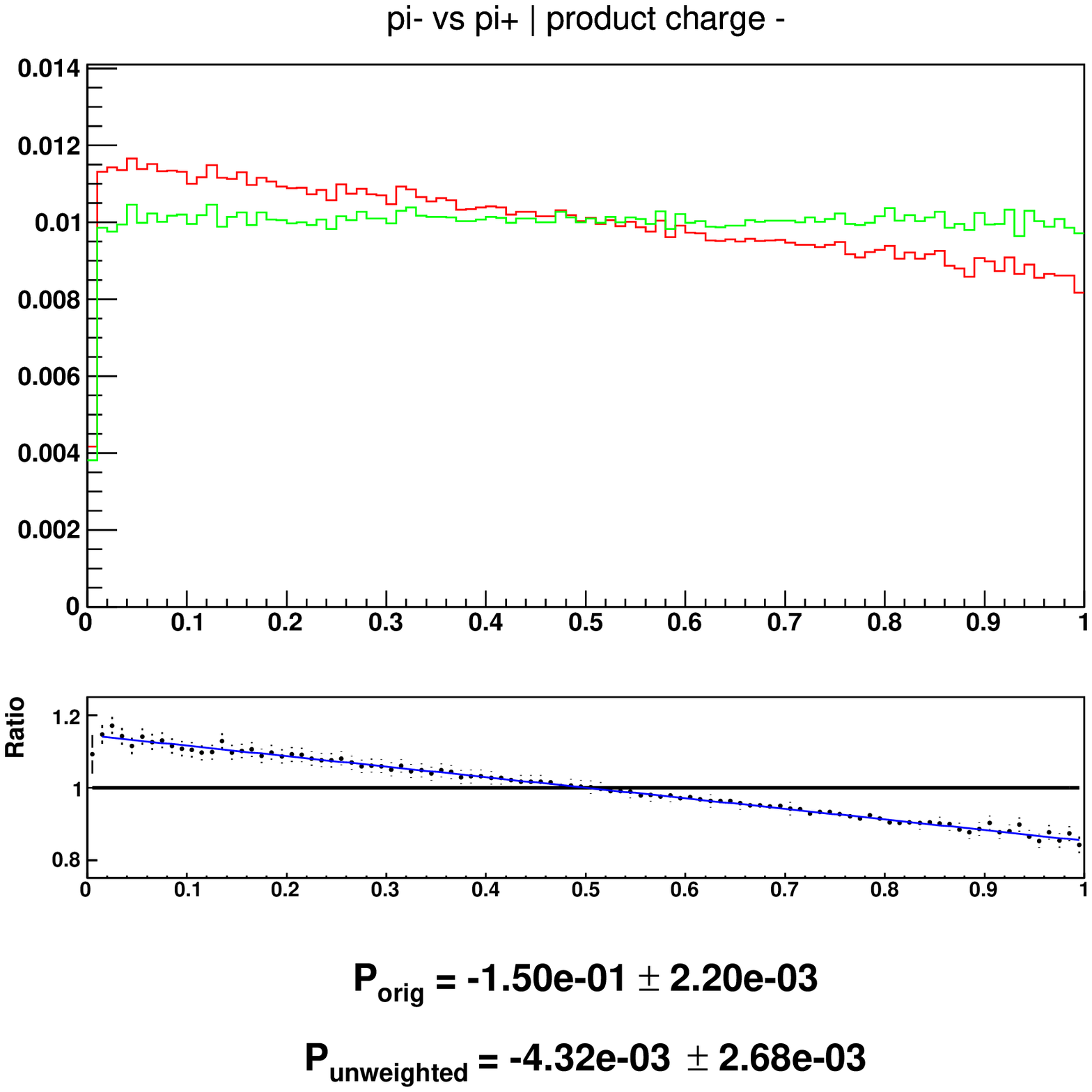}}
\resizebox*{0.49\textwidth}{!}{\includegraphics{\przedro 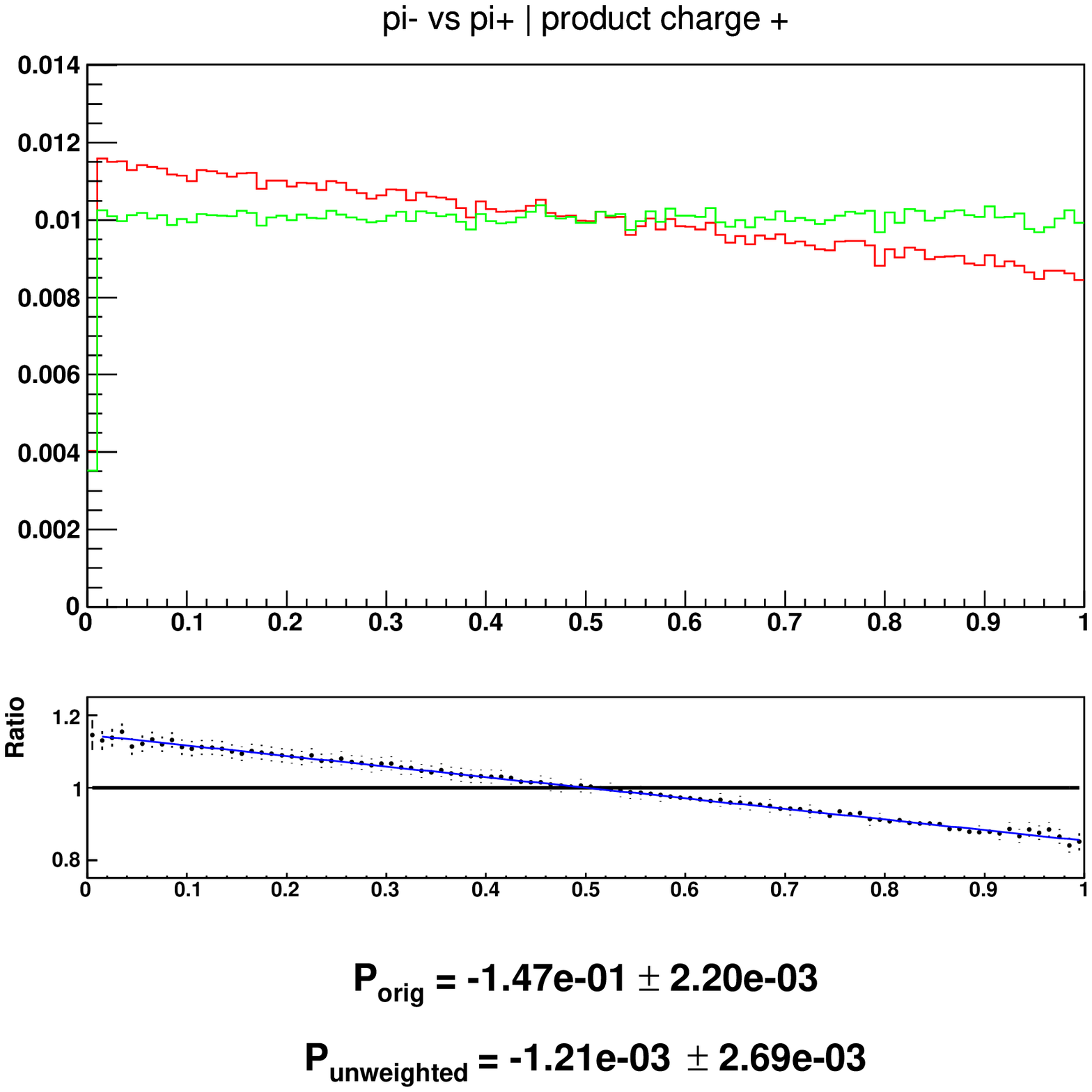}} \\
\caption{\small Fractions of  $\tau^+$ and $\tau^-$ energies carried by their visible  decay products:
two dimensional lego plots and one dimensional spectra$^{18}$.
\textcolor{red}{Red line} is  for original sample,
\textcolor{green}{green line} \greenlineis
black line is ratio \textcolor{red}{original}/\textcolor{green}{modified} with whenever available superimposed result for the
fitted functions.
}\label{Fig:spectra2}
\end{figure}

\newpage
\subsubsection{The energy spectrum: $\tau^- \to \mu^-, e^-$ {\tt vs } $\tau^+ \to \rho^+$}
\vspace{3\baselineskip}

\begin{figure}[h!]
\centering
\resizebox*{0.49\textwidth}{!}{\includegraphics{\przedro 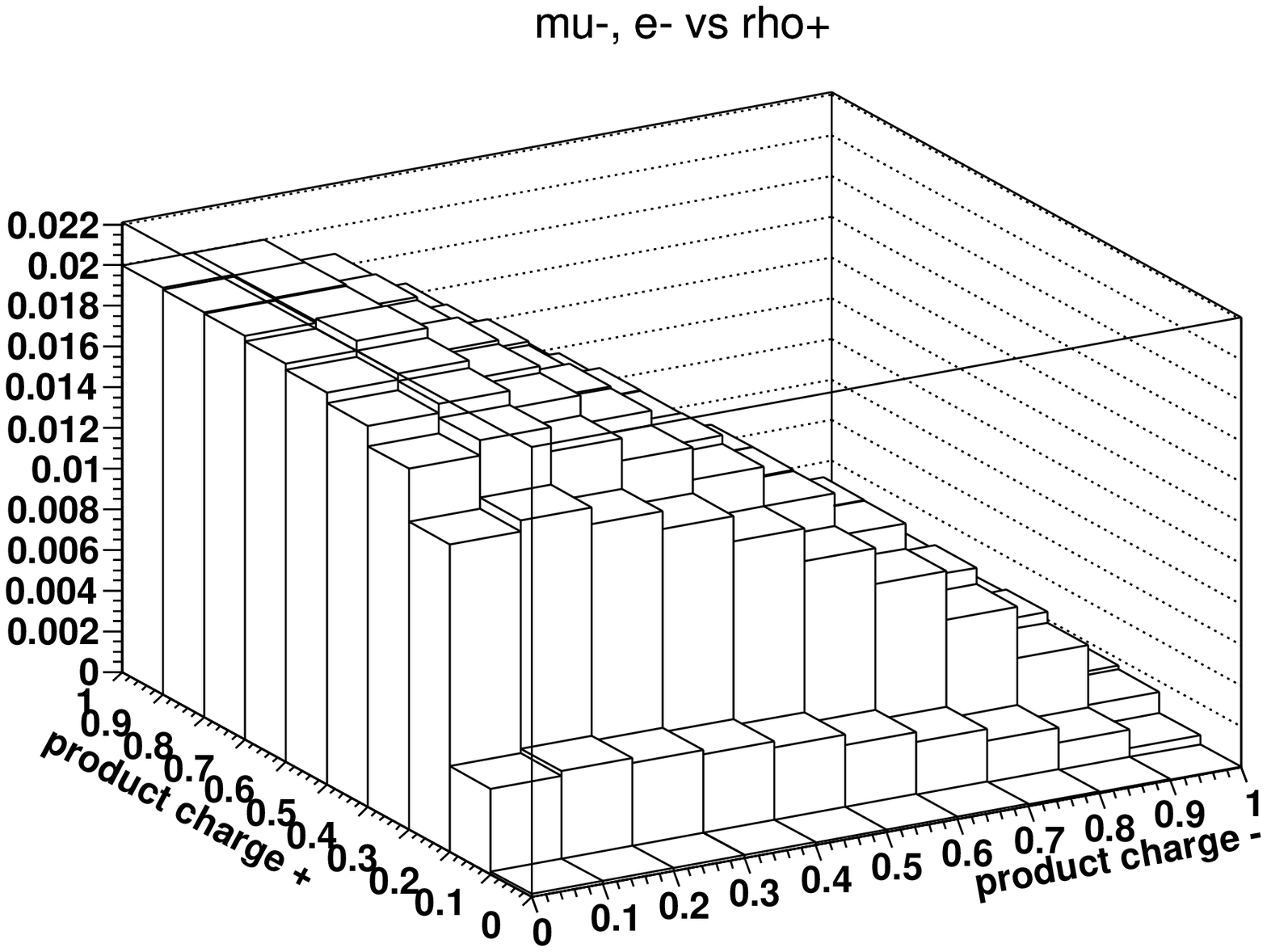}}
\resizebox*{0.49\textwidth}{!}{\includegraphics{\przedro 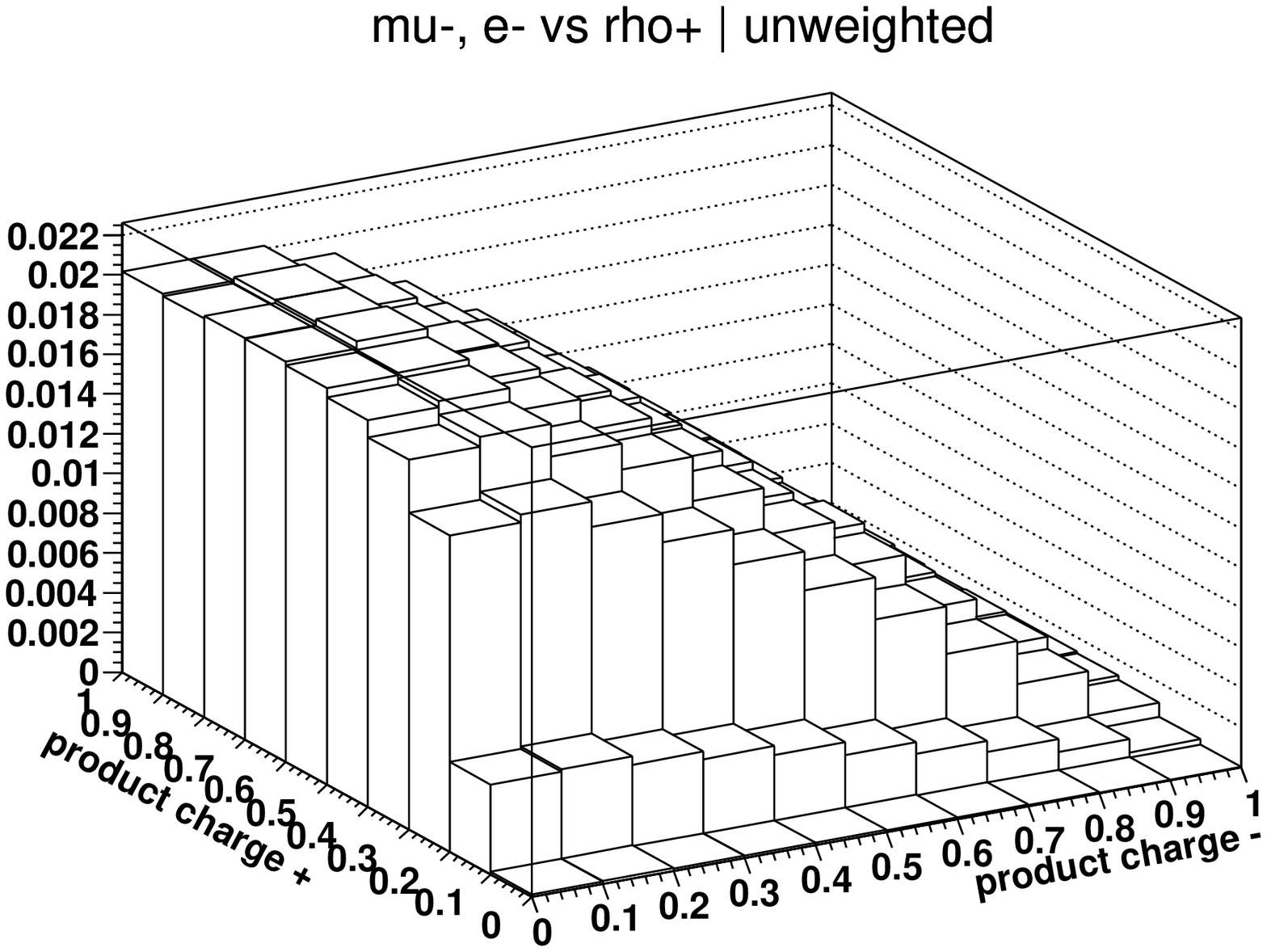}} \\
\resizebox*{0.49\textwidth}{!}{\includegraphics{\przedro 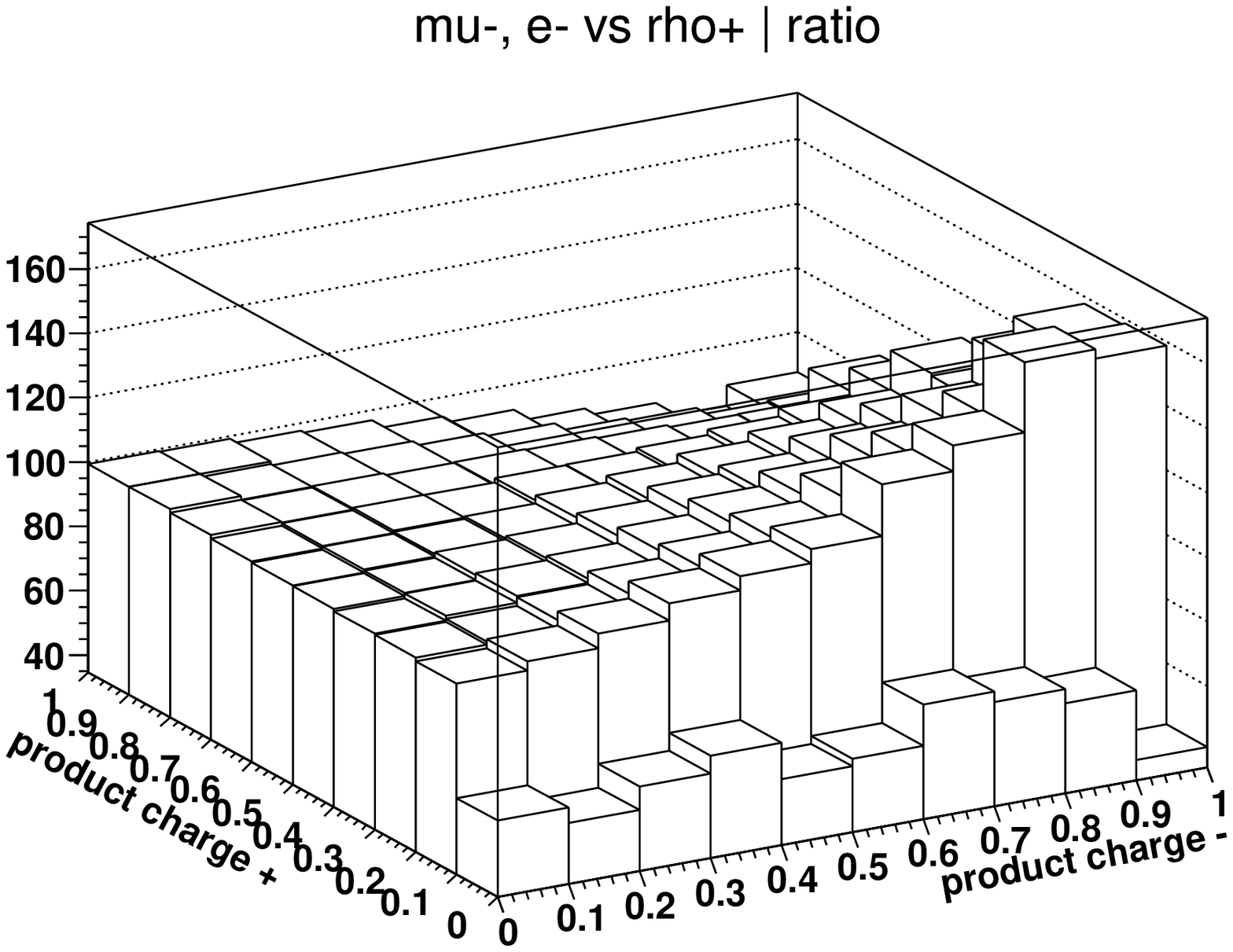}}
\resizebox*{0.49\textwidth}{!}{\includegraphics{\przedro 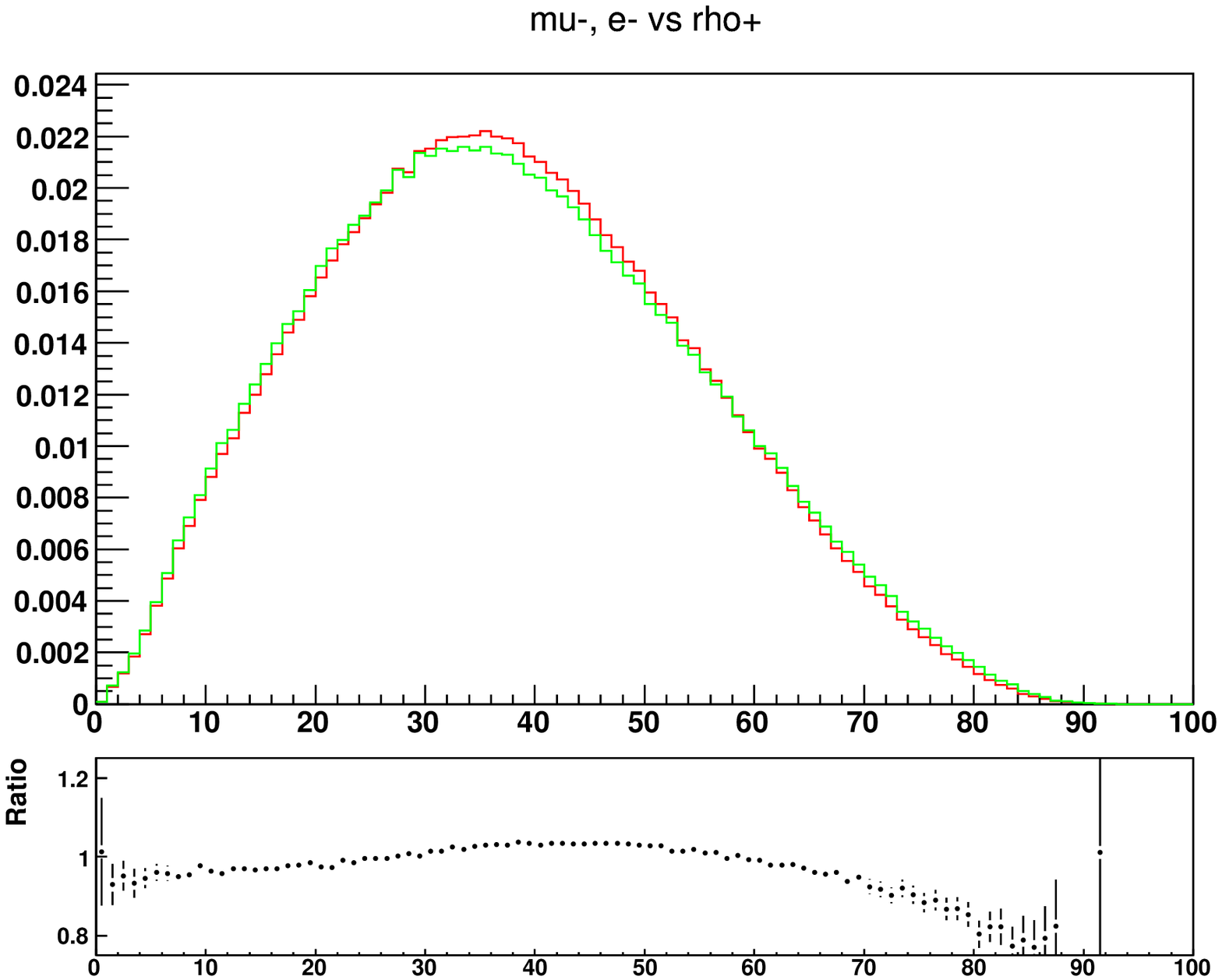}} \\
\resizebox*{0.49\textwidth}{!}{\includegraphics{\przedro 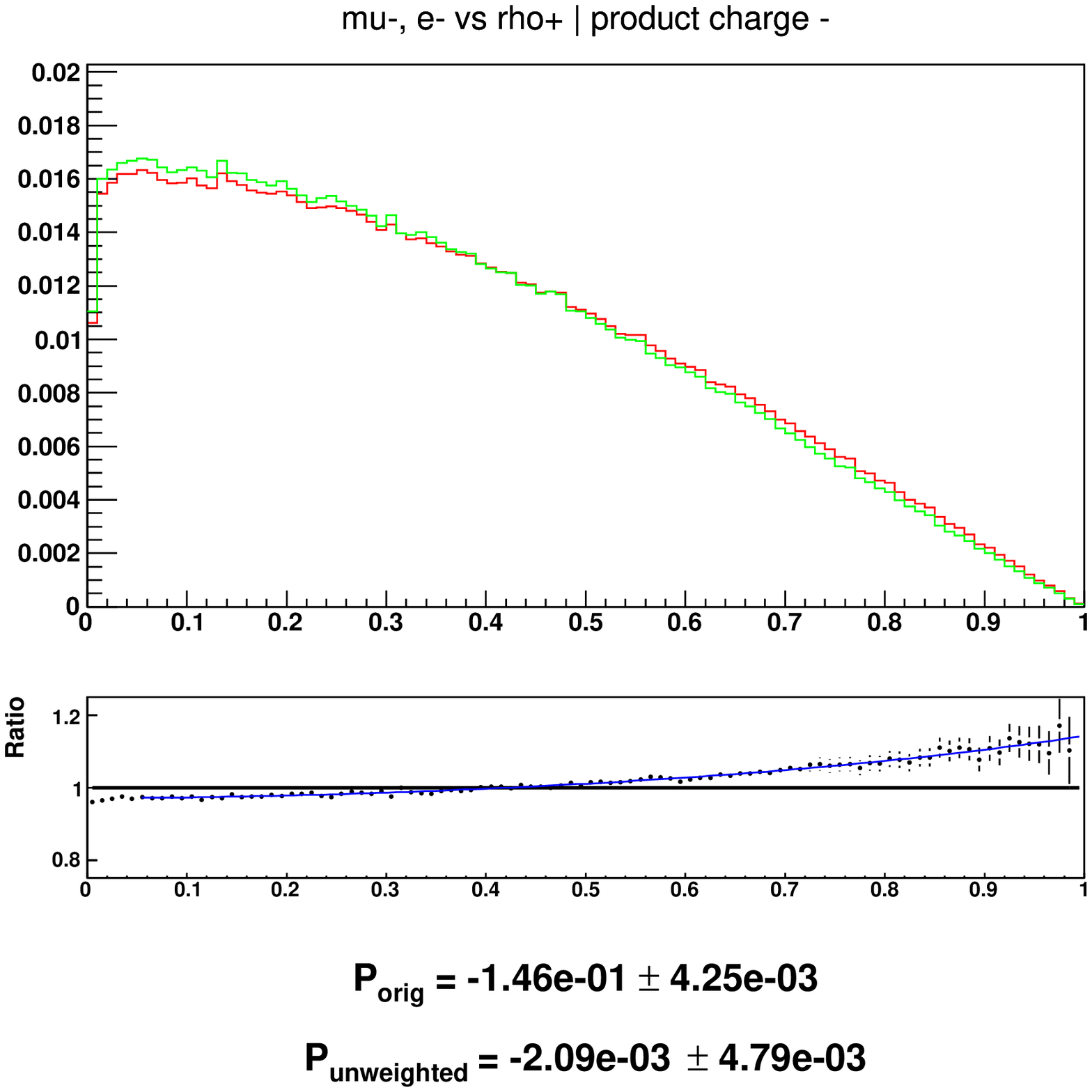}}
\resizebox*{0.49\textwidth}{!}{\includegraphics{\przedro 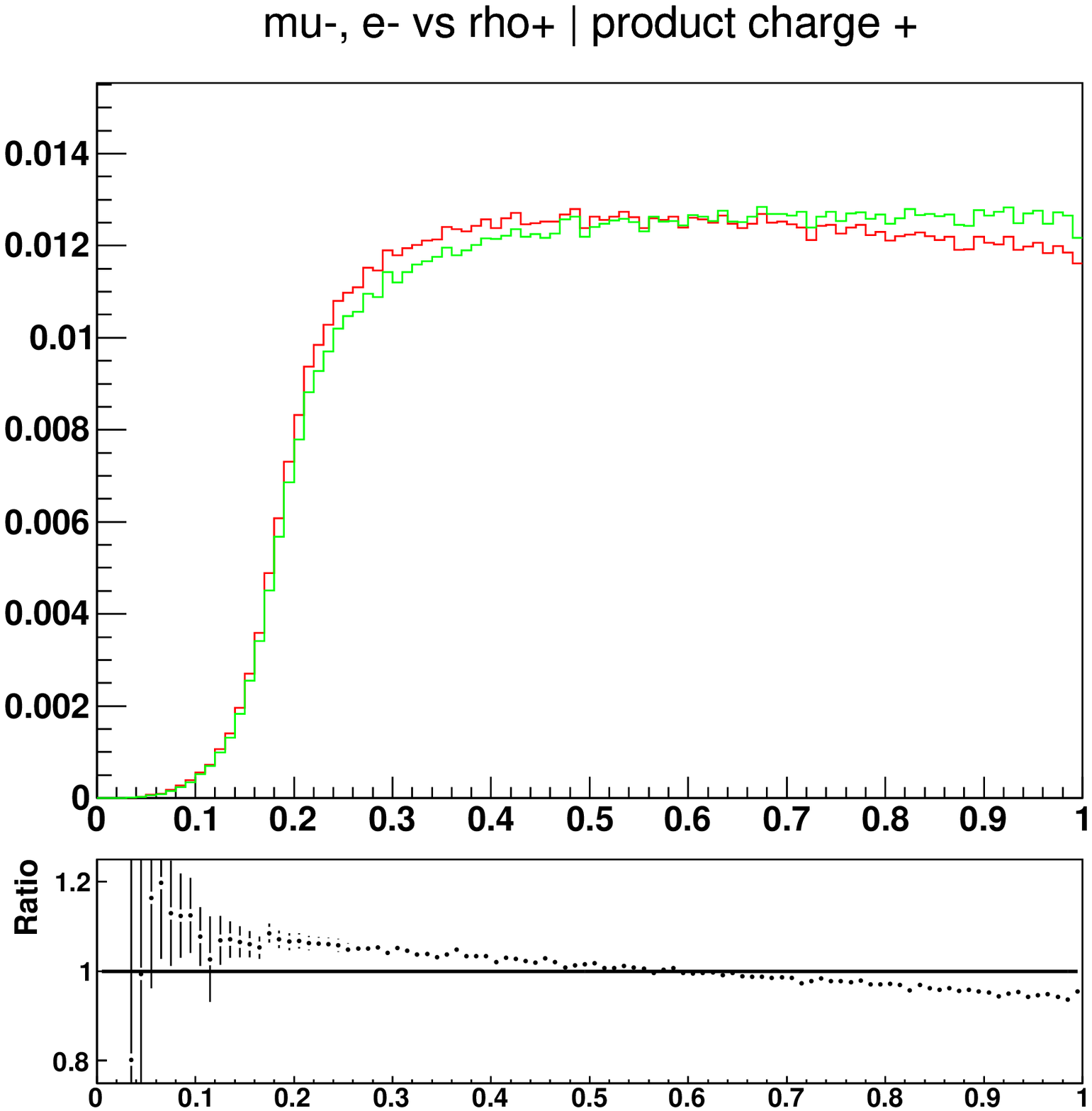}} \\
\caption{\small Fractions of  $\tau^+$ and $\tau^-$ energies carried by their visible  decay products:
two dimensional lego plots and one dimensional spectra$^{18}$.
\textcolor{red}{Red line} is  for original sample,
\textcolor{green}{green line} \greenlineis
black line is ratio \textcolor{red}{original}/\textcolor{green}{modified} with whenever available superimposed result for the
fitted functions.
}
\end{figure}

\newpage
\subsubsection{The energy spectrum: $\tau^- \to \rho^-$ {\tt vs } $\tau^+ \to \mu^+, e^+$}
\vspace{3\baselineskip}

\begin{figure}[h!]
\centering
\resizebox*{0.49\textwidth}{!}{\includegraphics{\przedro 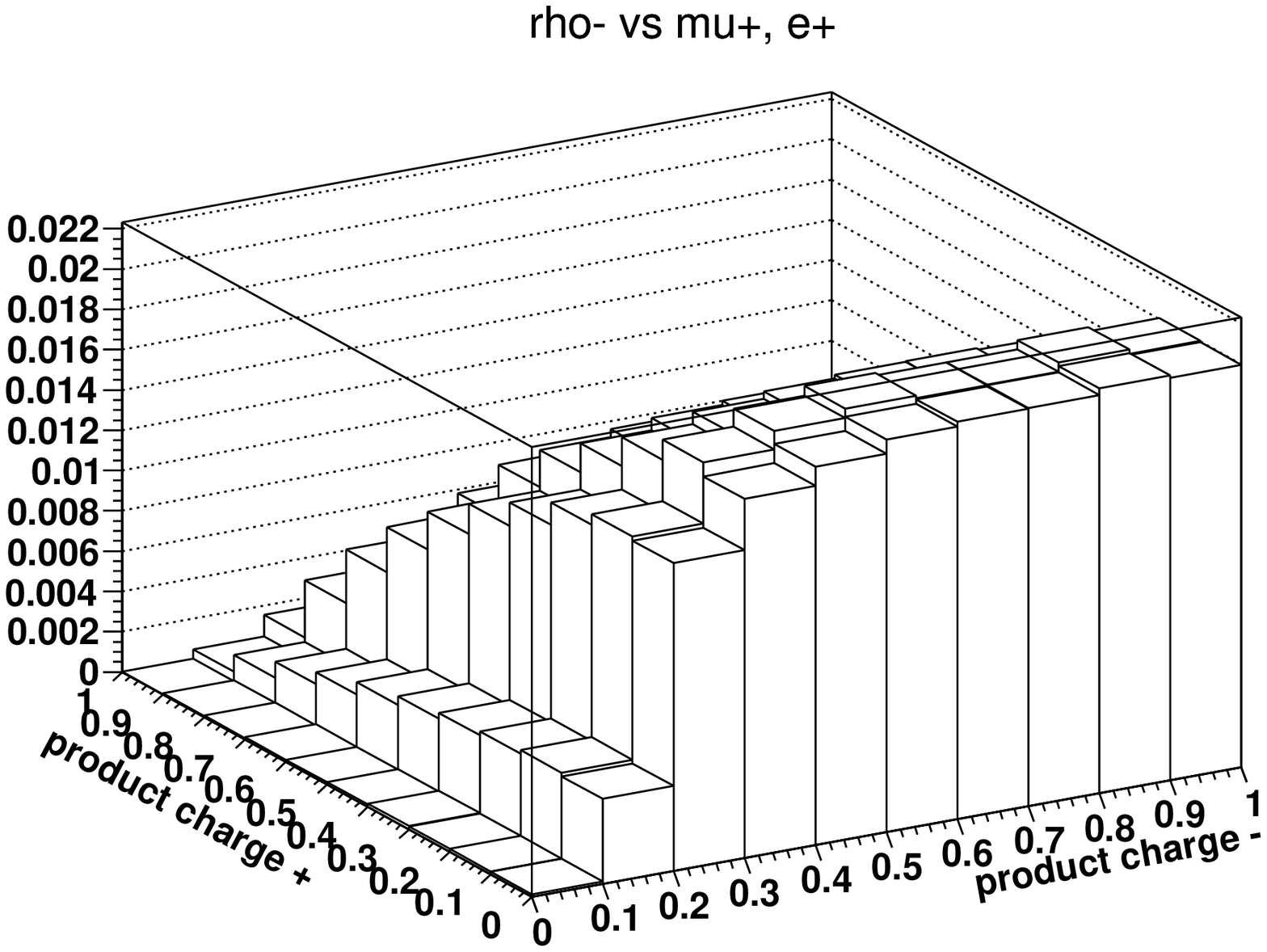}}
\resizebox*{0.49\textwidth}{!}{\includegraphics{\przedro 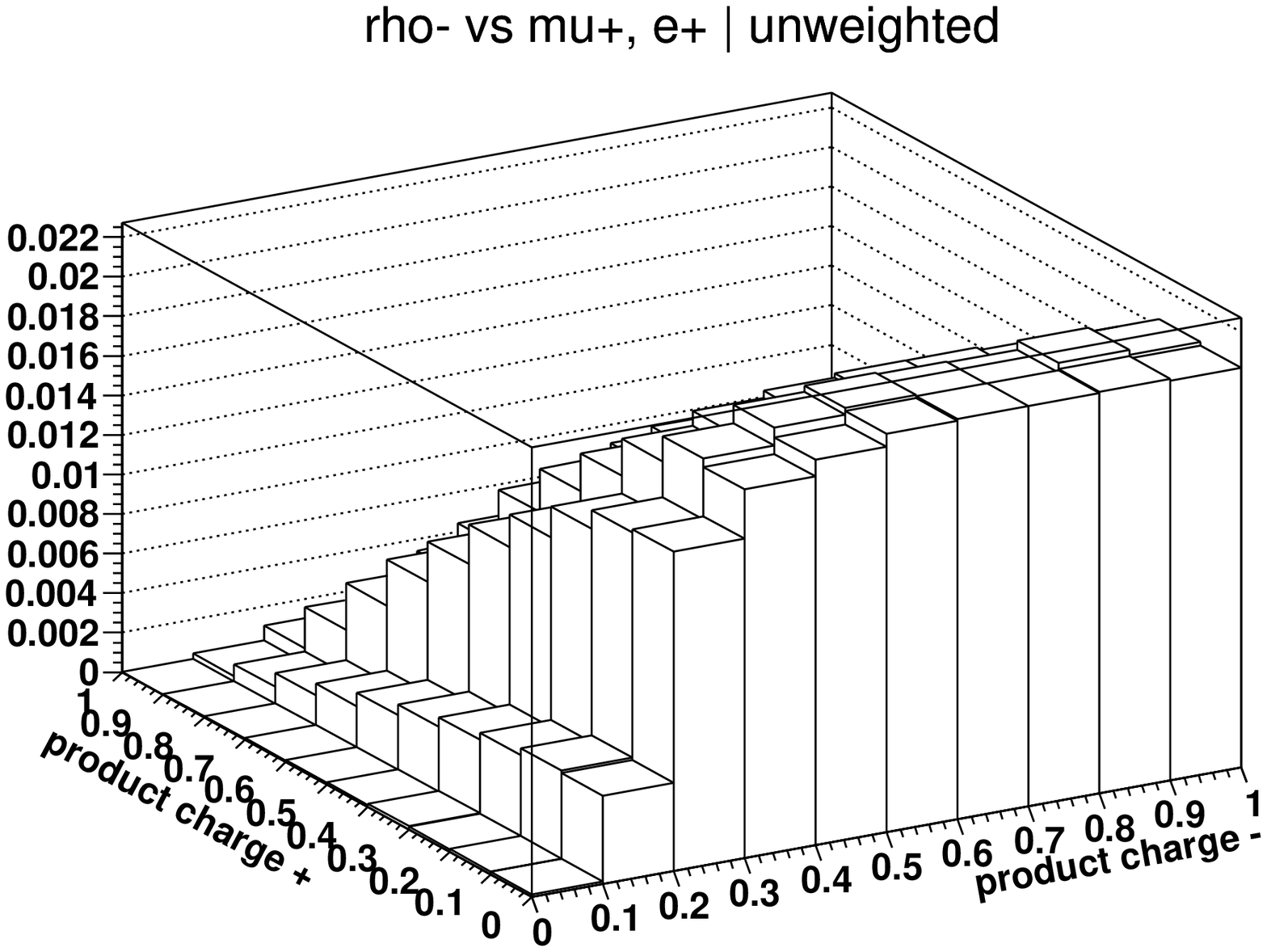}} \\
\resizebox*{0.49\textwidth}{!}{\includegraphics{\przedro 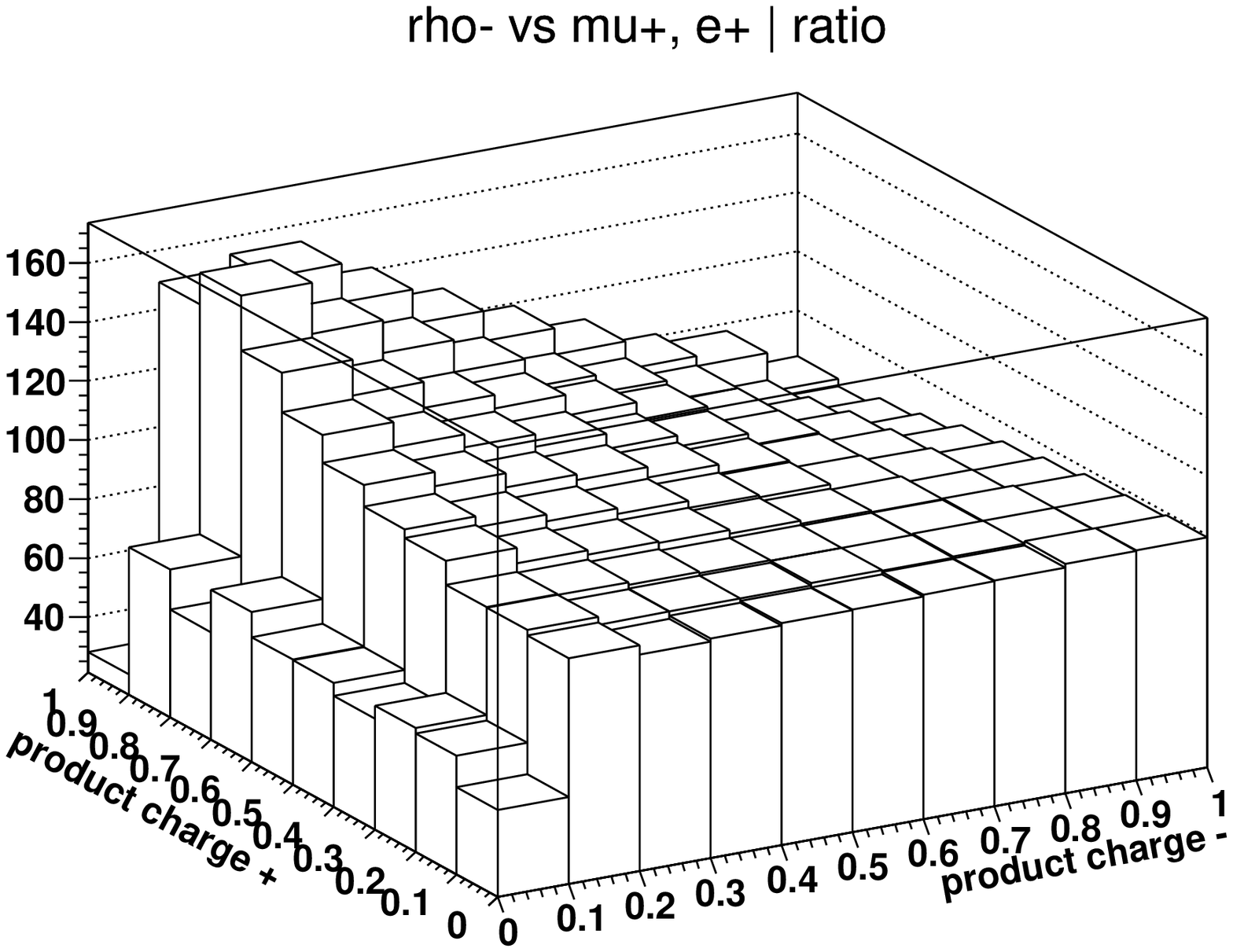}}
\resizebox*{0.49\textwidth}{!}{\includegraphics{\przedro 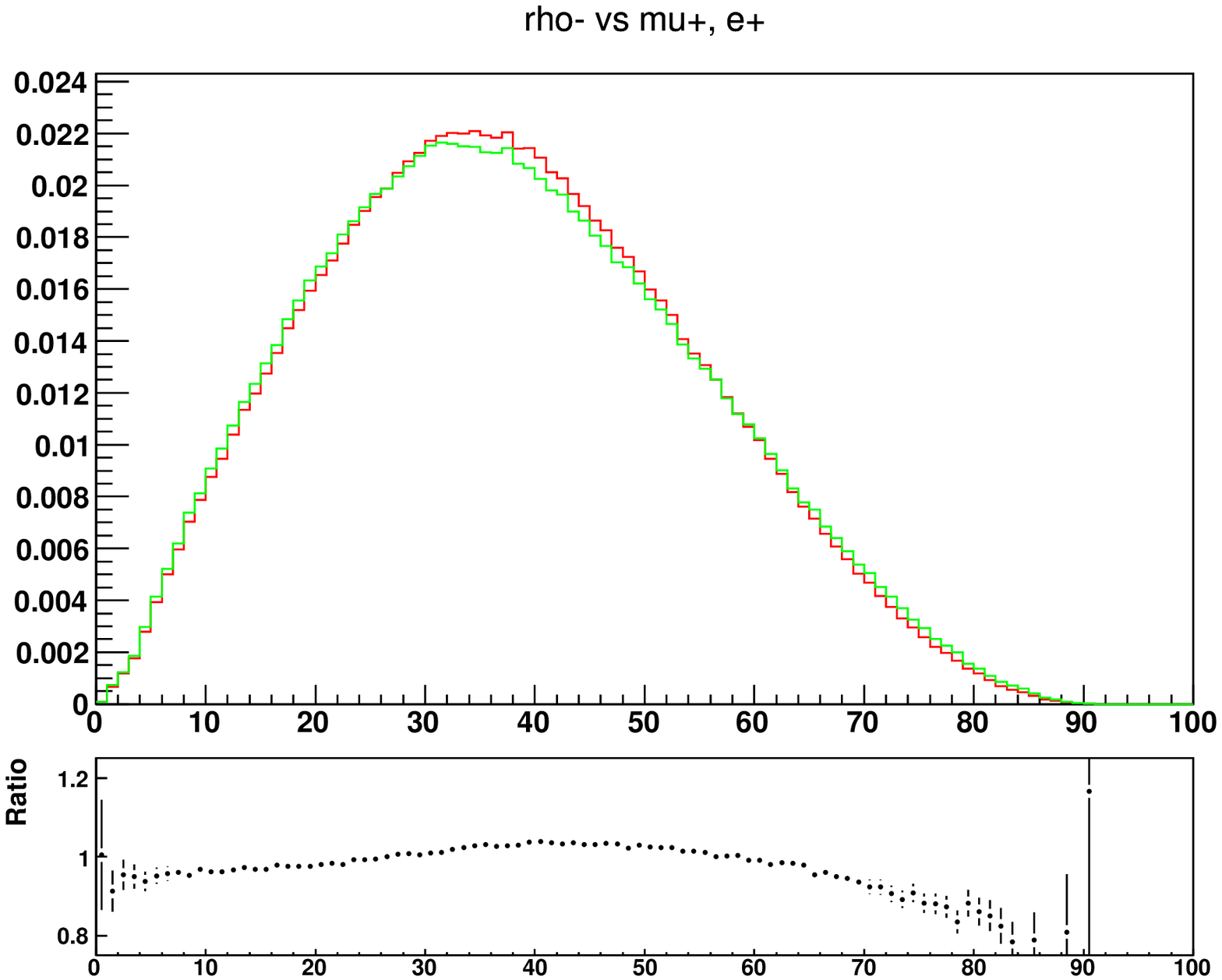}} \\
\resizebox*{0.49\textwidth}{!}{\includegraphics{\przedro 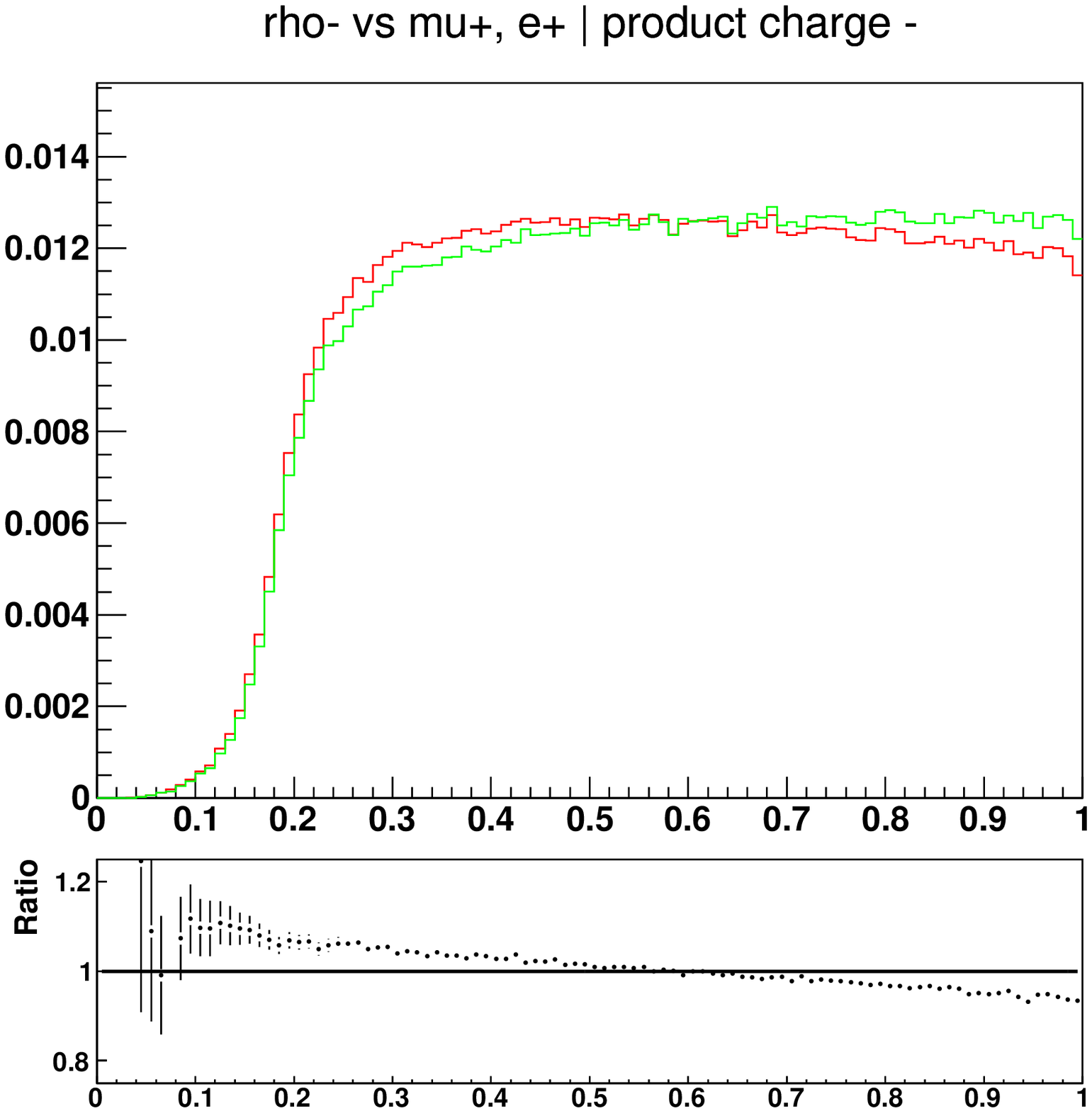}}
\resizebox*{0.49\textwidth}{!}{\includegraphics{\przedro 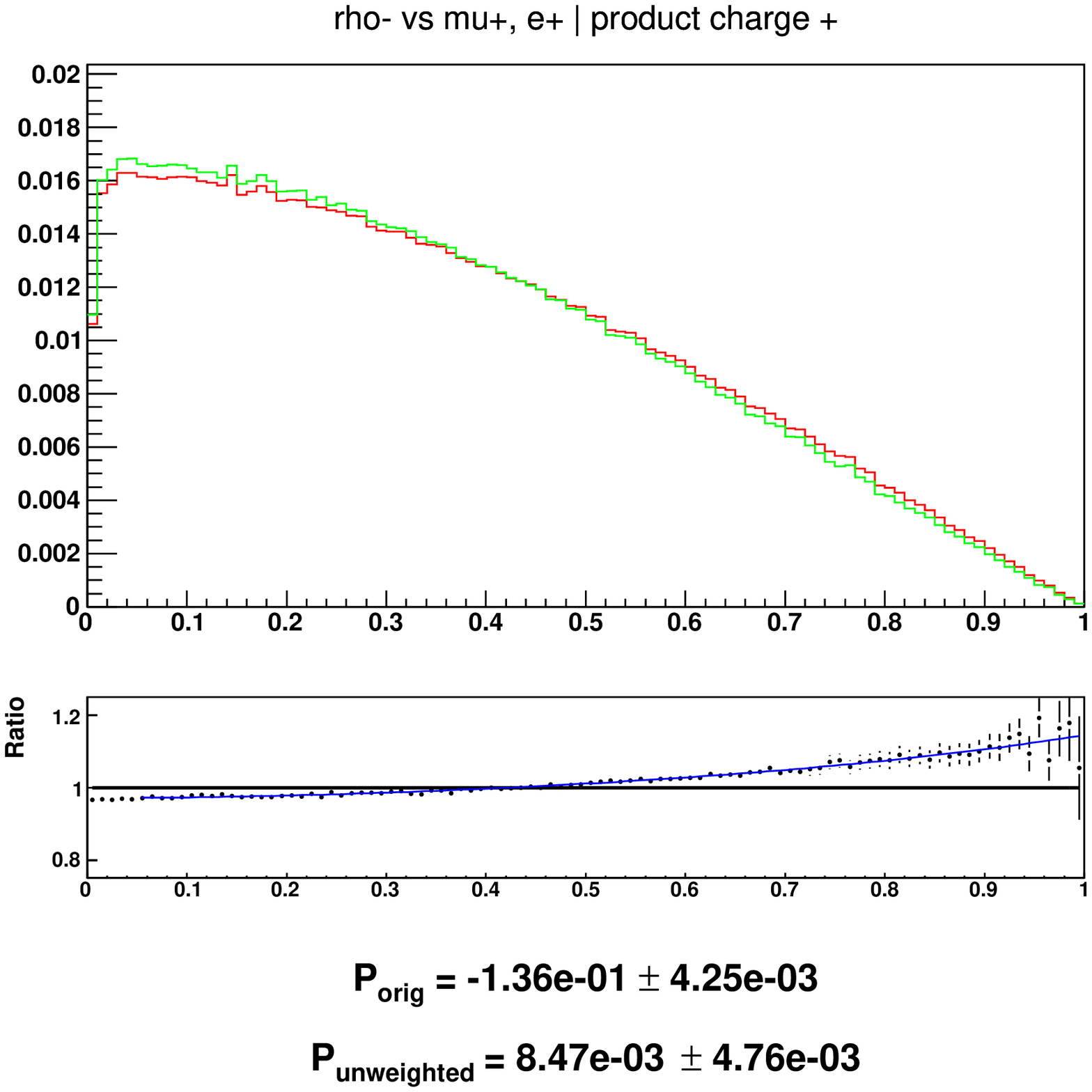}} \\
\caption{\small Fractions of  $\tau^+$ and $\tau^-$ energies carried by their visible  decay products:
two dimensional lego plots and one dimensional spectra$^{18}$.
\textcolor{red}{Red line} is  for original sample,
\textcolor{green}{green line} \greenlineis
black line is ratio \textcolor{red}{original}/\textcolor{green}{modified} with whenever available superimposed result for the
fitted functions.
}
\end{figure}

\newpage
\subsubsection{The energy spectrum: $\tau^- \to \pi^-$ {\tt vs } $\tau^+ \to \rho^+$}
\vspace{3\baselineskip}

\begin{figure}[h!]
\centering
\resizebox*{0.49\textwidth}{!}{\includegraphics{\przedro 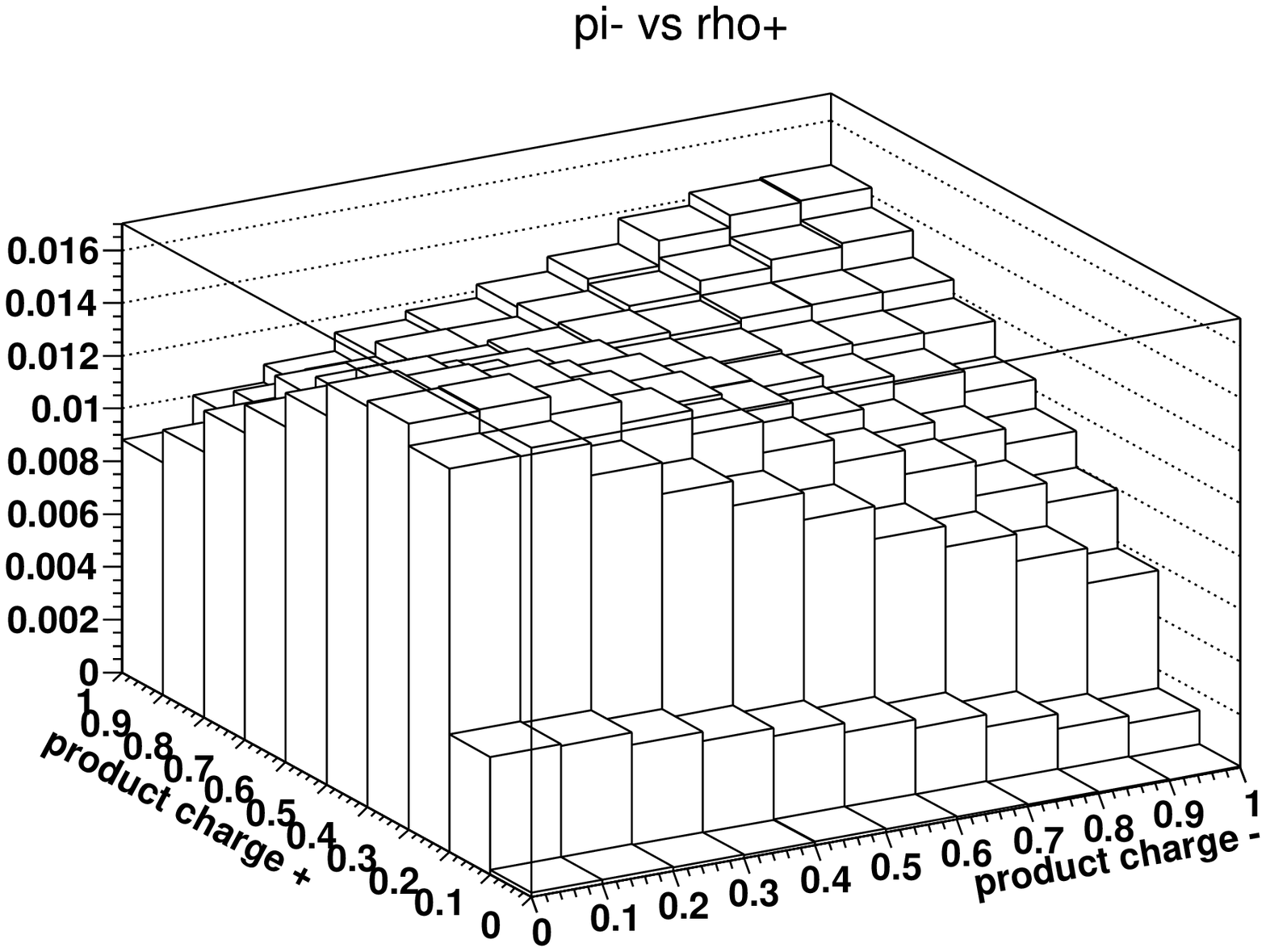}}
\resizebox*{0.49\textwidth}{!}{\includegraphics{\przedro 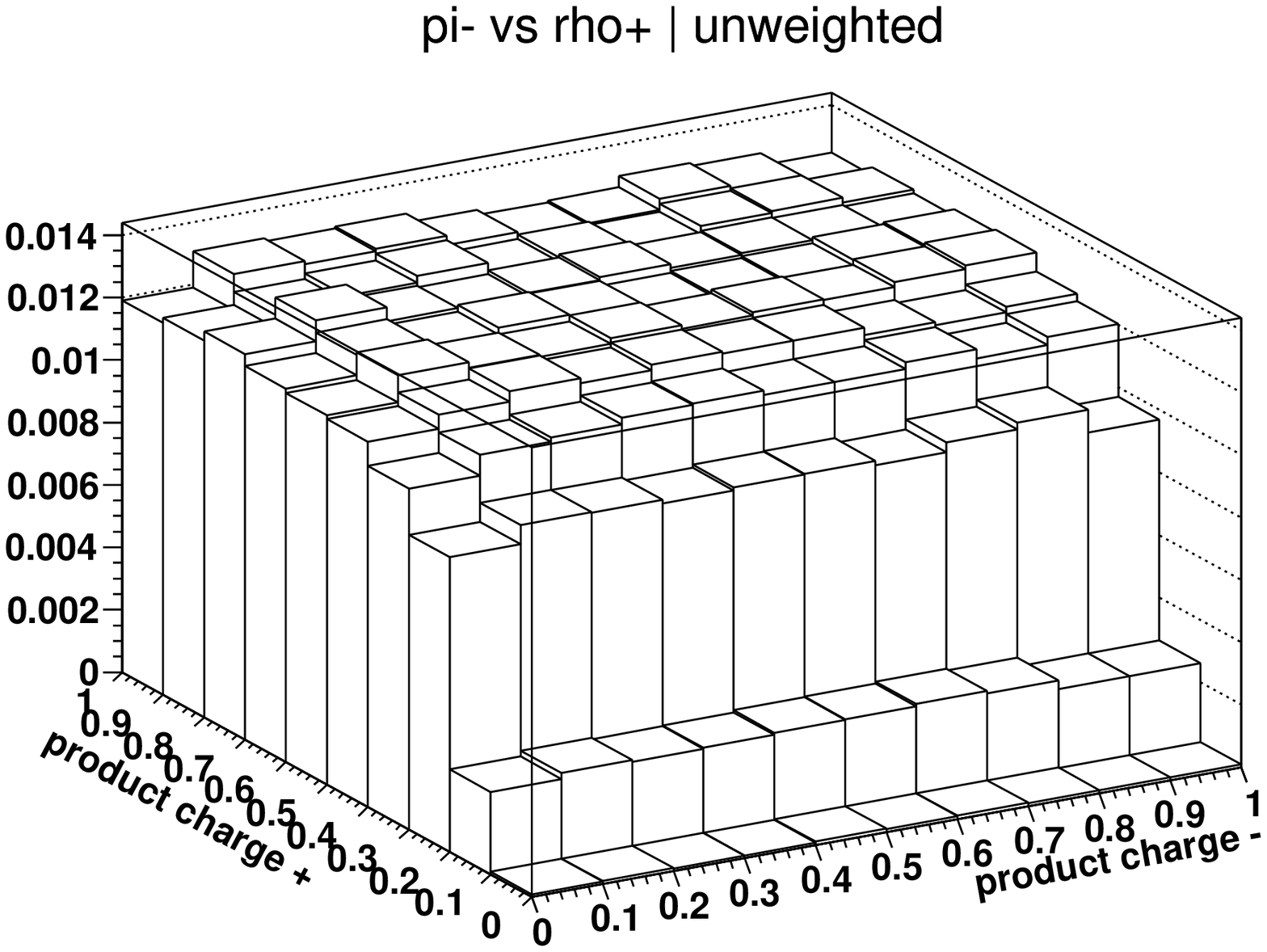}} \\
\resizebox*{0.49\textwidth}{!}{\includegraphics{\przedro 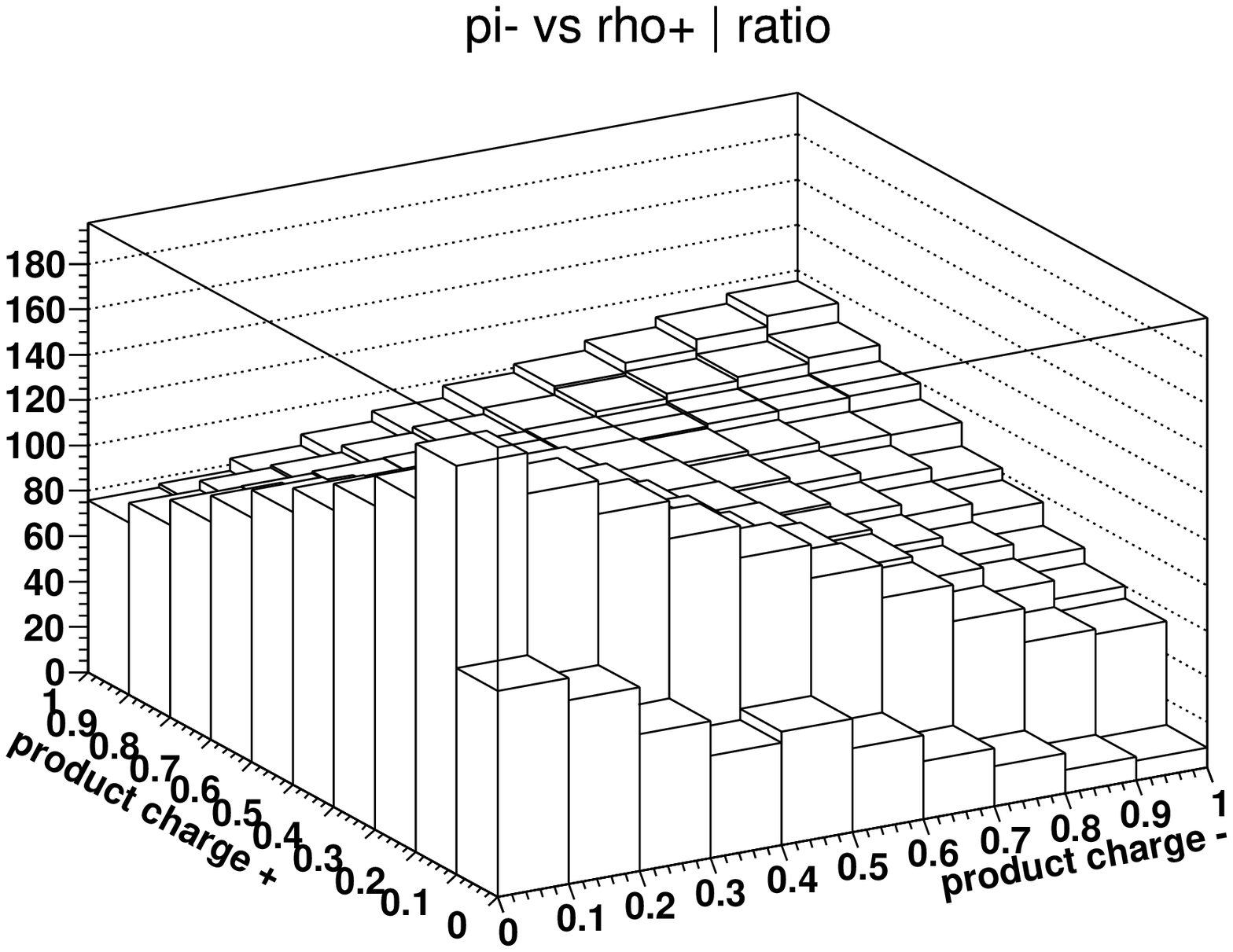}}
\resizebox*{0.49\textwidth}{!}{\includegraphics{\przedro 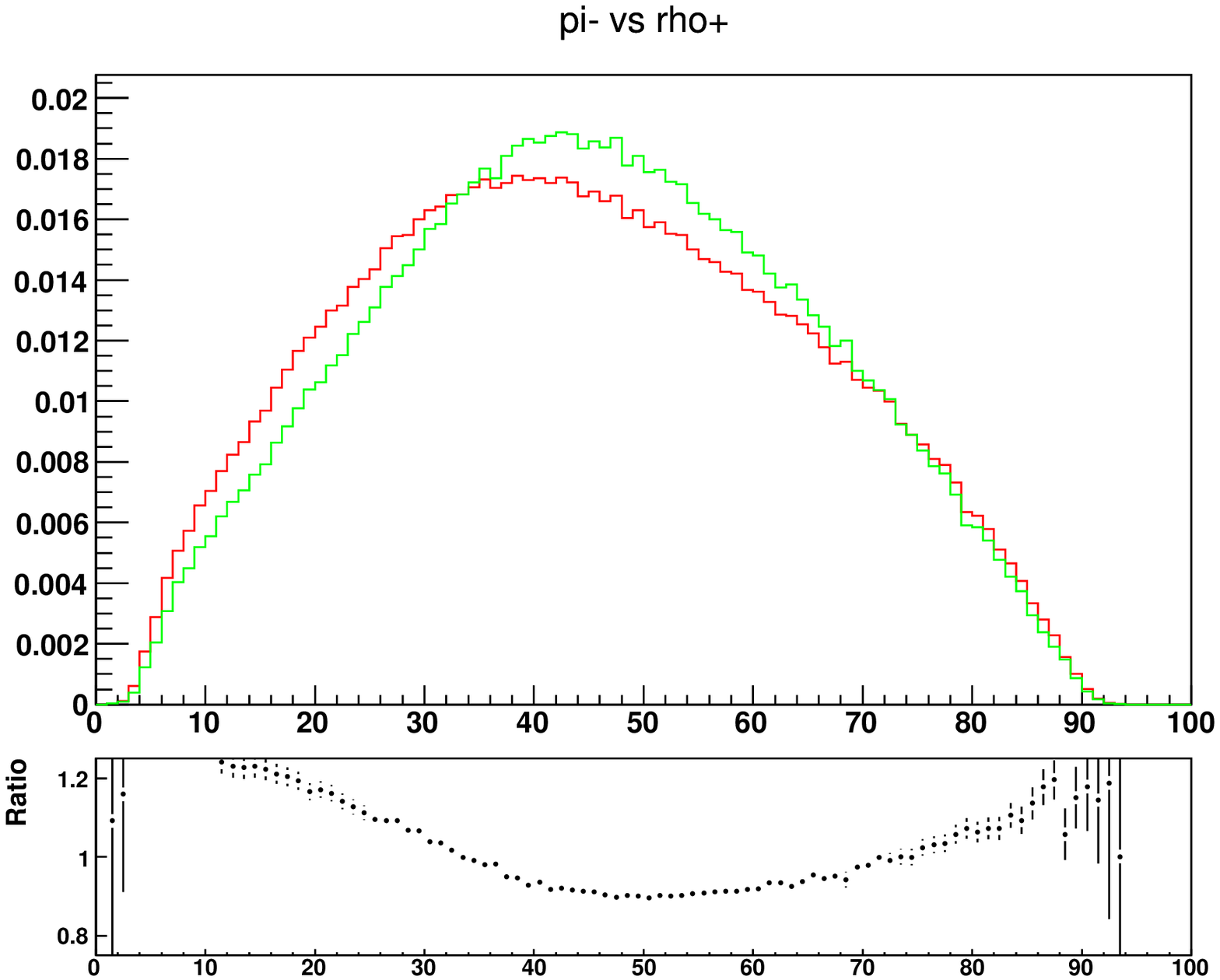}} \\
\resizebox*{0.49\textwidth}{!}{\includegraphics{\przedro 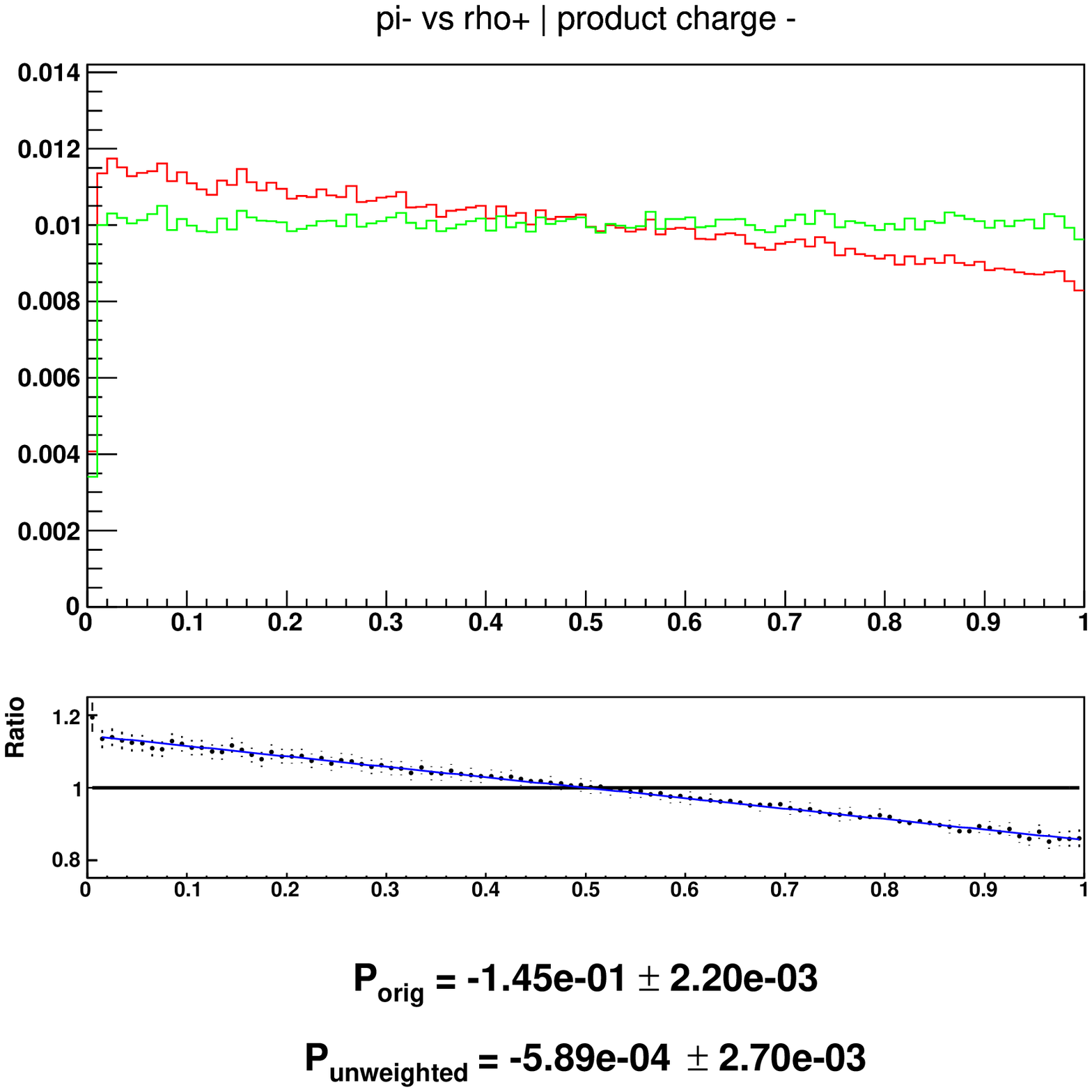}}
\resizebox*{0.49\textwidth}{!}{\includegraphics{\przedro 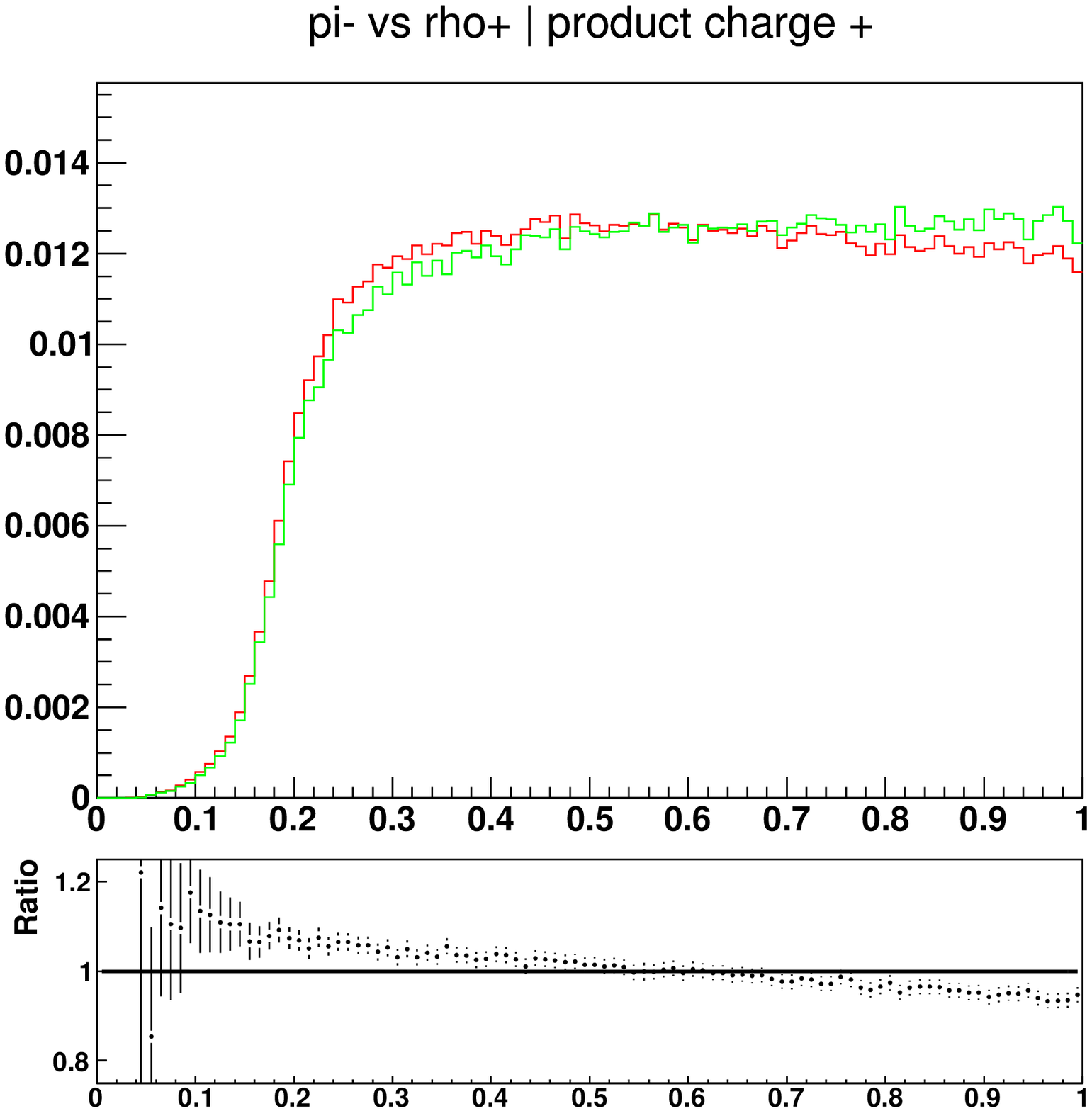}} \\
\caption{\small Fractions of  $\tau^+$ and $\tau^-$ energies carried by their visible  decay products:
two dimensional lego plots and one dimensional spectra$^{18}$.
\textcolor{red}{Red line} is  for original sample,
\textcolor{green}{green line} \greenlineis
black line is ratio \textcolor{red}{original}/\textcolor{green}{modified} with whenever available superimposed result for the
fitted functions.
}
\end{figure}

\newpage
\subsubsection{The energy spectrum: $\tau^- \to \rho^-$ {\tt vs } $\tau^+ \to \pi^+$}
\vspace{3\baselineskip}

\begin{figure}[h!]
\centering
\resizebox*{0.49\textwidth}{!}{\includegraphics{\przedro 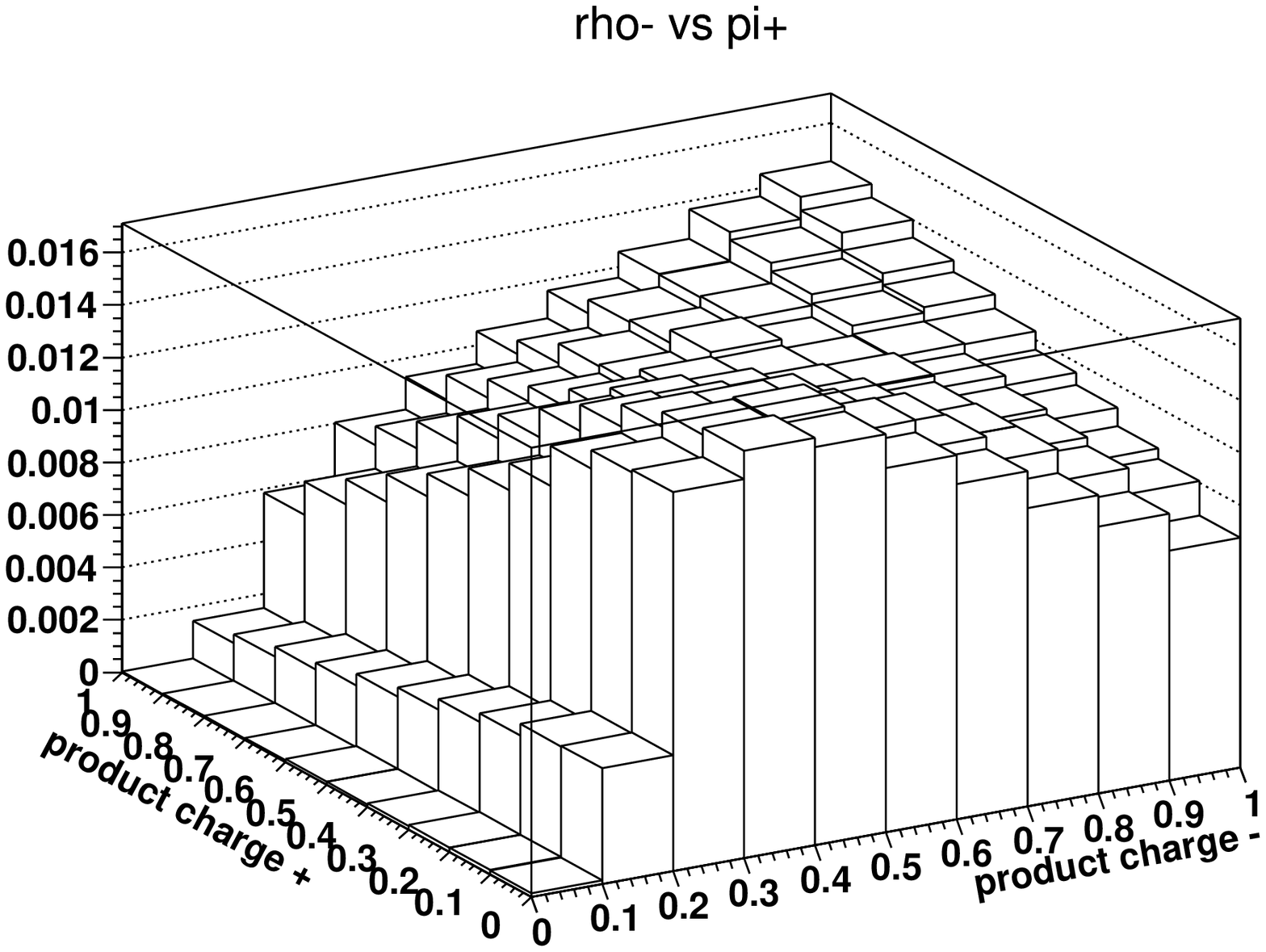}}
\resizebox*{0.49\textwidth}{!}{\includegraphics{\przedro 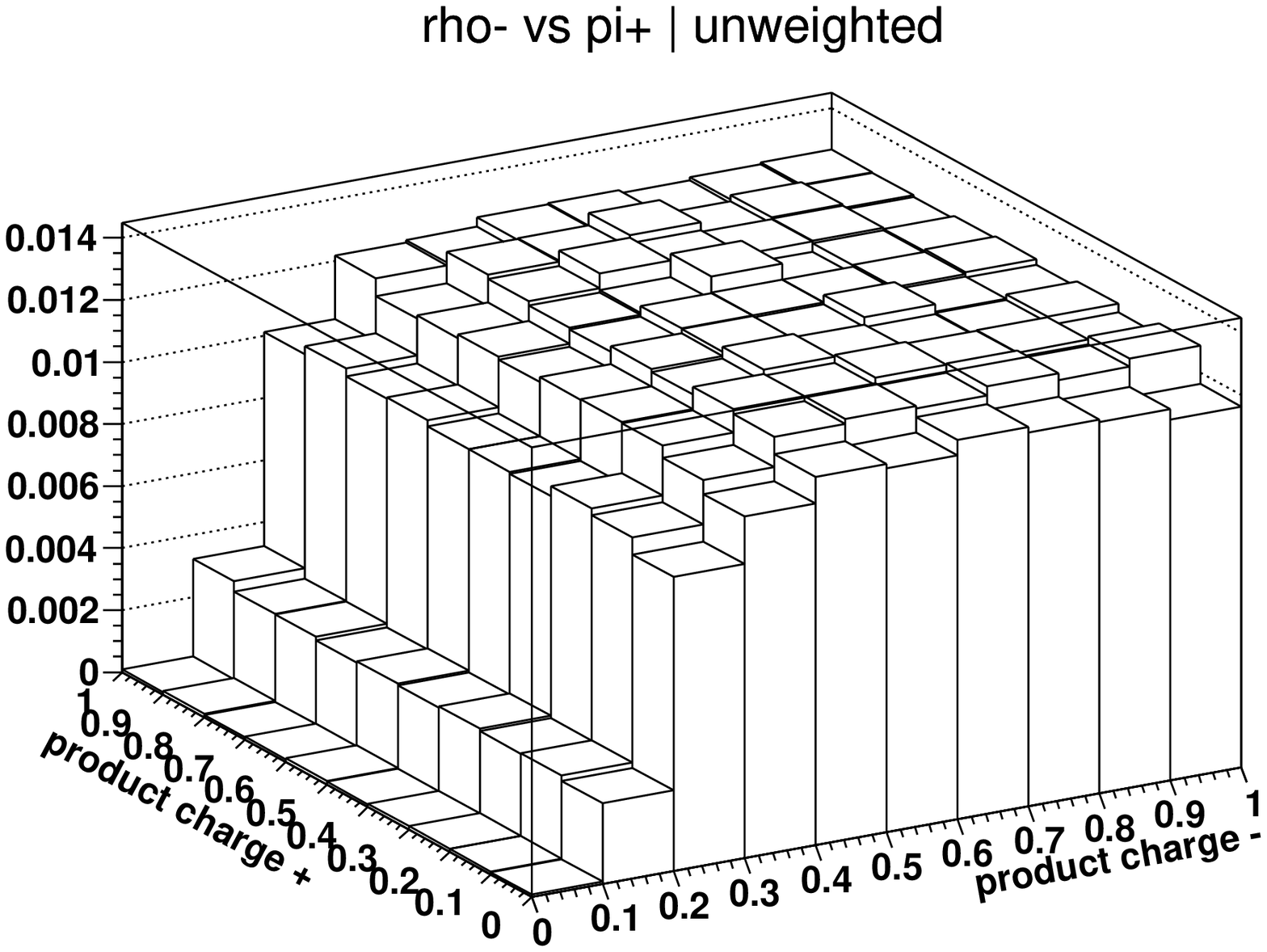}} \\
\resizebox*{0.49\textwidth}{!}{\includegraphics{\przedro 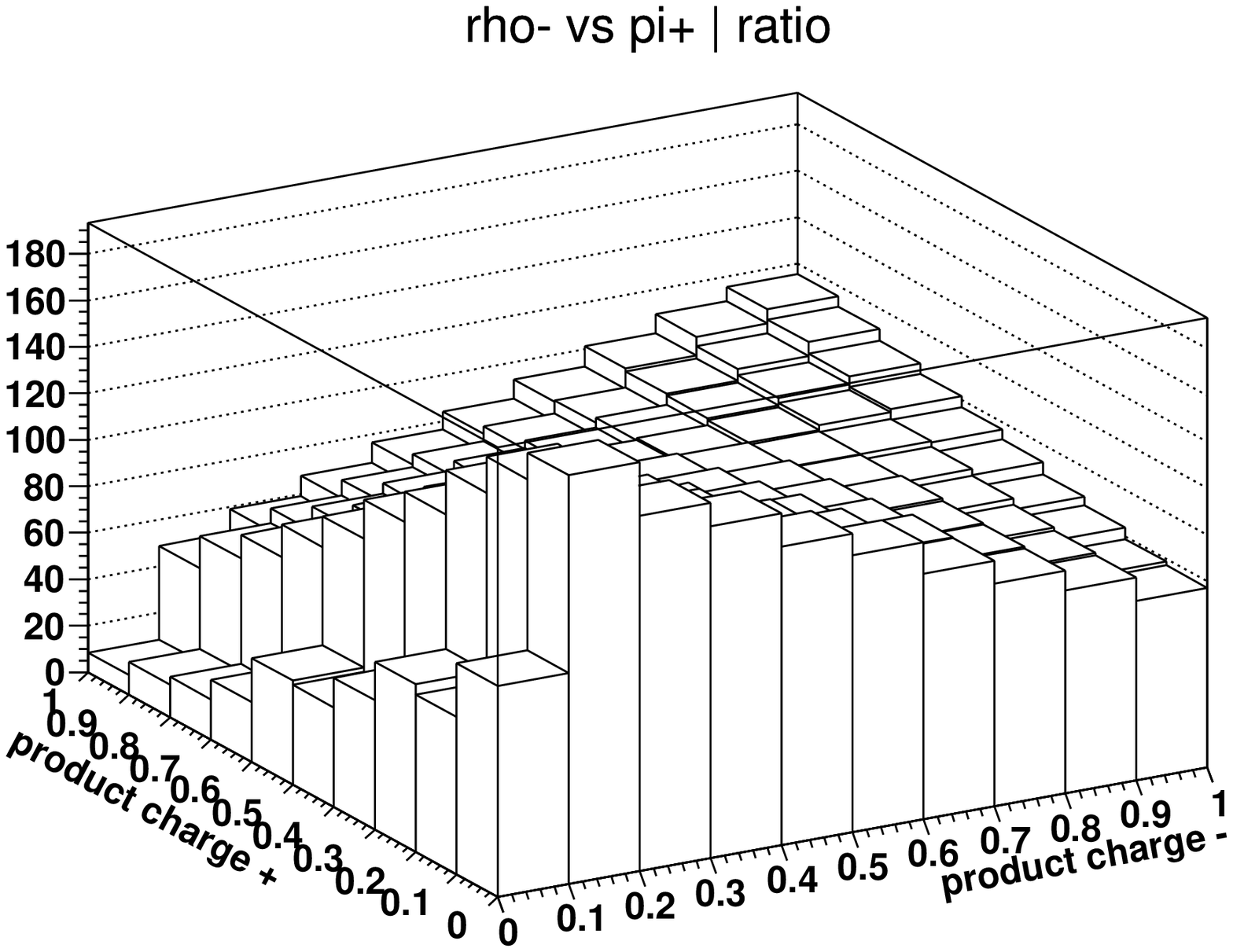}}
\resizebox*{0.49\textwidth}{!}{\includegraphics{\przedro 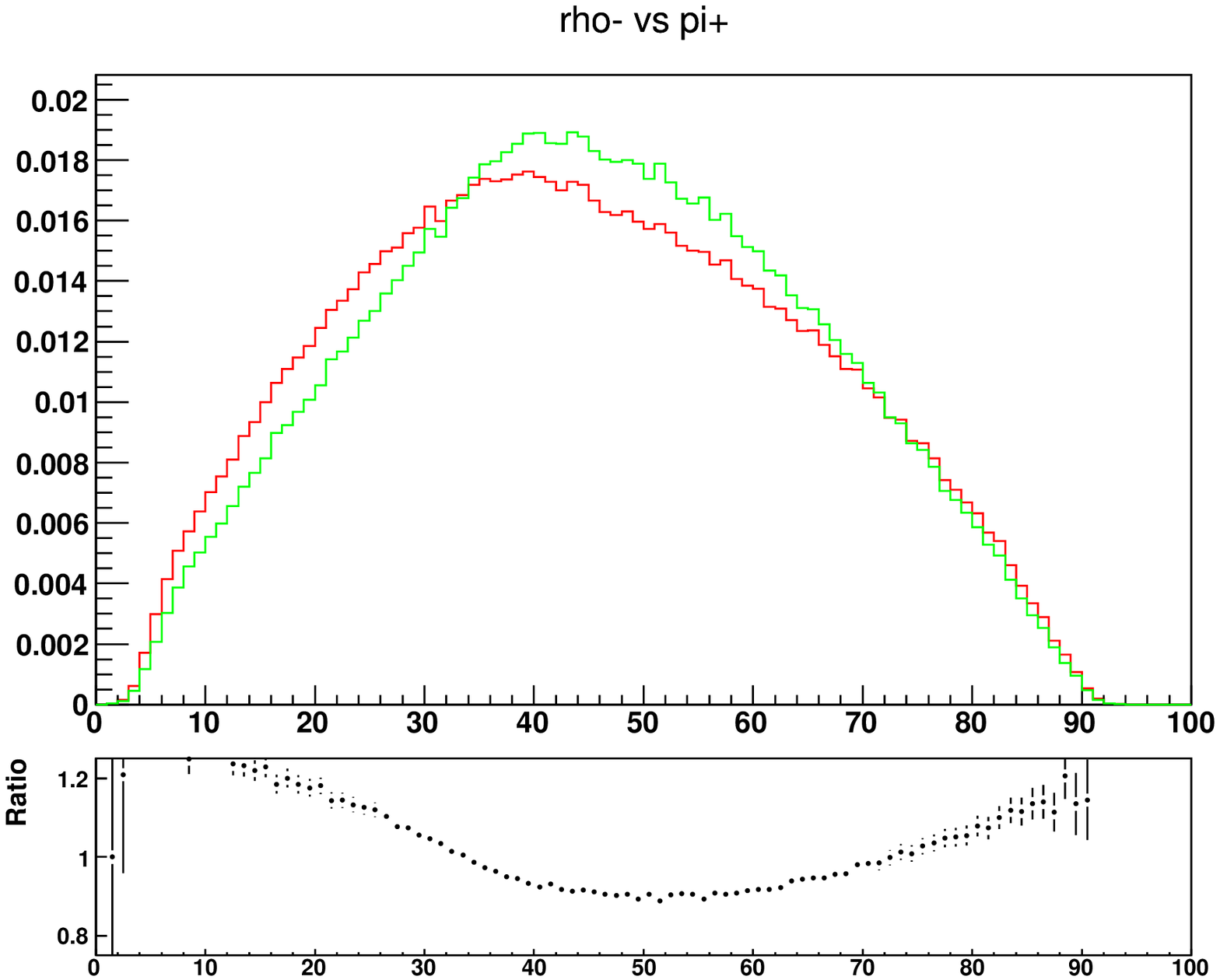}} \\
\resizebox*{0.49\textwidth}{!}{\includegraphics{\przedro 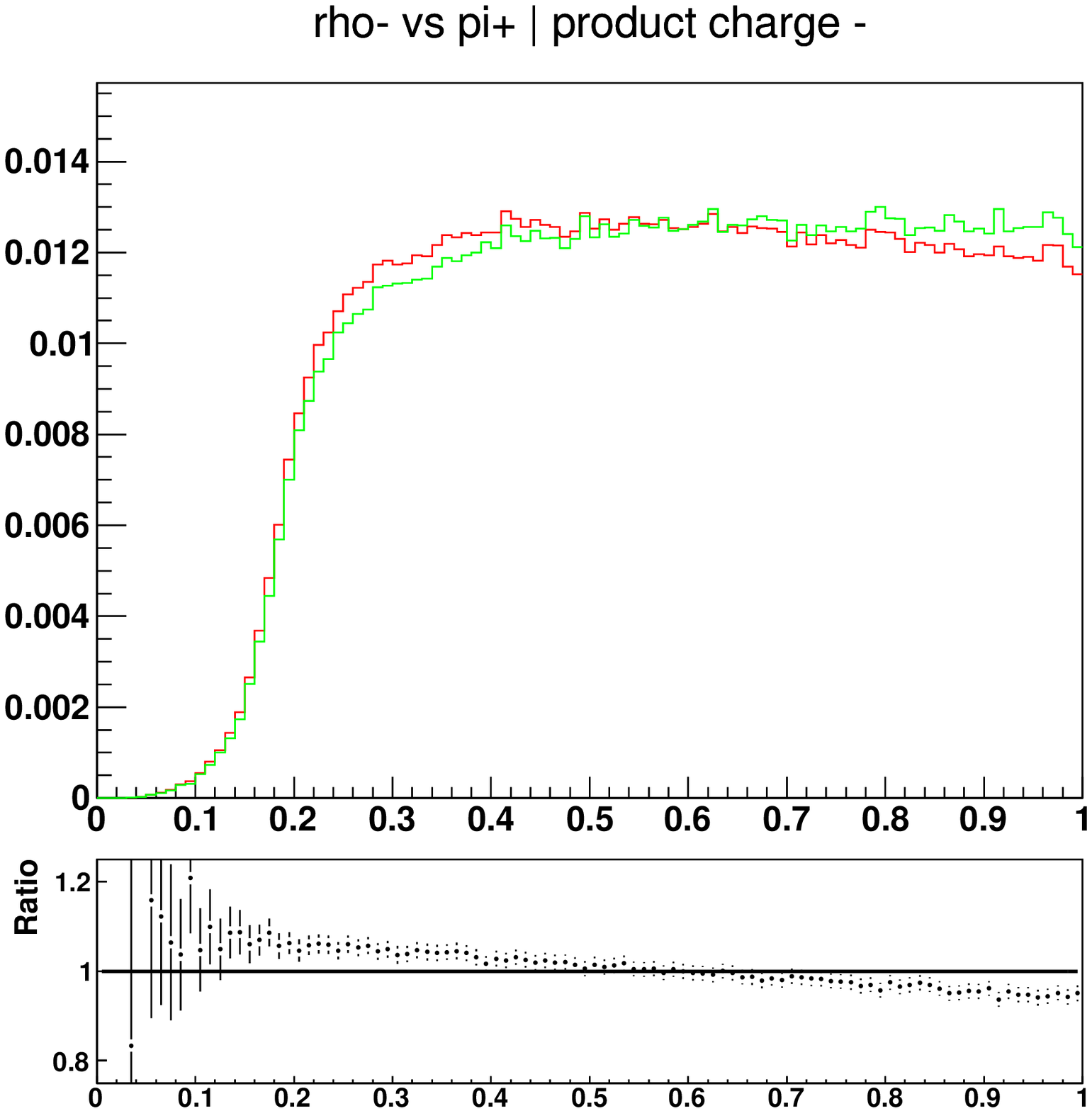}}
\resizebox*{0.49\textwidth}{!}{\includegraphics{\przedro 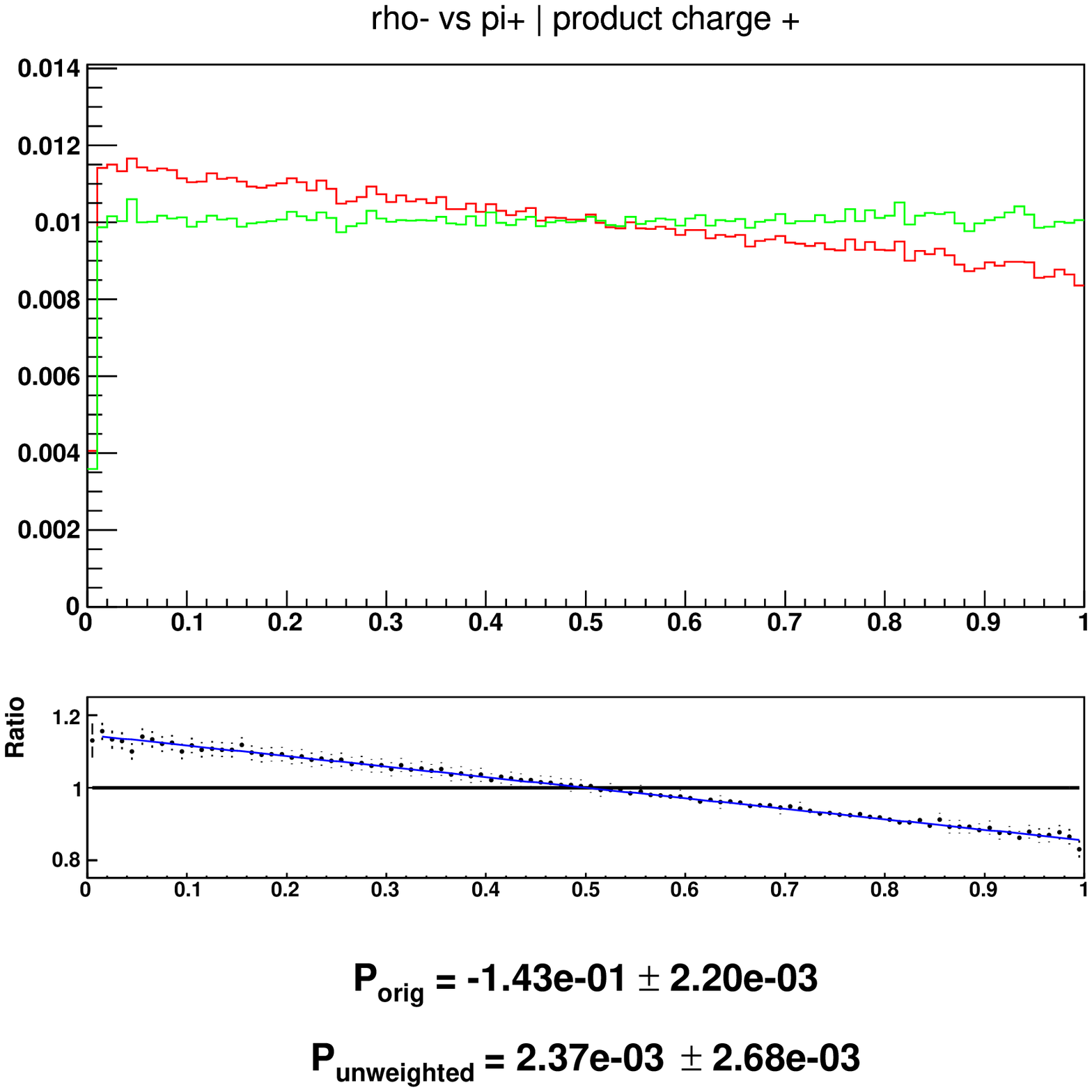}} \\
\caption{\small Fractions of  $\tau^+$ and $\tau^-$ energies carried by their visible  decay products:
two dimensional lego plots and one dimensional spectra$^{18}$.
\textcolor{red}{Red line} is  for original sample,
\textcolor{green}{green line} \greenlineis
black line is ratio \textcolor{red}{original}/\textcolor{green}{modified} with whenever available superimposed result for the
fitted functions.
}
\end{figure}

\newpage
\subsubsection{The energy spectrum: $\tau^- \to \rho^-$ {\tt vs } $\tau^+ \to \rho^+$}
\vspace{3\baselineskip}

\begin{figure}[h!]
\centering
\resizebox*{0.49\textwidth}{!}{\includegraphics{\przedro 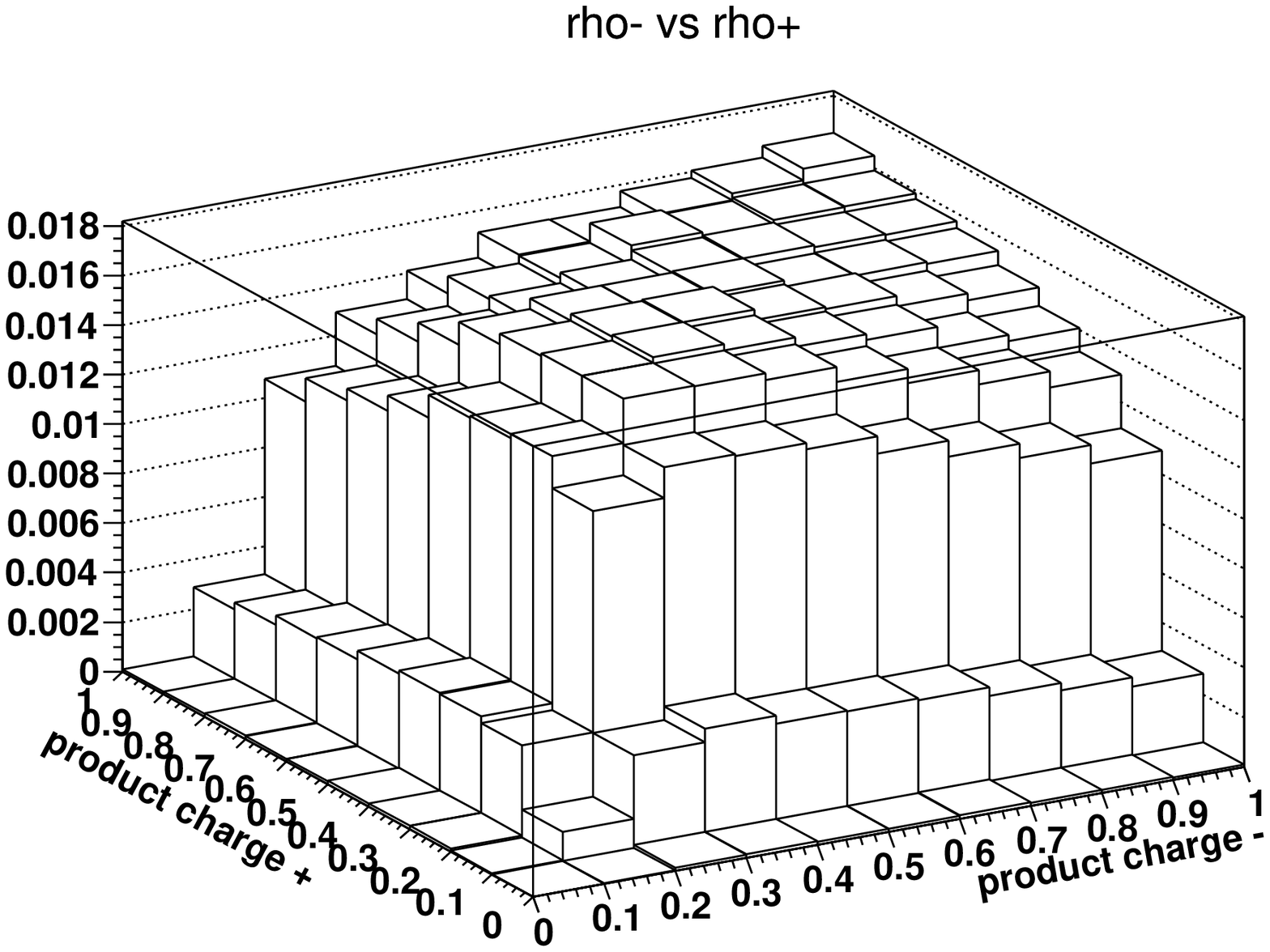}}
\resizebox*{0.49\textwidth}{!}{\includegraphics{\przedro 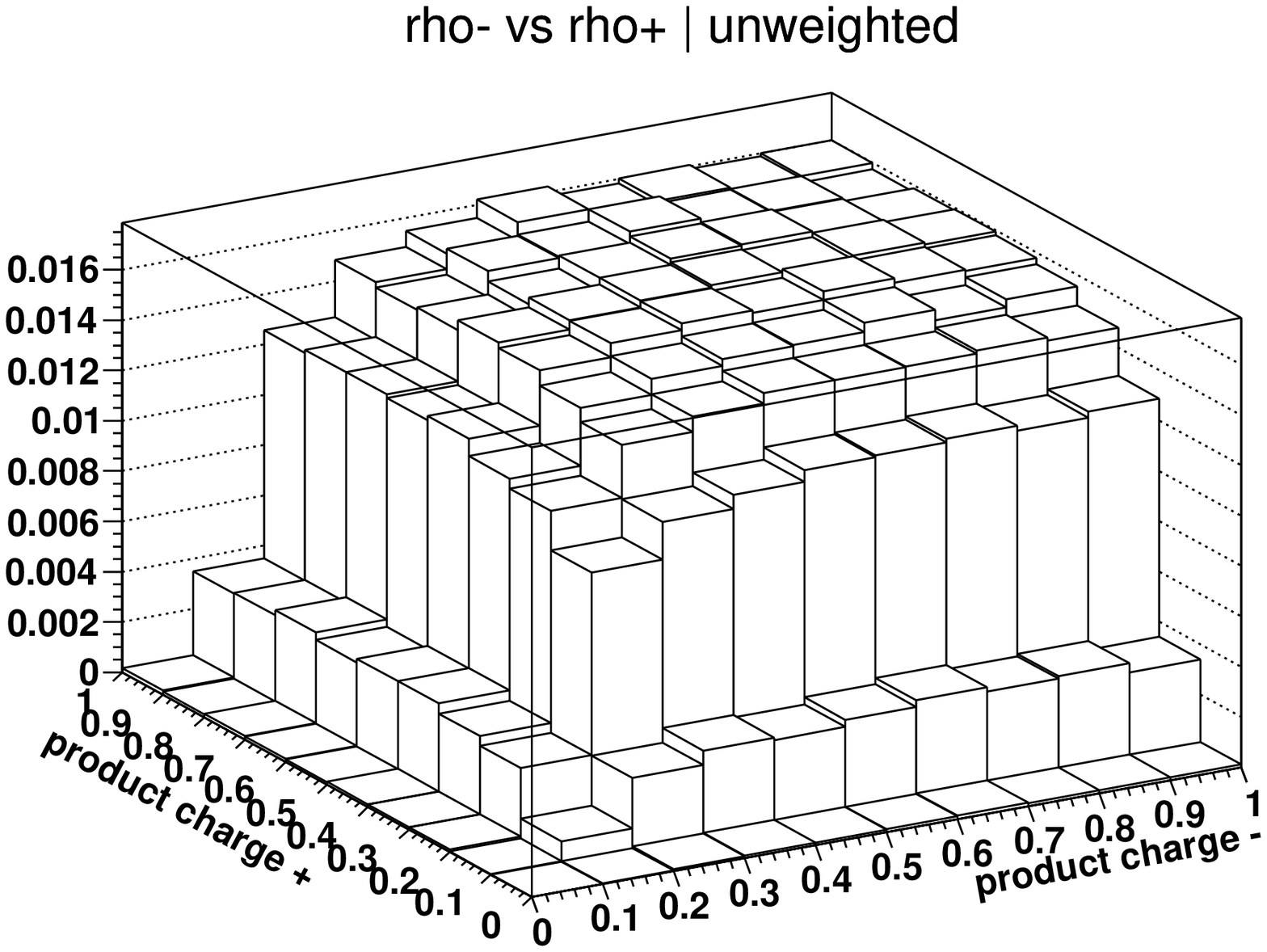}} \\
\resizebox*{0.49\textwidth}{!}{\includegraphics{\przedro 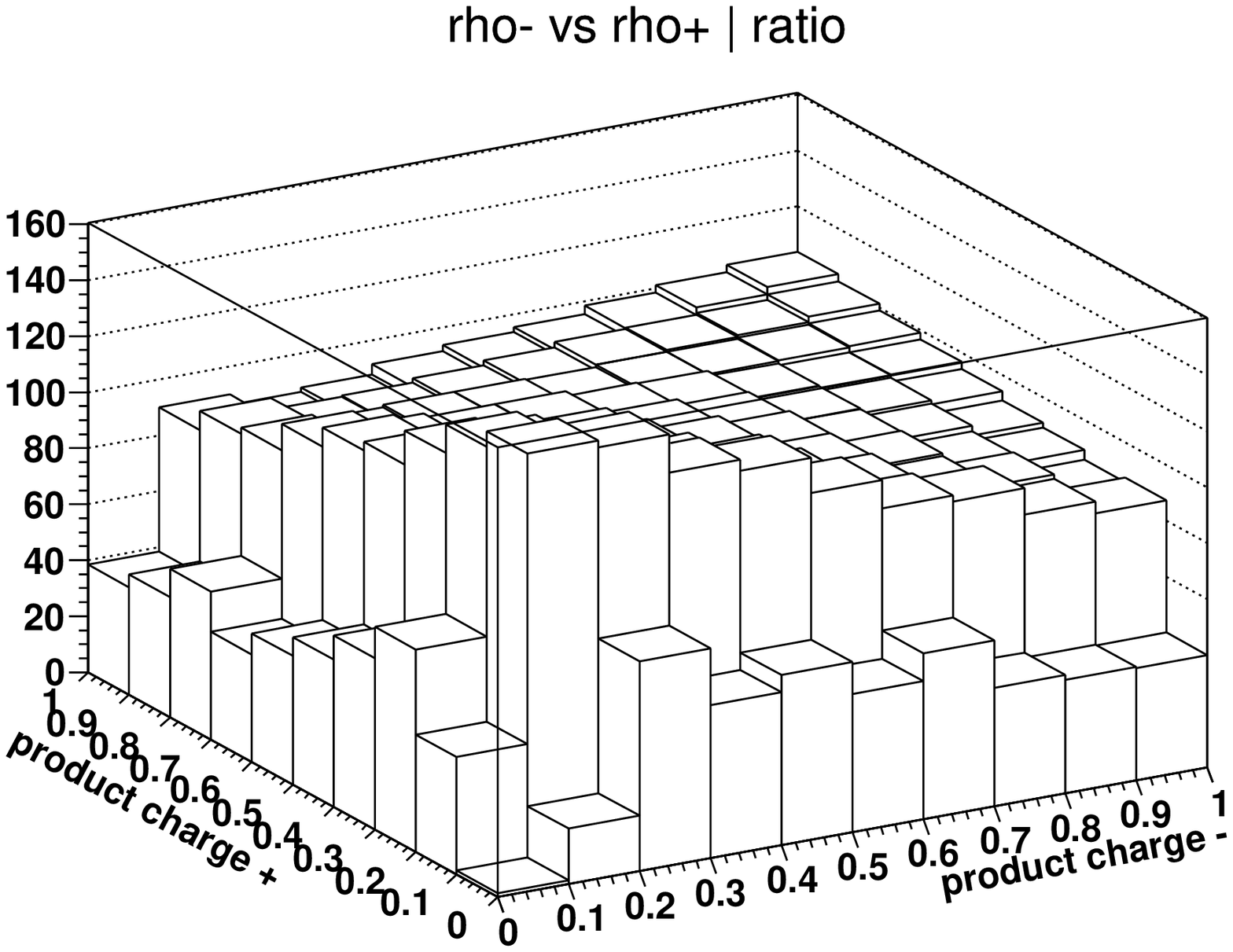}}
\resizebox*{0.49\textwidth}{!}{\includegraphics{\przedro 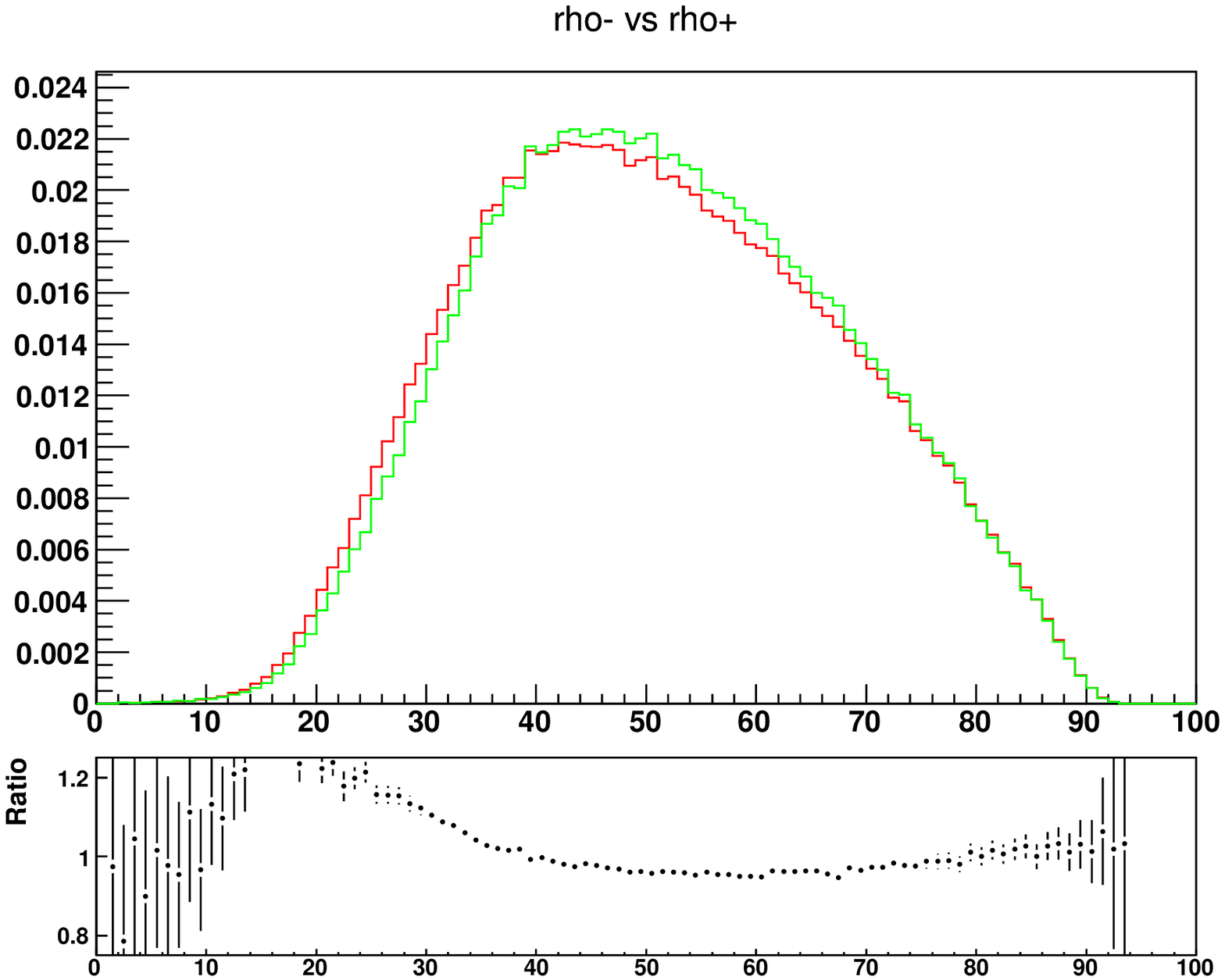}} \\
\resizebox*{0.49\textwidth}{!}{\includegraphics{\przedro 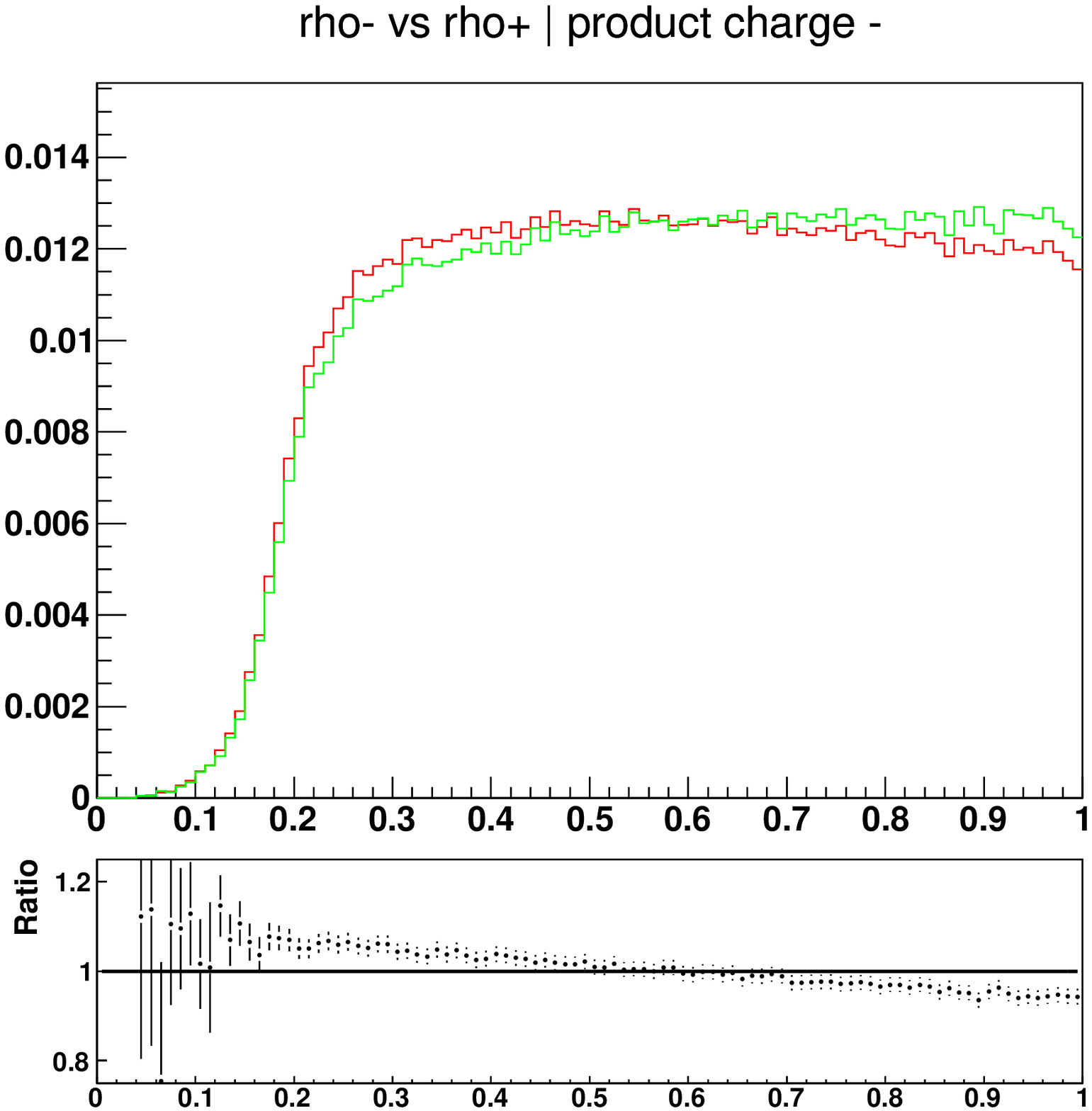}}
\resizebox*{0.49\textwidth}{!}{\includegraphics{\przedro 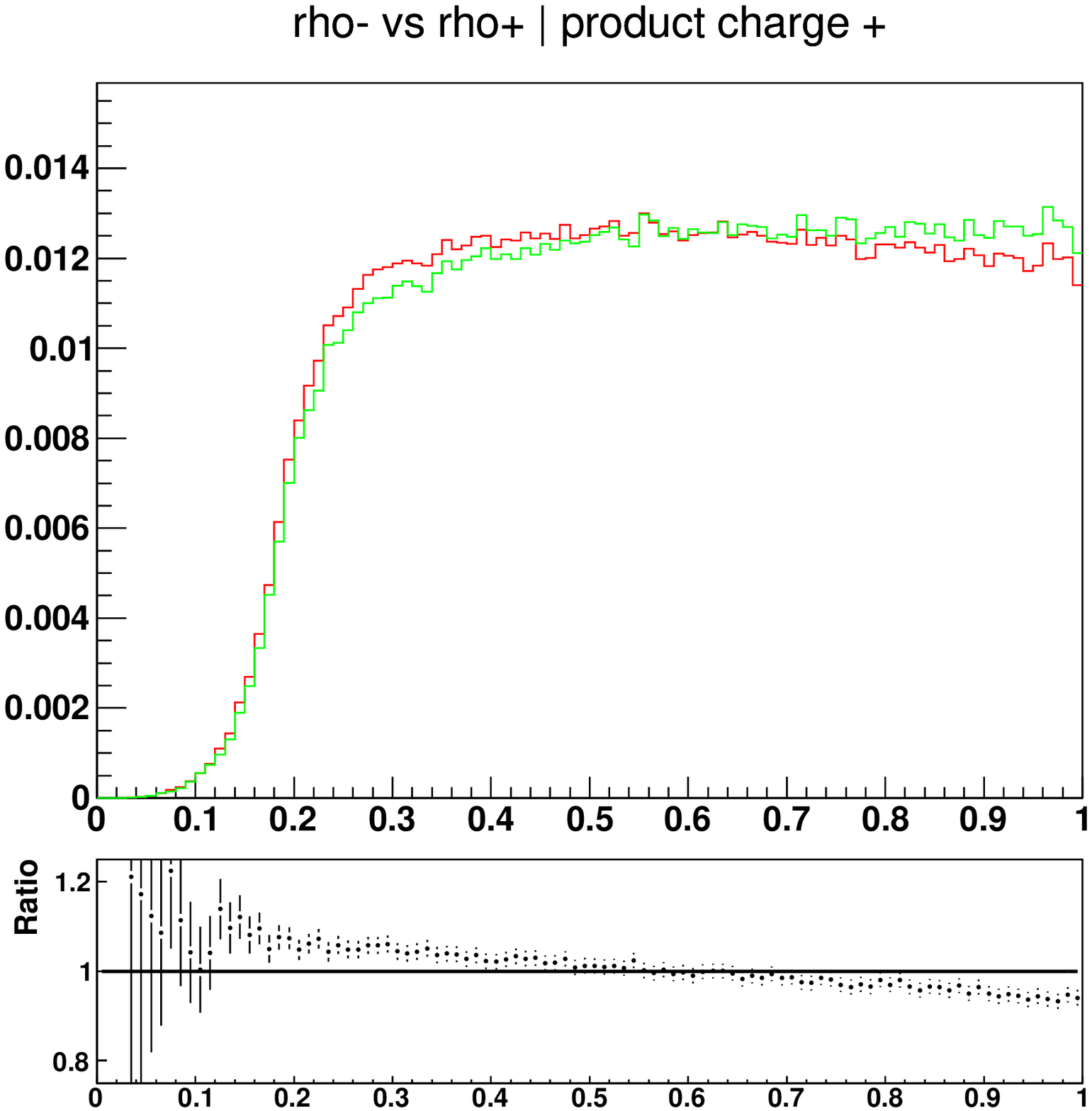}} \\
\caption{\small Fractions of  $\tau^+$ and $\tau^-$ energies carried by their visible  decay products:
two dimensional lego plots and one dimensional spectra.
\textcolor{red}{Red line} is  for original sample,
\textcolor{green}{green line} \greenlineis
black line is ratio \textcolor{red}{original}/\textcolor{green}{modified}.
}\label{Fig:spectra3}
\end{figure}

%% file: appendixB.tex

\newpage
\section{$\Phi$ decays: Z decay sample but with Higgs couplings used for  spin }\label{sec:OutputH} 
 
In this section, we monitor effects of Higgs couplings for spin correlations
and compare it with case of unpolarized $\tau$'s.
As expected, one can observe strong opposite sign as in $Z$ case, 
spin correlations, and there is  no effect on single $\tau$ decay product spectra.

\setcounter{figure}{0}

The invariant mass distribution and break-down on the $\tau$ decay channels are shown
for $\tau \nu_\tau$-pair originating from $\Phi$ decay.
For this purpose $Z \to \tau \tau$ events were used, with $\tau$ decay products removed 
and $\tau$ leptons decayed again using configuration of {\tt Tauola++} like for $H \to \tau \tau$ decay.
For the fits,  all 100 bins except 
the first five in $\tau \to l \nu_l \nu_\tau $ case  and first one in $\tau \to \pi \nu$ case,  were used.



\vskip 3 mm

\begin{figure}[h!]
\centering 
\resizebox*{0.45\textwidth}{!}{\includegraphics{\przedro 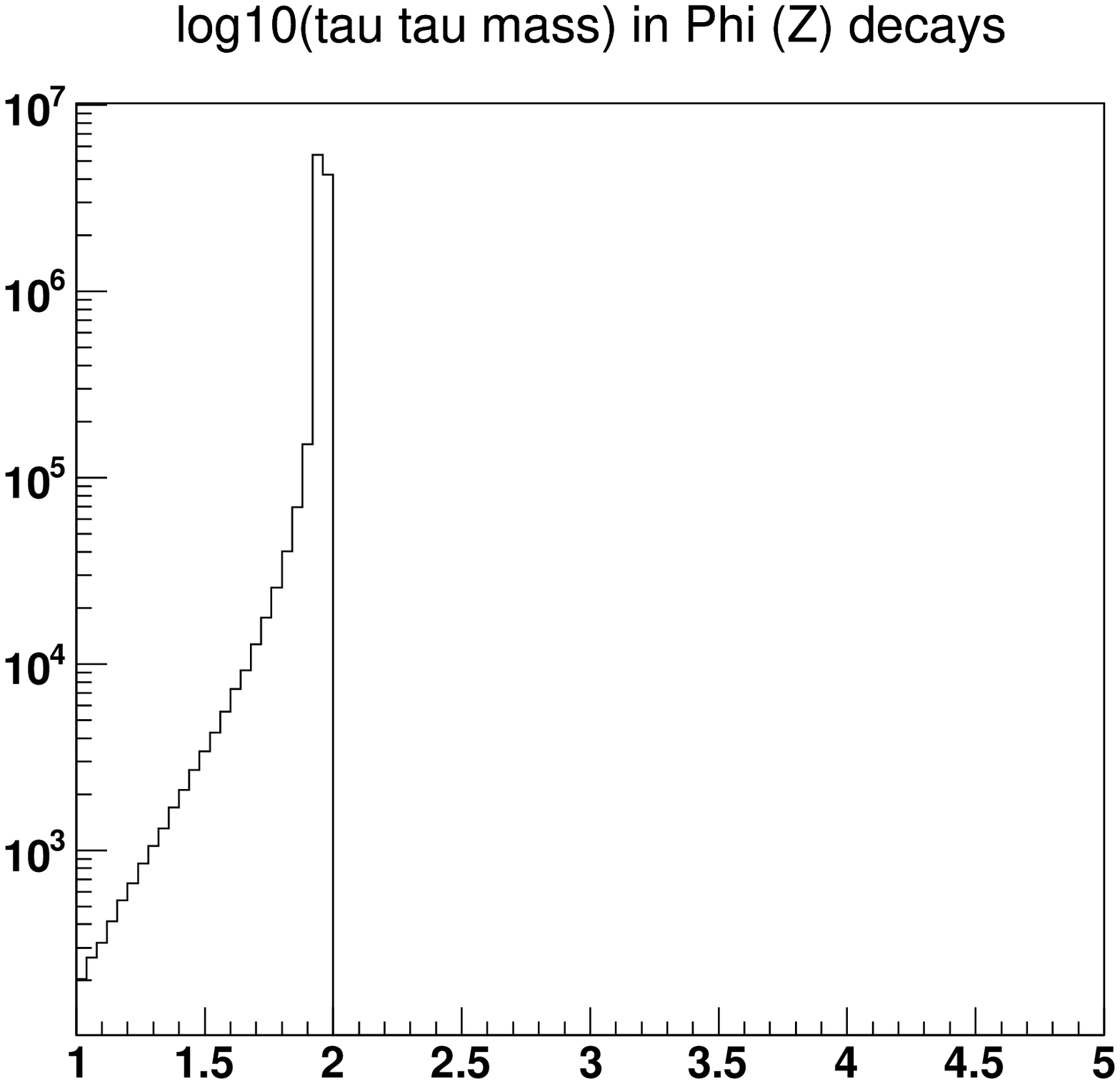}} \\
\end{figure}

{\small \verbinput{input-Phi-event-count.txt} }

\newpage

\subsection{The energy spectrum: $\tau^- \to \mu^-, e^-$ {\tt vs } $\tau^+ \to \mu^+, e^+$}
\vspace{1\baselineskip}

\begin{figure}[h!]
\centering
\resizebox*{0.49\textwidth}{!}{\includegraphics{\przedro 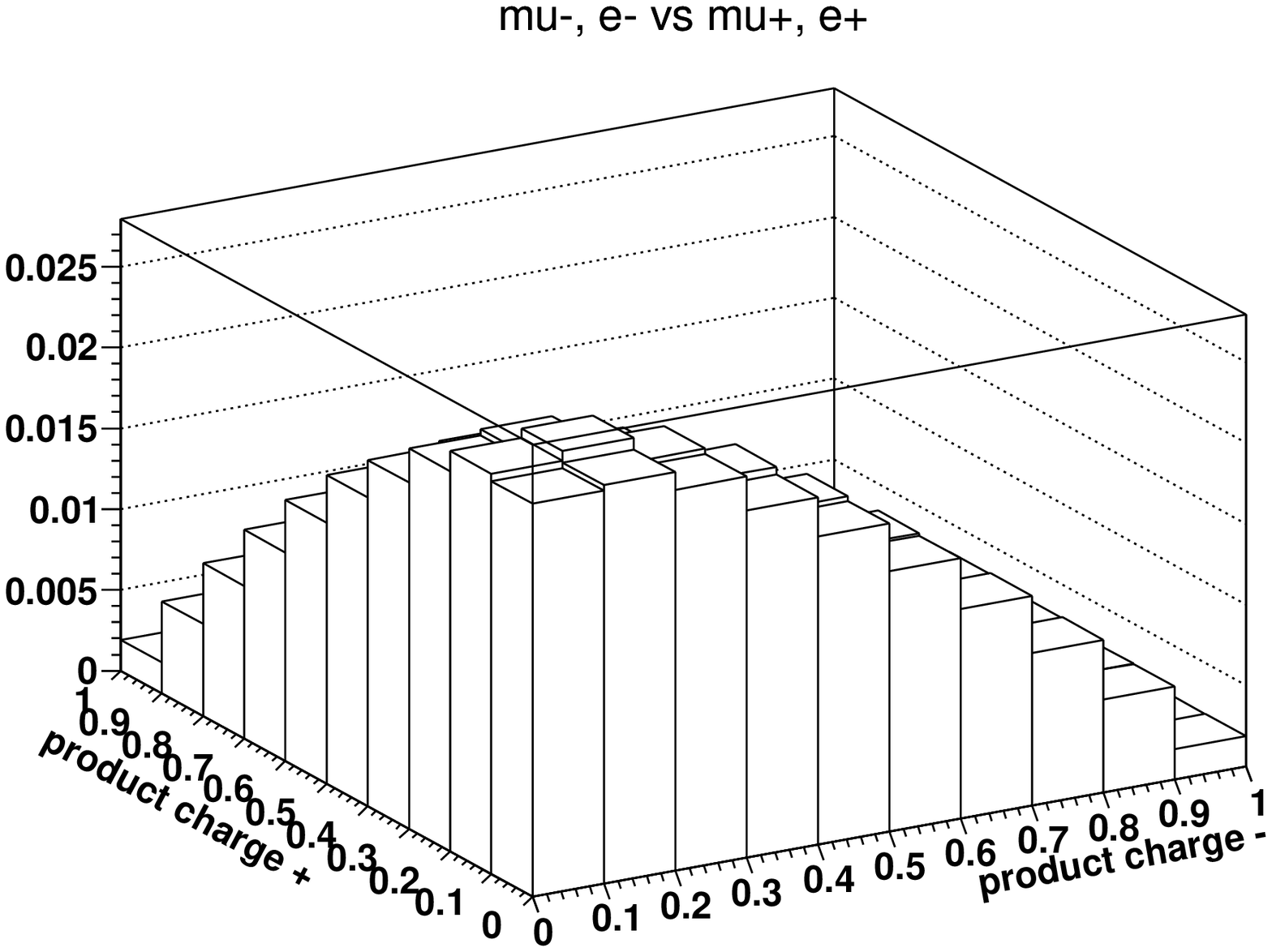}}
\resizebox*{0.49\textwidth}{!}{\includegraphics{\przedro 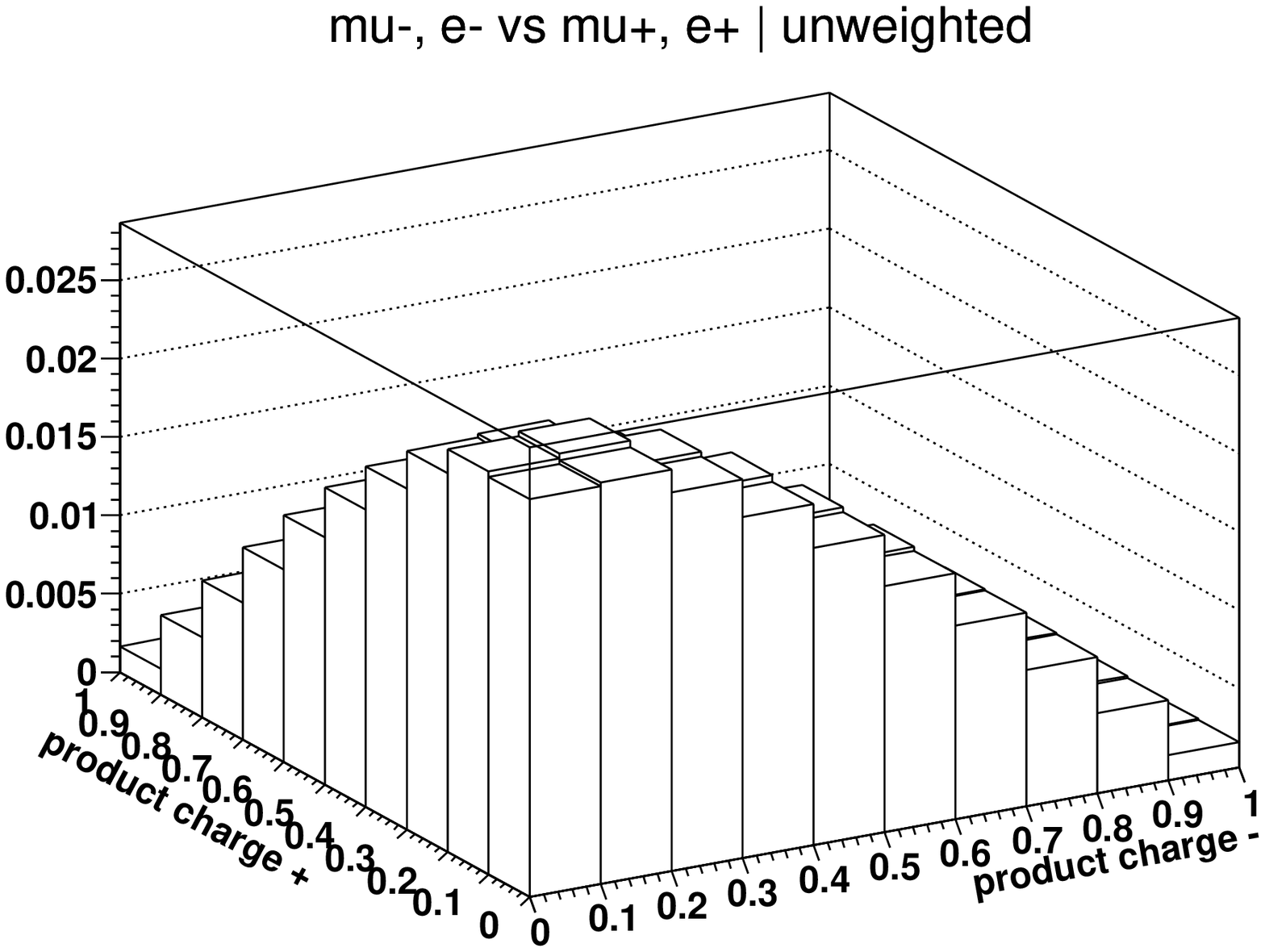}} \\
\resizebox*{0.49\textwidth}{!}{\includegraphics{\przedro 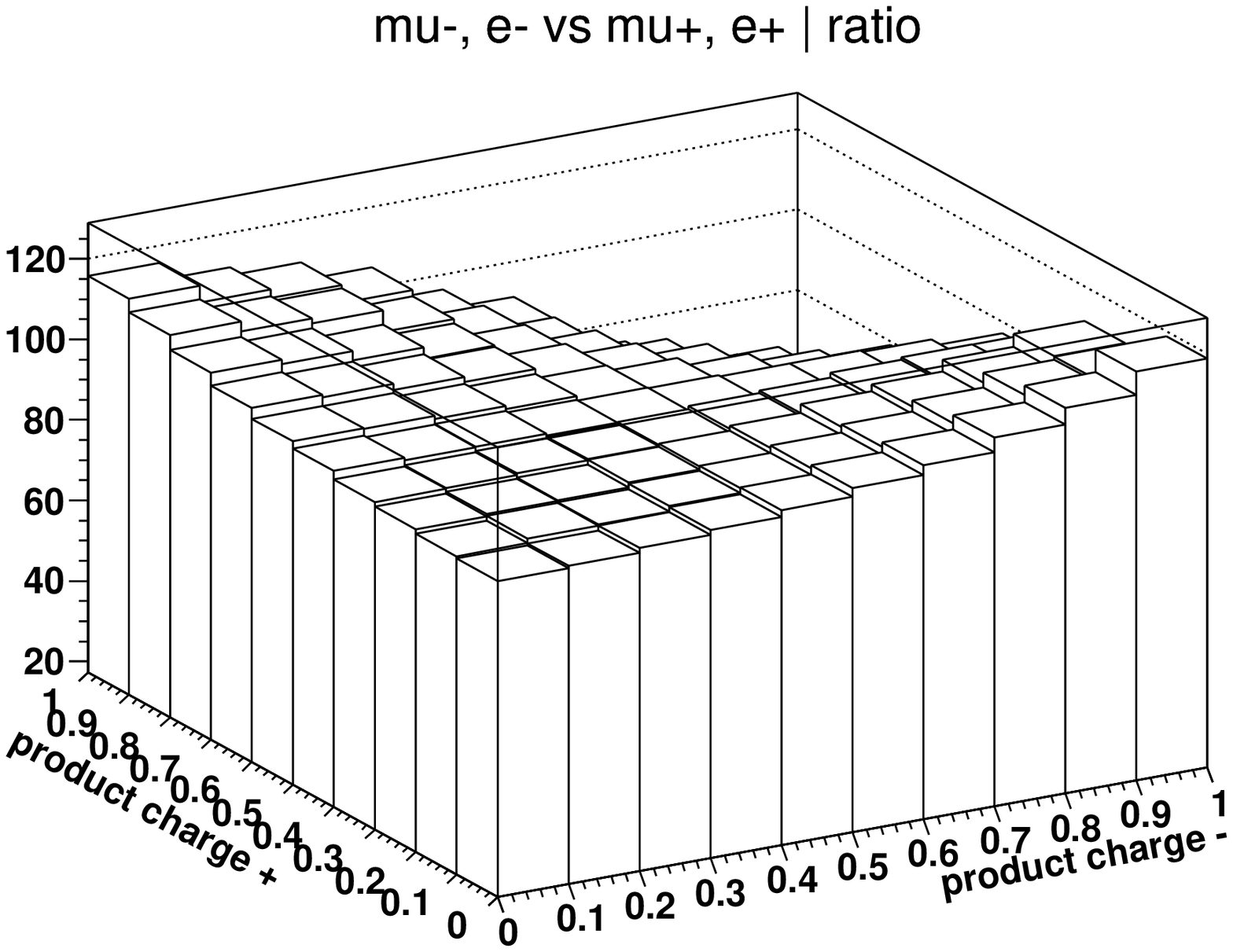}}
\resizebox*{0.49\textwidth}{!}{\includegraphics{\przedro 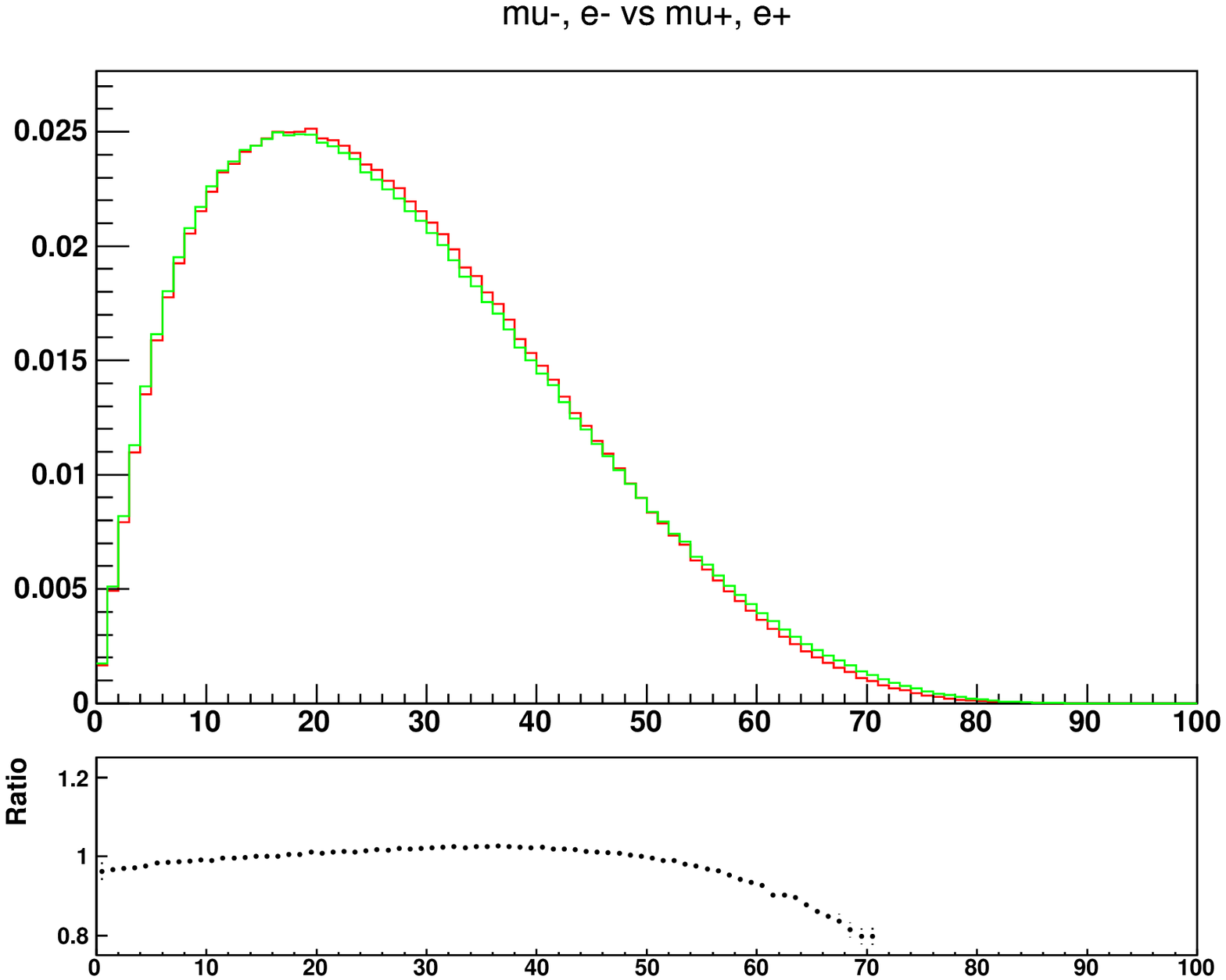}} \\
\resizebox*{0.49\textwidth}{!}{\includegraphics{\przedro 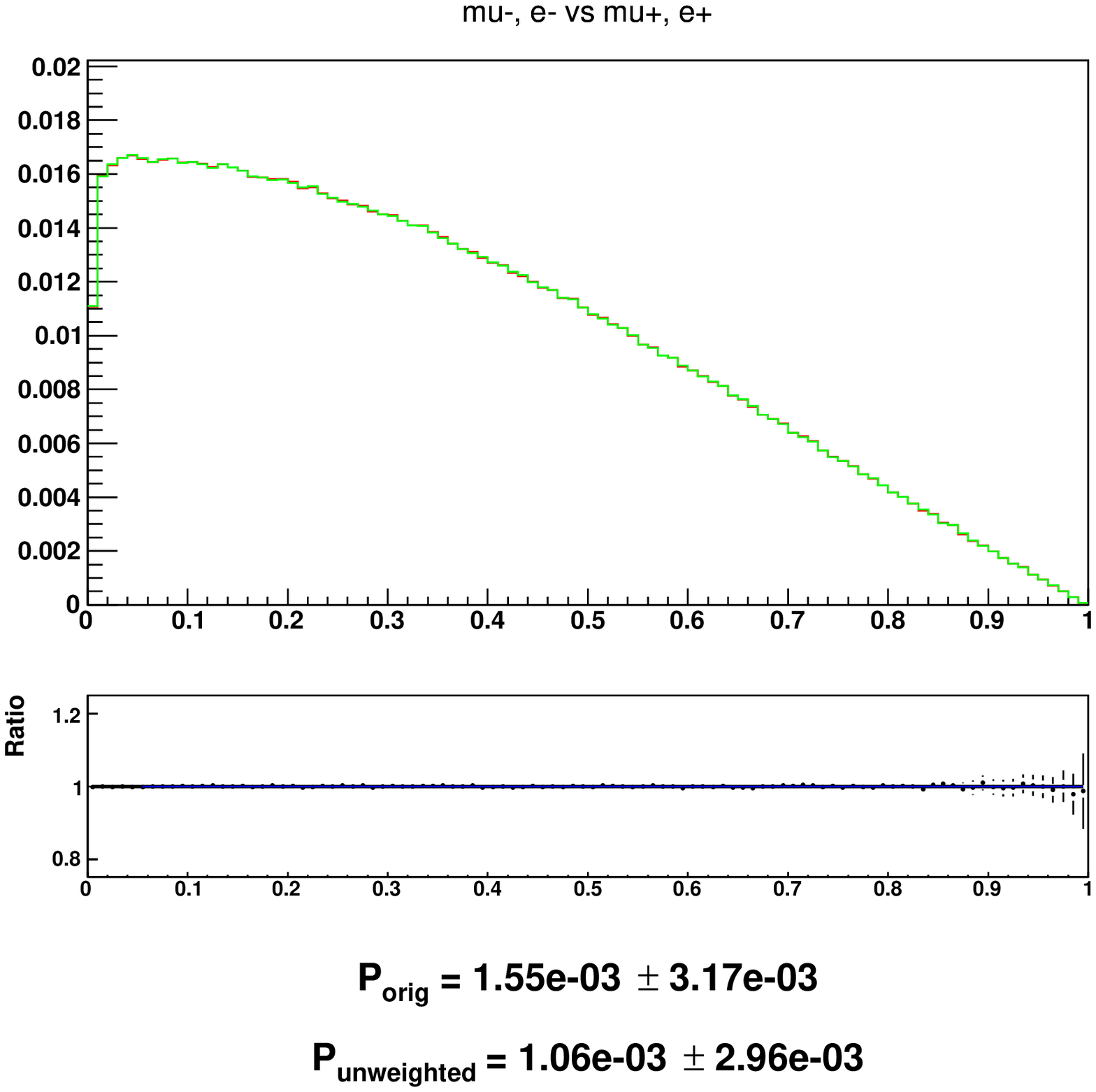}}
\resizebox*{0.49\textwidth}{!}{\includegraphics{\przedro 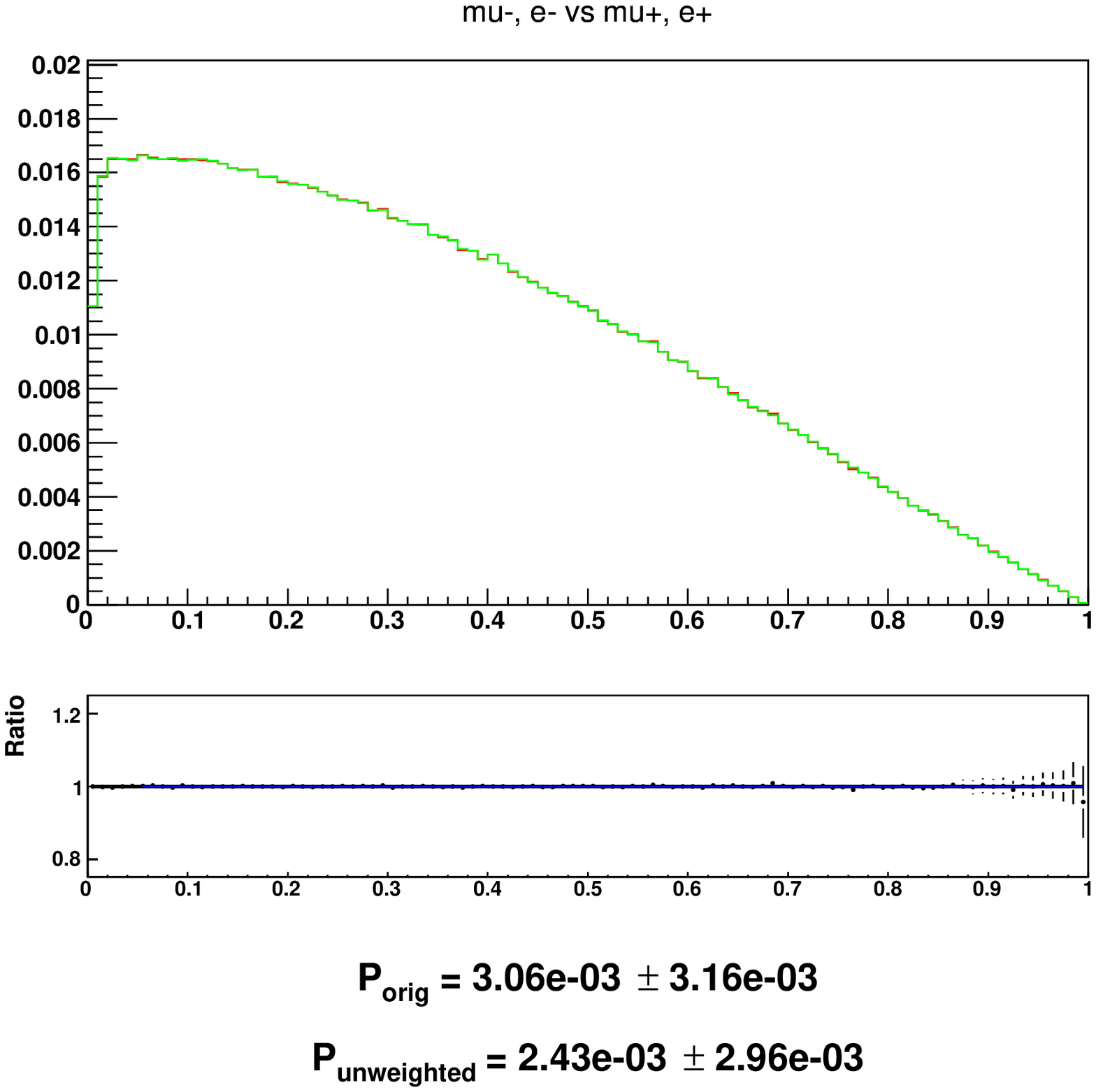}} \\
\caption{\small Fractions of  $\tau^+$ and $\tau^-$ energies carried by their visible  decay products:
two dimensional lego plots and one dimensional spectra$^{18}$.
\textcolor{red}{Red line} (and left scattergram) is sample with spin effects like of Higgs,
\textcolor{green}{green line} (and right scattergram) \greenlineis
black line is ratio \textcolor{red}{original}/\textcolor{green}{modified} with whenever available superimposed result for the
fitted functions.
}
\end{figure}

\newpage
\subsection{The energy spectrum: $\tau^- \to \mu^-, e^-$ {\tt vs } $\tau^+ \to \pi^+$}
\vspace{1\baselineskip}

\begin{figure}[h!]
\centering
\resizebox*{0.49\textwidth}{!}{\includegraphics{\przedro 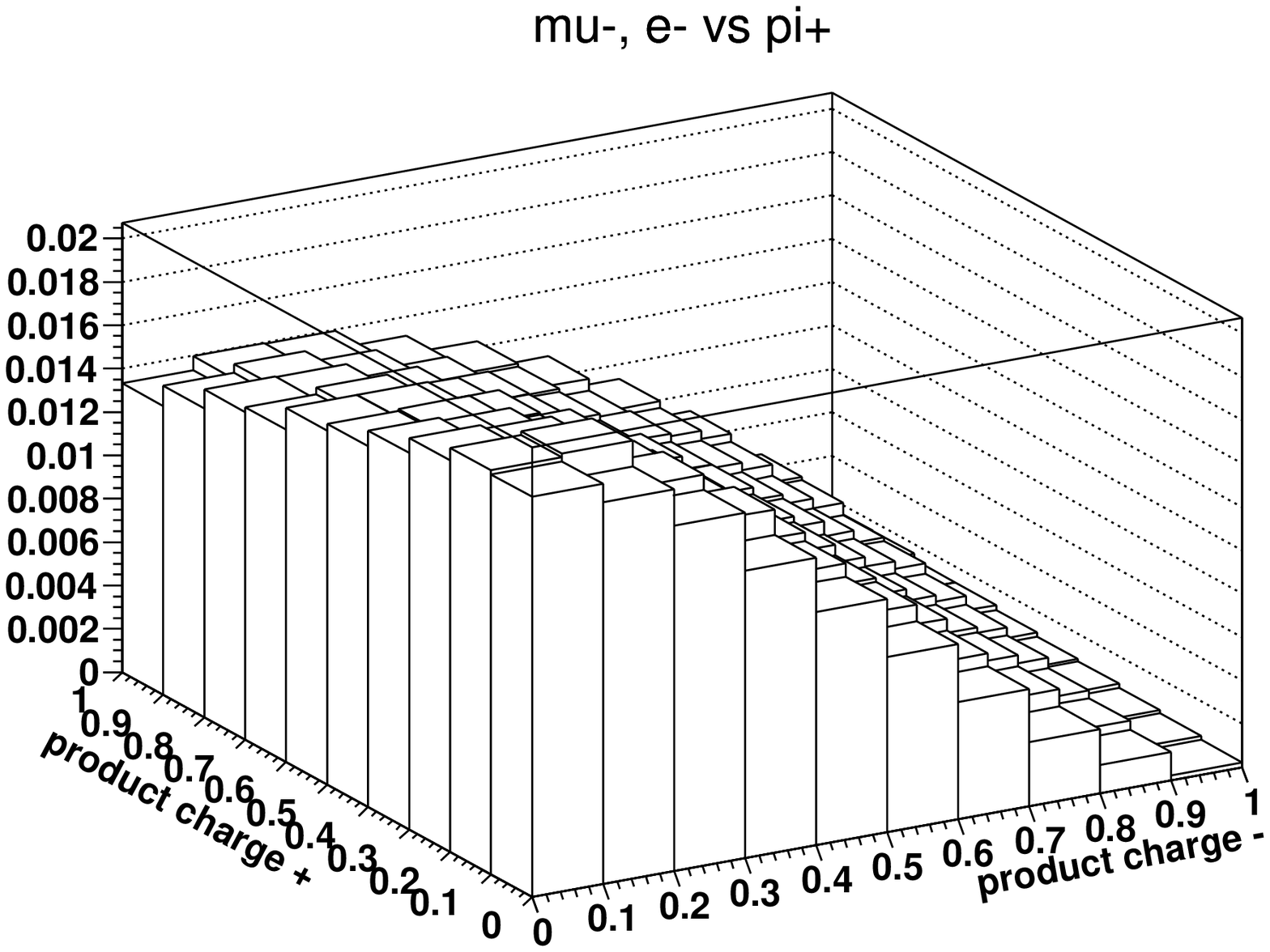}}
\resizebox*{0.49\textwidth}{!}{\includegraphics{\przedro 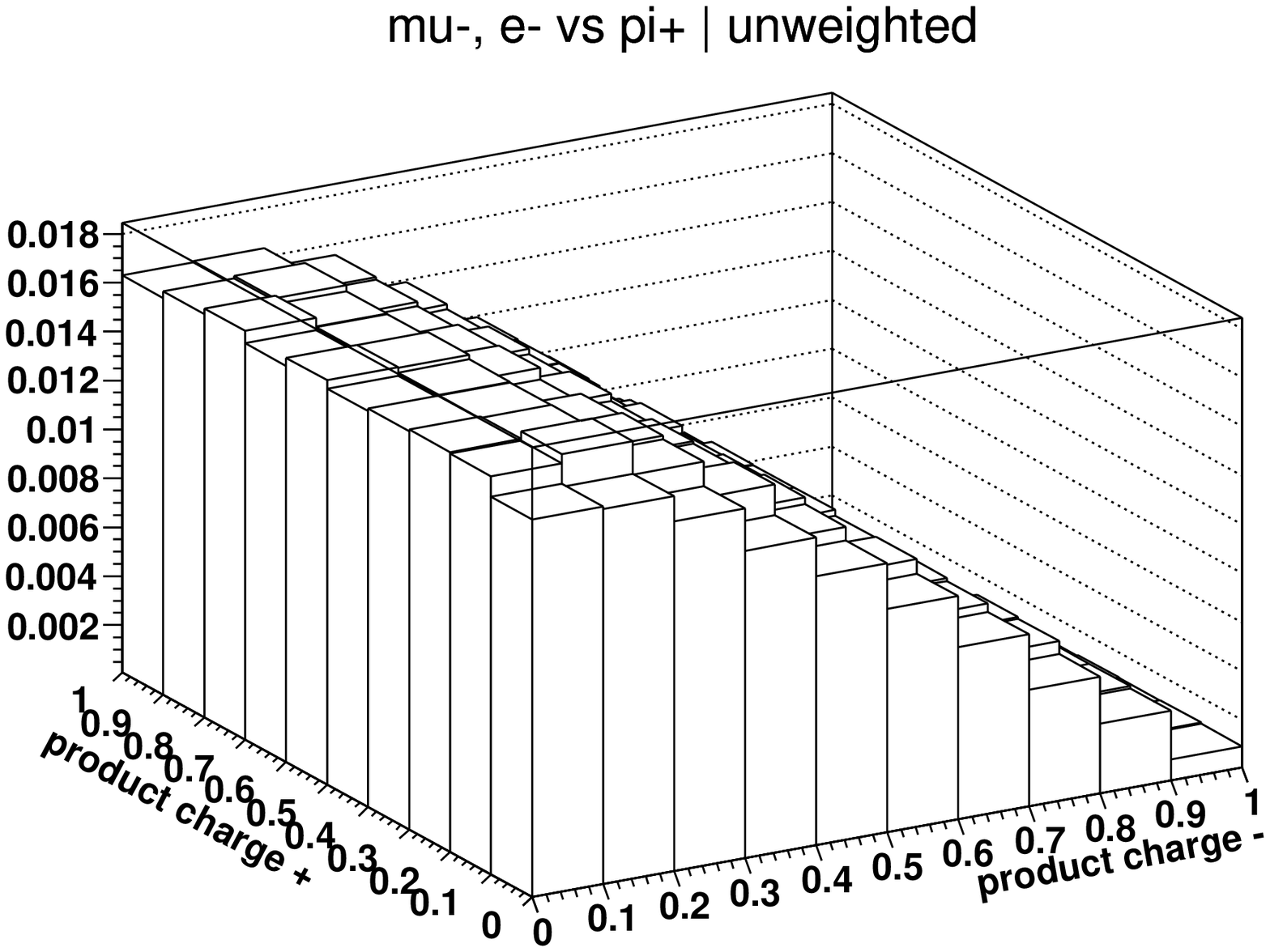}} \\
\resizebox*{0.49\textwidth}{!}{\includegraphics{\przedro 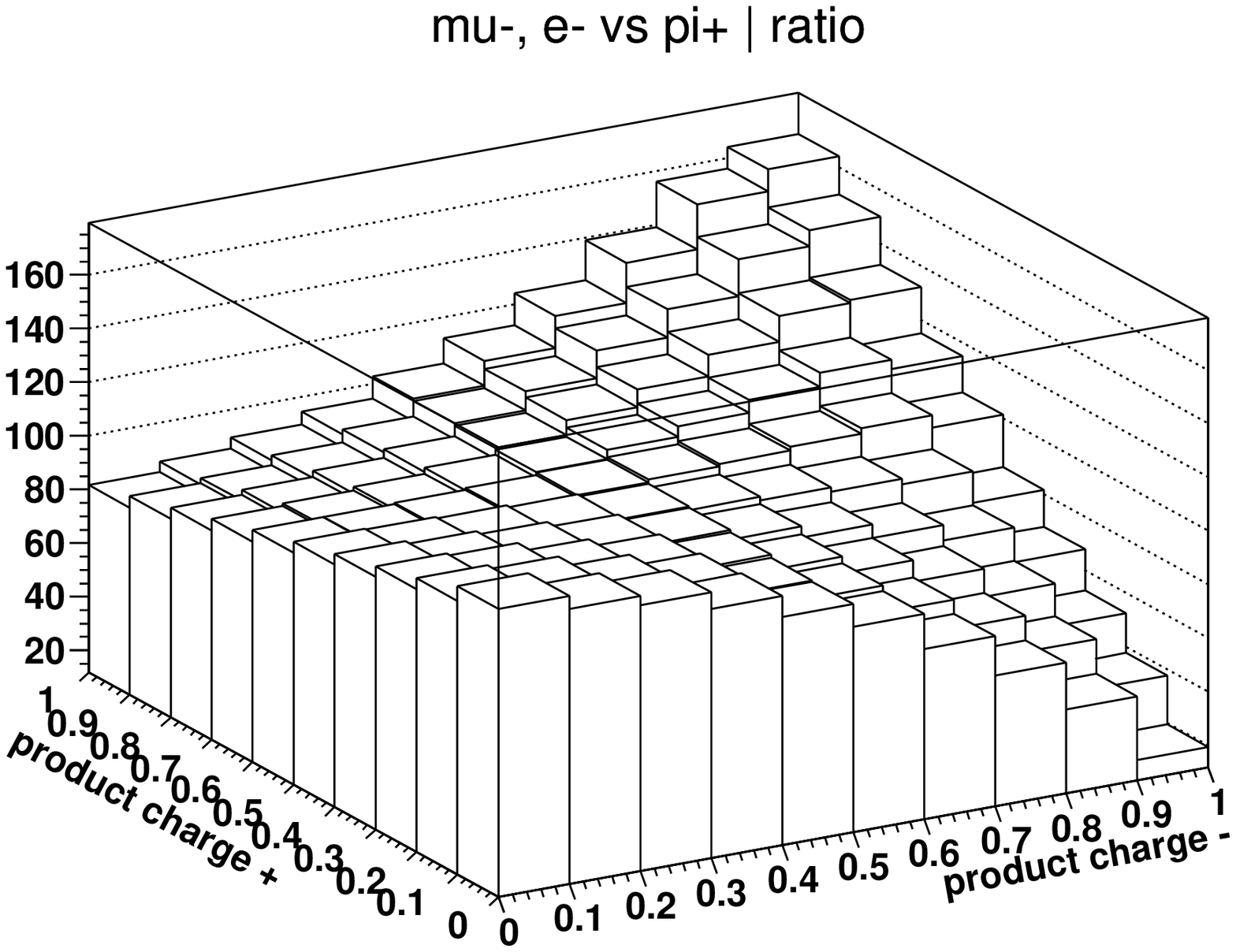}}
\resizebox*{0.49\textwidth}{!}{\includegraphics{\przedro 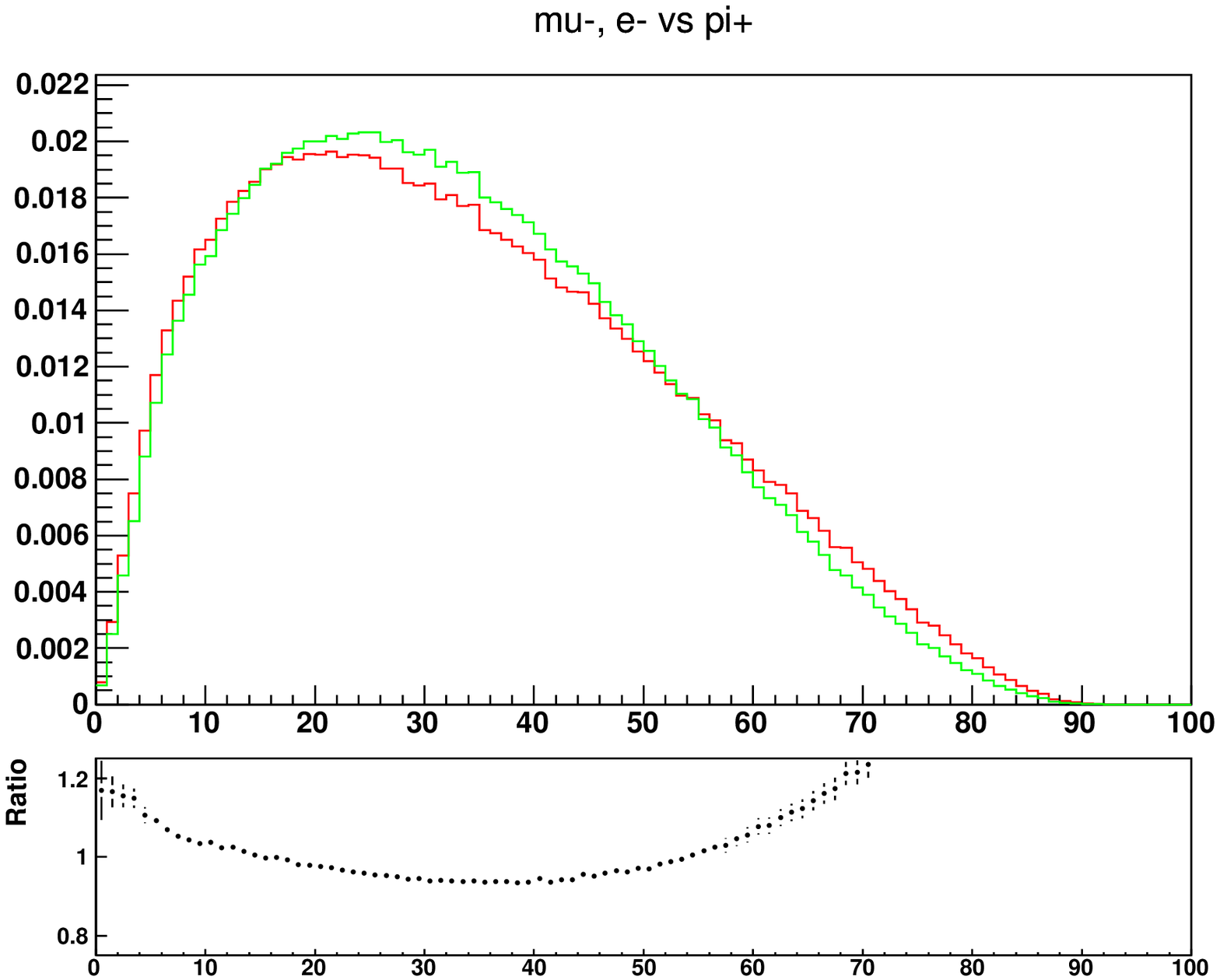}} \\
\resizebox*{0.49\textwidth}{!}{\includegraphics{\przedro 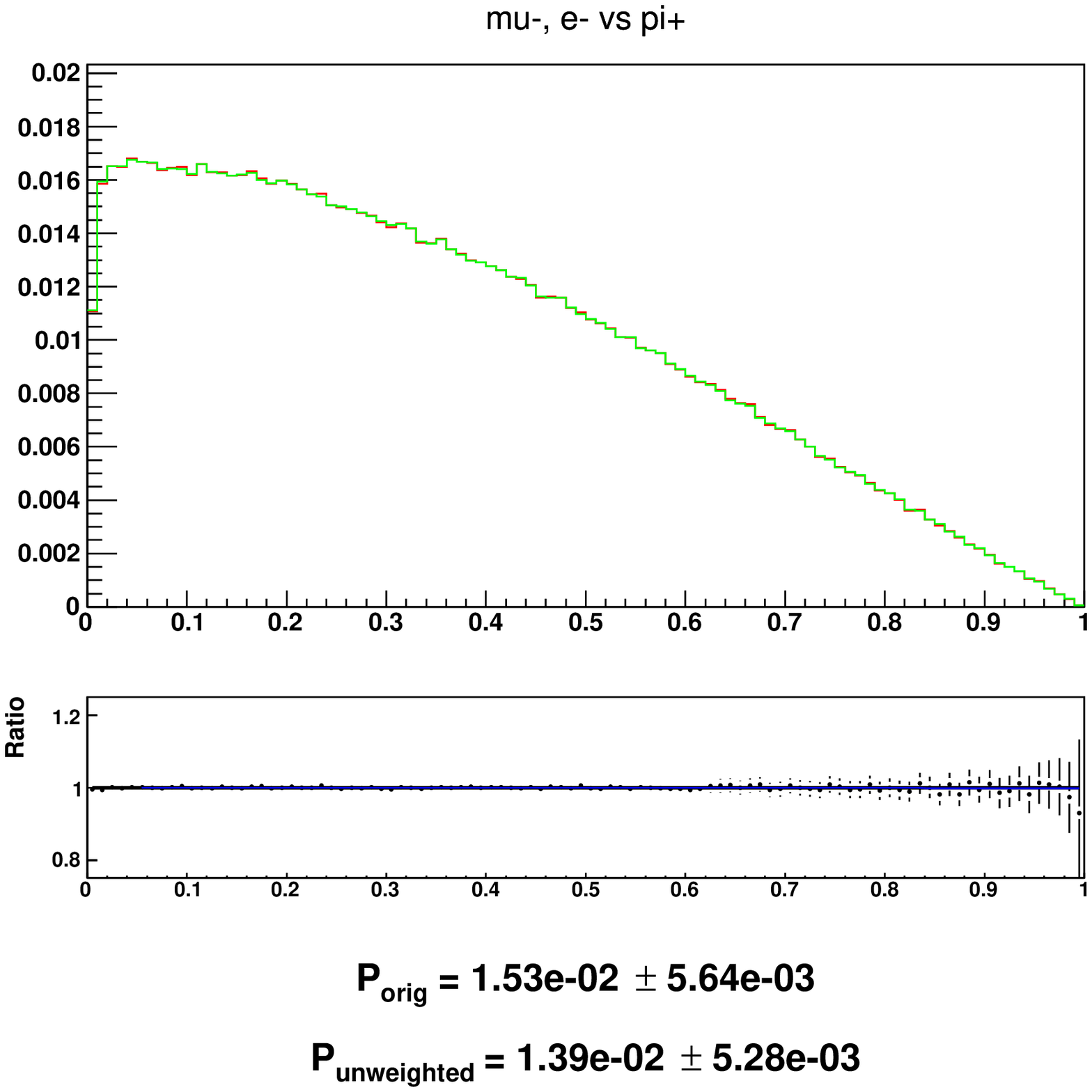}}
\resizebox*{0.49\textwidth}{!}{\includegraphics{\przedro 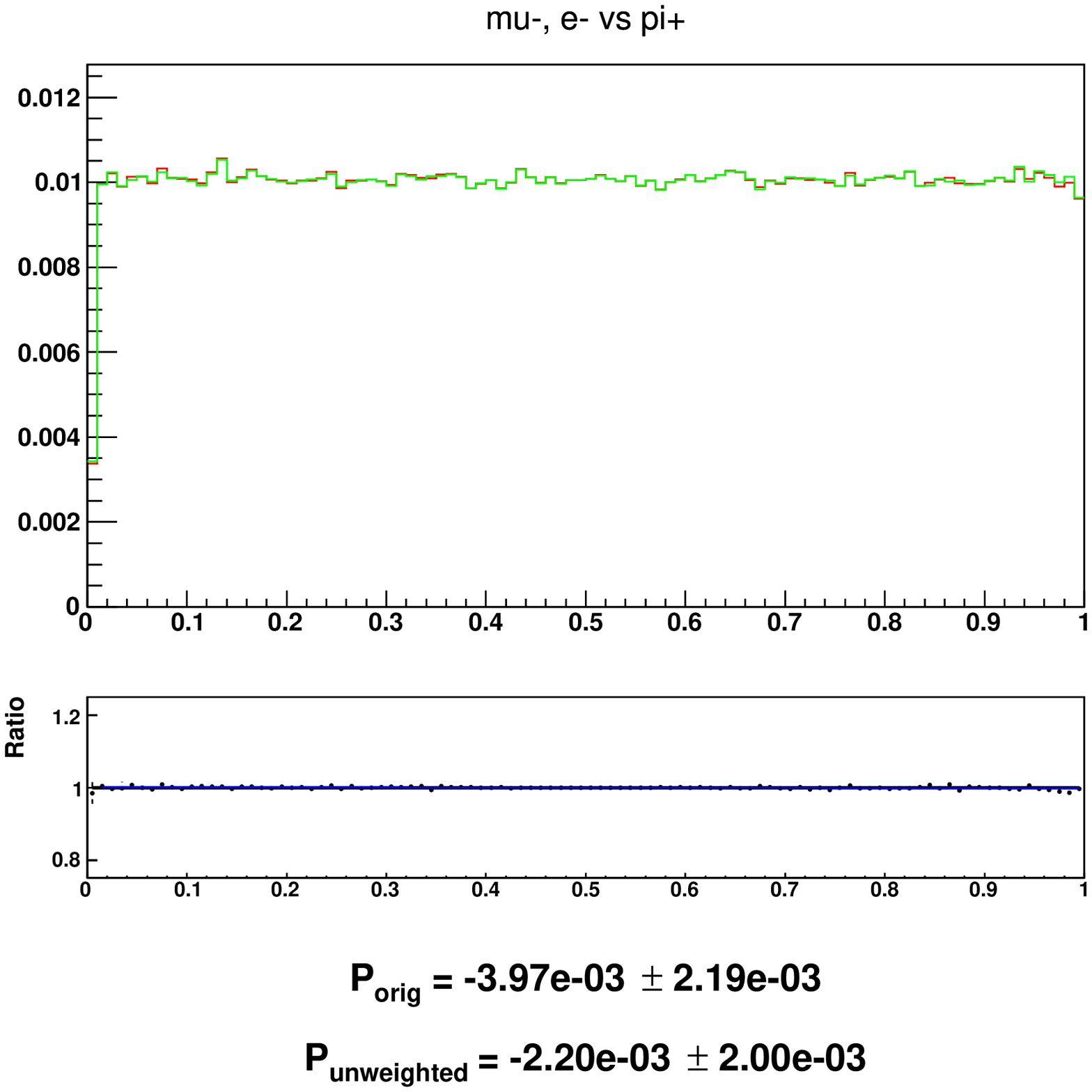}} \\
\caption{\small Fractions of  $\tau^+$ and $\tau^-$ energies carried by their visible  decay products:
two dimensional lego plots and one dimensional spectra$^{18}$.
\textcolor{red}{Red line} (and left scattergram) is sample with spin effects like of Higgs,
\textcolor{green}{green line} (and right scattergram) \greenlineis
black line is ratio \textcolor{red}{original}/\textcolor{green}{modified} with whenever available superimposed result for the
fitted functions.
}
\end{figure}

\newpage
\subsection{The energy spectrum: $\tau^- \to \pi^-$ {\tt vs } $\tau^+ \to \mu^+, e^+$}
\vspace{1\baselineskip}

\begin{figure}[h!]
\centering
\resizebox*{0.49\textwidth}{!}{\includegraphics{\przedro 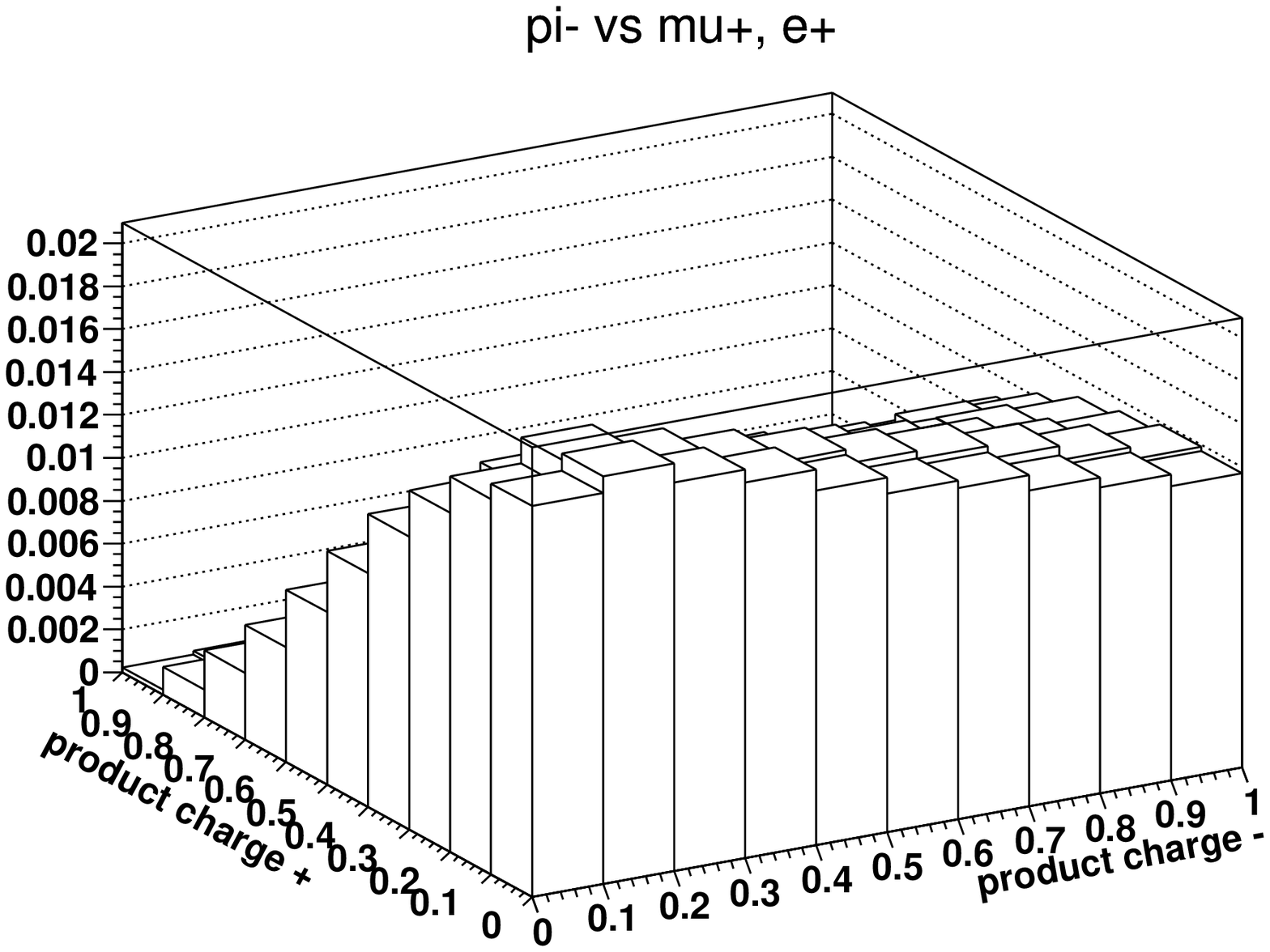}}
\resizebox*{0.49\textwidth}{!}{\includegraphics{\przedro 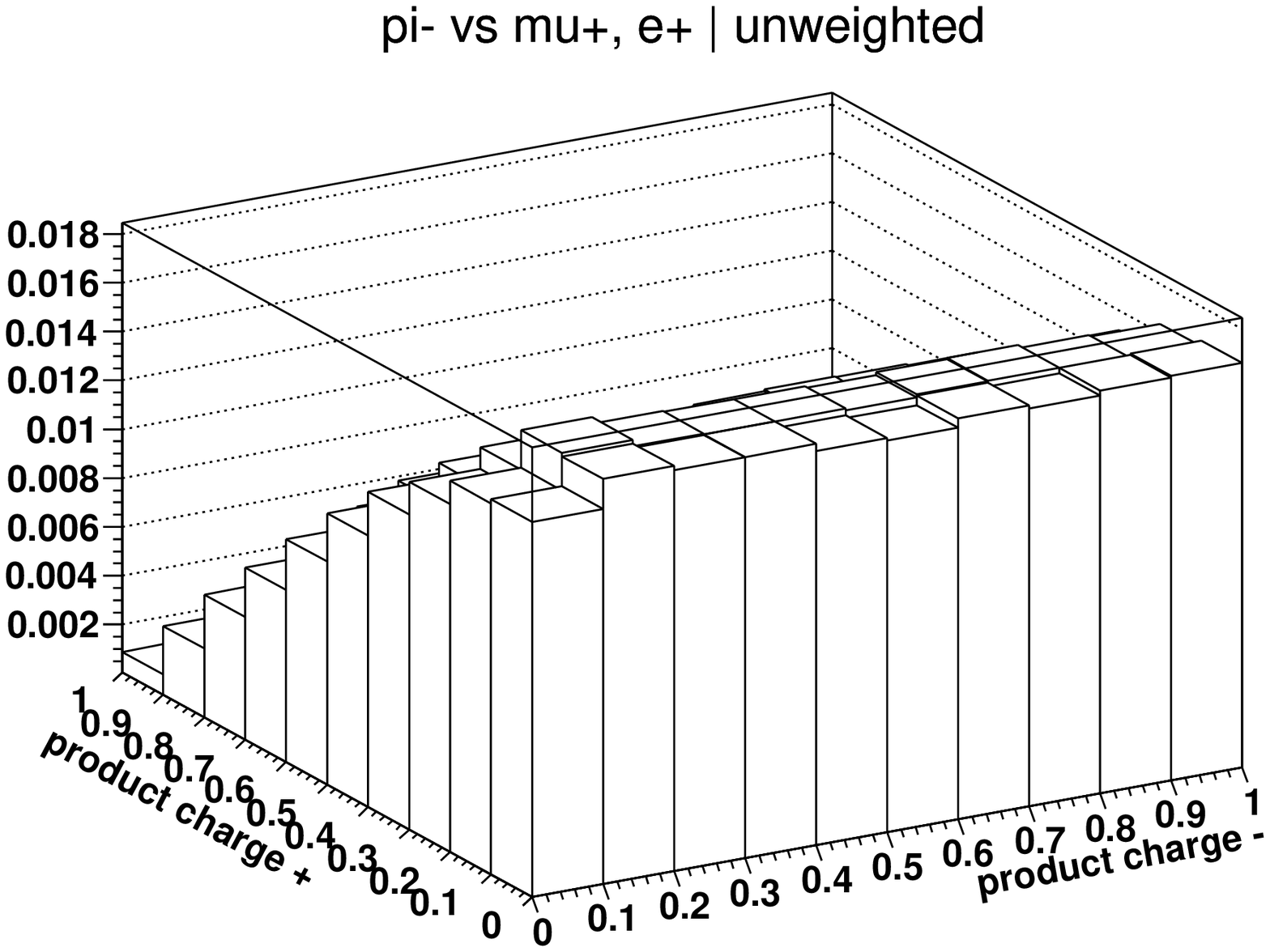}} \\
\resizebox*{0.49\textwidth}{!}{\includegraphics{\przedro 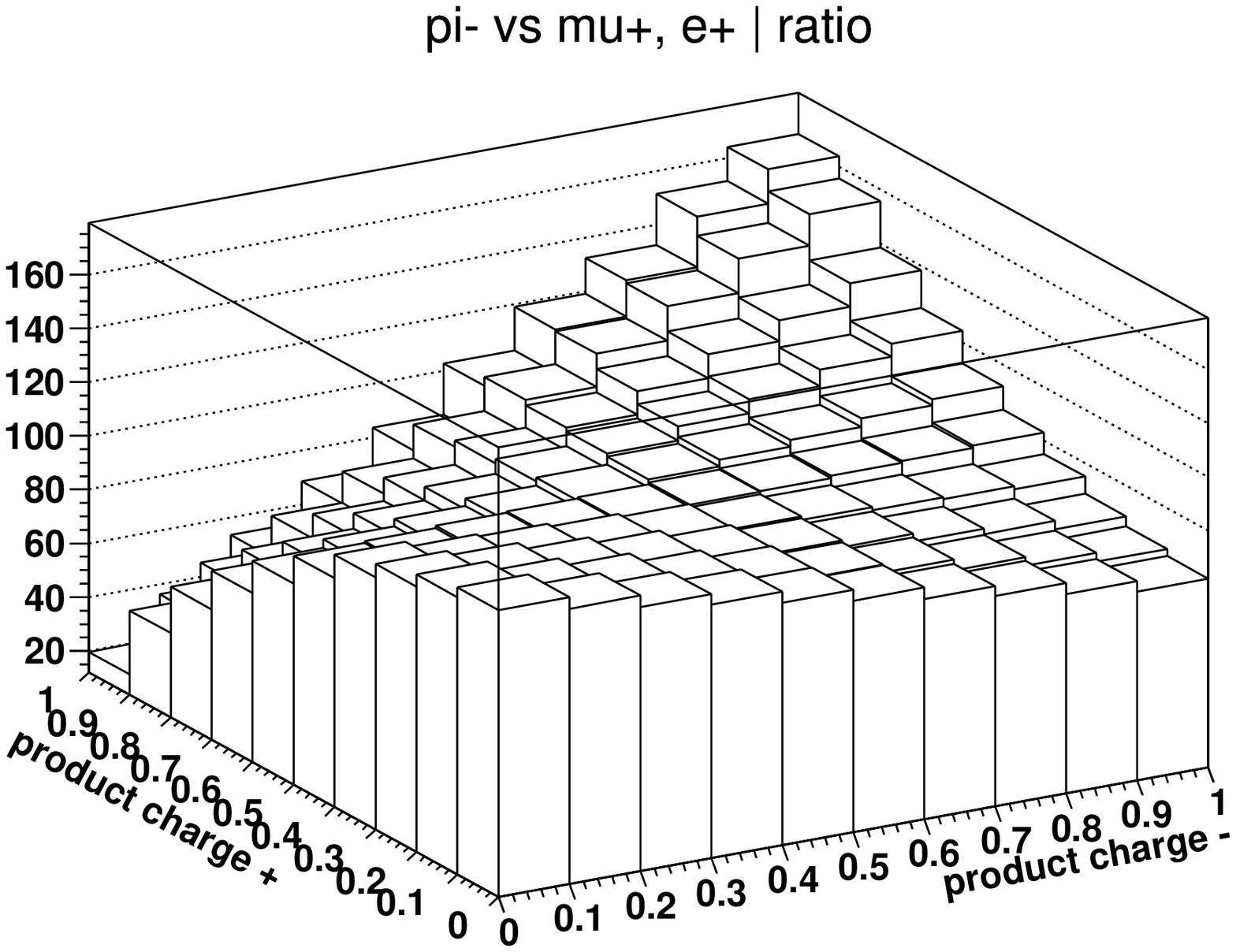}}
\resizebox*{0.49\textwidth}{!}{\includegraphics{\przedro 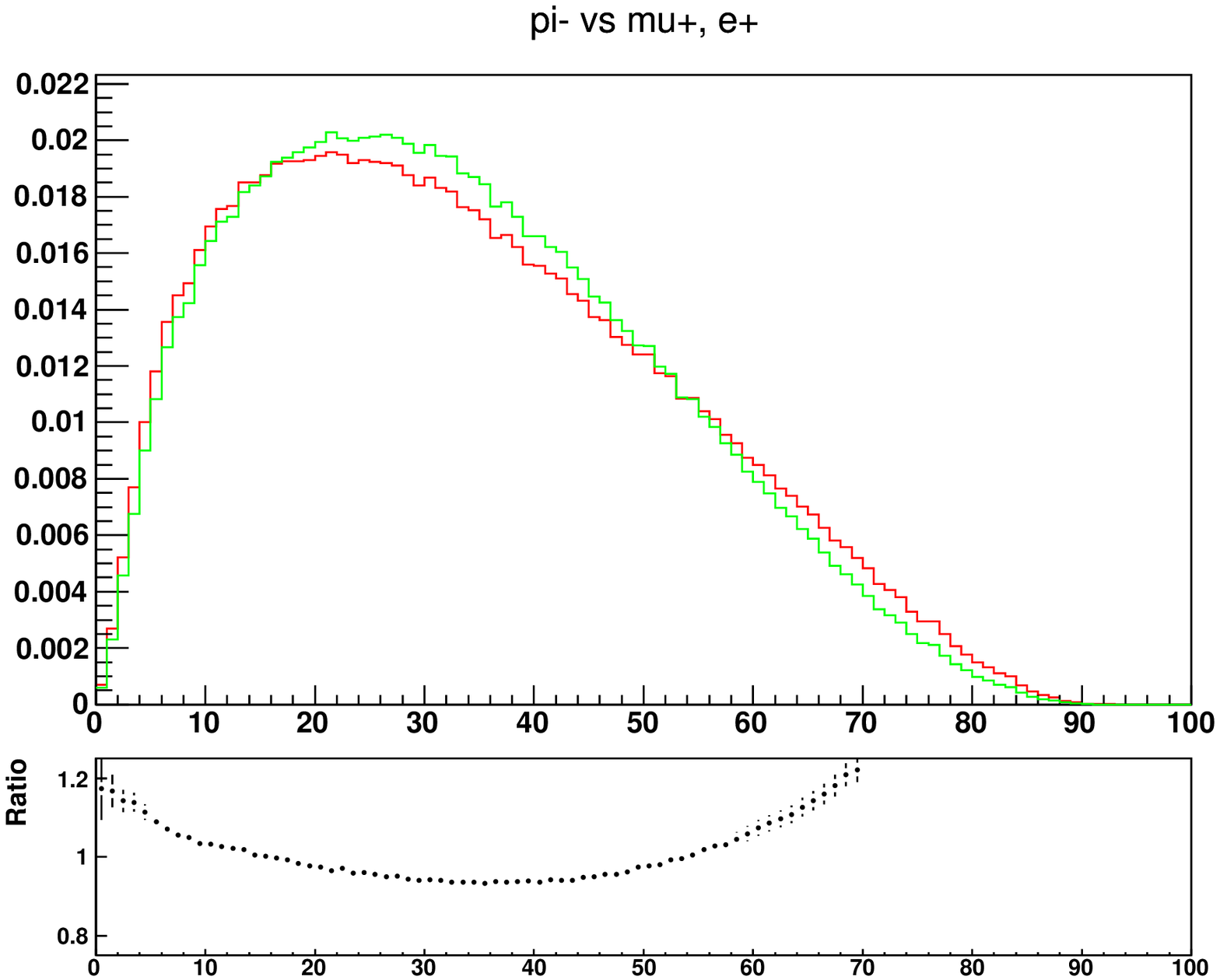}} \\
\resizebox*{0.49\textwidth}{!}{\includegraphics{\przedro 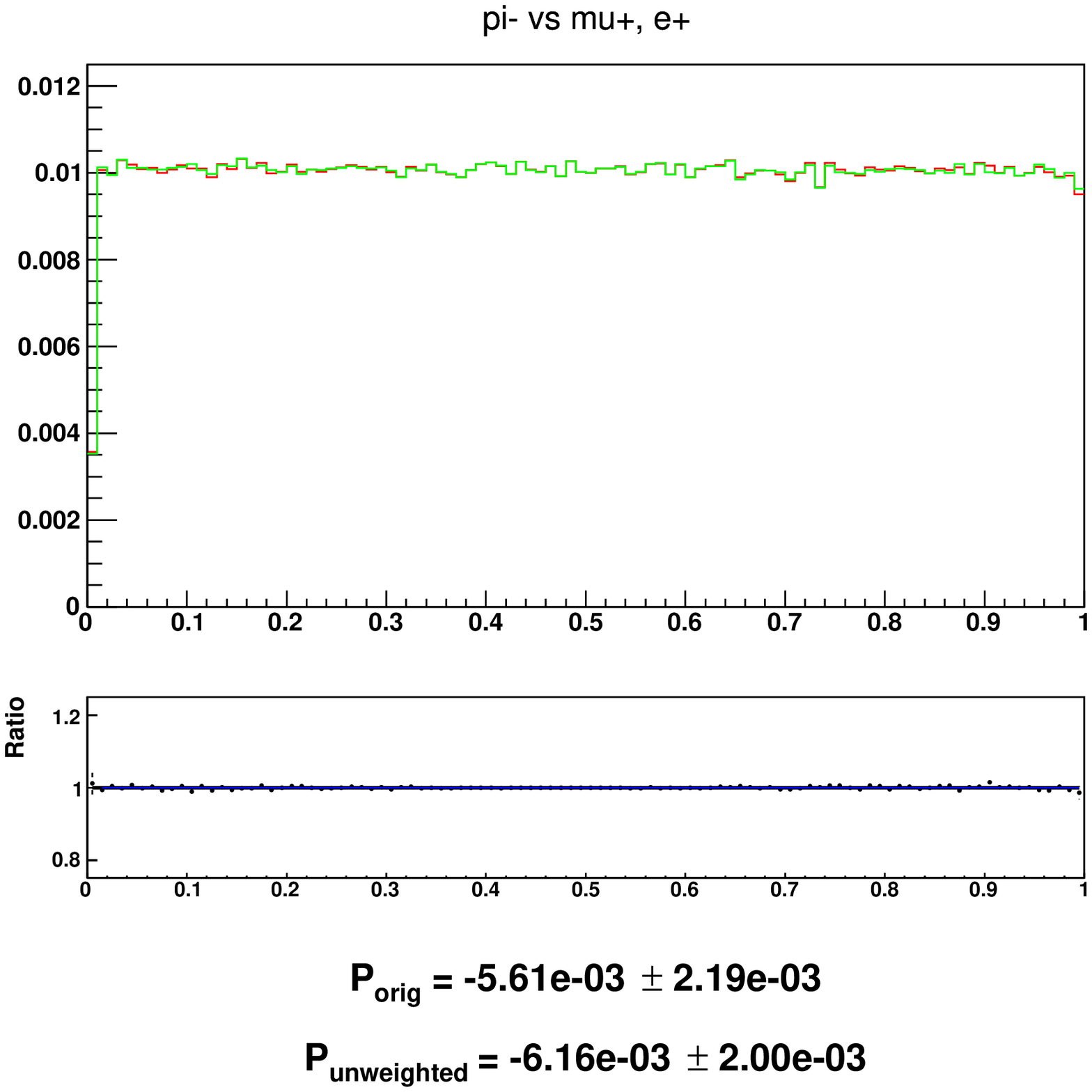}}
\resizebox*{0.49\textwidth}{!}{\includegraphics{\przedro 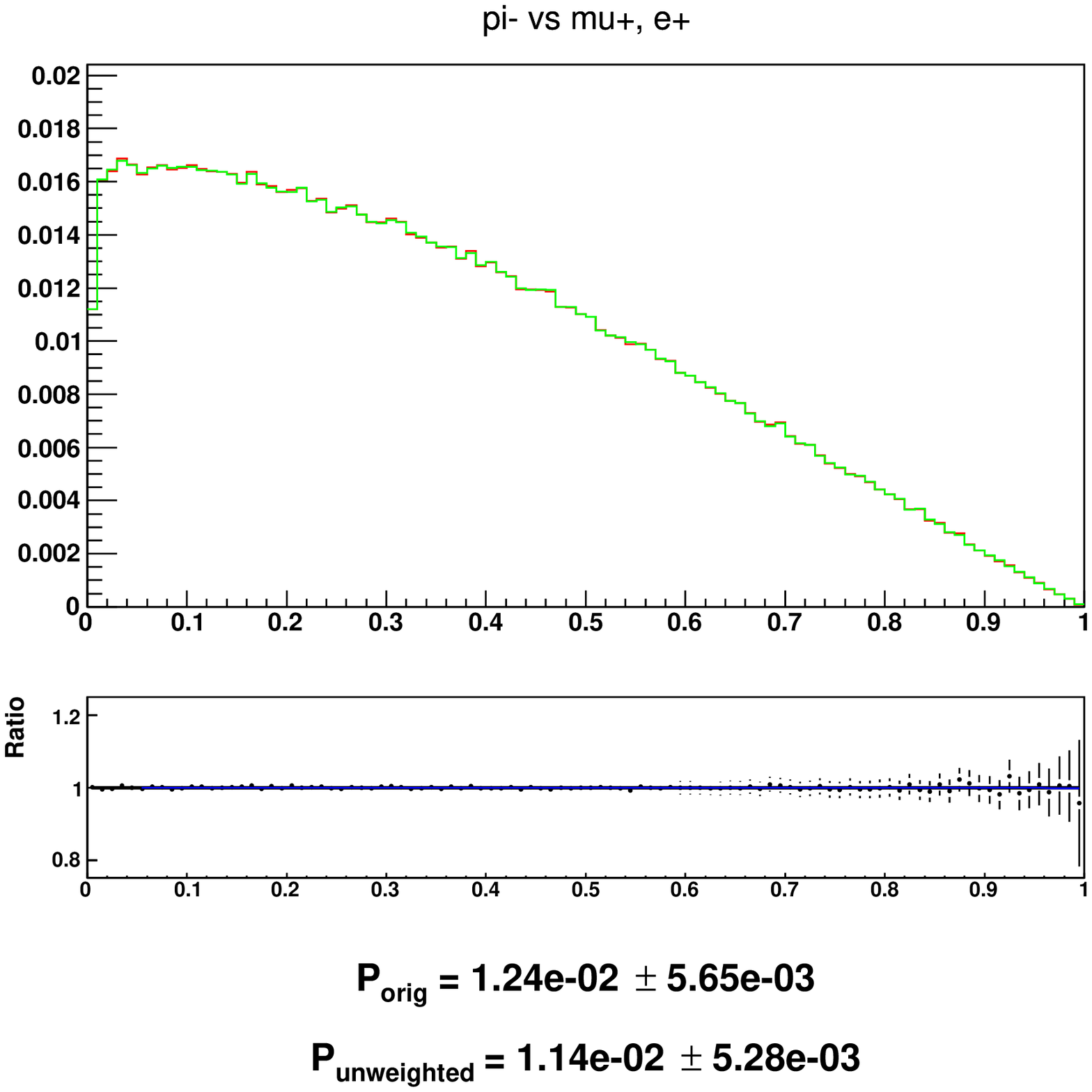}} \\
\caption{\small Fractions of  $\tau^+$ and $\tau^-$ energies carried by their visible  decay products:
two dimensional lego plots and one dimensional spectra$^{18}$.
\textcolor{red}{Red line} (and left scattergram) is sample with spin effects like of Higgs,
\textcolor{green}{green line} (and right scattergram) \greenlineis
black line is ratio \textcolor{red}{original}/\textcolor{green}{modified} with whenever available superimposed result for the
fitted functions.
}
\end{figure}

\newpage
\subsection{The energy spectrum: $\tau^- \to \pi^-$ {\tt vs } $\tau^+ \to \pi^+$}
\vspace{1\baselineskip}

\begin{figure}[h!]
\centering
\resizebox*{0.49\textwidth}{!}{\includegraphics{\przedro 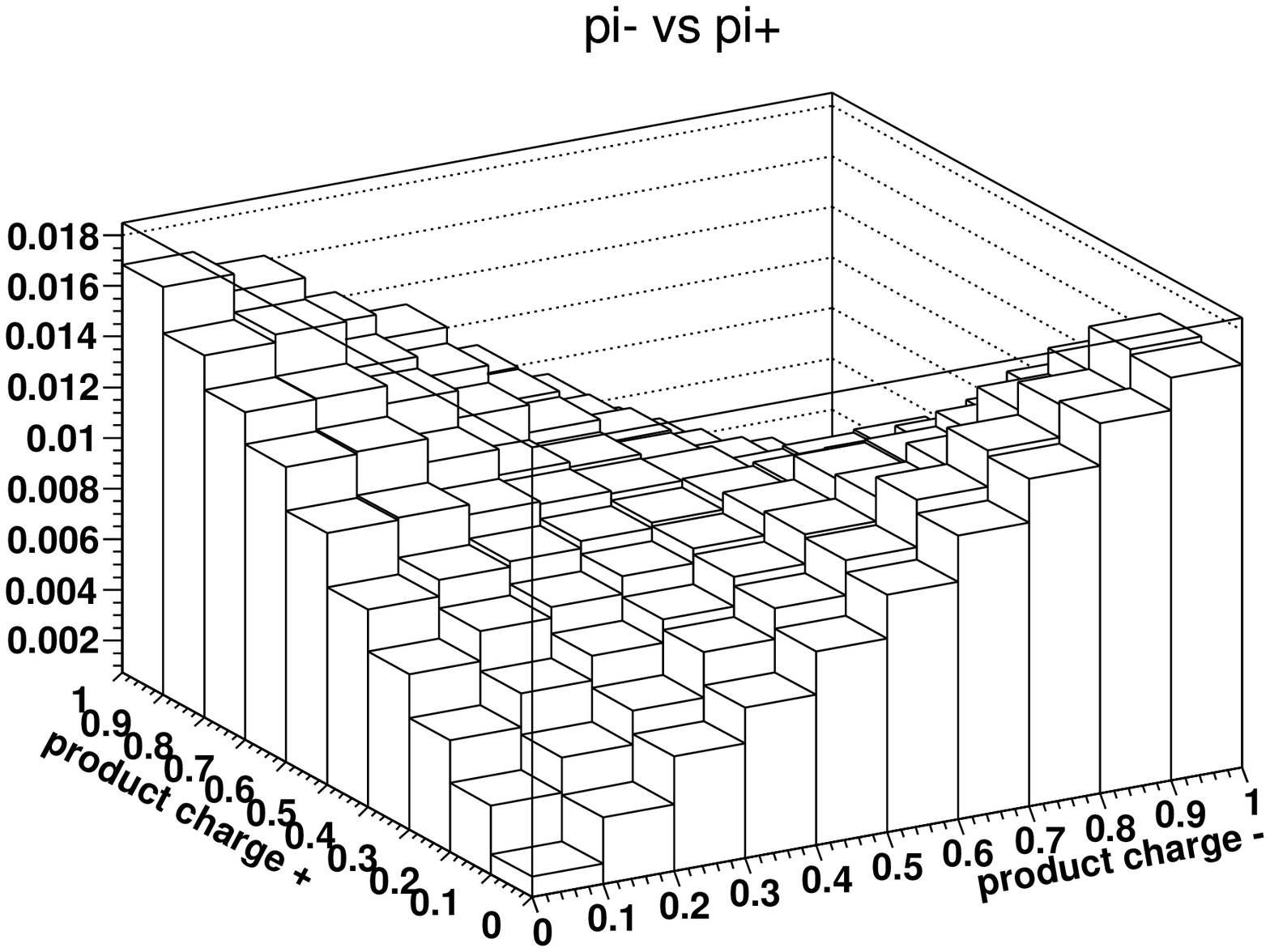}}
\resizebox*{0.49\textwidth}{!}{\includegraphics{\przedro 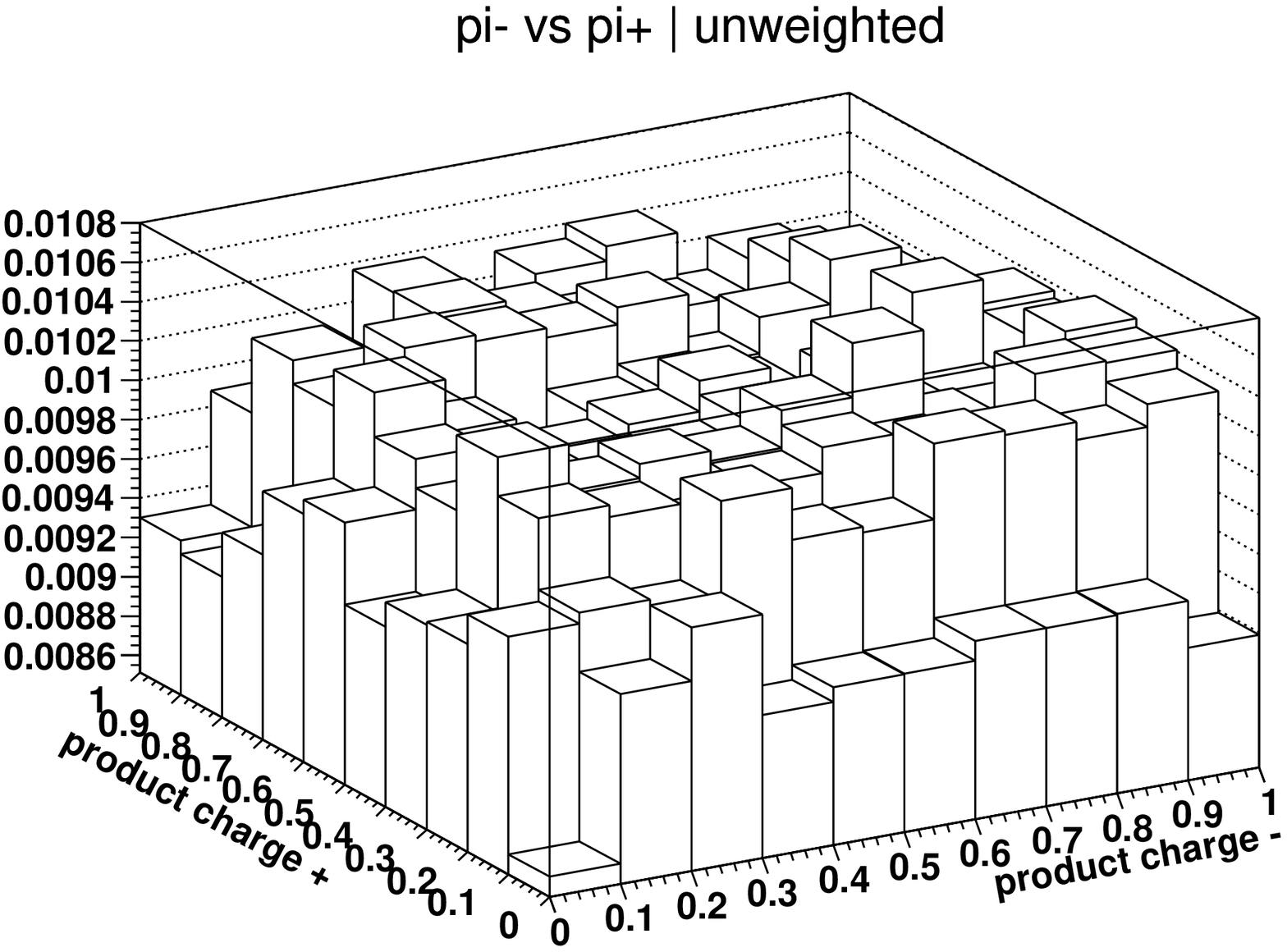}} \\
\resizebox*{0.49\textwidth}{!}{\includegraphics{\przedro 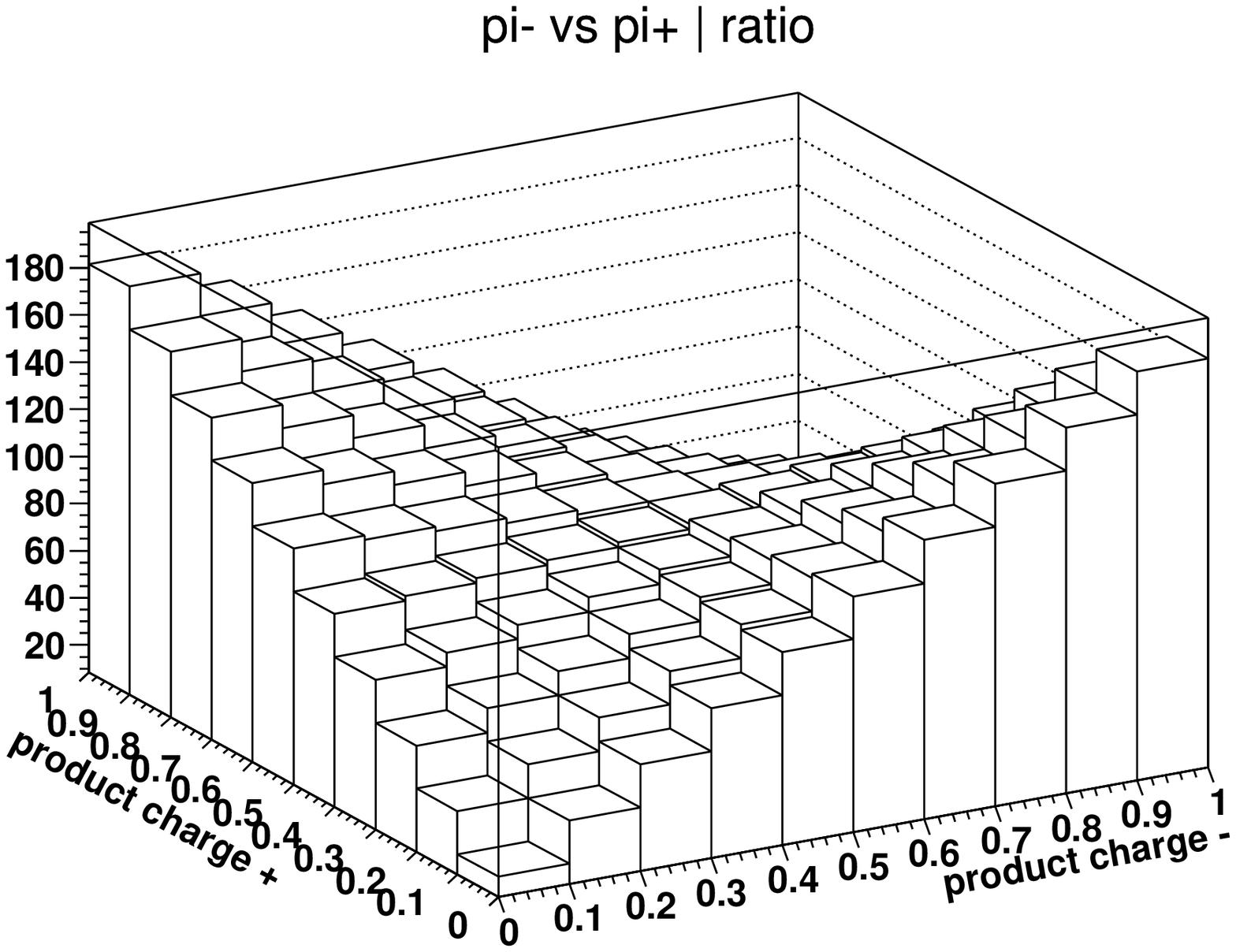}}
\resizebox*{0.49\textwidth}{!}{\includegraphics{\przedro 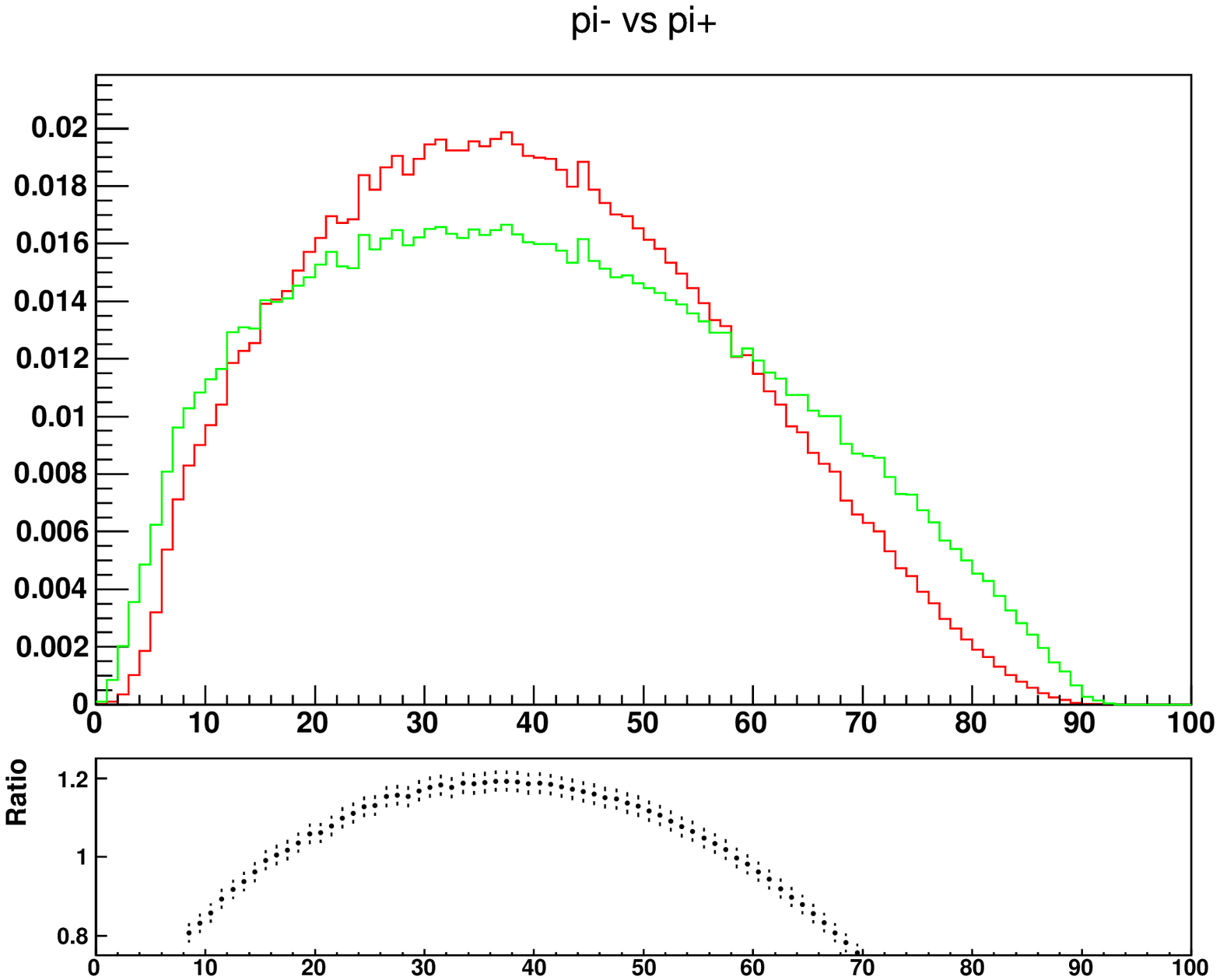}} \\
\resizebox*{0.49\textwidth}{!}{\includegraphics{\przedro 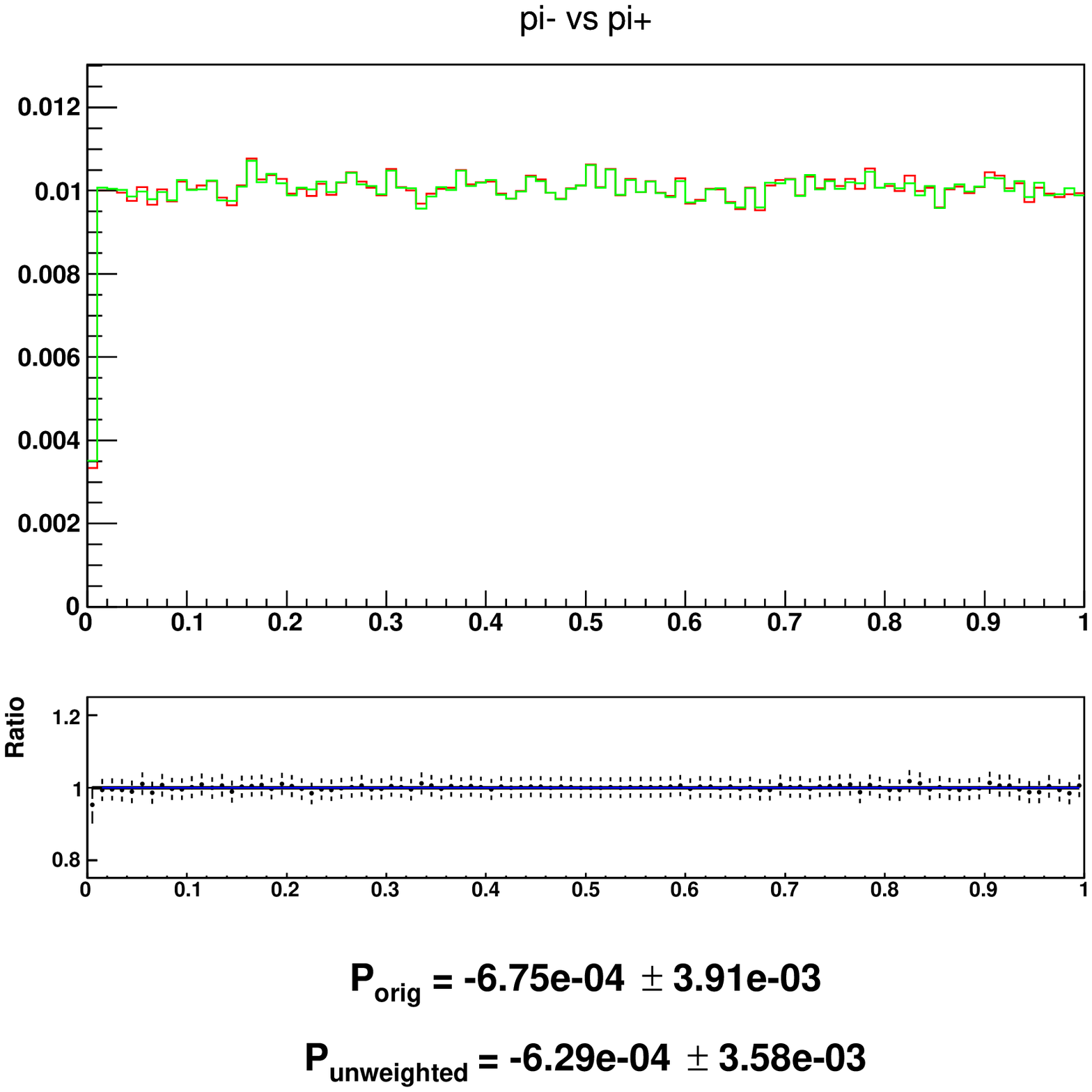}}
\resizebox*{0.49\textwidth}{!}{\includegraphics{\przedro 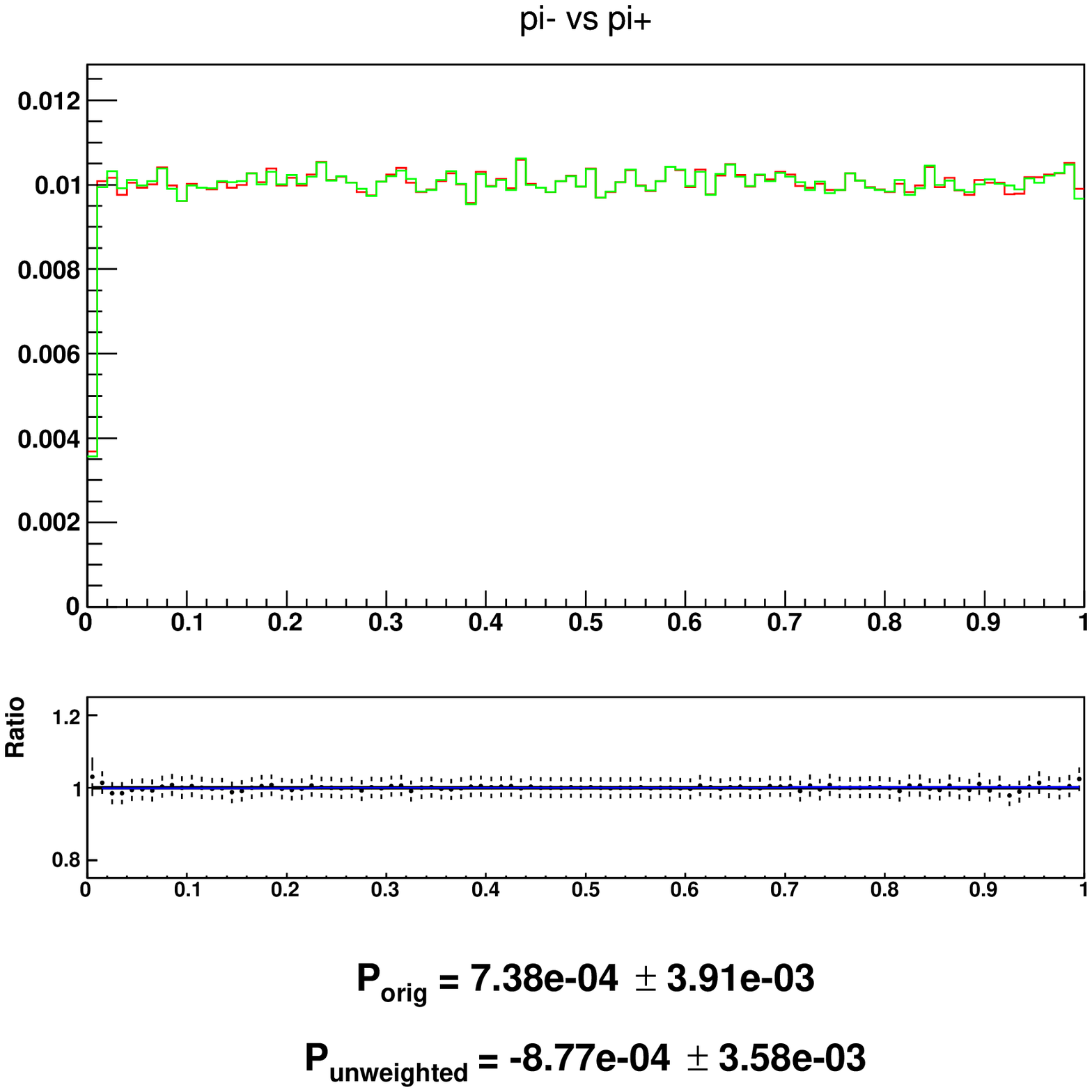}} \\
\caption{\small Fractions of  $\tau^+$ and $\tau^-$ energies carried by their visible  decay products:
two dimensional lego plots and one dimensional spectra$^{18}$.
\textcolor{red}{Red line} (and left scattergram) is sample with spin effects like of Higgs,
\textcolor{green}{green line} (and right scattergram) \greenlineis
black line is ratio \textcolor{red}{original}/\textcolor{green}{modified} with whenever available superimposed result for the
fitted functions.
}\label{Fig:scattergrams2}
\end{figure}

\newpage
\subsection{The energy spectrum: $\tau^- \to \mu^-, e^-$ {\tt vs } $\tau^+ \to \rho^+$}
\vspace{1\baselineskip}

\begin{figure}[h!]
\centering
\resizebox*{0.49\textwidth}{!}{\includegraphics{\przedro 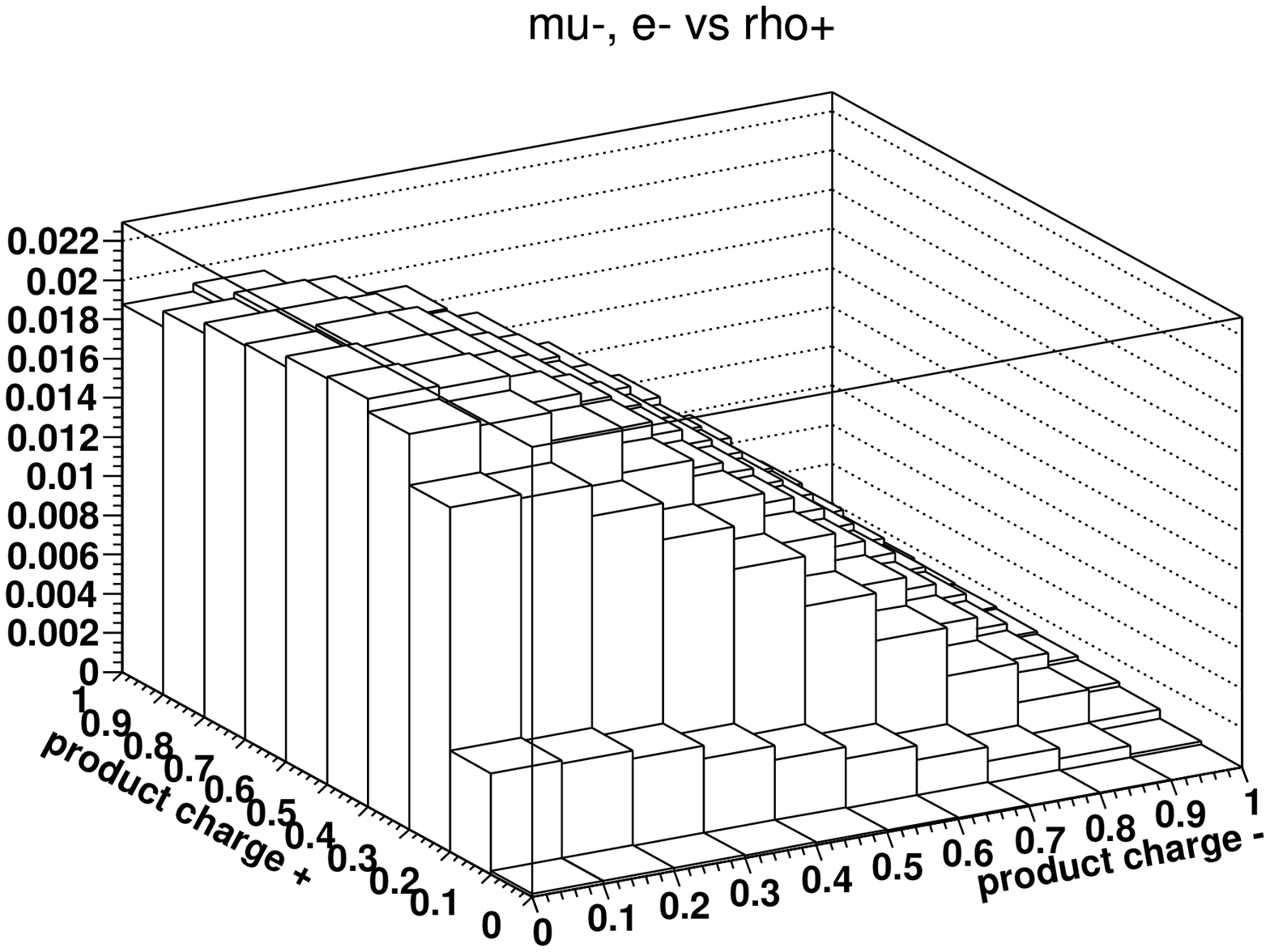}}
\resizebox*{0.49\textwidth}{!}{\includegraphics{\przedro 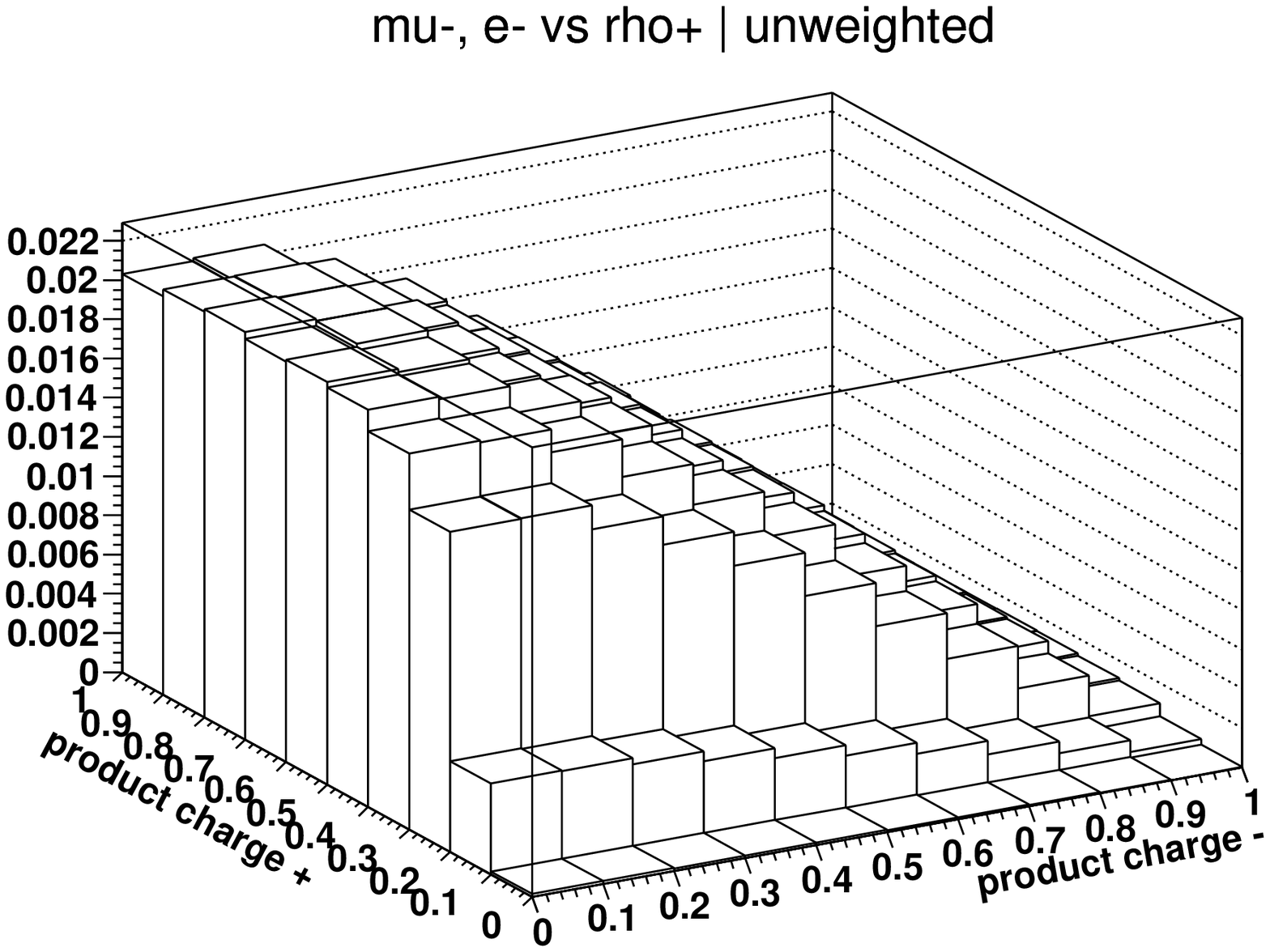}} \\
\resizebox*{0.49\textwidth}{!}{\includegraphics{\przedro 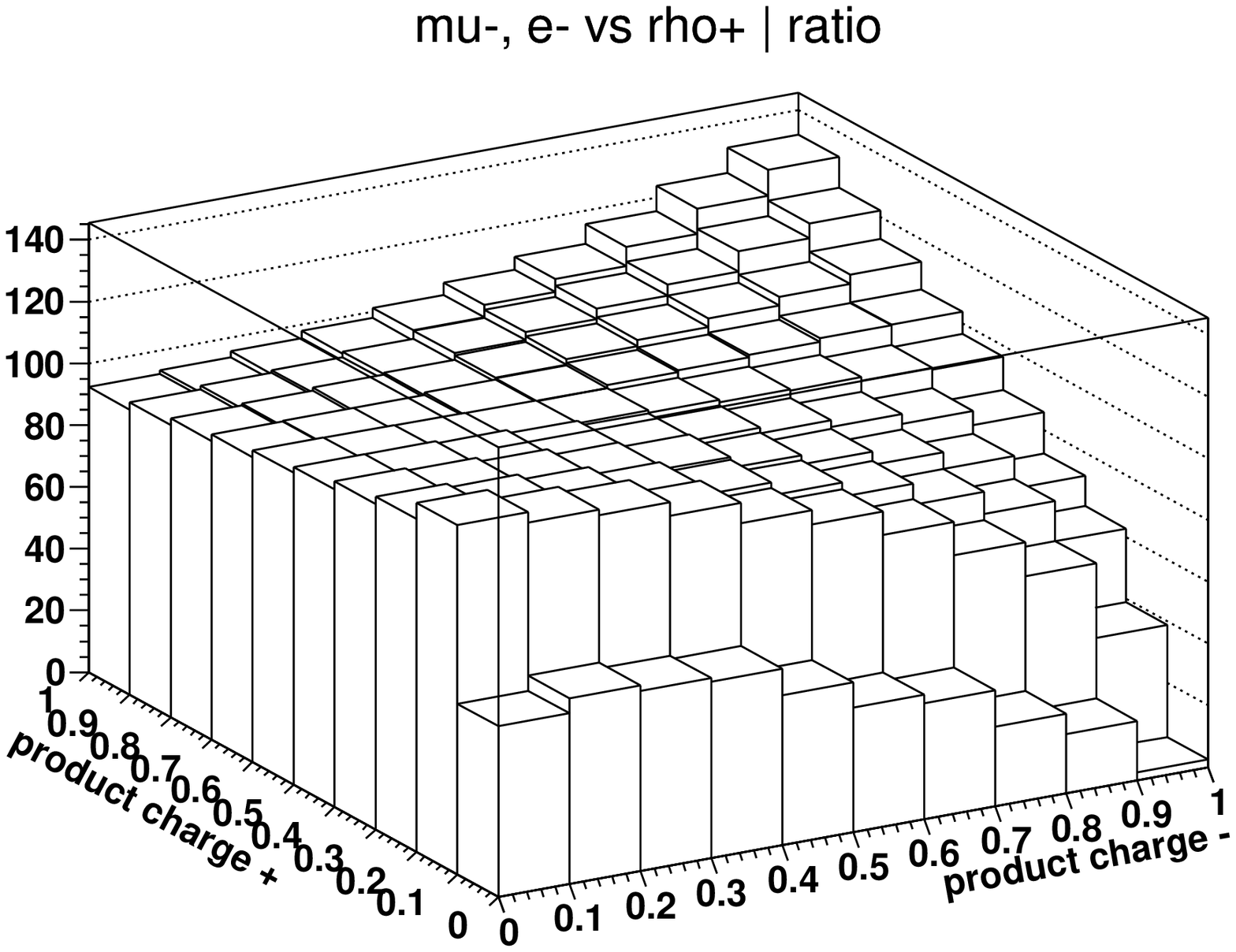}}
\resizebox*{0.49\textwidth}{!}{\includegraphics{\przedro 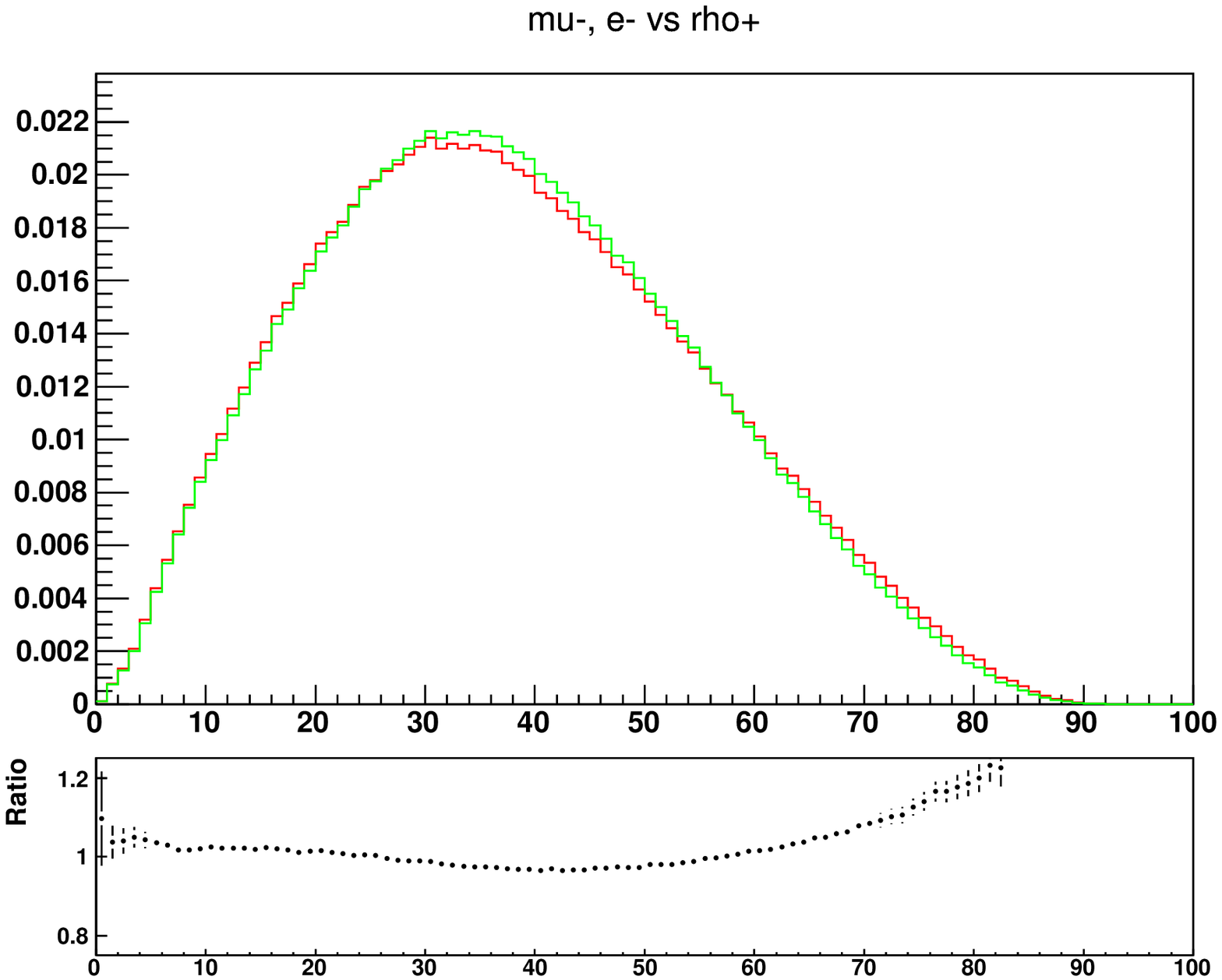}} \\
\resizebox*{0.49\textwidth}{!}{\includegraphics{\przedro 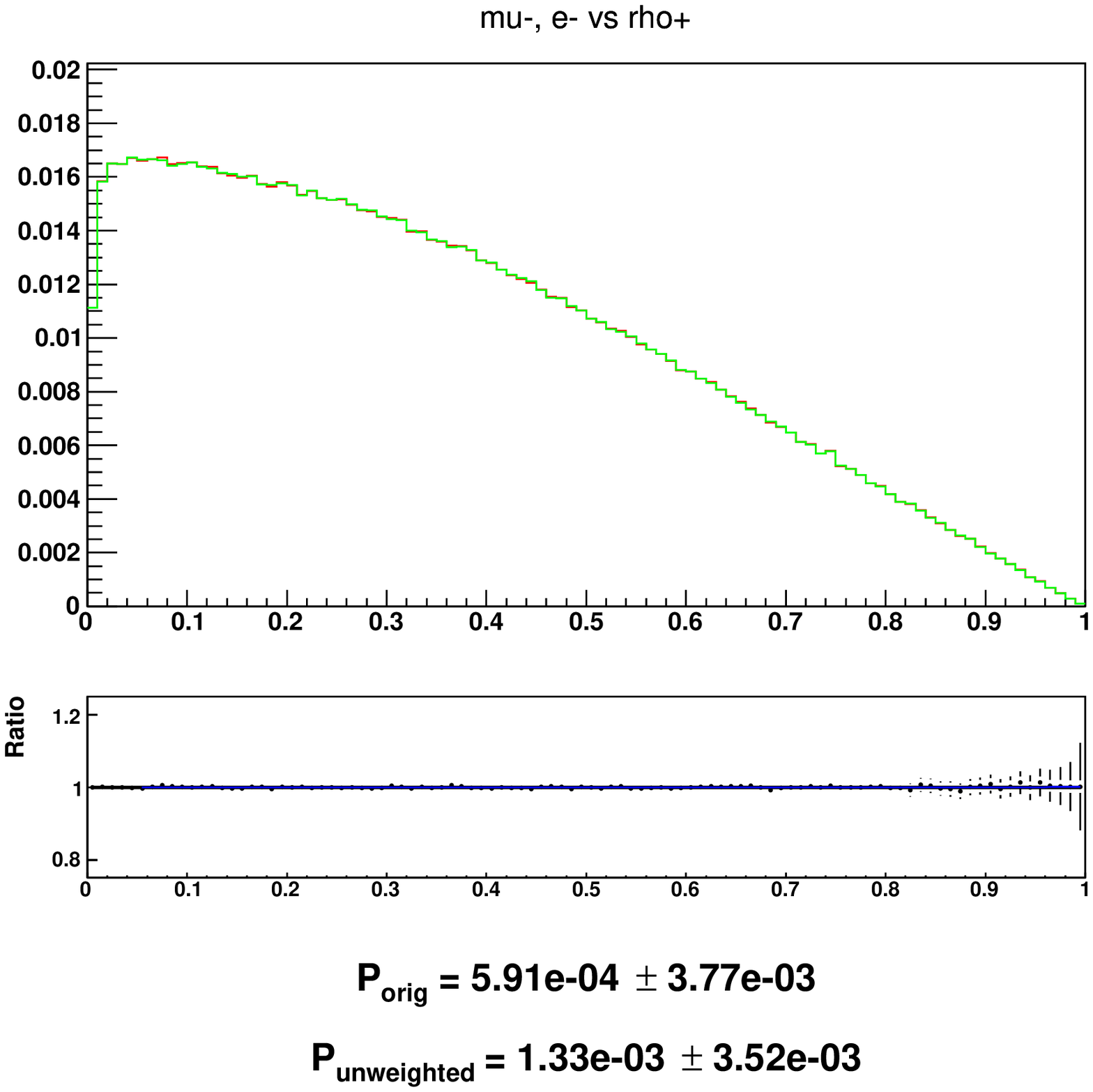}}
\resizebox*{0.49\textwidth}{!}{\includegraphics{\przedro 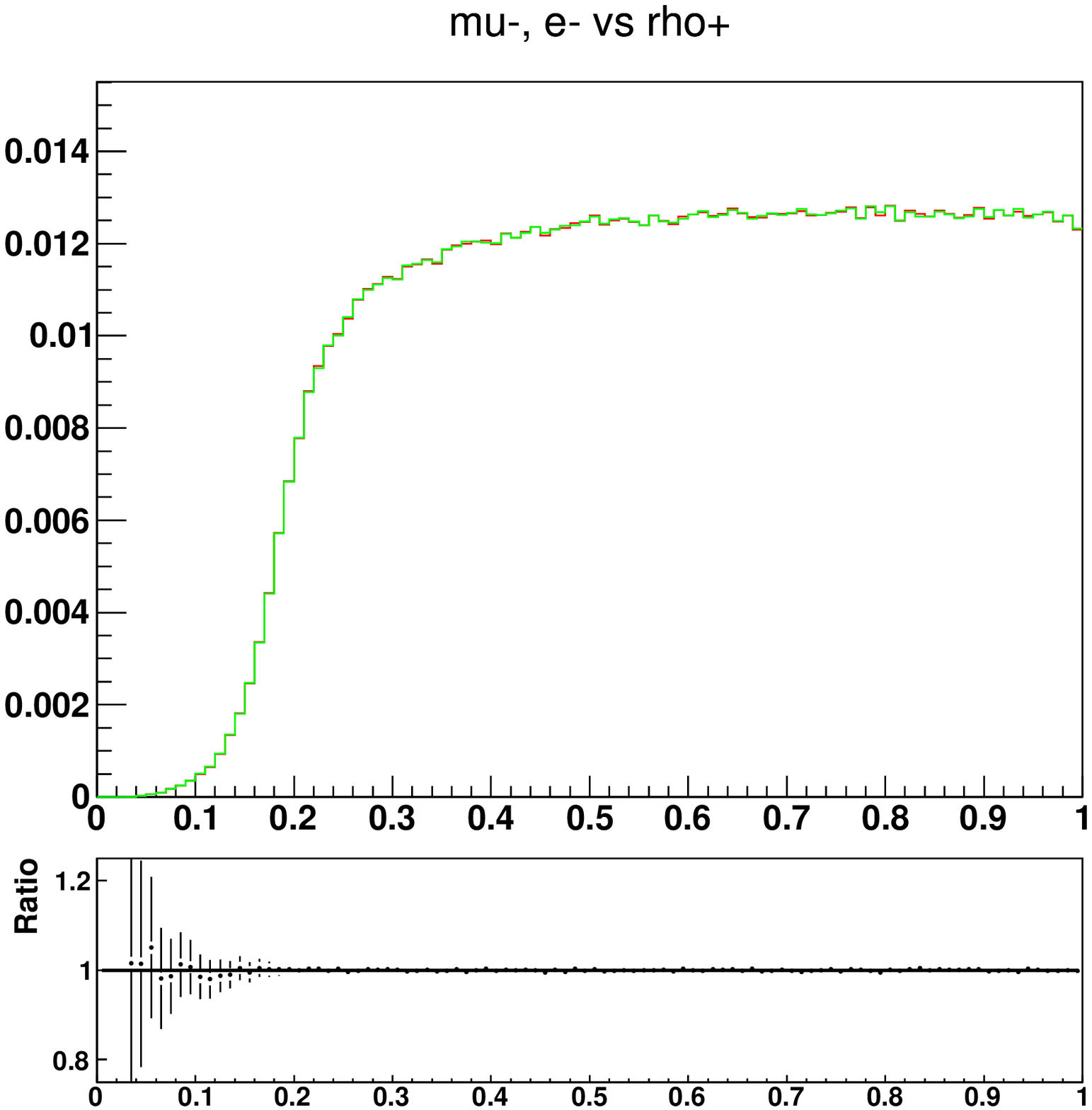}} \\
\caption{\small Fractions of  $\tau^+$ and $\tau^-$ energies carried by their visible  decay products:
two dimensional lego plots and one dimensional spectra$^{18}$.
\textcolor{red}{Red line} (and left scattergram) is sample with spin effects like of Higgs,
\textcolor{green}{green line} (and right scattergram) \greenlineis
black line is ratio \textcolor{red}{original}/\textcolor{green}{modified} with whenever available superimposed result for the
fitted functions.
}
\end{figure}

\newpage
\subsection{The energy spectrum: $\tau^- \to \rho^-$ {\tt vs } $\tau^+ \to \mu^+, e^+$}
\vspace{1\baselineskip}

\begin{figure}[h!]
\centering
\resizebox*{0.49\textwidth}{!}{\includegraphics{\przedro 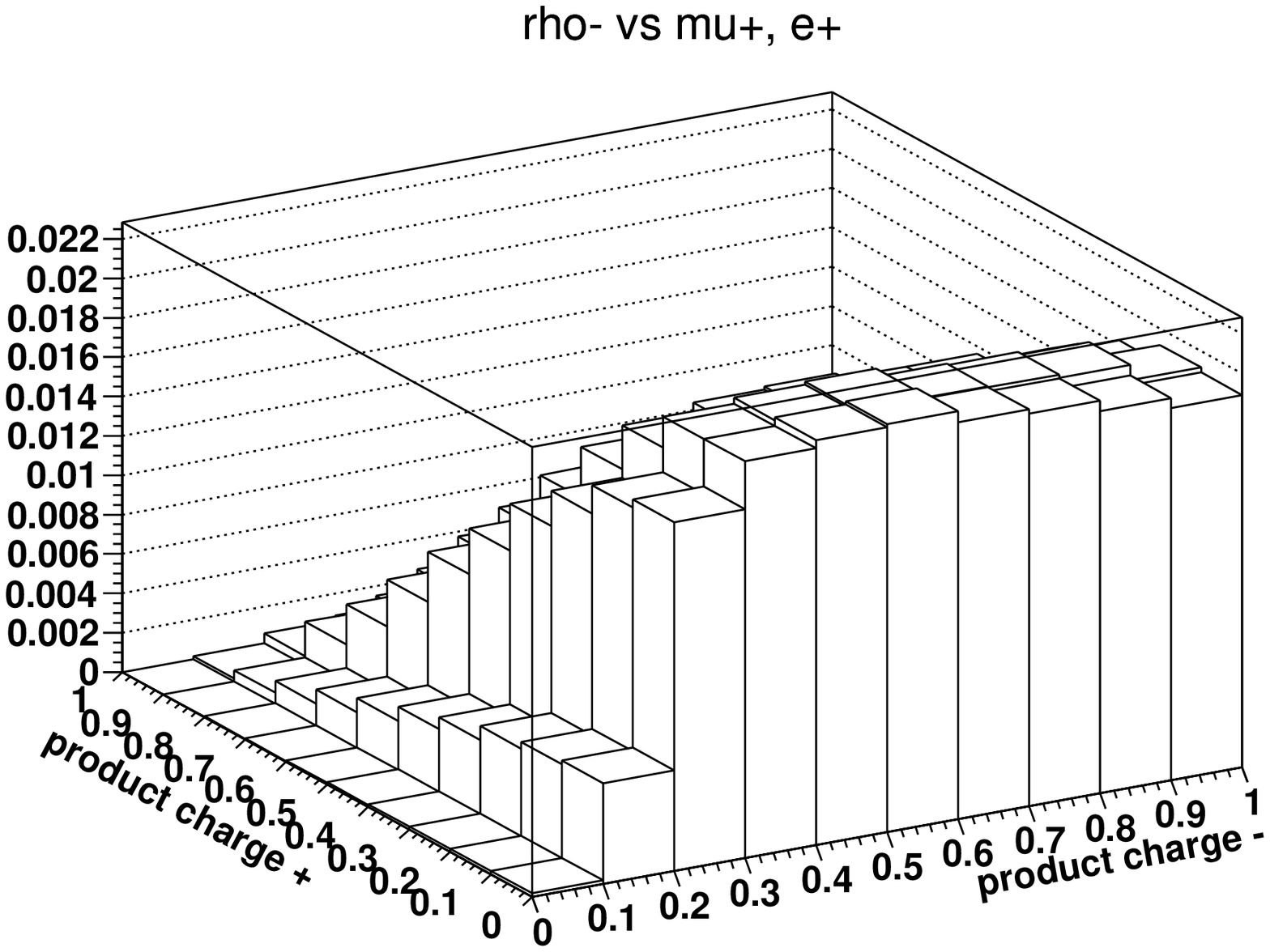}}
\resizebox*{0.49\textwidth}{!}{\includegraphics{\przedro 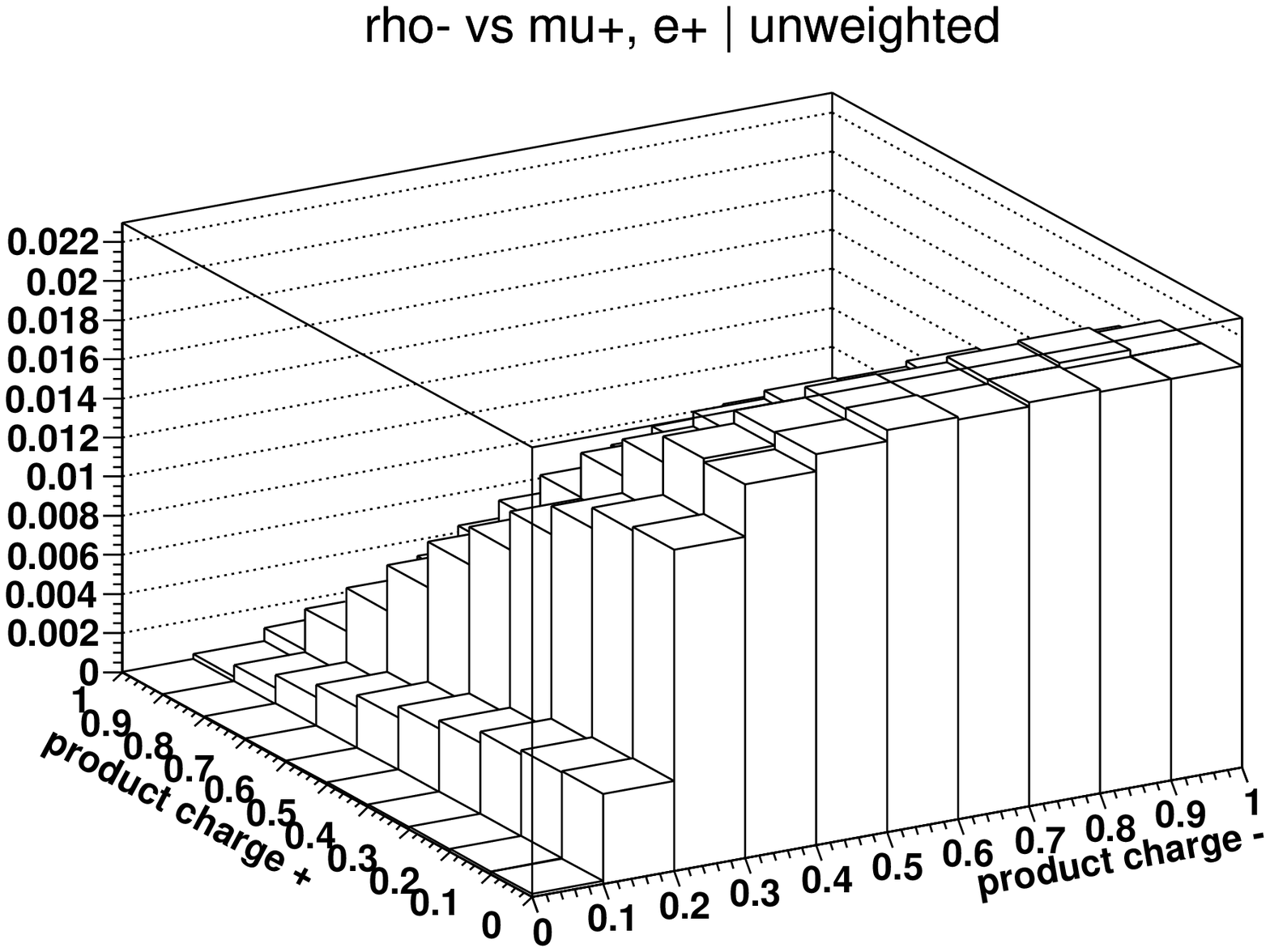}} \\
\resizebox*{0.49\textwidth}{!}{\includegraphics{\przedro 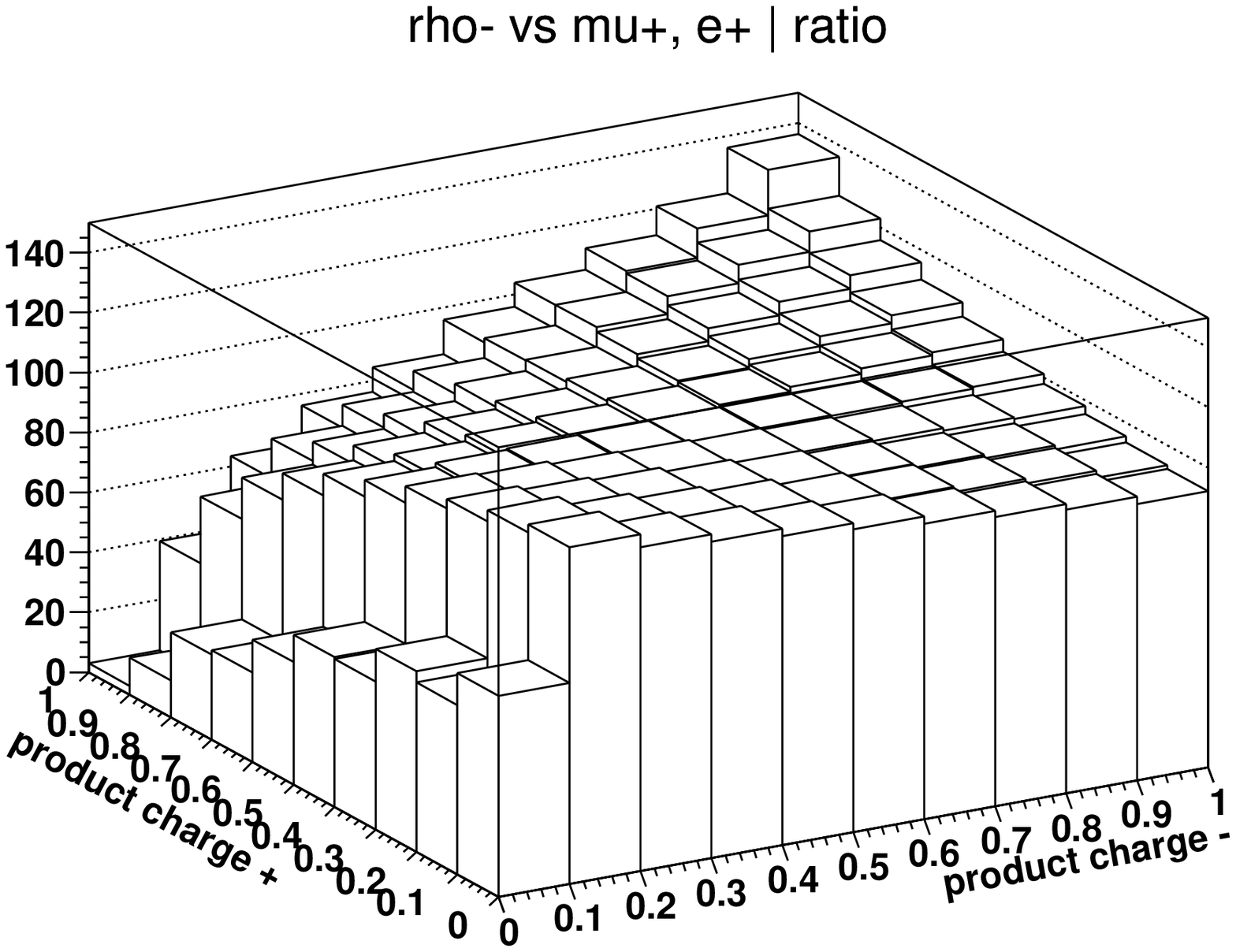}}
\resizebox*{0.49\textwidth}{!}{\includegraphics{\przedro 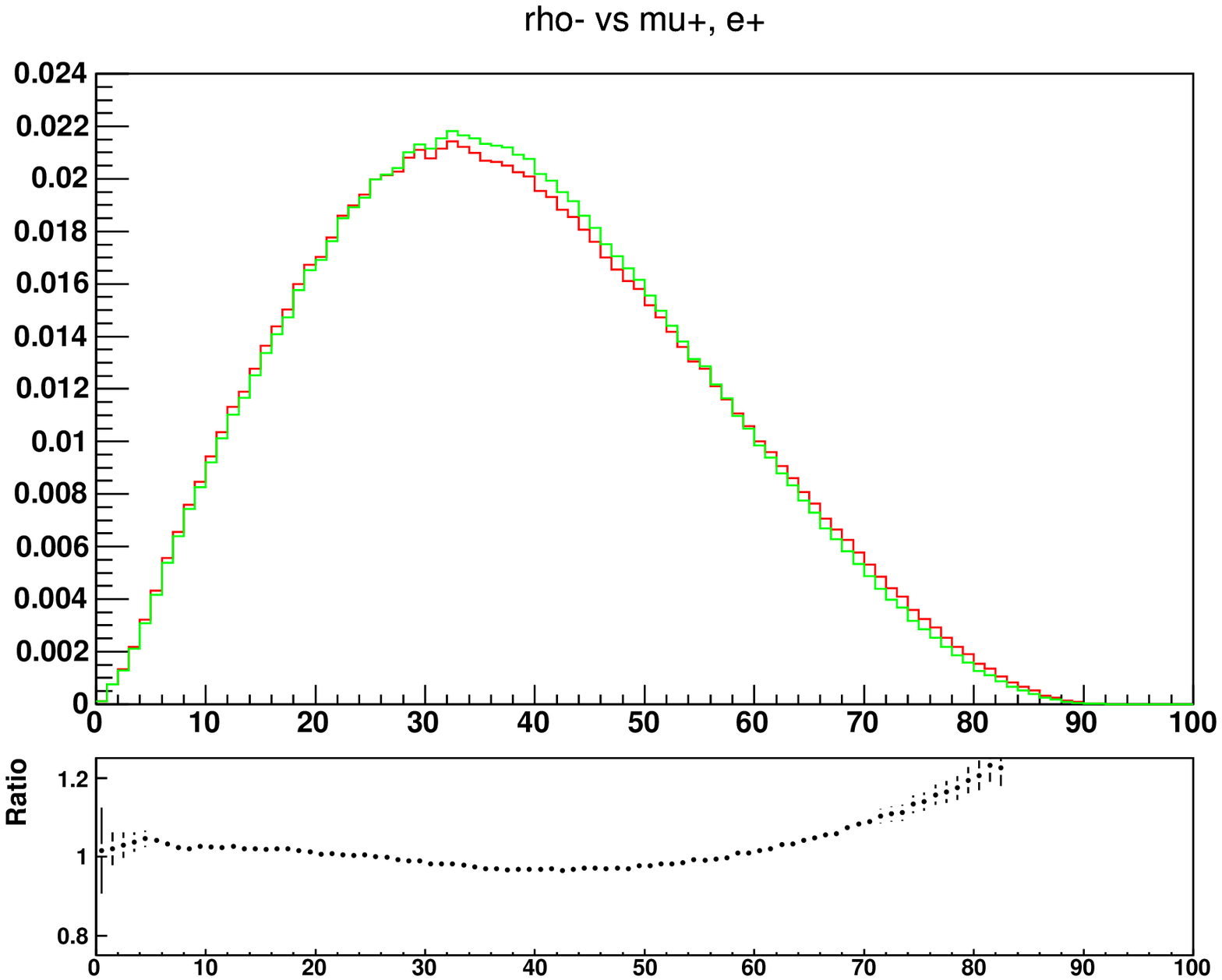}} \\
\resizebox*{0.49\textwidth}{!}{\includegraphics{\przedro 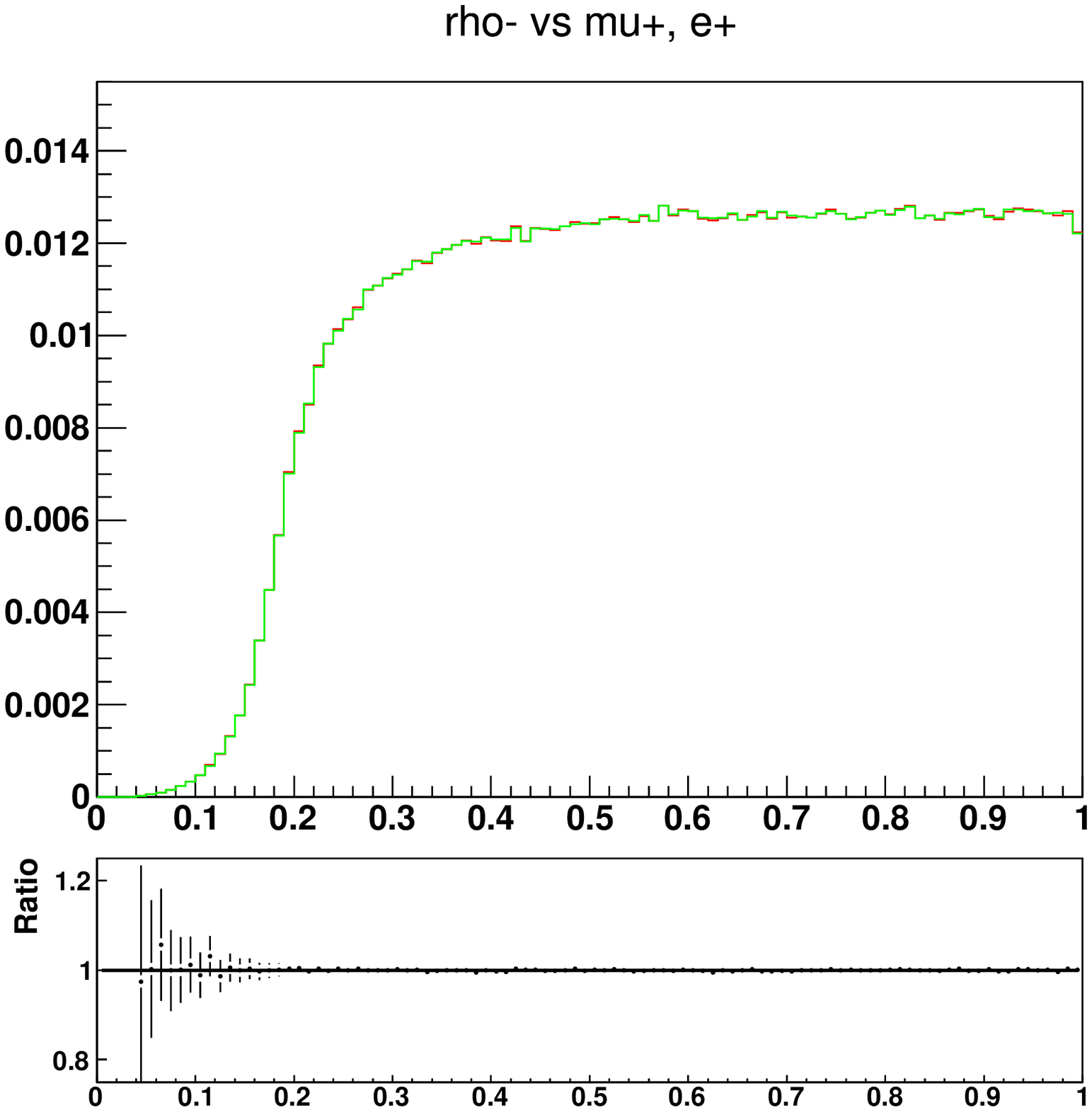}}
\resizebox*{0.49\textwidth}{!}{\includegraphics{\przedro 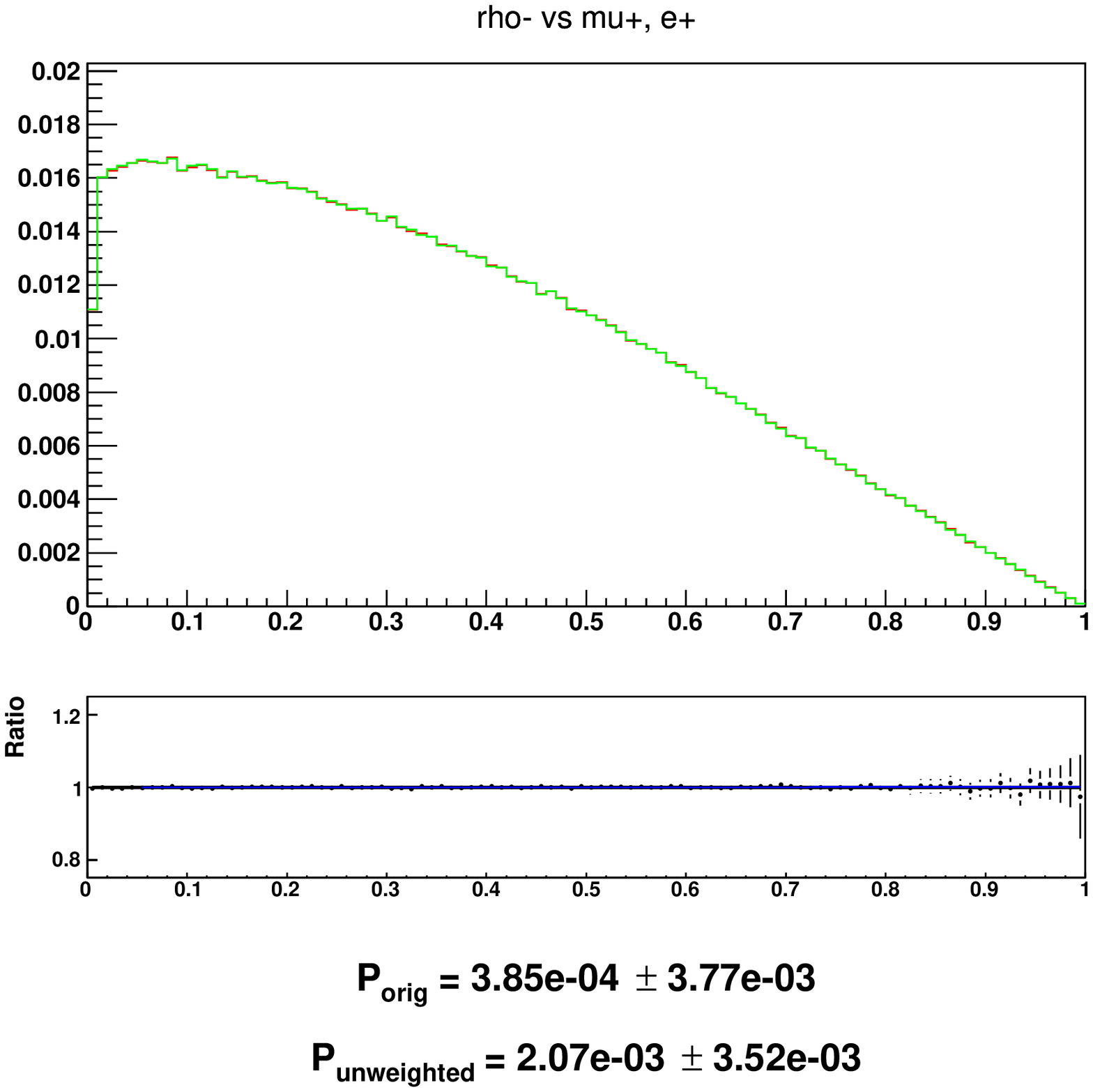}} \\
\caption{\small Fractions of  $\tau^+$ and $\tau^-$ energies carried by their visible  decay products:
two dimensional lego plots and one dimensional spectra$^{18}$.
\textcolor{red}{Red line} (and left scattergram) is sample with spin effects like of Higgs,
\textcolor{green}{green line} (and right scattergram) \greenlineis
black line is ratio \textcolor{red}{original}/\textcolor{green}{modified} with whenever available superimposed result for the
fitted functions.
}
\end{figure}

\newpage
\subsection{The energy spectrum: $\tau^- \to \pi^-$ {\tt vs } $\tau^+ \to \rho^+$}
\vspace{1\baselineskip}

\begin{figure}[h!]
\centering
\resizebox*{0.49\textwidth}{!}{\includegraphics{\przedro 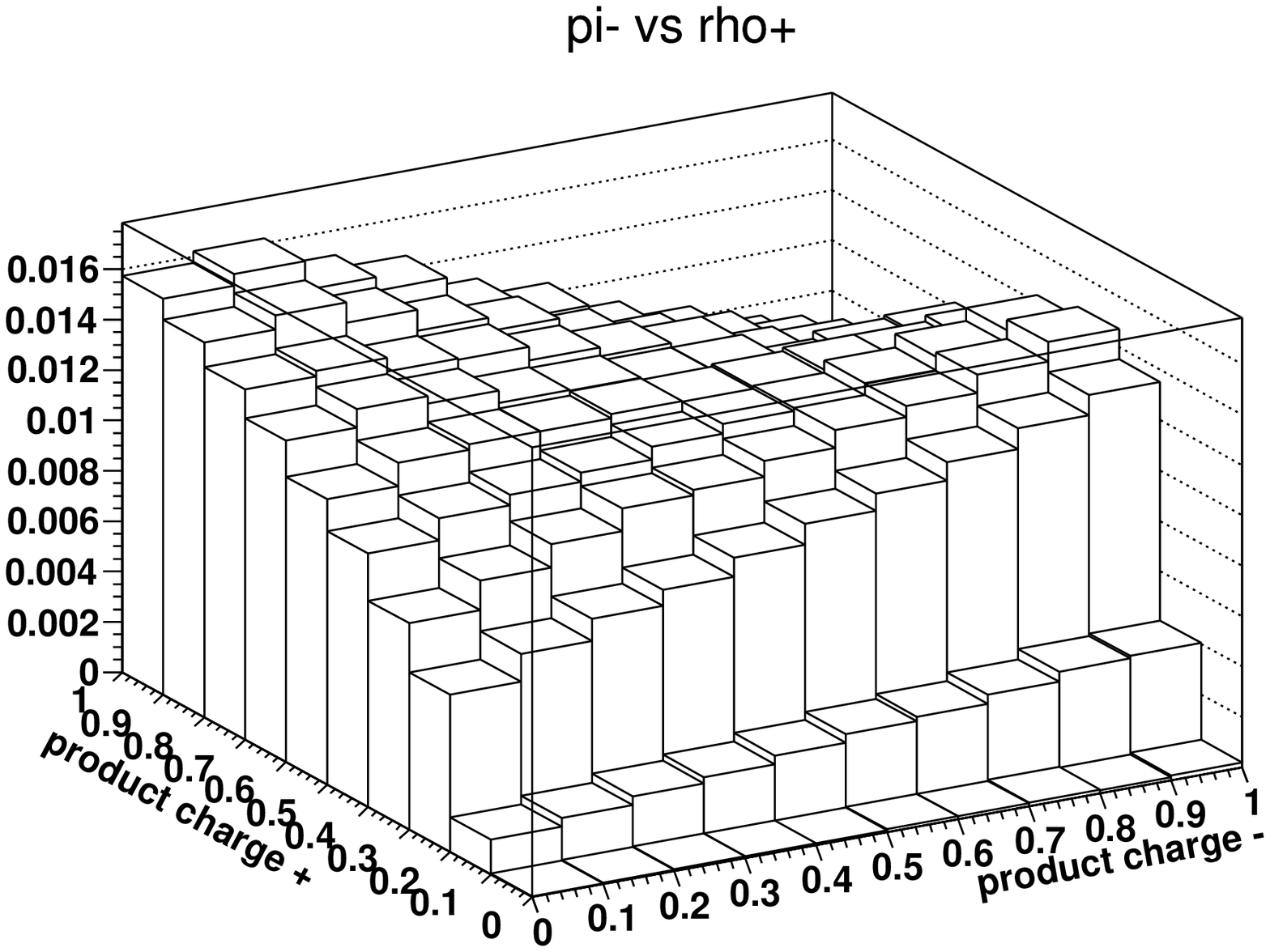}}
\resizebox*{0.49\textwidth}{!}{\includegraphics{\przedro 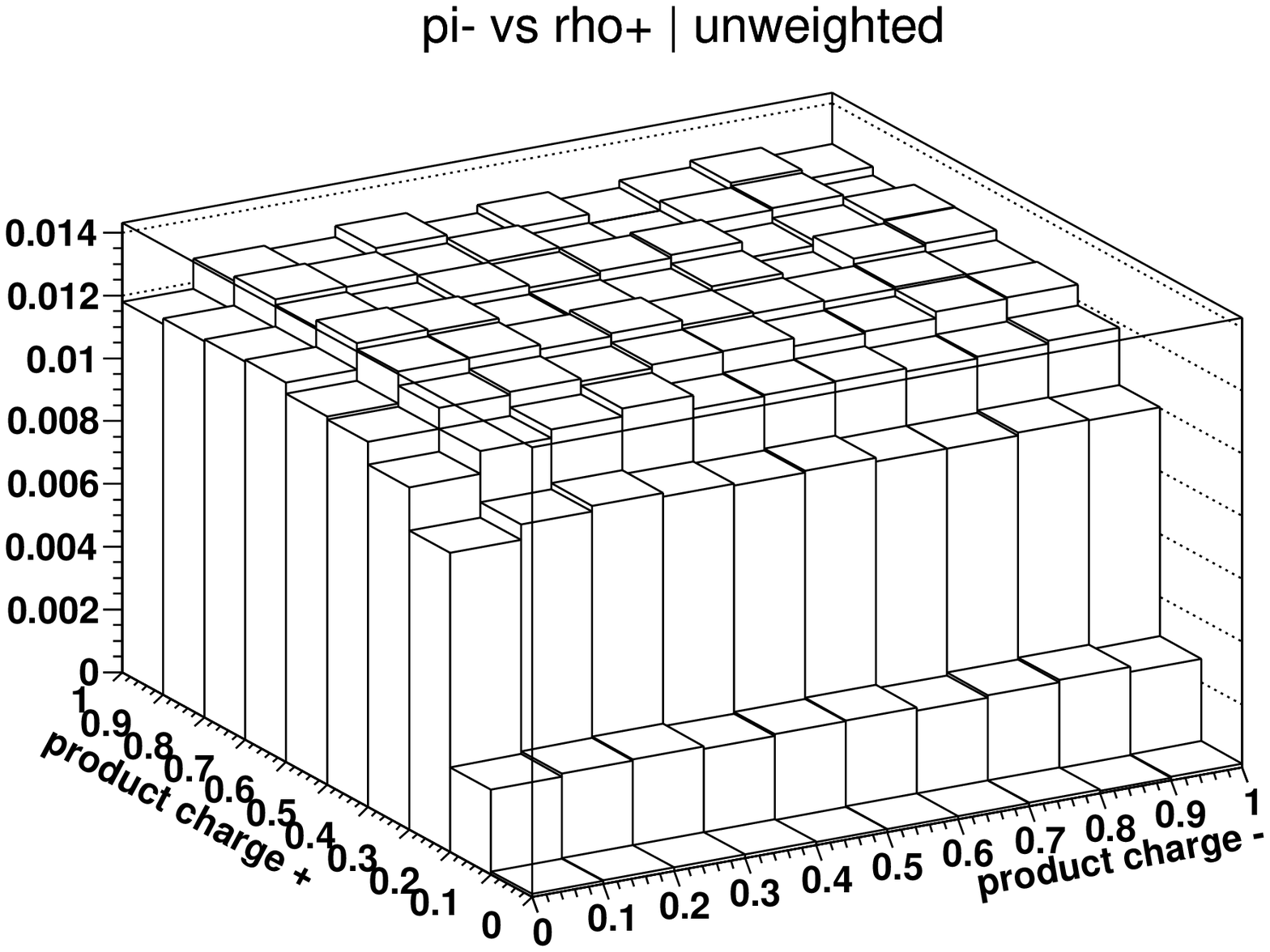}} \\
\resizebox*{0.49\textwidth}{!}{\includegraphics{\przedro 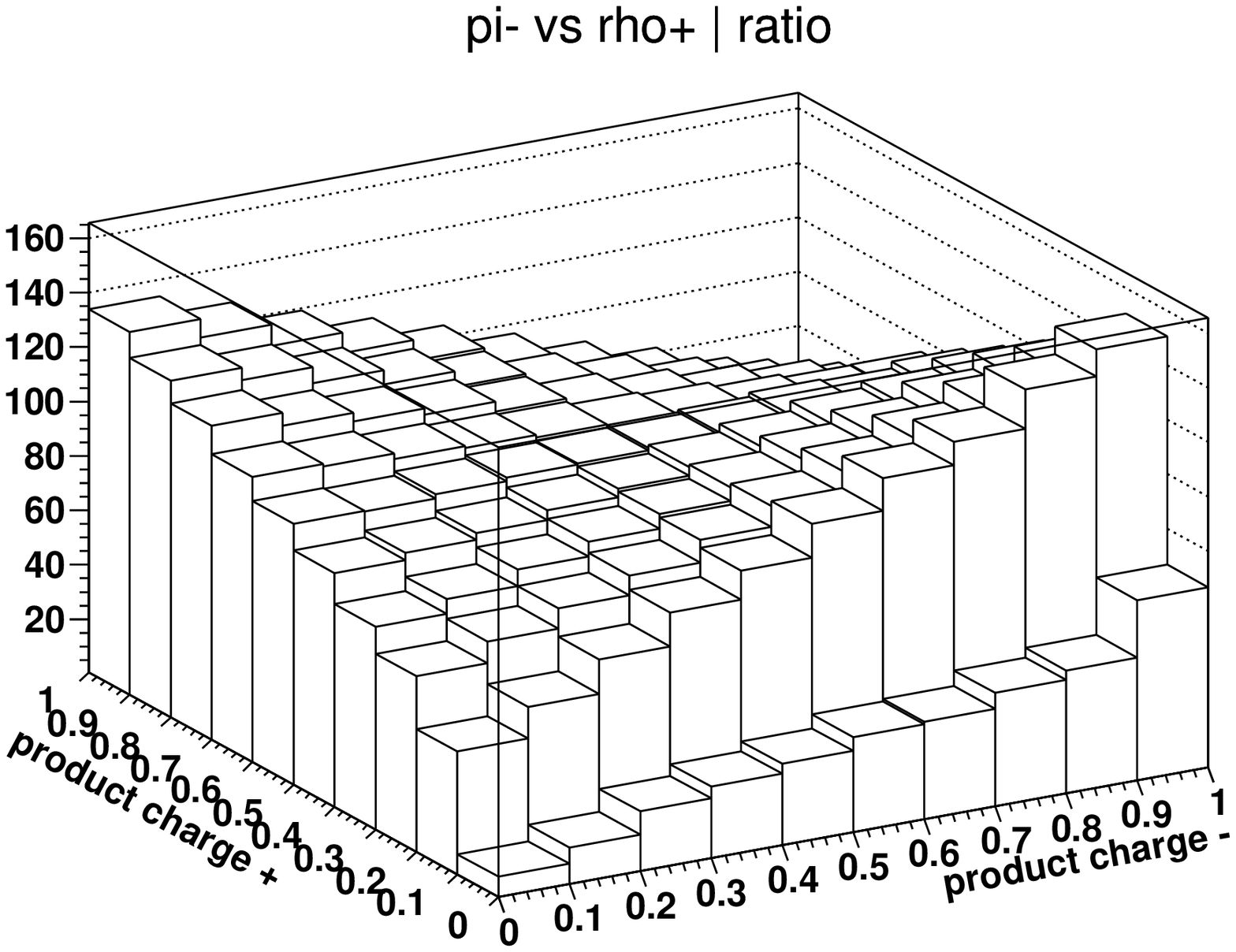}}
\resizebox*{0.49\textwidth}{!}{\includegraphics{\przedro 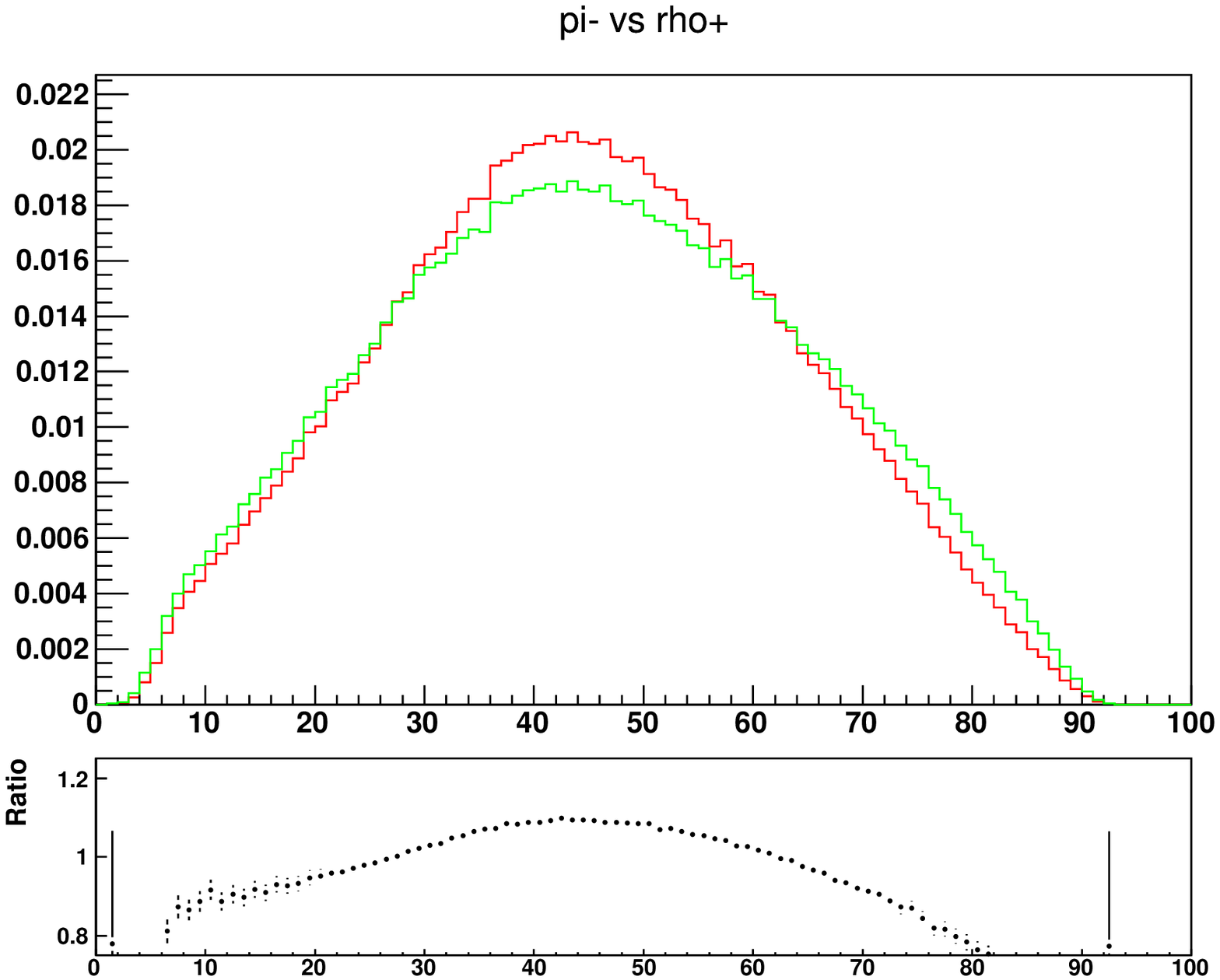}} \\
\resizebox*{0.49\textwidth}{!}{\includegraphics{\przedro 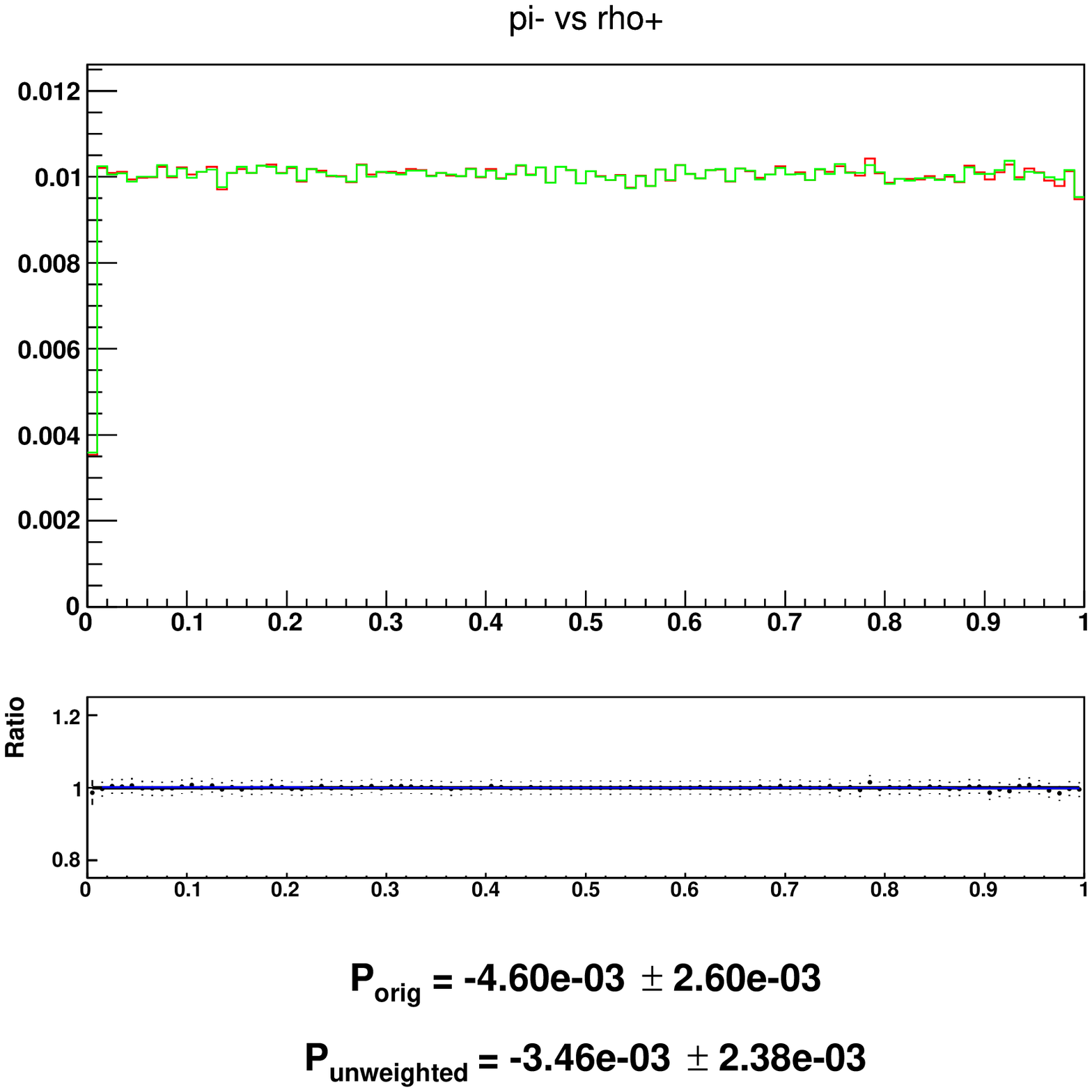}}
\resizebox*{0.49\textwidth}{!}{\includegraphics{\przedro 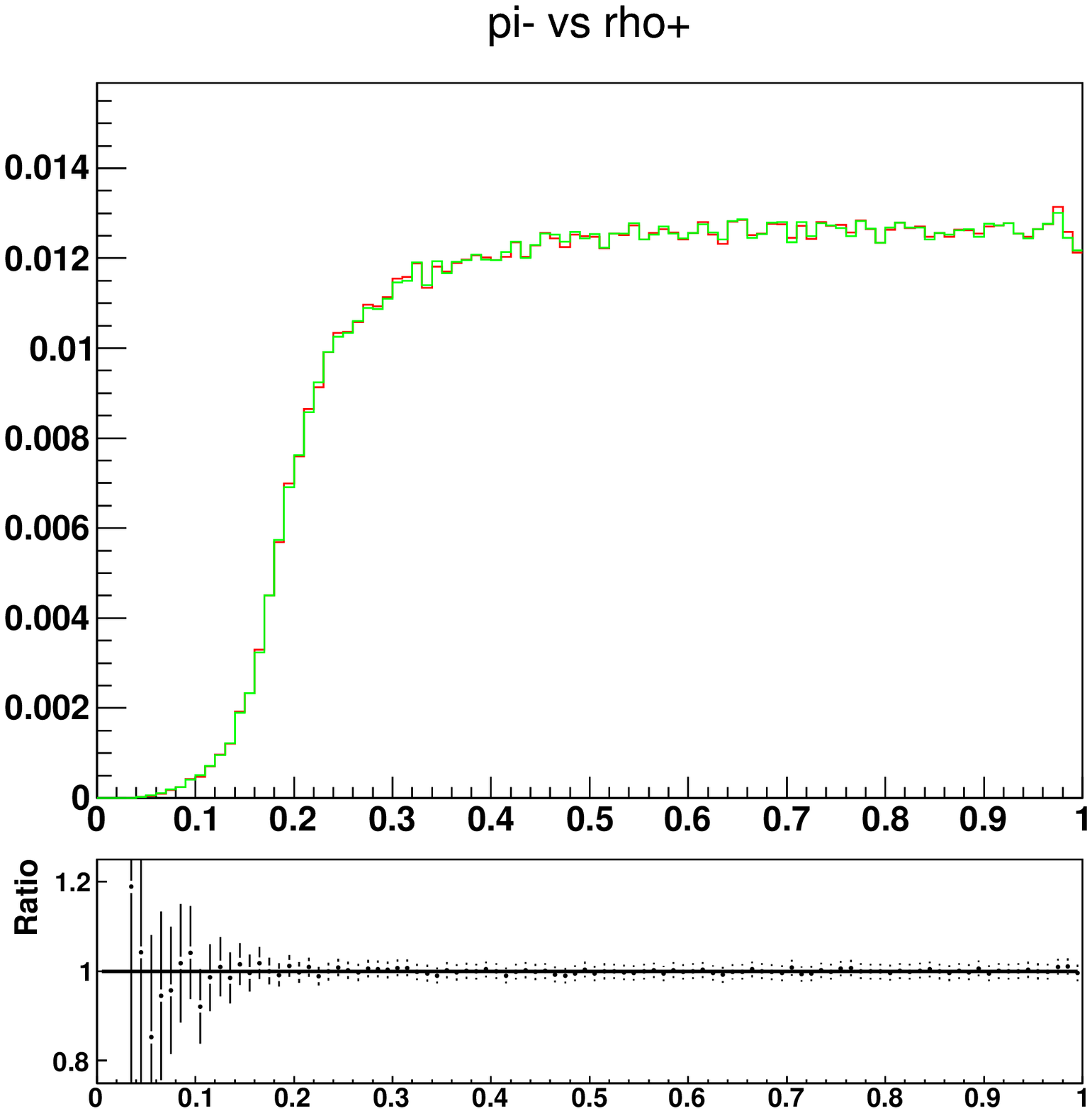}} \\
\caption{\small Fractions of  $\tau^+$ and $\tau^-$ energies carried by their visible  decay products:
two dimensional lego plots and one dimensional spectra$^{18}$.
\textcolor{red}{Red line} (and left scattergram) is sample with spin effects like of Higgs,
\textcolor{green}{green line} (and right scattergram) \greenlineis
black line is ratio \textcolor{red}{original}/\textcolor{green}{modified} with whenever available superimposed result for the
fitted functions.
}
\end{figure}

\newpage
\subsection{The energy spectrum: $\tau^- \to \rho^-$ {\tt vs } $\tau^+ \to \pi^+$}
\vspace{1\baselineskip}

\begin{figure}[h!]
\centering
\resizebox*{0.49\textwidth}{!}{\includegraphics{\przedro 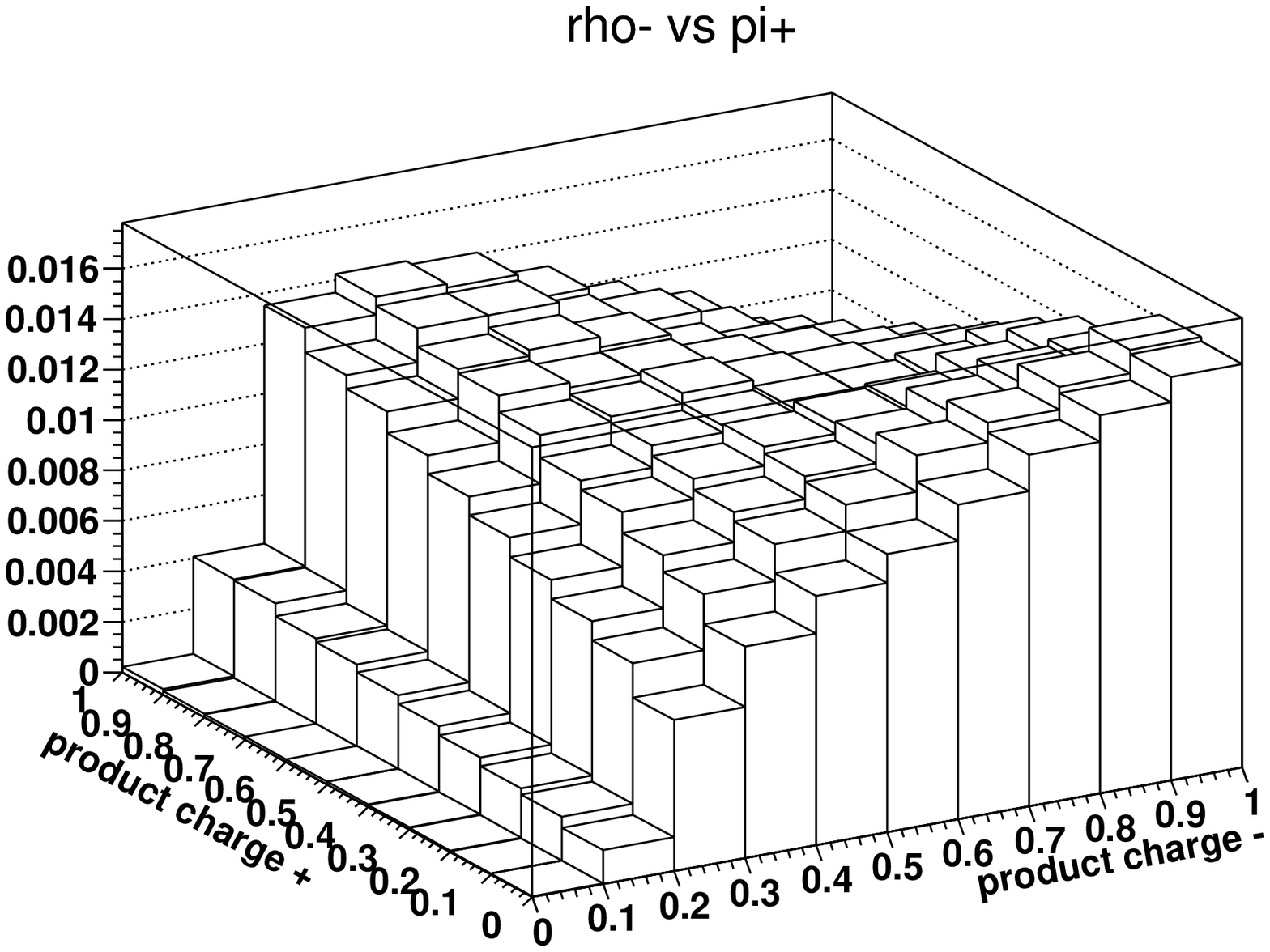}}
\resizebox*{0.49\textwidth}{!}{\includegraphics{\przedro 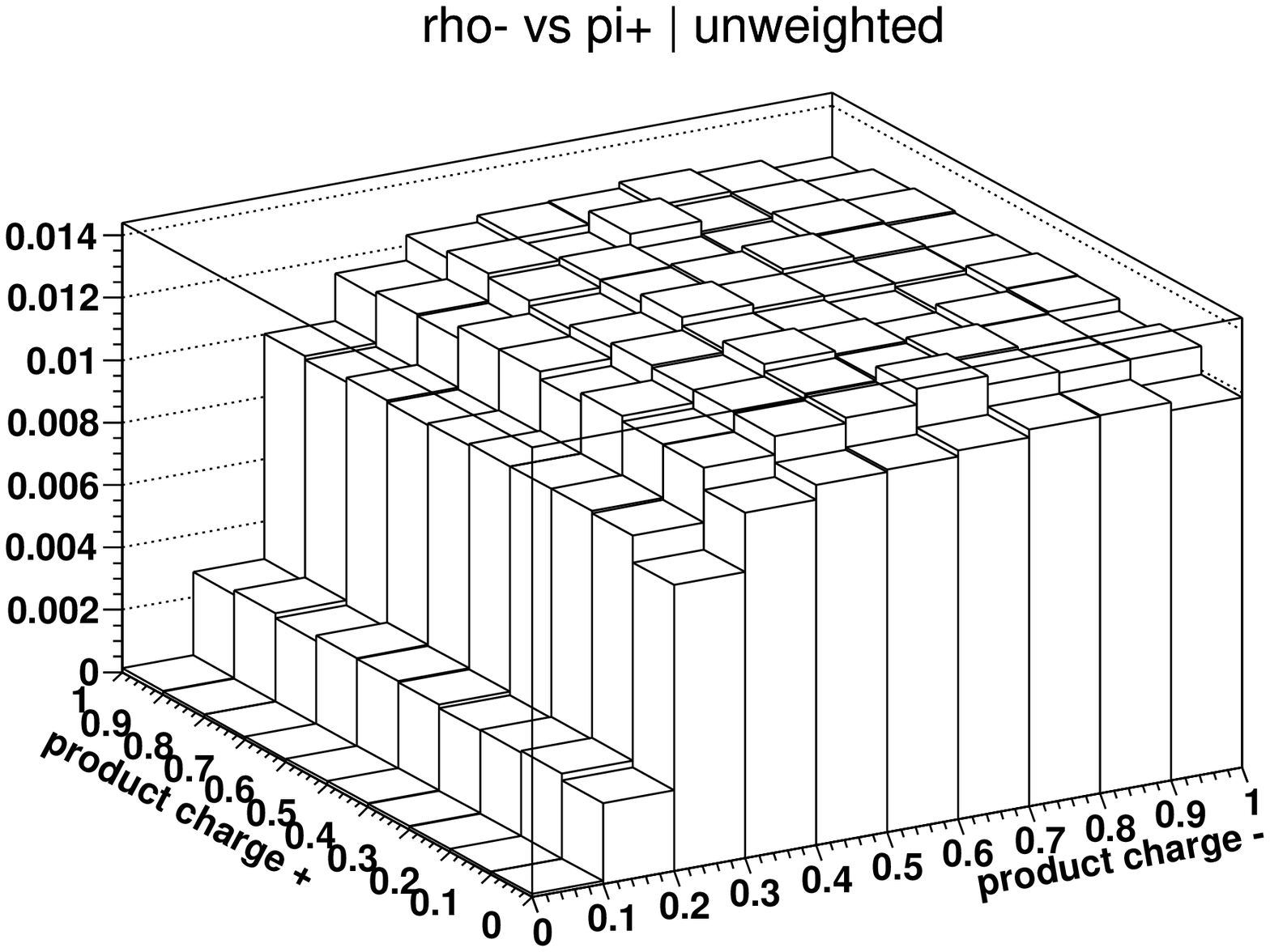}} \\
\resizebox*{0.49\textwidth}{!}{\includegraphics{\przedro 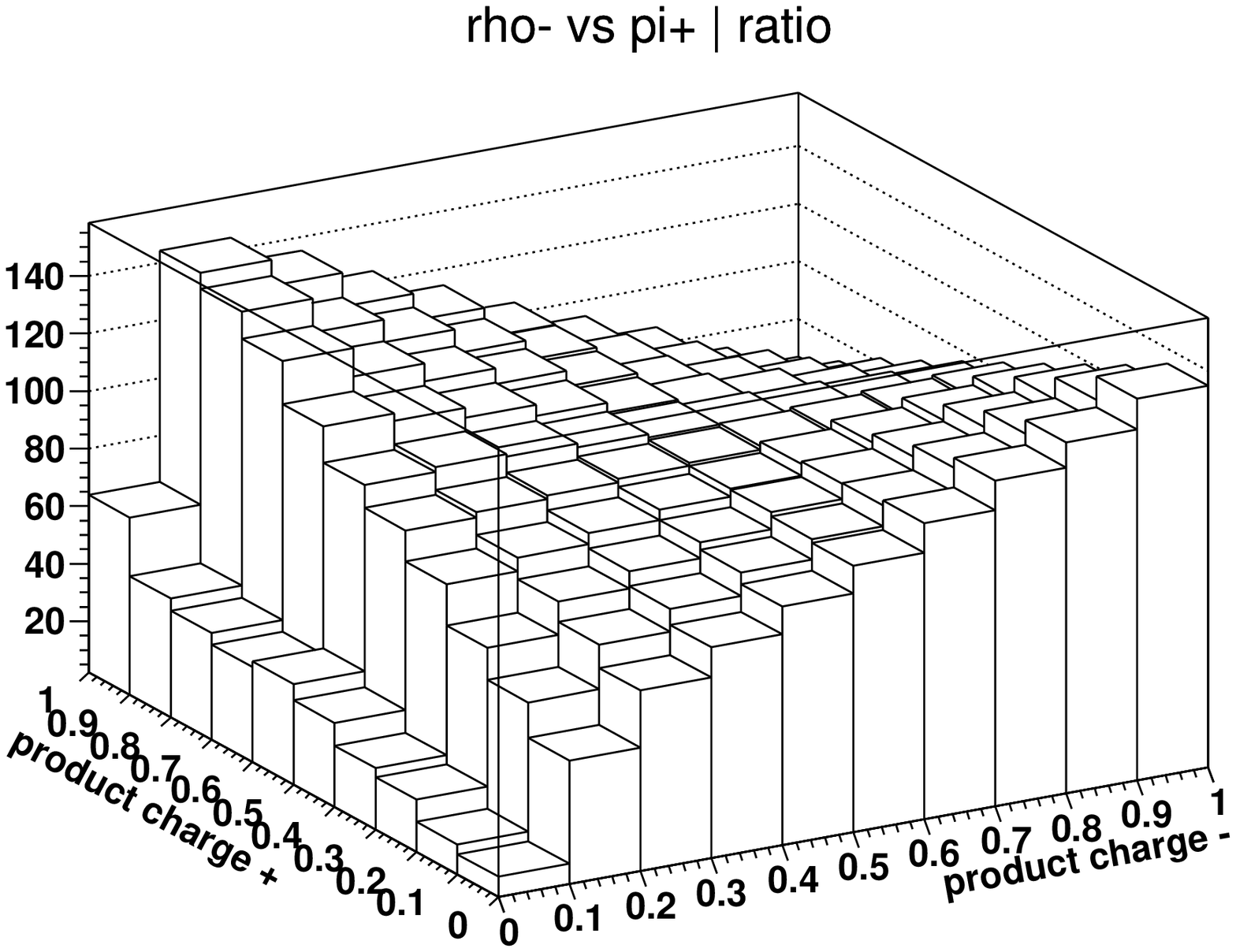}}
\resizebox*{0.49\textwidth}{!}{\includegraphics{\przedro 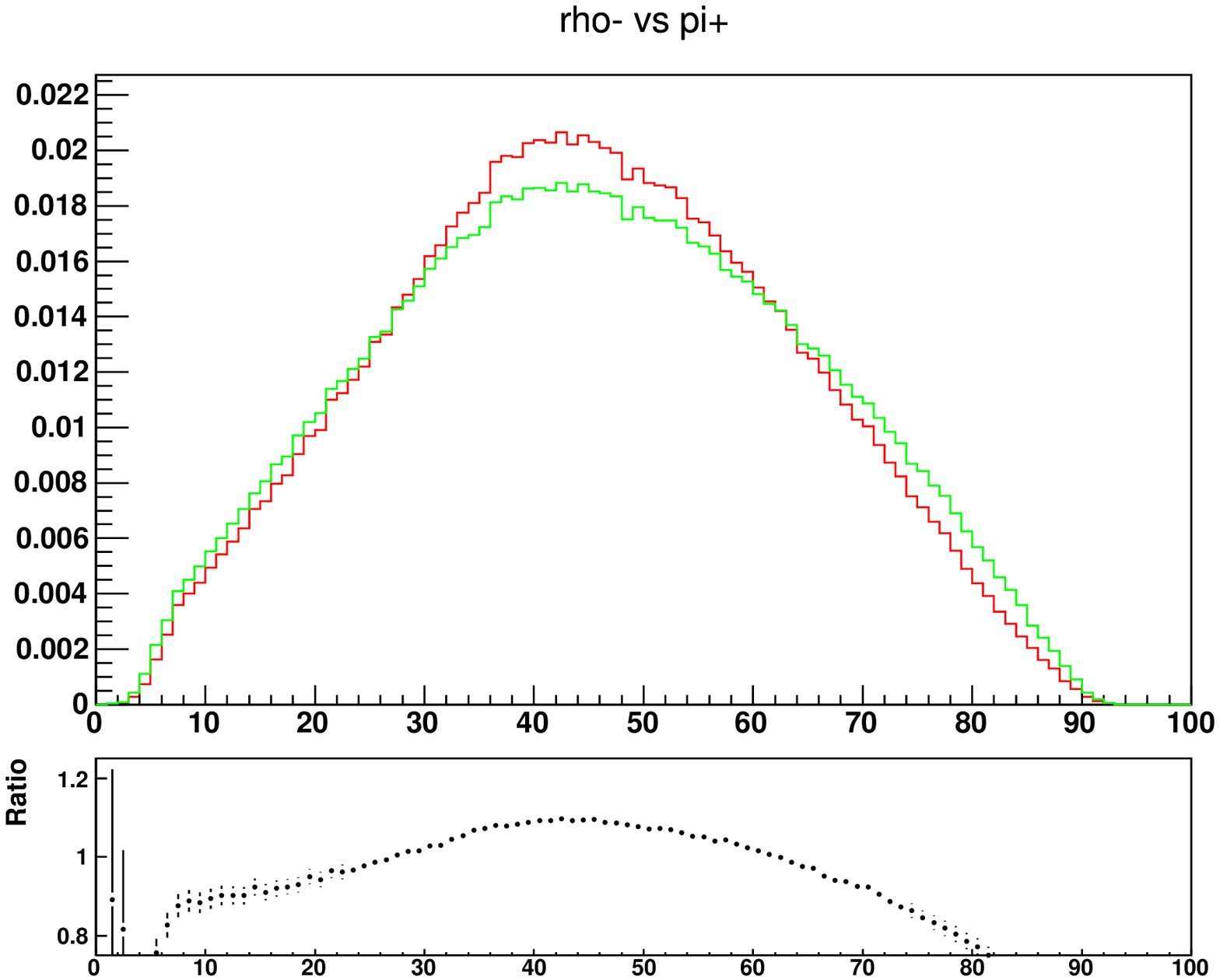}} \\
\resizebox*{0.49\textwidth}{!}{\includegraphics{\przedro 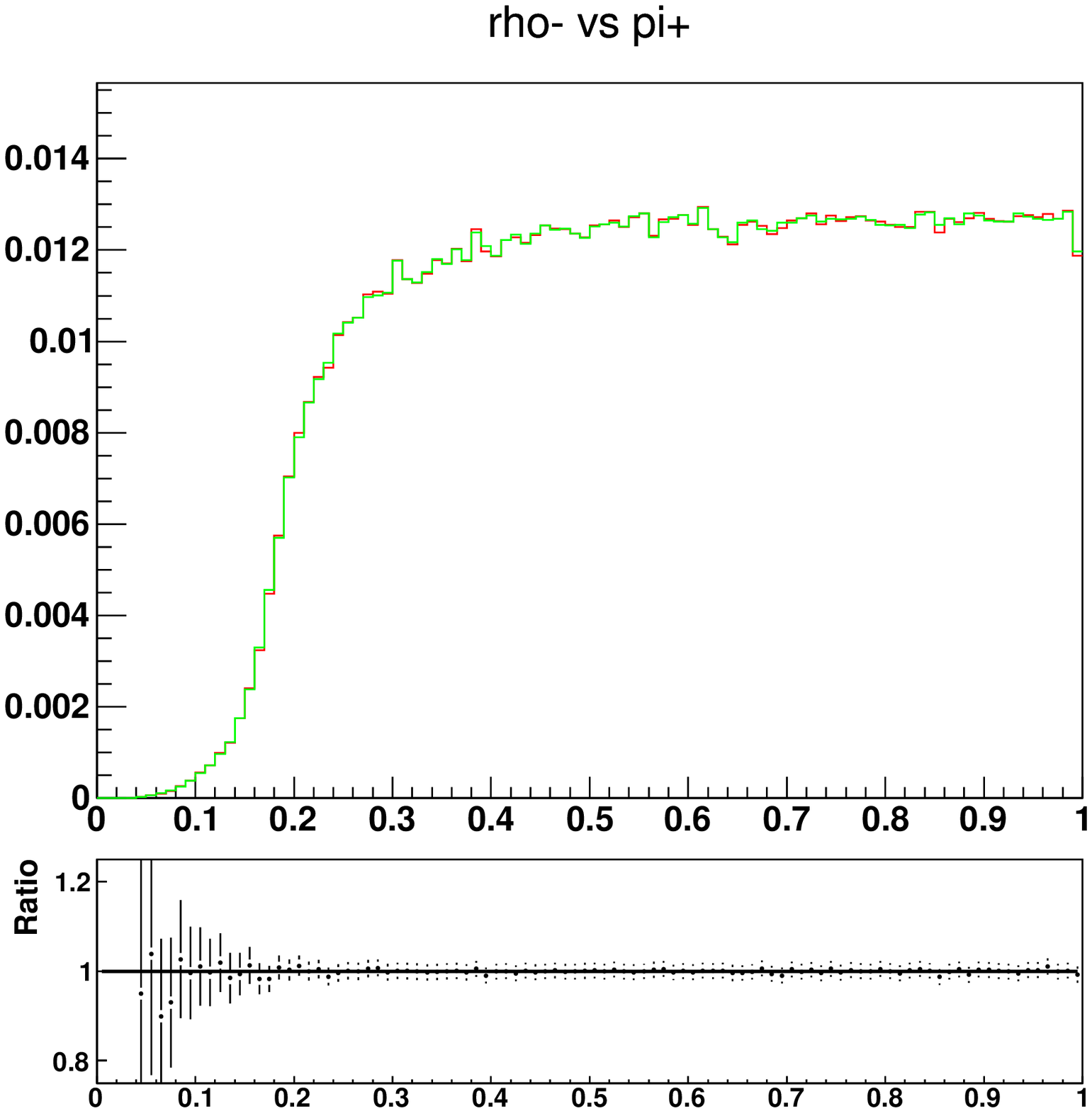}}
\resizebox*{0.49\textwidth}{!}{\includegraphics{\przedro 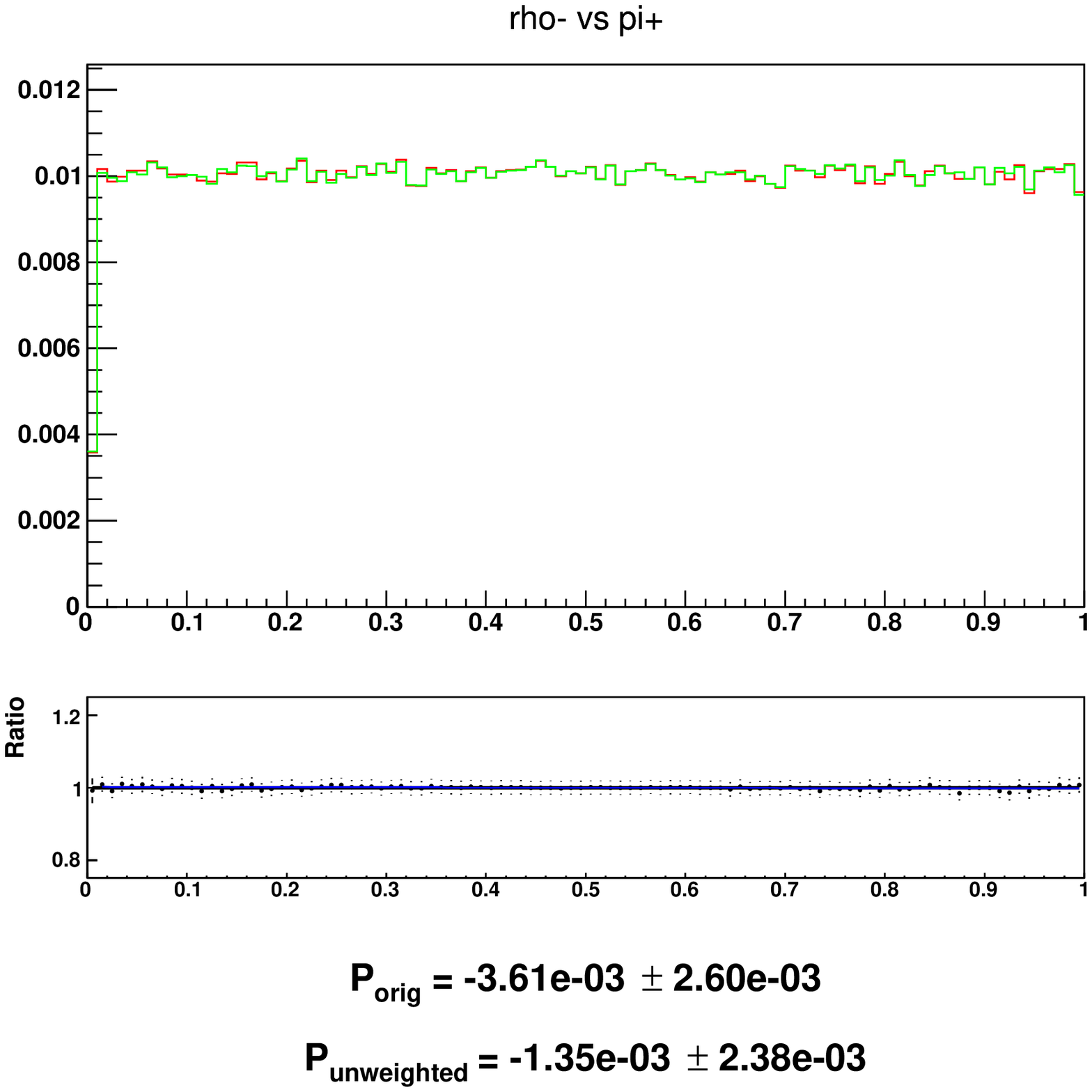}} \\
\caption{\small Fractions of  $\tau^+$ and $\tau^-$ energies carried by their visible  decay products:
two dimensional lego plots and one dimensional spectra$^{18}$.
\textcolor{red}{Red line} (and left scattergram) is sample with spin effects like of Higgs,
\textcolor{green}{green line} (and right scattergram) \greenlineis
black line is ratio \textcolor{red}{original}/\textcolor{green}{modified} with whenever available superimposed result for the
fitted functions.
}
\end{figure}

\newpage
\subsection{The energy spectrum: $\tau^- \to \rho^-$ {\tt vs } $\tau^+ \to \rho^+$}
\vspace{1\baselineskip}

\begin{figure}[h!]
\centering
\resizebox*{0.49\textwidth}{!}{\includegraphics{\przedro 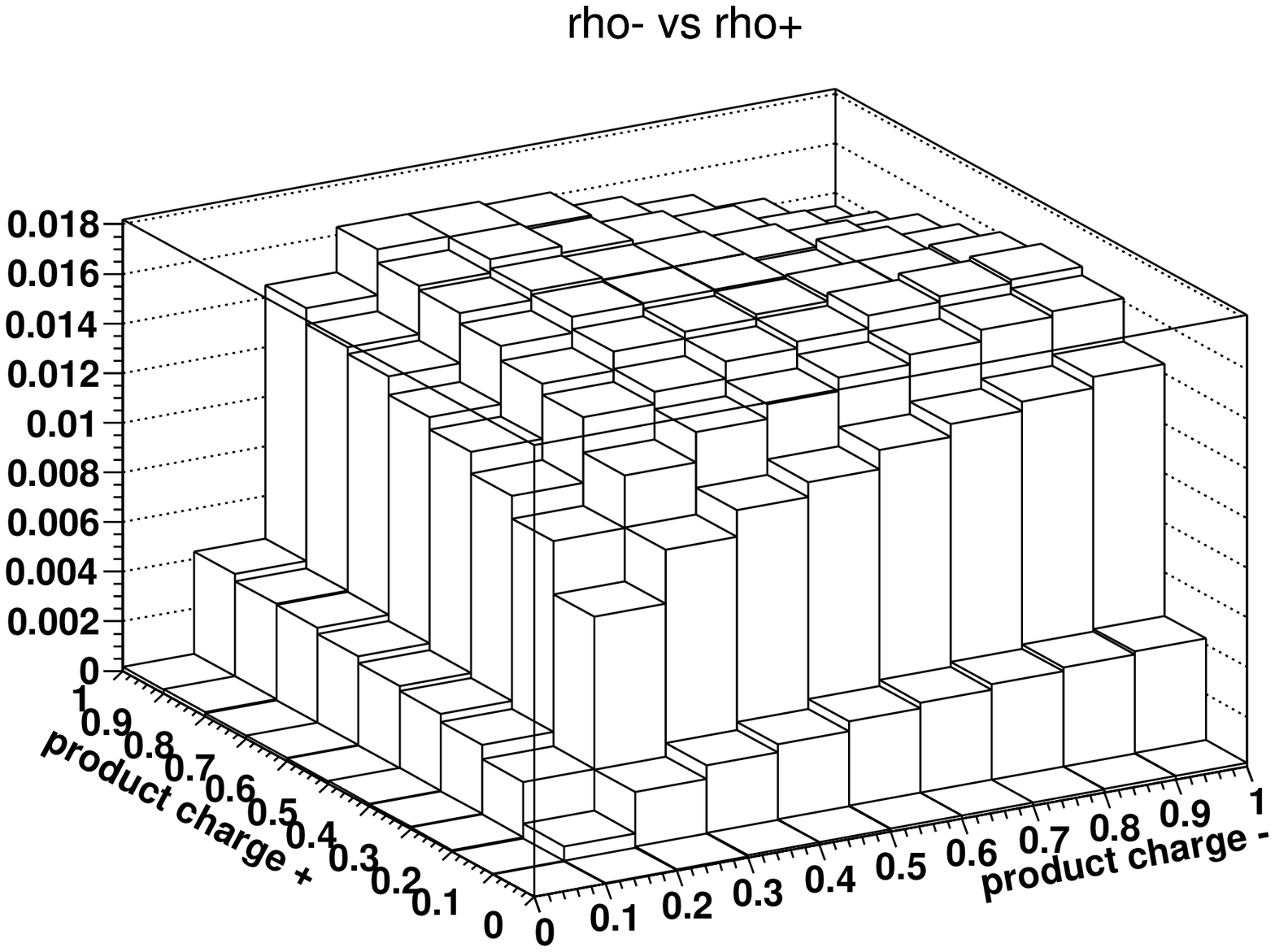}}
\resizebox*{0.49\textwidth}{!}{\includegraphics{\przedro 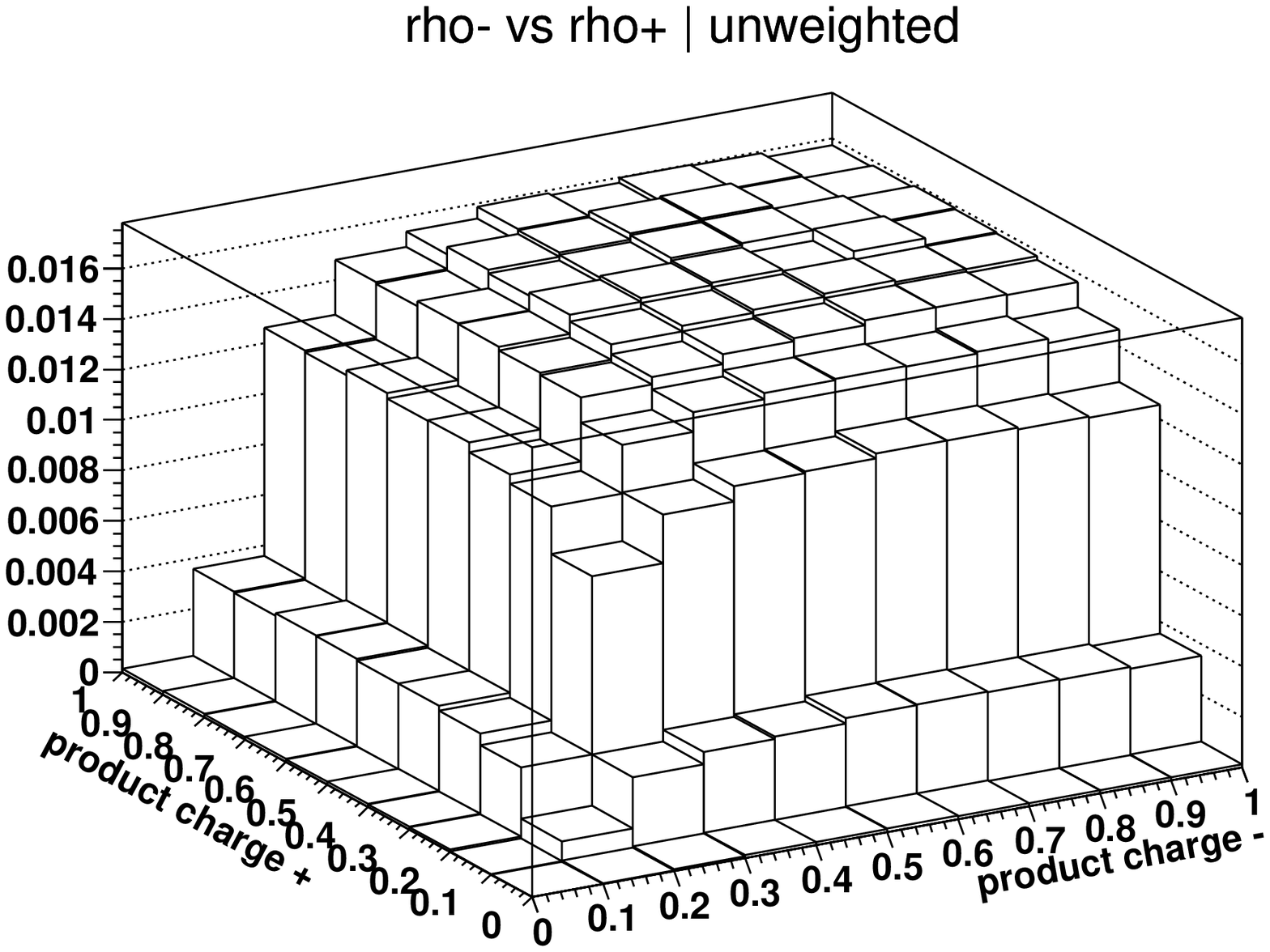}} \\
\resizebox*{0.49\textwidth}{!}{\includegraphics{\przedro 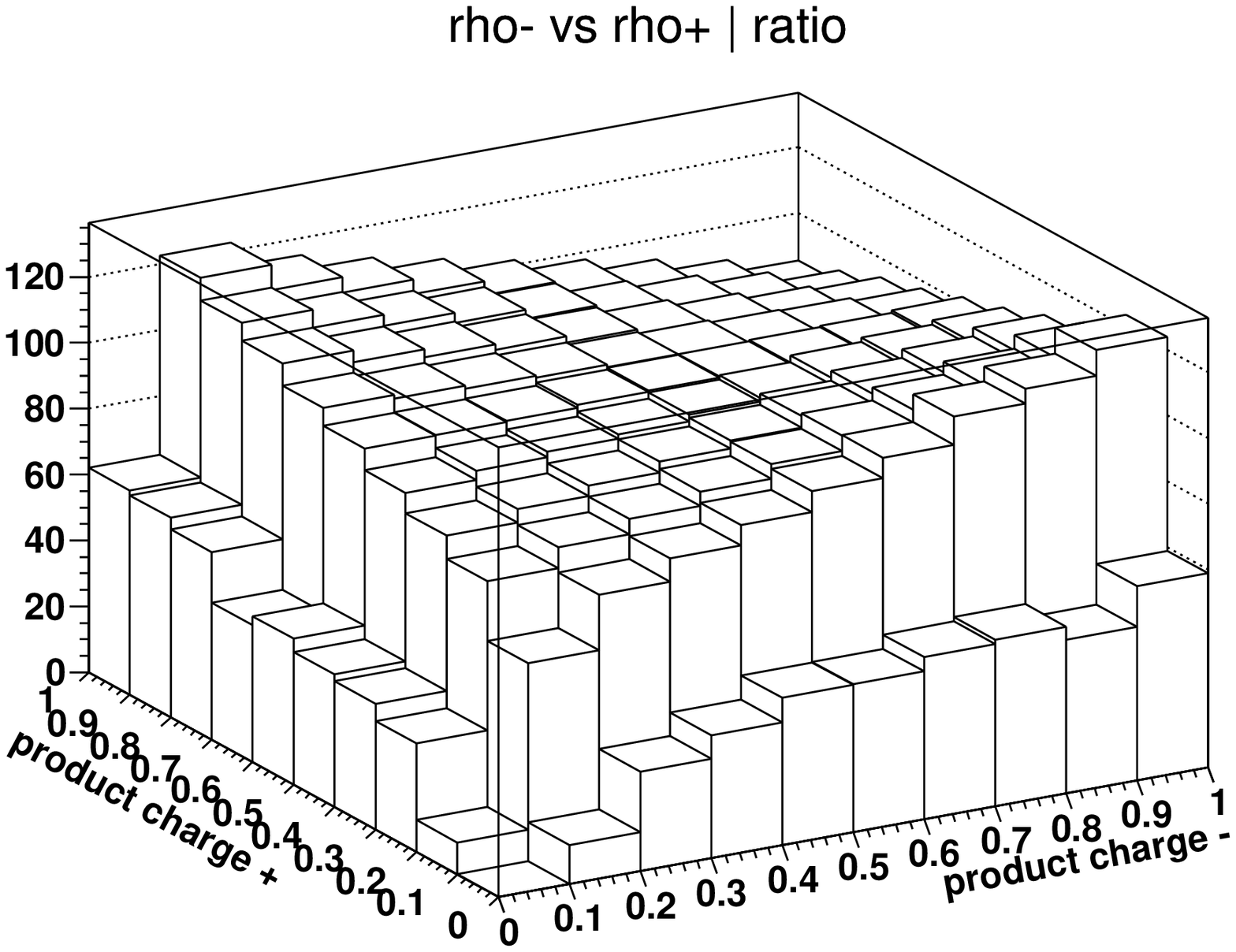}}
\resizebox*{0.49\textwidth}{!}{\includegraphics{\przedro 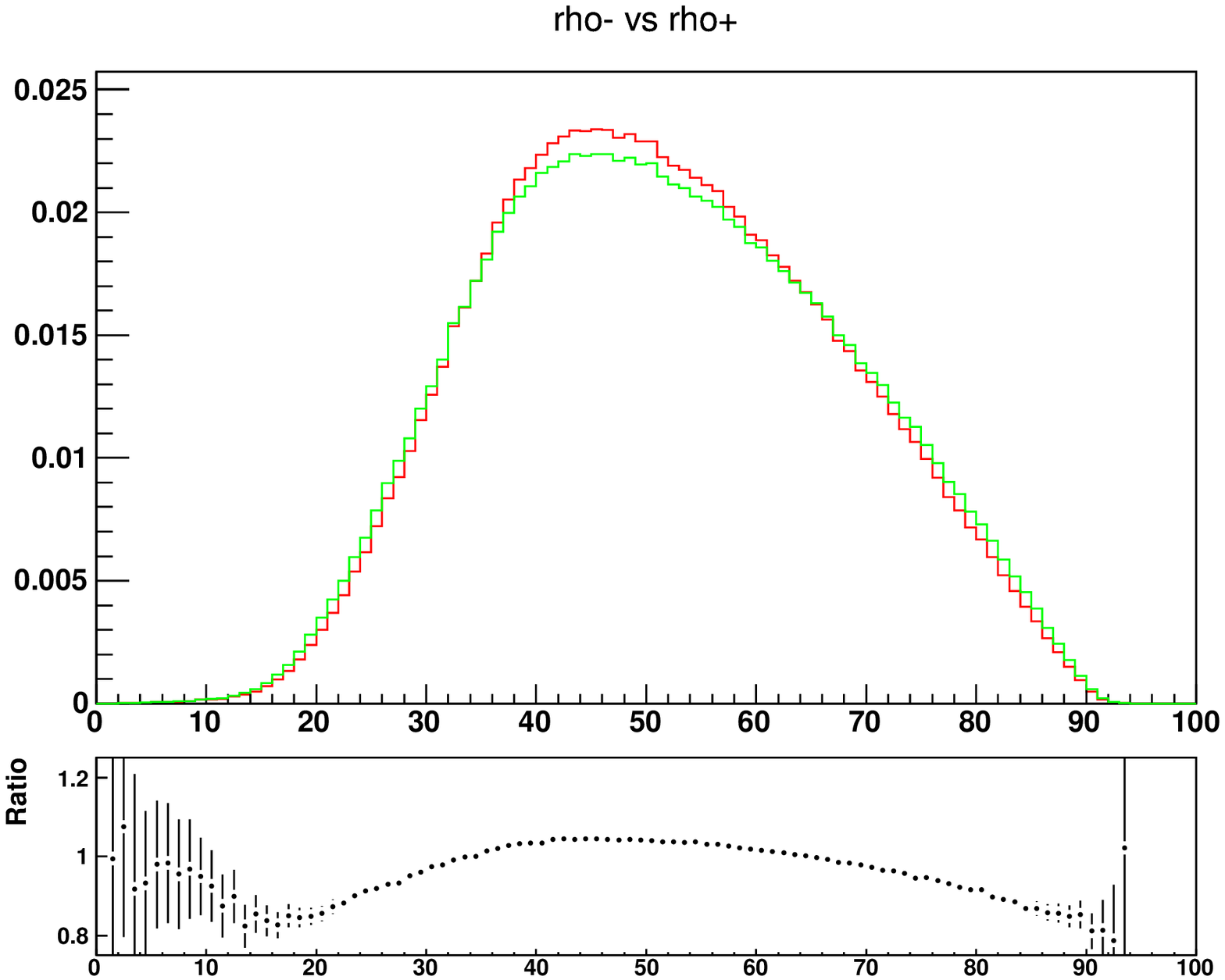}} \\
\resizebox*{0.49\textwidth}{!}{\includegraphics{\przedro 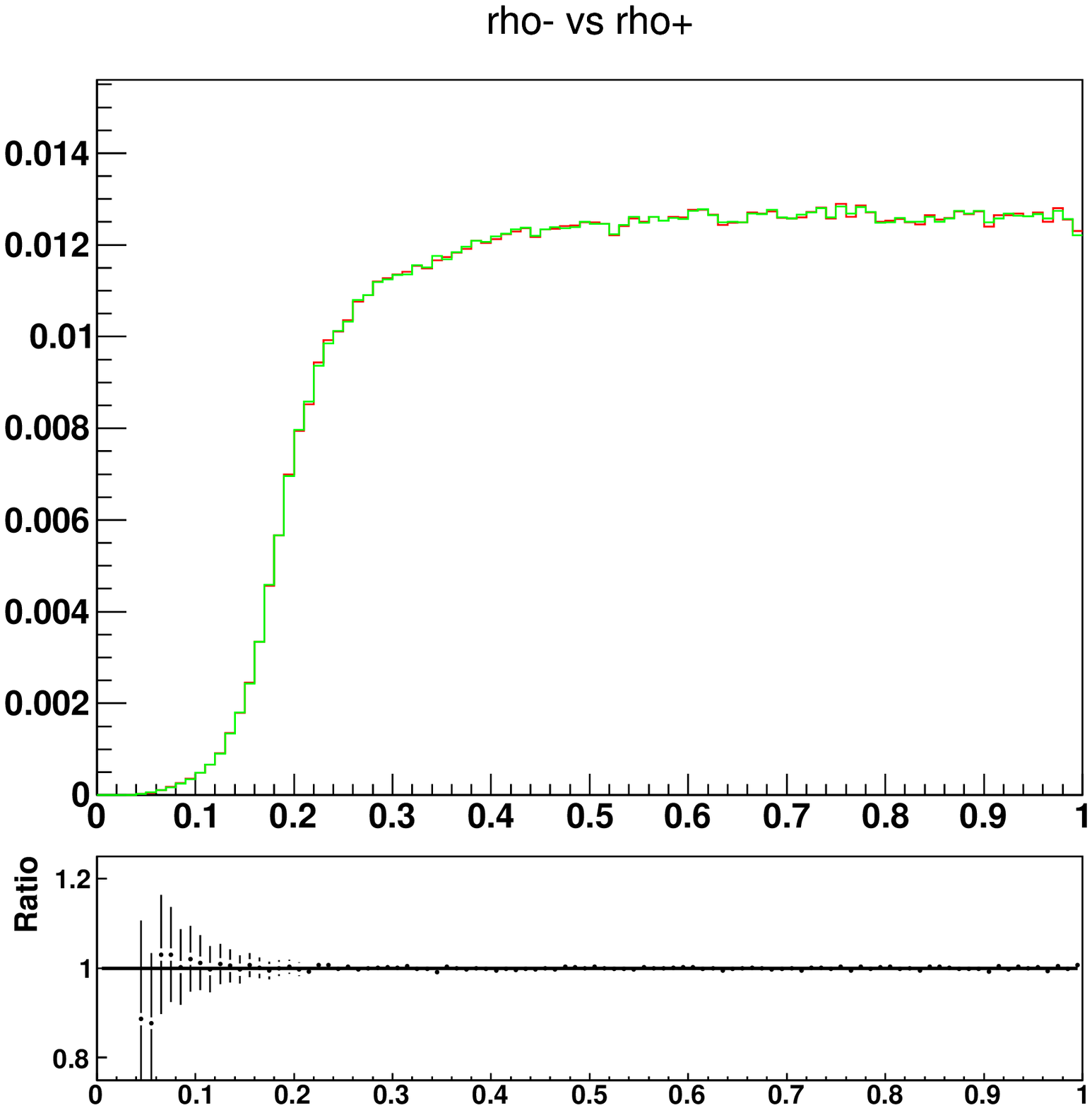}}
\resizebox*{0.49\textwidth}{!}{\includegraphics{\przedro 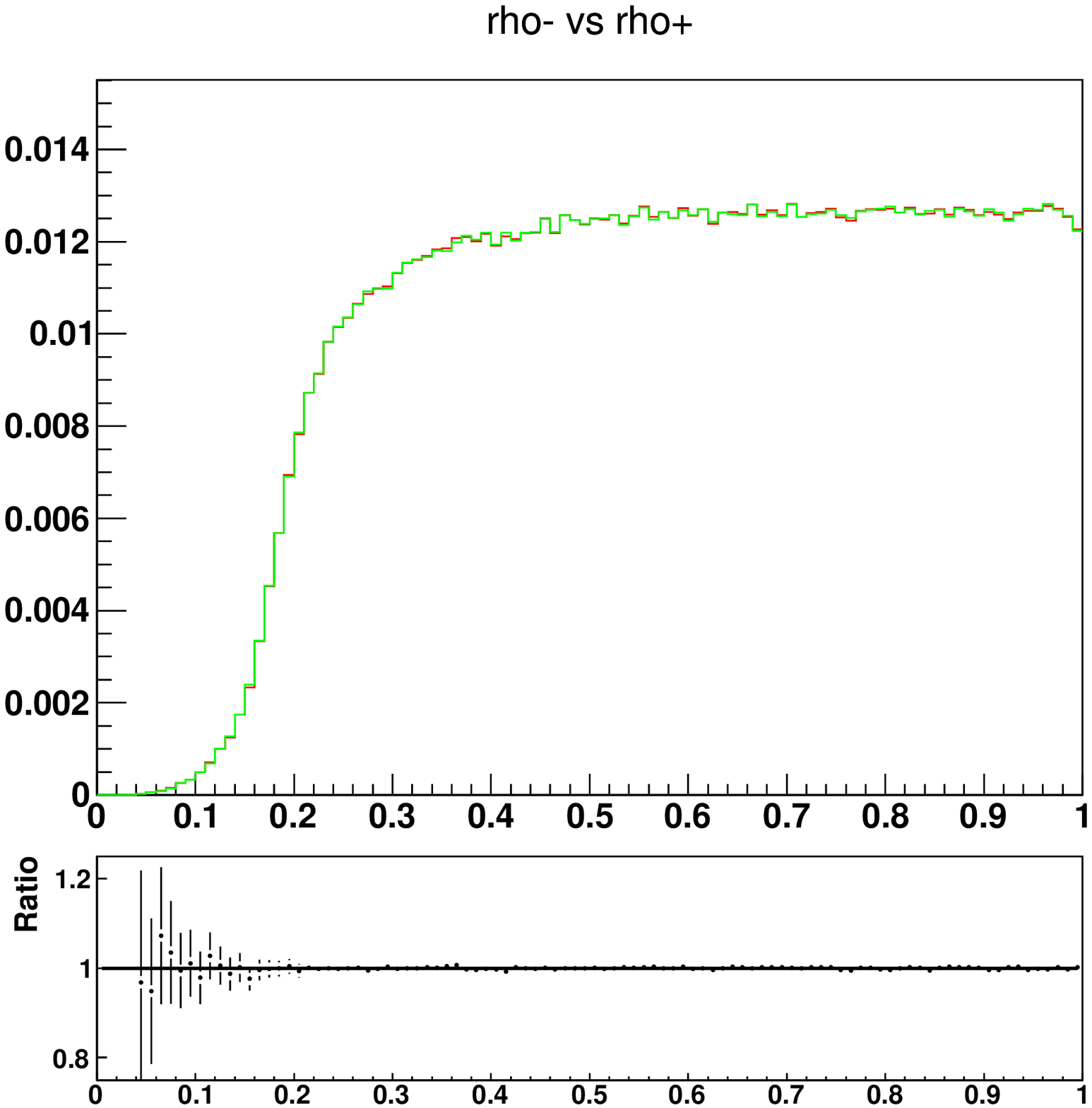}} \\
\caption{\small Fractions of  $\tau^+$ and $\tau^-$ energies carried by their visible  decay products:
two dimensional lego plots and one dimensional spectra.
\textcolor{red}{Red line} (and left scattergram) is sample with spin effects like of Higgs,
\textcolor{green}{green line} (and right scattergram) \greenlineis
black line is ratio \textcolor{red}{original}/\textcolor{green}{modified}.
}
\end{figure}